%% file: Thesis-Manuscript_Main.tex
\documentclass[12pt, twoside, a4paper]{report}
\usepackage[dvips]{graphicx}
\usepackage{latexsym}
\usepackage{epsf}
\usepackage{epsfig}
\usepackage{color}
\usepackage{a4}
\usepackage{amsmath} 
\usepackage{amssymb}
\usepackage{amsxtra}
\usepackage{fancyhdr}
\usepackage{enumitem}

\usepackage[T1]{fontenc}
\usepackage{setspace}

\usepackage[nottoc,notlot,notlof]{tocbibind}

\usepackage{amsmath}
\usepackage[cmintegrals]{newtxmath}
\usepackage{graphicx}
\graphicspath{{Figures/}}
\usepackage{url}
\usepackage{cite}
\usepackage{algorithm}
\usepackage{algpseudocode}

\makeatletter
\def\BState{\State\hskip-\ALG@thistlm}
\makeatother

\usepackage{epstopdf}
\usepackage{amsmath}
\usepackage{amsxtra}
\usepackage{amstext}
\usepackage{amssymb}
\usepackage{latexsym}
\usepackage{dsfont} 
\usepackage{color}
\usepackage{multirow}
\usepackage{float}
\usepackage{hyperref}

\usepackage{tabularx,colortbl}
\usepackage[table]{xcolor}

\usepackage{lscape}

\usepackage{algorithm}
\usepackage{algpseudocode}

\usepackage{units}
\usepackage{cases}
\usepackage[ subrefformat=parens,labelformat=parens, caption=false, font=footnotesize]{subfig}
\usepackage{mathtools}

\newcommand{\mbf}[1]{\mathbf{#1}}
\newcommand{\blue}[1]{{\color{blue}{#1}}} 

\newcommand{\nth}[1]{{#1}{\text{th}}}

\newcommand{\abs}[1]{\left|{#1}\right|}
\newcommand{\norm}[1]{\left\|{#1}\right\|}
\newcommand{\red}[1]{{\color{red}{#1}}} 

\newcommand\Tstrut{\rule{0pt}{2.3ex}}         
\newcommand\Bstrut{\rule[-1ex]{0pt}{0pt}}   

\DeclareMathOperator*{\argmax}{arg\,max}   
\DeclareMathOperator*{\argmin}{arg\,min}

\usepackage{mathtools}

\newcommand{\ML}{\mathrm{ML}}
\newcommand{\PML}{\mathrm{PML}}
\newcommand{\NC}{\mathrm{N/C}}
\newcommand{\PNC}{\mathrm{PN/C}}
\newcommand{\CD}{\mathrm{CD}}
\newcommand{\PCD}{\mathrm{PCD}}
\newcommand{\cML}{\mathrm{cML}}
\newcommand{\LORD}{\mathrm{LORD}}
\newcommand{\SSD}{\mathrm{SSD}}
\newcommand{\VSSD}{\mathrm{VSSD}}
\newcommand{\SLORD}{\mathrm{SLORD}}
\newcommand{\SSSD}{\mathrm{SSSD}}
\newcommand{\IRC}{\mathrm{IRC}}
\newcommand{\WR}{\mathrm{WR}}

\newcommand{\RML}{\mathrm{RML}}
\newcommand{\RAD}{\mathrm{RAD}}
\newcommand{\MMSE}{\mathrm{MMSE}}
\newcommand{\ZF}{\mathrm{ZF}}
\newcommand{\user}{\mathrm{user}}
\newcommand{\inter}{\mathrm{inter}}
\newcommand{\Prb}{\mathrm{Pr}}
\newcommand{\err}{\mathrm{err}}
\newcommand{\A}{\mathrm{A}}
\newcommand{\B}{\mathrm{B}}
\newcommand{\C}{\mathrm{C}}
\newcommand{\D}{\mathrm{D}}

\newcommand{\F}{\mathrm{F}}

\newcommand{\Tr}{\mathrm{Tr}}
\newcommand{\Rp}{\mathring{\mbf{R}}}

\newcommand{\rp}{\mathring{r}}
\newcommand{\rrp}{\mathring{\mbf{r}}}

\newcommand{\TLORD}{\mathrm{T-LORD}}
\newcommand{\AIR}{\mathrm{AIR}}
\newcommand{\LogMAP}{\mathrm{Log-MAP}}
\newcommand{\MaxLogMAP}{\mathrm{Max-Log-MAP}}
\newcommand{\sgn}{\mathrm{sgn}}
\newcommand{\tr}{\mathrm{tr}}

\newcommand{\CP}{\mathrm{CP}}

\newcommand{\LORP}{\mathrm{LORP}}
\newcommand{\NCP}{\mathrm{NCP}}
\newcommand{\GP}{\mathrm{GP}}

\newcommand{\Opt}{\mathrm{Opt}}
\newcommand{\var}{\emph{var}}
\newcommand{\muu}{\emph{mu}}
\newcommand{\Detector}{\mathrm{Detector}}

\begin{document}

\pagestyle{empty}
\include{cover_new}

\pagenumbering{roman} \setcounter{page}{5}

\pagestyle{empty} \tableofcontents \listoffigures \listoftables
\newpage

\pagestyle{plain} \pagenumbering{arabic} \setcounter{page}{1}
{\setstretch{1.5}
\include{abst}

\include{abbreviations}
\include{notation}
\include{Chapter1_INTRO}

\include{Chapter2_REF}

\include{Chapter3_TURBO}
\include{Chapter4_PRQRD}

\include{Chapter5_PUNDET}
\include{Chapter6_ANALYSIS}
\include{Chapter7_MUMIMO2x2}

\include{Chapter8_MUMIMOLARGE}

\include{Chapter9_MC_MIMO}

\include{Chapter10_MASSIVE}
\include{Chapter11_CONCLUSION}


}


\end{document}

%% file: cover_new.tex
\begin{titlepage}

\begin{center}

\LARGE{AMERICAN UNIVERSITY OF BEIRUT\\} \vspace{3cm}
\LARGE{Large Multiuser MIMO Detection: \\Algorithms and Architectures\\} \vspace{3cm}
\normalsize{by\\} \Large{HADI AKRAM SARIEDDEEN\\} \vspace{3cm}

\normalsize{A dissertation\\}
\normalsize{submitted in partial fulfillment of the requirements\\}
\normalsize{for the degree of Doctor of Philosophy\\}
\small{to the Department of Electrical and Computer Engineering\\}
\normalsize{of the Maroun Semaan Faculty of Engineering and Architecture\\}
\normalsize{at the American University of Beirut\\} \vspace{4cm}
\normalsize{Beirut, Lebanon\\
 March 2018\\}



%
%
%
%
%
%

\end{center}

\end{titlepage} 

%% file: abst.tex

\chapter*{An Abstract of the Dissertation of}
\addcontentsline{toc}{chapter}{Abstract}

\begin{tabbing}
\underline{\normalsize{Hadi Akram Sarieddeen}}\quad\quad
for\quad\quad\=\underline{Doctor of Philosophy}\\
\>\underline{Major}: Electrical and Computer Engineering
\end{tabbing}
\vspace{0.5cm} Title: \underline{Large Multiuser MIMO Detection: Algorithms and Architectures}\\

\vspace{0.5cm}
After decades of research on multiple-input multiple-output (MIMO) technology, including paradigm shifts from point-to-point to multiuser MIMO (MU-MIMO), an ample literature exists on techniques to exploit the spatial dimension to increase link throughput and network capacity of wireless communication systems. \emph{Massive} MIMO, which supports hundreds of antennas at the base station (BS), is celebrated as the key enabling technology of the upcoming fifth generation (5G) wireless communication standard. However, the use of \emph{large} MIMO systems in the future is also indispensable, especially for high-speed wireless backhaul connectivity. \emph{Large} MIMO systems use tens of antennas in communication terminals, and can afford a large number of antennas on both the transmitter and the receiver sides. While favorable propagation in \emph{massive} MIMO ensures that reliable performance can be achieved by simple linear processing, the inherent symmetry in \emph{large} MIMO renders the computational complexity of near-optimal signal processing schemes exponential in the number of antennas.

In this thesis, we investigate the problem of efficient data detection in \emph{large} MIMO and high order MU-MIMO systems. First, near-optimal low-complexity detection algorithms are proposed for regular MIMO systems. Then, a family of low-complexity hard-output and soft-output detection schemes based on channel matrix puncturing targeted for \emph{large} MIMO systems is proposed. The performance of these schemes is characterized and analyzed mathematically, and bounds on capacity, diversity gain, and probability of bit error are derived. After that, efficient high order MU-MIMO detectors are proposed, based on joint modulation classification and subspace detection, where the modulation type of the interferer is estimated, while multiple decoupled streams are individually detected. Hardware architectures are designed for the proposed algorithms, and the promised gains are verified via simulations. Finally, we map the studied search-based detection schemes to low-resolution precoding at the transmitter side in \emph{massive} MIMO and report the performance-complexity tradeoffs.

%% file: abbreviations.tex
\chapter*{Abbreviations}
\label{chapter:Abbreviations}

\begin{tabbing}

5G \quad\quad\quad\quad\quad\quad\quad \= fifth generation\\
ADC \> analog-to-digital converter\\
AIR \> achievable information rate\\
AIR-PM \> achievable-information-rate-based partial marginalization\\
ALRT \> average likelihood ratio test\\
BER \> bit error rate\\
BPSK \> binary phase shift keying\\
BRF \> breadth-first\\
BS \> base station\\
BSF \> best-first\\
CC \> cyclic cumulant\\
CCR \> correct classification ratio\\
CD \> chase detector\\
CDF \> cumulative distribution function\\
CFER \> coded frame error rate\\
CG \> center generator\\
CMLD \> counter-ML-distance-based MC \\
CYLD \> cyclic layered orthogonal lattice detector\\
CYSD \> cyclic subspace detection\\
DAC \> digital-to-analog converter\\
DF \>  depth-first\\
DI \> distance increment\\
DP \> diagonal-process\\
FER \> frame error rate\\
FLOP \> floating-point operation\\
GLRT \> generalized likelihood ratio test\\
GR \> Givens rotation\\
GS \> Gram-Shmidt\\
HLRT \> hybrid likelihood ratio test\\
HO \> hard-output\\
HT \> Householder transformation\\
IA \> interference-aware\\
II \> interference-ignoring\\
IRC \> interference rejection combining\\
Iter-LC-LORD \>iterative LC-LORD\\
LC-LORD \> low-complexity layered orthogonal lattice detector\\
LLR \> log-likelihood ratio\\
LO-LC-LORD \>layer-ordered LC-LORD\\
LORD \> layered orthogonal lattice detector\\
LORP \> layered orthogonal lattice precoder\\
LQP \> linear quantized precoder\\
LSD \> list sphere decoding\\
LTE \> long term evolution\\
MAP \> maximum-a-posteriori\\
MC \> modulation classification\\
MGS \> modified Gram-Schmidt\\
MIMO \> multiple-input multiple-output\\
ML \> maximum likelihood\\
MMSE \> minimum mean square error\\
MMSE-LC-LORD \> minimum mean square error LC-LORD\\
MR \> minimum cumulative residual\\
MRC \> maximal ratio combining\\
MRQRD \> minimum cumulative residual QR decomposition\\
MT \> modulation type\\
MU-MIMO \> multiuser multiple-input multiple-output\\
N/C \> nulling-and-cancellation\\
NLQP \> nonlinear quantized precoder\\
OFDM \> orthogonal frequency-division multiplexing\\
PCD \> punctured chase detector\\
PED \> partial Euclidean distance\\
PEP \> pairwise error probability\\
PM \> partial marginalization\\
PML \> punctured maximum likelihood\\
PN/C \> punctured nulling-and-cancellation\\
PPCD \> partially-punctured chase detector\\
PPN/C \> partially-punctured nulling-and-cancellation\\
PR-QRD \> permutation-robust QR decomposition\\
PSV \> partial symbol vector\\
PWLD \> pairwise layered orthogonal lattice detector\\
PWSD \> pairwise subspace detector\\
QAM \> quadrature amplitude modulation\\
QPSK \> quadrature phase shift keying\\
QRD \> QR decomposition\\
RAD \> real addition\\
RegTh-LC-LORD \> region-thresholding LC-LORD\\
RF \> radio frequency\\
RML \> real multiplication\\
SD \> sphere decoder\\
SE \>  Schnorr-Euchner\\
SIC \> successive interference cancellation\\
SINR \> signal-to-interference-plus-noise ratio\\
SL-MMSE \> single-layer minimum mean square error\\
SNR \> signal-to-noise ratio\\
SO \> soft-output\\
SP \> sphere precoder\\
SPLD \> single-permutation layered orthogonal lattice detector\\
SPSD \> single-permutation subspace detector\\
SQRD \> sorted QR decomposition\\
SQUID \> squared-infinity norm Douglas-Rachford splitting\\
SSD \> subspace detector\\
SSSD \> symbol-based subspace detector\\
T-LORD \>turbo layered orthogonal lattice detector\\
TP \> triangular-process\\
TSA \> triangular systolic array\\
UE \> user equipment\\
UTM \> upper-triangular matrix\\
V-BLAST \> vertical Bell Labs layered space time\\
VSSD \> vector-based subspace detector\\
WiFi \> wireless fidelity\\
WRD \> WR decomposition\\
ZF \> zero forcing\\
ZFDF \> ZF with decision-feedback\\

\end{tabbing}

%% file: notation.tex
\chapter*{Symbols and Notation}
\label{chapter:Notation}

Bold upper case, bold lower case, and lower case letters correspond to matrices, vectors, and scalars, respectively. Unless otherwise stated, all variables are complex. In what follows we list key symbols in this thesis and detail the notation.

\begin{tabbing}

\textbf{Latin Alphabet} \\~\\
$\mbf{A}$ \quad\quad\quad \= upper-triangular sub-matrix of $\mbf{R}$\\
$\mathring{\mbf{A}}$ \> upper-triangular sub-matrix of $\Rp$\\

$\bar{\mbf{B}}$ \> matrix used in AIR computations\\
$B$ \> number of BS antennas in massive MIMO\\
$\mbf{b}$ \> first $N-1$ elements of the last column of $\mbf{R}$\\
$\mathring{\mbf{b}}$ \> first $N-1$ elements of the last column of $\Rp$\\
$\mbf{b}_{n}$ \> coded bit-representation of a symbol $x_n$\\
$b_{n,k}$ \> $\nth{k}$ element of $\mbf{b}_{n}$\\
$\bar{b}_{ij}$ \> element of $\bar{\mbf{B}}$ at the $\nth{i}$ row and $\nth{j}$ column\\

$\mbf{C}_r$ \> receive antenna correlation matrix\\
$\mbf{C}_t$ \> transmit antenna correlation matrix\\
$C_{\mbf{H}}$ \> capacity of regular channel\\
$C_{\Rp}$ \> achievable rate under channel puncturing\\
$C_{\Rp,\Opt}$ \> capacity under channel puncturing\\
$c$ \> equivalent to $r_{N,N}$\\
$\mathring{c}$ \> equivalent to $\rp_{N,N}$\\

$\mbf{d}$ \> difference between transmitted and erroneously detected symbol vector\\
$d^{\ML}$ \> Euclidean distance metric of hard-output ML solution\\
$d^{\overline{\ML}}_{n,k}$ \> counter-ML Euclidean distance metric corresponding to $b_{n,k}$\\

$\mbf{G}_r$ \> modified Gram matrix for AIR computations\\

$\mbf{H}$ \> channel matrix under rich scattering\\
$\bar{\mbf{H}}$ \> augmented channel matrix in massive MIMO\\
$\mbf{H}_1$ \> first $N-1$ columns of $\mbf{H}$\\
$\mbf{H}_c$ \> correlated channel matrix\\
$\mbf{H}_r$ \> modified channel matrix for AIR computations\\
$\mbf{h}_{n}$ \> $\nth{n}$ column of $\mbf{H}$\\

$I_{\AIR}$ \> achievable information rate\\

$J$ \> number of frames retaining the modulation classification output\\
$\bar{J}$ \> number of iterations in Iter-LC-LORD algorithm\\

$L$ \> number of OFDM symbols in modulation classification\\
$\bar{L}$ \> number of possible quantization labels\\
$L_{\mbf{H}}$ \> lower bound on capacity of regular channel\\
$L_{\Rp}$ \> lower bound on achievable rate under channel puncturing\\
$\mathcal{L}$ \> set of possible quantization labels\\
$\bar{\mathcal{L}}$ \> set of possible quantization symbols\\

$M$ \> number of receive antennas\\
$\mathcal{M}$ \> specific modulation constellation\\

$N$ \> number of transmit antennas\\
$N_{\user}$ \> number of user antennas\\
$N_{\inter}$ \> number of interfering antennas\\
$\mbf{n}$ \> noise vector\\

$\mbf{P}$ \> precoding matrix\\
$\bar{\mbf{P}}$ \> permutation matrix\\
$\mbf{P}^{\dag}$ \> orthogonal projection onto the column space of $\mbf{H}$\\
$\mbf{P}^{\perp}$ \> orthogonal projection onto the left nullspace of $\mbf{H}$\\
$P$ \> maximum allocated power\\
$\bar{\mbf{p}}_n$ \> $\nth{n}$ column of $\bar{\mbf{P}}$\\
$P^{\Detector}$ \> BER of a specific $\Detector$\\
$\acute{P}^{\Detector}$ \> BER of a specific channel-punctured $\Detector$\\
$P^{\A}$ \> probability of error value used in BER analysis\\
$P^{\B}$ \> probability of error value used in BER analysis\\
$P^{\C}$ \> probability of error value used in BER analysis\\
$P^{\D}$ \> probability of error value used in BER analysis\\
$\acute{P}^{\B}$ \> probability of error value used in BER analysis\\
$\acute{P}^{\C}$ \> probability of error value used in BER analysis\\
$\acute{P}^{\D}$ \> probability of error value used in BER analysis\\
$P_n$ \> BER at layer $n$ of a N/C detector\\
$\acute{P}_n$ \> BER at layer $n$ of a PN/C detector\\
$\mathcal{P}$ \> list of candidate symbol vectors in PCD\\
$\mathcal{P}^{(1)}_{n,k}$ \> subset of $\mathcal{P}$ where the bit $b_{n,k}$ is 1\\
$\mathcal{P}^{(0)}_{n,k}$ \> subset of $\mathcal{P}$ where the bit $b_{n,k}$ is 0\\

$\mbf{Q}$ \> unitary matrix generated by the QRD of $\mbf{H}$\\
$\mbf{q}_{n}$ \> $\nth{n}$ column of $\mbf{Q}$\\
$q$ \> number of bits per symbol\\

$\mbf{R}$ \> UTM generated by the QRD of $\mbf{H}$\\
$\tilde{\mbf{R}}$ \> scaled $\mbf{R}$ in massive MIMO\\
$\Rp$ \> punctured UTM generated by the WRD of $\mbf{H}$\\
$\mbf{R}_1$ \> first $N-1$ columns of $\mbf{R}$\\
$\Rp_1$ \> first $N-1$ columns of $\Rp$\\
$\mbf{r}_n$ \> $\nth{n}$ column of $\mbf{R}$\\
$\rrp_n$ \> $\nth{n}$ column of $\Rp$\\
$r_{ij}$ \> element of $\mbf{R}$ at the $\nth{i}$ row and $\nth{j}$ column\\
$\rp_{ij}$ \> element of $\Rp$ at the $\nth{i}$ row and $\nth{j}$ column\\

$S$ \> number of possible modulation types\\
$\mathcal{S}$ \> list of candidate symbol vectors in CD\\
$\mathcal{S}^{(1)}_{n,k}$ \> subset of $\mathcal{S}$ where the bit $b_{n,k}$ is 1\\
$\mathcal{S}^{(0)}_{n,k}$ \> subset of $\mathcal{S}$ where the bit $b_{n,k}$ is 0\\
$\mathsf{SNR}$ \> SNR value\\
$\mbf{s}$ \> symbol vector before quantization\\
$\bar{\mbf{s}}$ \> augmented symbol vector in massive MIMO\\
$\tilde{\mbf{s}}$ \> modified $\bar{\mbf{s}}$ after QRD\\

$T$ \> number of observations (tones) in modulation classification\\
$\bar{T}$ \> number of detection/decoding iterations\\

$\mbf{U}$ \> matrix used in AIR computations\\
$U$ \> number of users in massive MIMO\\
$\mathcal{U}_j$ \> modulation type of interferer $j$\\
$u_{ij}$ \> element of $\mbf{U}$ at the $\nth{i}$ row and $\nth{j}$ column\\

$\mbf{W}$ \> normalized matrix generated by the WRD of $\mbf{H}$\\
$\mbf{w}_{n}$ \> $\nth{n}$ column of $\mbf{W}$\\

$\mathcal{X}$ \> finite $N$-dimensional lattice\\
$\mathcal{X}_{n}$ \> normalized constellation at layer $n$\\
$\mathcal{\bar{X}}_j$ \> $N$-dimensional lattice corresponding to hypothesis $j$ of modulation types\\
$\mathcal{X}_{n,k}^{(0)}$ \> subset of $\mathcal{X}$ where the bit $b_{n,k}$ is 0\\
$\mathcal{X}_{n,k}^{(1)}$ \> subset of $\mathcal{X}$ where the bit $b_{n,k}$ is 1\\
$\mathcal{X}_{n,n,k}^{(0)}$ \> subset of $\mathcal{X}_n$ where the bit $b_{n,k}$ is 0\\
$\mathcal{X}_{n,n,k}^{(1)}$ \> subset of $\mathcal{X}_n$ where the bit $b_{n,k}$ is 1\\
$\mbf{x}$ \> transmitted symbol vector\\
$\mbf{x}_{1}$ \> first $N-1$ elements of $\mbf{x}$\\
$\mbf{x}^{(1)}$ \> true transmitted symbol vector\\
$\mbf{x}^{(2)}$ \> erroneously detected symbol vector\\
$\hat{\mbf{x}}^{\Detector}$ \> hard-output vector solution of a specific $\Detector$\\
$x_n$ \> $\nth{n}$ element of $\mbf{x}$\\
$\hat{x}^{\Detector}_n$  \> $\nth{n}$ element of $\hat{\mbf{x}}^{\Detector}$\\

$\mbf{y}$ \> received vector\\
$\tilde{\mbf{y}}$ \> modified received vector after QRD\\
$\bar{\mbf{y}}$ \> modified received vector after WRD\\
$\tilde{\mbf{y}}_{1}$ \> first $N-1$ elements of $\tilde{\mbf{y}}$\\
$\bar{\mbf{y}}_{1}$ \> first $N-1$ elements of $\bar{\mbf{y}}$\\
$\hat{\mbf{y}}^{\Detector}$ \> equalized output vector of a specific $\Detector$\\
$\hat{y}^{\Detector}_n$ \> $\nth{n}$ element of $\hat{\mbf{y}}^{\Detector}$\\~\\

\textbf{Greek Alphabet} \\~\\

$\alpha$ \quad\quad\quad \= transmit correlation factor\\
$\bar{\alpha}$ \> receive correlation factor\\

$\beta$ \> precoding factor\\

$\gamma$ \> branch SNR\\

$\mbf{\Delta}$ \> used in diversity analysis for regular channels\\
$\mathring{\mbf{\Delta}}$ \> used in diversity analysis for punctured channels\\

$\Theta$ \> modulation constellation of reduced size\\
$\theta_1$ \> complex multiplications saved under puncturing\\
$\theta_2$ \> number of FLOPS required for QRD\\
$\theta_3$ \> number of FLOPS required for puncturing\\

$\bar{\Lambda}$ \> modulation type of the user of interest\\
$\Lambda_j$ \> modulation type $j$ of interferer\\

$\lambda_{n,k}^{\Detector}$ \> the LLR of bit $b_{n,k}$ of a specific $\Detector$\\

$\mu$ \> variable used in BER analysis\\

$\xi_j$ \> a priori LLRs for bits corresponding to $x_j$\\

$\varrho$ \> minimum error distance on a constellation\\
$\bar{\varrho}$ \> variable used in BER analysis\\

$\sigma^2$ \> noise variance\\

$\Upsilon_j^2$ \> scaled chi-squared distributed random variable with $j$ degrees of freedom\\

$\Phi$ \> empty constellation\\
$\varphi^{\Detector}$ \> cost function of a specific $\Detector$\\

$\chi_j^2$ \> chi-squared distributed random variable with $j$ degrees of freedom\\~\\

\textbf{Notation} \\~\\
$\mbf{0}$ \quad\quad\quad\quad\quad \= column vector of zeroes\\

$\abs{\cdot}$ \> scalar norm or cardinality of a set\\
$\norm{\cdot}$ \> vector $\text{L}_2$ norm\\
$\norm{\cdot}_{\F}$ \> matrix Frobenius norm\\
$\lfloor\cdot\rceil$ \> slicing operation\\

$\mathcal{C}$ \> set of complex numbers\\
$\mathcal{CN}(\muu,\var)$ \> complex Gaussian distribution with mean $\muu$ and variance $\var$\\

$d(\mbf{x})$ \> Euclidean distance metric as a function of $\mbf{x}$\\
$\bar{d}(\mbf{x})$ \> Euclidean distance metric under puncturing\\

$\mathsf{E}[\cdot]$ \> expected value\\

$(\cdot)^\mathcal{H}$ \> conjugate transpose\\

$\mbf{I}_N$ \> identity matrix of size $N$\\
$\Im(\cdot)$ \> imaginary part\\

$\log(\cdot)$ \> natural logarithm\\

$\mathcal{N}(\cdot)$ \> normal distribution\\

$\Prb(\cdot)$ \> probability density function\\
$\mathcal{P_r}(\cdot)$ \> precoding function\\

$Q(\cdot)$ \> Q-function\\
$\mathcal{Q}(\cdot)$ \> quantizer-mapping function\\

$\mathcal{R}$ \> set of real numbers\\
$\Re(\cdot)$ \> real part\\

$(\cdot)^{T}$ \> transpose function\\
$\Tr(\cdot)$ \> trace function\\

\end{tabbing}

%% file: Chapter1_INTRO.tex

\chapter{Introduction}\label{chapter:introduction}

%
\section{MIMO Wireless Technology}
\label{sec:MIMO_Technology}

Wireless data usage continues to increase with the enhancements in smartphones and broadband-enabled portables, leading to an exponential growth in mobile data traffic. Such increasing demands are met by optimized network architectures, and one of the most important optimizations is taking advantage of the spatial dimension to improve reliability, spectral efficiency, and spatial separation of users. Towards that end, multiple-input multiple-output (MIMO) technology has been successfully used in several wireless communications standards \cite{802.11,802.16,WCDMA,LTE_36.211}. MIMO technology \cite{2003_Paulraj} is a technique by which more antennas are added to increase link throughput and network capacity. However, conventional MIMO configurations fall short of providing the required spatial diversity in the upcoming fifth generation (5G) mobile communication standard, which promises to connect billions of devices and achieve several gigabit-per-second data rates.

After decades of research on MIMO technology~\cite{foschini1998limits,telatar1999capacity,hochwald2003achieving,Marzetta746779,2003_Paulraj}, including a paradigm shift from point-to-point to multiuser MIMO (MU-MIMO)~\cite{Gesbert4350224}, \emph{massive} MIMO~\cite{Larsson_6736761,Bjornson7402270,Bjornson7031971,marzetta2016fundamentals,bjornson2017massive} is currently being celebrated as a key enabling technology for 5G. With massive MIMO, a base station (BS) can simultaneously accommodate a large number (100 or more) of co-channel users. This allows for fine-grained beamforming to serve hundreds of user equipments (UEs) in the same time-frequency resources, resulting in an order-of-magnitude increase in capacity~\cite{Ngo_2013,Yang6415389,Wagner6172680,Couillet6375928,Hoydis6415388}. However, many challenges have to be addressed in order to achieve the promised theoretical advantages. For example, pilot contamination is a fundamental limitation in a multi-cell system~\cite{Ngo6493983,Bjornson7294693}, and non-ideal hardware is an inevitable constraint~\cite{Bjornson6891254,Bjornson7080890}.

Despite the extensive work on massive MIMO, \emph{large} MIMO will also play an important role in the future. Large MIMO systems~\cite{chockalingam2014large} use tens of antennas in communication terminals, and can afford a large number of antennas on both the transmitter and the receiver sides, such as for example $8\times8$, $16\times16$, $32\times32$, and $64\times64$ configurations. Large point-to-point MIMO wireless links are of specific interest in 5G for high-speed wireless backhaul connectivity between BSs. Also, multipoint-to-point large MU-MIMO can be used in 5G in the uplink when the number of served transmitting users is less than, but comparable to, the number of BS antennas. Nevertheless, large MIMO can also be considered for point-to-multipoint downlink MU-MIMO~\cite{2011_Duplicy}, whether in enhanced versions of current wireless communications standards, or in 5G, where users sharing the same physical resource blocks are chosen based on the degree of orthogonality of their cascaded precoder and channel.

Furthermore, the order of modulation types (MTs) is rising to increase capacity. For example, quadrature amplitude modulations (QAMs) of size 1024 (1024-QAM) and beyond are currently being accommodated. Such modulations have previously found use in low-noise high-performance infrastructures, and they are now paving their way into future wireless communication standards. At the receiver side, the main disadvantage of employing such MTs is the scalability of existing data detection schemes.

%
\subsection{Detection in Single-User MIMO}
\label{sec:MIMO_Detection}

After being traditionally driven by diversity-multiplexing tradeoffs, recent wireless communication system designs have been driven by two factors; system performance in terms of throughput and bit error rate (BER), and system complexity in terms of processing latency and computational complexity.

The performance of MIMO systems is largely determined by the detection scheme at the receiver side; various schemes provide different performance and complexity tradeoffs~\cite{bai2014low}. Linear detectors, such as zero forcing (ZF) and minimum mean square error (MMSE), are the least-complex, but the least-optimal as well. On the other hand, maximum likelihood (ML) detectors are optimal but most computationally intensive, with a complexity that grows exponentially with the number of antennas. Several sub-optimal detectors fill the spectrum in between, including sphere decoders (SDs) and their variants~\cite{771234Viterbo,Hassibi,Vikalo,2014_sphereP1_mansour,2014_sphereP2_mansour,Barbero4543065,Barbero4400127,Jalden4801667}. Moreover, in addition to conventional hard-output (HO) detectors, soft-output (SO) detectors play an important role in near-capacity achieving systems, but are more complex because they require processing significantly more signal combinations to generate reliability information.

In massive MIMO systems, linear detectors achieve near-optimal performance by exploiting the channel hardening effect \cite{Ngo_2013}, and approximate matrix inversions via Neumann series approximations \cite{Rosario_2016} are often used for practical implementations. However, large MIMO systems do not have very large receive-to-transmit antenna ratios. Hence, they cannot achieve the performance gains of asymmetric massive MIMO systems, and they do not allow for similar practical implementations, where Neumann series expansions fail to converge. For large MIMO systems, the detection schemes in the literature are grouped into several areas: detection based on local search \cite{Vardhan_2008,Li_2010}; detection based on meta-heuristics \cite{Datta_2010,Srinidhi_2011}; detection via message passing on graphical models \cite{Goldberger_2011,Narasimhan_2014}; lattice reduction (LR) aided detection \cite{Wubben_2011,Zhou_LR_2013}; and detection using Monte Carlo sampling \cite{Datta_Gibbs_2013}. However, for these schemes to achieve a near-ML performance with high orders of antennas and modulation constellations, the entailed complexity would be prohibitive.

A popular family of MIMO detectors that achieves good performance and complexity tradeoffs employs nonlinear subset-stream detection. The nulling-and-cancellation (N/C) detector \cite{choi2006nulling} is a low-complexity member of this family; it consists of linear nulling followed by successive interference cancellation (SIC). The chase detector (CD) \cite{waters2004chase,waters2008chase} is a more complex member of this family; it first creates a list of candidate decision vectors, and then chooses the best candidate from this list as a final decision. Chase detection is considered a special case of list detection. However, it differs from list sphere decoding (LSD) \cite{hochwald2003achieving}, for example, in the way the list is generated and administered; in LSD, list admission is based on proximity to an initial solution, while in CD, list generation is deterministic, and is done by spanning all possible sub-tree symbols emanating from the root symbol in a specific layer of interest. Furthermore, other popular subset-stream detectors exist (e.g.,~\cite{jiang2005joint,jiang2005uniform,Ariyavisitakul}), that decompose the channel matrix into lower order sub-channels to reduce the number of jointly detected streams.

All aforementioned subset-stream detectors make use of QR decomposition (QRD). However, the SO subspace detector (SSD) \cite{ojard2008method}, transforms the channel matrix via a punctured QRD, which we refer to in this thesis as WR decomposition (WRD). In~\cite{2014_mansour_SPL_WLD,2014_mansour_eurasip_WLD,7565040Sarieddeen}, WRD-based SSD is generalized to allow for joint detection of arbitrary-sized subsets of decoupled streams, and efficient implementation methods are presented. The QRD-based version of this detector is called the layered orthogonal lattice detector (LORD) \cite{Siti-1,tomasoni2012hardware}, and both are special cases of the CD. To the best of our knowledge, the use of punctured QRD in MIMO detectors has not been studied analytically in the literature, and its applicability to large MIMO systems has not been addressed.

%
\subsection{Detection in Multiuser MIMO}
\label{sec:MUMIMO_Detection}

MU-MIMO technology \cite{2003_Paulraj,2011_Duplicy} allows simultaneous transmissions to multiple users over the same time-frequency resource elements, by using multiple antennas at the transmitter and the receiver. The main issue in the multi-user scenario is interference. Intra-cell interference occurs when a BS sends information to multiple users within a cell, over spatially almost-orthogonal channels. At the receiving side, the desired user knows its channel and tries to estimate the interference without knowing the MT of the interfering signal.

Different interference mitigation proposals have led to different receiver designs. Conventional linear processing techniques only use the channel estimate of the co-scheduled user, without requiring the knowledge of its MT. Such techniques include \cite{Lee_2011} interference-ignoring (II), interference rejection combining (IRC), and single-layer MMSE (SL-MMSE), with the latter two having the exact coded performance \cite{2011_Bai}. However, if the detectors explicitly take into account the modulation formats of the desired and interference signals, remarkable performance gains can be achieved. Such interference-aware (IA) detectors, ML and minimum distance detectors \cite{Lee_2011} for example, are noise limited, rather than interference limited, and are not prone to error floors like conventional detectors.

Since current communication standards do not provide information about the interfering MT in the downlink, several techniques emerged, that decide on a specific interfering MT. In~\cite{2011_Ghaffar_a,2011_Ghaffar_b}, the constellation of the interfering user's signal is presumed to be 16-QAM, regardless of its actual size, and without making any attempt to estimate it. A better approach, however, is to add an interference modulation classification (MC) routine, followed by a regular IA detector \cite{7450603Gomaa,Bae}. MC is the task of recognizing the MT employed at the transmitter of a detected signal, which is required for various military and civilian applications. In particular, cognitive radio with adaptive MTs \cite{arslan2007cognitive} is a promising future application of MC. In such scenario, the transmitter dynamically adjusts the data rate by switching the modulation order depending on channel conditions. By employing automatic (blind) MC at the receiver, the communication overhead can be significantly reduced.

MC techniques can be classified into two categories\cite{Dobre}: feature-based and likelihood-based. With feature-based classification, inherent characteristics of the received waveform are exploited, such as higher order correlations, hierarchical cumulants, zero-crossing rates, and power estimations. Such characteristics are regarded as discriminant features and decisions are made based on their observed values. With likelihood-based classification, on the other hand, the decision is made on the modulation format that has the highest probability within multiple hypotheses. This is achieved by computing complex likelihood functions. In this thesis, we consider a combination of both.

The two main likelihood-based MC approaches \cite{Wei,Panagiotou,Hameed2009} are the average likelihood ratio test (ALRT) and the generalized likelihood ratio test (GLRT). While ALRT treats the signal and channel parameters as unknown random variables with known distributions, GLRT treats them as deterministic but unknown. The hybrid likelihood ratio test (HLRT) is a combination of the previous two. These approaches were extended to multiuser and MIMO scenarios \cite{Choqueuse,Shim,Ramezani2013,Eldemerdash2016}.

The most popular feature-based approach exploits the higher-order cyclic cumulants (CCs) of the baseband intercepted signal as powerful features for linear digital MC \cite{Dobre2010,Muhlhaus2012,Muhlhaus2013}. Calculating the higher-order cumulants of the sum of independent processes is mathematically convenient, and the intrinsic cyclostationarity of communication signals makes the CCs robust to interference and stationary noise. Moreover, without perfect channel state information (CSI), independent component analysis has been used \cite{Muhlhaus2013b} to blindly estimate the channel in conjunction with either likelihood-based or feature-based MC.

%
\subsection{Low-Resolution Precoding in Massive MIMO}
\label{sec:MMIMO_Precoding}

As the number of antennas increases, and if each antenna element has its own radio frequency (RF) chain at the BS, the hardware complexity and system costs will significantly increase, as well as the circuit power consumption, especially in the context of mmWave systems~\cite{bogale2016massive,swindlehurst2014millimeter} with high sampling rates. The dominant sources of power consumption at a BS with massive antenna arrays are analog-to-digital converters (ADCs) in the uplink and digital-to-analog converters (DACs) in the downlink. For instance, the dissipated power in ADCs scales exponentially in the number of resolution bits and linearly in the sampling rate~\cite{Walden761034}. Moreover, a massive number of antennas puts extreme capacity requirements on the fronthaul interconnect link between the baseband processing unit and the radio unit (RF components), especially when these two units are separated by a large distance, such as in a cloud radio access network architecture~\cite{Park6924850}, where the baseband processing is migrated from the BSs to a centralized unit.

The challenge is to jointly reduce system costs, power consumption, and interconnect bandwidth with minimal performance degradation. Recent research trends aim at either reducing the number of converters, by partitioning the signal processing operations between analog and digital domains using hybrid beamforming~\cite{Han_2015}, or reducing their bit resolutions~\cite{mo2017hybrid}. The latter employs coarse quantization, which has the extra benefit of lowering the linearity and noise requirements, because quantization noise may dominate the noise introduced by mixers, oscillators, filters, and low-noise amplifiers, which further reduces the RF circuit power. It was argued in~\cite{Sarajlic7990217} that the energy efficiency is maximized at intermediate ADC resolutions, typically in the range of $4$ to $8$ bits. In the extreme case of 1-bit quantization~\cite{Mo_2014,Mo_2014_2}, only simple low-complexity comparators are required~\cite{hoyos2005monobit}, and there is no need for automatic gain control circuitry to match the dynamic range of the ADCs. It is known that quadrature phase shift keying (QPSK) is capacity achieving over complex-valued Gaussian channels in the 1-bit case~\cite{5351659Singh}, as well as with Rayleigh-fading assuming perfect CSI~\cite{5592653Krone}.

With low-resolution ADCs in the uplink~\cite{mollen2017achievable,Fan7307134}, a special design of signaling schemes and receiver algorithms is required at the BS to combat the resultant nonlinearity. Note that by exploiting the time division duplex reciprocity, only uplink channels need to be estimated. However, channel estimation on the basis of quantized observations is challenging~\cite{708938Lok}, especially with fast fading channels. In such scenarios, QPSK is optimal only when the signal-to-noise ratio (SNR) exceeds a coherence-time-dependant threshold~\cite{4594988Mezghani}. In~\cite{risi2014massive}, a system employing 1-bit ADCs with QPSK is shown to achieve large sum-rate throughputs when the BS employs a least squares channel estimator, followed by a linear maximal ratio combining (MRC) or ZF detector. This study is extended to high-order modulations in~\cite{7439790Choi}. Bussgang's decomposition~\cite{bussgang1952crosscorrelation} is used for channel estimation in other studies~\cite{Li7931630,jacobsson2017throughput}, and a joint channel- and data-estimation algorithm is presented in~\cite{wen2016bayes}, which outperforms separate channel estimation and data detection at the expense of high complexity. Furthermore, since implementing ZF or MMSE requires the computation of matrix inversions, computationally efficient approximations based on truncated polynomial expansions~\cite{Kammoun6812124,Hoydis6034083} or conjugate-gradient techniques~\cite{Yin7037382} have been proposed. Nevertheless, efficient nonlinear detection schemes are viable alternatives that can boost performance if the BS can afford a marginal increase in complexity. In~\cite{Zhang8234637} 1-bit massive MIMO detection based on variational approximate message passing was proposed. 

With low-resolution DACs in the downlink, conventional low-complexity linear quantized precoders (LQPs), such as ZF or MMSE, followed by quantization, can achieve good performance, but only at high transmit-to-receive antenna ratios and low-to-moderate SNRs~\cite{Saxena7946265,Xu7953567}. To compensate the performance loss in 1-bit massive MIMO systems with linear processing, $2.5$ times more antennas need to be deployed at the BS~\cite{Li7887699}. However, reliable data transmission can be retained under quantization if sophisticated precoding algorithms that can mitigate both multi-user interference and quantization artifacts are employed. In~\cite{jacobsson2017quantized}, two nonlinear quantized precoders (NLQPs) are proposed; the first is based on semi-definite relaxation and squared-infinity norm Douglas-Rachford splitting (SQUID), while the second adapts the SD to a quantized sphere precoder (SP). In~\cite{castaneda20171,castaneda2017pokemon}, two low-complexity nonlinear 1-bit precoding algorithms based on biconvex relaxation are presented. They achieve better error-rate performance compared to linear precoding followed by quantization. Heuristic nonlinear precoding schemes can also provide a good performance-complexity tradeoff, such as subset-codebook precoding~\cite{tirkkonen2017subset}. Furthermore, two-stage spatio-temporal precoding structures~\cite{Gokceoglu7742960} can be used to suppress interuser interference.

There are several other notable studies in the literature on massive MIMO with coarse quantization. Mixed resolution architectures~\cite{liang2016mixed,Zhang7886292,Ding8069011,Pirzadeh8063410} and non-uniform resolutions~\cite{Kong7896590} are considered to increase system performance. Solutions in the context of frequency-selective wideband channels that use orthogonal frequency division multiplexing (OFDM) have also been studied, such as in~\cite{jacobsson2017linear,jacobsson2017massive} for the downlink, and in~\cite{6987288Wang,7458830Studer} for the uplink. Moreover, while most studies assume Nyquist-rate sampling at the receiver, which is not optimal in the presence of quantization~\cite{6109374Zhang}, it is shown in~\cite{6222713Krone} that high-order constellations such as 16-QAM can be supported with 1-bit quantization when oversampling is applied at the receiver.

To sum up, most of the reference studies consider linear precoding and detection at the BS in massive MIMO systems. The performance of these linear solutions has been bounded analytically, including the case of coarse quantization, and efficient architectures have been proposed. Moreover, nonlinear precoding and detection solutions promise significant performance enhancements, especially in the presence of 1-bit ADCs and DACs. However, these solutions are not adequately addressed in the literature. There are mainly three gaps in recently proposed nonlinear solutions: they usually entail high complexity, their performance is not characterized analytically, and they are often studied disjointly as either precoding or detection schemes.

%
\section{Contributions and Outline}
\label{sec:Contributions}

The purpose of this thesis is to design efficient algorithms and architectures for MIMO, large MIMO, and MU-MIMO detection, as well as massive MIMO precoding. Using theoretical analysis and empirical simulations, the proposed algorithms are proven to be high-performance and low-complexity solutions. The structure of this thesis is as follows:

\textbf{Chapter 2} introduces the detection problem in spatial multiplexing and presents reference linear and nonlinear receivers.

\textbf{Chapter 3} presents early results on dual-layer MIMO systems as a starter. Several approaches are proposed to reduce the complexity of iterative  detection and decoding when high order MTs are employed. It is argued that low-complexity LORD (LC-LORD) introduces significant performance degradation, especially with high channel correlation. We propose improving the location of a reduced region of search within a 1024-QAM constellation, as well as enhancing the bit log-likelihood ratio (LLR) approximation. The proposed schemes are studied in the context of non-iterative and iterative detection and decoding, and significant gains are achieved in both cases.

\textbf{Chapter 4} presents a permutation-robust QRD (PR-QRD) technique, using the modified Gram-Schmidt (GS) orthogonalization procedure and elementary matrix operations. This technique is then used to reduce the complexity of two popular detectors in the literature. First, computationally efficient subspace detection schemes based on special layer ordering, followed by PR-QRD are proposed. A hardware architecture is designed, which allows building an 8-layer detector from 4-layer and 2-layer constituent detector blocks. Second, PR-QRD is used in low-complexity SD, where an optimized layer-ordering scheme based on the minimum cumulative residual (MR) criterion is considered.

\textbf{Chapter 5} presents a family of low-complexity detection schemes based on channel matrix puncturing targeted for large MIMO systems. It is well-known that the computational cost of MIMO detection based on QRD is directly proportional to the number of non-zero entries involved in back-substitution and slicing operations in the triangularized channel matrix, which can be too high for low-latency applications involving large MIMO dimensions. By systematically puncturing the channel to have a specific structure, it is demonstrated that the detection process can be accelerated by employing standard schemes such as CD, LSD, N/C detection, and SSD on the transformed matrix. The difference between optimal channel shortening and efficient channel puncturing is also highlighted in this chapter. Simulations of coded and uncoded scenarios certify that the proposed schemes scale up efficiently, both in the number of antennas and constellation size, as well as in the presence of correlated channels.

\textbf{Chapter 6} introduces a theoretical analysis. The performance of the proposed channel-punctured detectors is characterized and analyzed mathematically, and bounds on the capacity, diversity gain, and probability of bit error are derived. Surprisingly, it is shown that puncturing does not negatively impact the receive diversity gain in HO detectors. The analysis is extended to SO detection when computing per-layer bit LLRs; it is shown that significant performance gains are attainable by ordering the layer of interest to be at the root when puncturing the channel.

\textbf{Chapter 7} presents near-optimal data detection schemes for dual-layer MU-MIMO systems. Joint likelihood-based MC of the co-scheduled user and data detection receivers are developed. By expanding the Max-Log- maximum-a-posteriori (MAP) MC approach to include distances of counter ML hypothesis symbols, the decision metric for MC is shown to be an accumulation over a set of tones of Euclidean distance computations, that are also used by the detectors for bit LLR soft decision generation. With a small complexity overhead, the proposed approaches achieve near-optimal performance. Efficient hardware architectures are presented for the proposed approaches. \\

\textbf{Chapter 8} extends the work on MU-MIMO to higher antenna orders. A detector that employs joint MC and low-complexity subspace detection is proposed, by which the MT of the interferer is estimated, while multiple decoupled streams are individually detected. A hierarchical MC scheme is proposed, comprising feature-based and near-optimal likelihood-based classifiers, as well as a classifier that always assumes the interfering MT to be a fixed high order QAM. An efficient hardware architecture that realizes the proposed algorithms is presented. Simulations demonstrate that depending on the channel condition, one of the proposed schemes can achieve near IA performance with a minimum complexity overhead.

\textbf{Chapter 9} presents a novel near-optimal low-complexity likelihood-based MC scheme for MIMO systems with adaptive MTs. First, the channel matrix is decomposed employing subspace decomposition, and then the MT on the partially decoupled stream of interest gets detected using a modified likelihood metric. A joint MC and subspace detection receiver is also presented.

\textbf{Chapter 10} extends the study to address the problem of efficient precoding in the downlink of massive MIMO systems that use 1-bit DACs. By adapting the procedures of popular search-based detection algorithms to 1-bit quantized precoding, two families of nonlinear precoders are proposed. The first employs QRD combined with tree-based search techniques, and the second uses Gibbs sampling for search enumerations without decomposing the channel. Simulations demonstrate that some of the proposed schemes outperform reference nonlinear precoders, both in performance and complexity with low order MIMO, and in performance with a graceful increase in computations in the context of massive MIMO with high order modulation types.

\textbf{Chapter 11} concludes the presented work and specifies future directions. \\

%
\section{Thesis-Related Publications}
\label{sec:publications}

At the time of writing, this thesis work had been incrementally published in three journal papers and eight conference papers as follows: \\

\textbf{Journal Papers}:

\begin{enumerate}
  \item H.~Sarieddeen, M.~M. Mansour, and A.~Chehab, ``Large {MIMO} detection schemes based on channel puncturing: Performance and complexity analysis,'' \emph{{IEEE} Trans. Commun.}, no.~99, pp. 1--1, 2017.
  \item H.~Sarieddeen, M.~M. Mansour, L.~Jalloul, and A.~Chehab, ``High order   multi-user {MIMO} subspace detection,'' \emph{J. of Signal Process. Syst.},   vol.~90, no.~3, pp. 305--321, Mar. 2017.
  \item H.~Sarieddeen, M.~M. Mansour, and A.~Chehab, ``Modulation classification via subspace detection in {MIMO} systems,'' \emph{{IEEE} Commun. Lett.}, vol.~21, no.~1, pp. 64--67, Jan. 2017.
\end{enumerate}

\textbf{Conference Papers}:	
\begin{enumerate}
  \item H.~Sarieddeen, M.~M. Mansour, and A.~Chehab, ``Channel-Punctured Large MIMO Detection,'' in \emph{Proc. IEEE Int. Symp. Inf. Theory (ISIT)}, Vail, CO, USA, Jun. 2018 (to appear).
  \item H.~Sarieddeen, M.~M. Mansour, and A.~Chehab, ``Hard-output chase detectors for large {MIMO}: {BER} performance and complexity analysis,'' in \emph{Proc. IEEE Int. Symp. Personal Indoor and Mobile Radio Commun. (PIMRC)}, Montreal, Canada, Oct. 2017, pp. 1--5.
  \item H.~Sarieddeen and M.~M. Mansour, ``Enhanced low-complexity layer-ordering for {MIMO} sphere detectors,'' in \emph{Proc. IEEE Int. Conf. Commun. (ICC)}, Kuala Lumpur, Malaysia, May 2016, pp. 1--6.
  \item H.~Sarieddeen, M.~M. Mansour, L.~M.~A. Jalloul, and A.~Chehab, ``Low-complexity joint modulation classification and detection in {MU-MIMO},'' in \emph{Proc. IEEE Wireless Commun. and Netw. Conf. (WCNC)}, Doha, Qatar, Apr. 2016, pp. 1--6.
  \item H.~Sarieddeen, M.~M. Mansour, and A.~Chehab, ``Efficient near-optimal 8x8 {MIMO} detector,'' in \emph{Proc. IEEE Wireless Commun. and Netw. Conf. (WCNC)}, Doha, Qatar, Apr. 2016, pp. 1--6.
  \item H.~Sarieddeen, M.~M. Mansour, and A.~Chehab, ``Efficient subspace detection for high-order {MIMO} systems,'' in \emph{Proc. IEEE Int. Conf. Acoustics, Speech, and Signal Process. (ICASSP)}, Shanghai, China, Mar. 2016, pp. 1001--1005.
  \item H.~Sarieddeen, M.~Mansour, L.~Jalloul, and A.~Chehab, ``Efficient near optimal joint modulation classification and detection for {MU-MIMO} systems,'' in \emph{Proc. IEEE Int. Conf. Acoustics, Speech, and Signal Process. (ICASSP)}, Shanghai, China, Mar. 2016, pp. 3706--3710.
  \item H.~Sarieddeen, M.~M. Mansour, L.~M.~A. Jalloul, and A.~Chehab, ``Low-complexity {MIMO} detector with 1024-{QAM},'' in \emph{Proc. IEEE Global Conf. on Signal and Inform. Process. (GlobalSIP)}, Orlando, FL, USA, Dec. 2015, pp. 883--887.
  \item H.~Sarieddeen, M.~M. Mansour, L.~M.~A. Jalloul, and A.~Chehab, ``Likelihood-based modulation classification for {MU-MIMO} systems,'' in \emph{Proc. IEEE Global Conf. on Signal and Inform. Process. (GlobalSIP)}, Orlando, FL, USA, Dec. 2015, pp. 873--877.
\end{enumerate}

%% file: Chapter2_REF.tex

\chapter{System Model and Reference Detectors}\label{chapter:ref_detectors}

%
\section{System Model}
\label{sec:ch2_model}

We consider spatial multiplexing in a MIMO system with $N$ transmit antennas and $M = N$ receive antennas. The equivalent complex baseband input-output system relation is given by
\begin{equation}\label{eq:ch2_sysmodel}
  \mbf{y} = \mbf{H}\mbf{x} + \mbf{n},
\end{equation}
where $\mbf{y}\in\mathcal{C}^{M\times1}$ is the received complex vector, $\mbf{H}=[\mbf{h}_{1}\cdots\mbf{h}_{n}\cdots \mbf{h}_{N}^{}]\in\mathcal{C}^{M\times N}$ is the channel matrix with entries that are assumed to be $\mathcal{CN}(0,1)$ i.i.d. random variables, $\mbf{x}=[x_{1}\cdots x_{n}\cdots x_{N}^{}]^{T}\in\mathcal{C}^{N\times1}$ is the transmitted symbol vector, and $\mbf{n}\in\mathcal{C}^{M\times1}$ is the noise vector with $\mathcal{CN}(0,\sigma^2)$ entries $\left(\mathsf{E}[\mbf{n}\mbf{n}^\mathcal{H}]=\sigma^{2}\mbf{I}_M\right)$.

Each symbol $x_{n}$, $n\in\{1,\cdots,N\}$, belongs to a normalized complex constellation ($\mathsf{E}[x_{n}^\mathcal{H}x_{n}^{}]\!=\!1$), and we have $\mbf{x}\!\in\! \mathcal{X}\!\triangleq\!\mathcal{X}_{1}\times \cdots \times \mathcal{X}_{n}\times \cdots \times\mathcal{X}_{N}^{} \subset \mathcal{C}^{N\times1}$, where $\mathcal{X}$ is the finite set of points on a $N$-dimensional lattice generated by all possible symbol vectors. For simplicity, we assume a uniform modulation constellation $\mathcal{M}$ on all layers, and hence $\mathcal{X}=\mathcal{M}^{N}$. The coded bit-representation of a symbol $x_n$ is denoted by $\mbf{b}_{n}\!=\!(b_{n,1},\cdots,b_{n,k},\cdots,b_{n,q})$, where $q\!=\!\lceil{\log(\abs{\mathcal{M}})\rceil}$ and $b_{n,k} \in \{0,1\}$ for $k=1,\cdots,q$. The SNR is defined in terms of the noise variance as
\begin{equation}\label{eq:ch2_sysmodel1}
  \mathsf{SNR}=\frac{N}{\sigma^{2}}.
\end{equation}

At the receiver side, and assuming perfect knowledge of the channel, QRD decomposes $\mbf{H}$ as $\mbf{H}\!=\!\mbf{Q}\mbf{R}$, where $\mbf{Q}\!=\![\mbf{q}_{1}\cdots\mbf{q}_{n}\cdots \mbf{q}_{N}^{}]\!\in\!\mathcal{C}^{M \times N}$ has orthonormal columns $\mbf{q}_n\in\mathcal{C}^{M\times 1}$ ($\mbf{Q}^\mathcal{H}\mbf{Q}\!=\!\mbf{I}_N$), and $\mbf{R}\!=\![r_{ij}^{}]\!\in\!\mathcal{C}^{N\times N}$ is a square upper-triangular matrix (UTM) with real and positive diagonal entries. The transformed receive symbol vector can then be equivalently expressed as
\begin{equation}\label{eq:ch2_sysmodel2}
  \tilde{\mbf{y}} = \mbf{Q}^\mathcal{H}\mbf{y} = \mbf{R}\mbf{x} + \mbf{Q}^\mathcal{H}\mbf{n},
\end{equation}
where $\mbf{Q}^\mathcal{H}\mbf{n}$ and $\mbf{n}$ are statistically identical since $\mbf{Q}$ is orthonormal.

%
\section{ML Detector}
\label{sec:ch2_ml}

An ``exhaustive'' log-max ML detector searches the complete lattice $\mathcal{X}$, computing $\abs{\mathcal{M}}^N$ Euclidean distance metrics, to solve for
\begin{equation}\label{eq:ch2_ML_dist}
  \hat{\mbf{x}}^{\ML} = \argmin_{\mbf{x} \in \mathcal{X}}\norm{\mbf{y} - \mbf{H}\mbf{x}}^{2} = \argmin_{\mbf{x} \in \mathcal{X}}\norm{\tilde{\mbf{y}} - \mbf{R}\mbf{x}}^{2},
\end{equation}
where equality  holds since $\mbf{Q}$ is unitary. The LLR of bit $b_{n,k}$, is generated as
\begin{equation}\label{eq:ch2_LLR1}
\lambda_{n,k}^{\ML} = \frac{1}{\sigma^{2}} \left(\min_{\mbf{x}\in \mathcal{X}_{n,k}^{(0)}}\norm{\mbf{y} - \mbf{H}\mbf{x}}^{2} - \min_{\mbf{x}\in \mathcal{X}_{n,k}^{(1)}} \norm{\mbf{y} - \mbf{H}\mbf{x}}^{2}\right),
\end{equation}
where the sets $\mathcal{X}_{n,k}^{(0)}\!\triangleq\!\{\mbf{x} \in \mathcal{X}: b_{n,k}\!=\!0\}$ and $\mathcal{X}_{n,k}^{(1)}\!\triangleq\!\{\mbf{x} \in \mathcal{X}: b_{n,k}\!=\!1\}$ correspond to subsets of symbol vectors in $\mathcal{X}$, having in the corresponding $\nth{k}$ bit of the $\nth{n}$ symbol a value of $0$ and $1$, respectively.

%
\section{MMSE and ZF Detectors}
\label{sec:ch2_mmse}

The MMSE detector generates an equalized output
\begin{equation}\label{eq:ch2_mmse}
\hat{\mbf{y}}^{\MMSE} = [\hat{y}_{1}^{\MMSE}\cdots \hat{y}_{n}^{\MMSE}\cdots \hat{y}_{N}^{\MMSE}]^{T} = \left(\mbf{H}^\mathcal{H}\mbf{H}\!+\!(1/\mathsf{SNR})\mbf{I}_{N}\right)^{-1} \mbf{H}^\mathcal{H}\mbf{y},
\end{equation}
and the LLRs can then be calculated as
\begin{equation}\label{eq:ch2_LLR3}
\lambda_{n,k}^{\MMSE}= \frac{1}{\sigma_{\MMSE,n}^{2}} \left(\min_{x_n\in \mathcal{X}_{n,n,k}^{(0)}} \abs{\hat{y}_n^{\MMSE} - x_n}^{2} - \min_{x_n\in \mathcal{X}_{n,n,k}^{(1)}} \abs{\hat{y}_n^{\MMSE} - x_n}^{2}\right),
\end{equation}
where the sets $\mathcal{X}_{n,n,k}^{(0)}\!\triangleq\!\{x_n \in \mathcal{X}_n: b_{n,k}\!=\!0\}$ and $\mathcal{X}_{n,n,k}^{(1)}\!\triangleq\!\{x_n \in \mathcal{X}_n: b_{n,k}\!=\!1\}$ correspond to subsets of symbols in $\mathcal{X}_n$, having in the corresponding $\nth{k}$ bit a value of $0$ and $1$, respectively, and $\sigma_{\MMSE,n}^{2}\!=\!\sigma^{2}\tau_n$ is a scaled variance with $\tau_n$ being the $\nth{n}$ diagonal element of the matrix $\left(\mbf{H}^\mathcal{H} \mbf{H}\!+\!(1/\mathsf{SNR}) \mbf{I}_{N}\right)^{-1}$.

Note that unbiased SO MMSE detection  \cite{Studer_MMSE} slightly outperforms this conventional detector. However, the performance gap is negligible, and thus this detector serves as a good reference. Similarly, ZF solves for an equalized output $\hat{\mbf{y}}^{\ZF} = \left(\mbf{H}^\mathcal{H}\mbf{H}\right)^{-1} \mbf{H}^\mathcal{H}\mbf{y}$, and the rest of the derivation remains intact.

%
\section{Sphere Decoder}
\label{sec:ch2_sphere}

The SD achieves exact log-max ML performance with less computations, by executing a tree-based search on a subset of $\mathcal{X}$, skipping vectors in the space whose partial distance already exceeds the current best distance. Note that for each bit, one of the two minima in \eqref{eq:ch2_LLR1} corresponds to the distance $d^{\ML}$ of the hard ML solution
\begin{equation}\label{eq:ch2_cml}
d^{\ML}=\min_{\mbf{x}\in \mathcal{X}} \norm{\tilde{\mbf{y}} - \mbf{R}\mbf{x}}^{2},
\end{equation}
having the bit representations $\mbf{b}_n^{\ML}$s associated with the ML solution $\mbf{x}^{\ML}$. The other minima corresponds to the distance $d^{\overline{\ML}}_{n,k}$ of the counter ML ($\overline{\ML}$) hypothesis, having in the same bit position the complement of $b_{n,k}^{\ML}$, $\overline{b_{n,k}^{\ML}}$, where:
\begin{equation}\label{eq:ch2_cml1}
d^{\overline{\ML}}_{n,k}=\min_{\mbf{x}\in \mathcal{X}_{n,k}^{(\overline{\mbf{b}_{n,k}^{\ML}})}} \norm{\tilde{\mbf{y}} - \mbf{R}\mbf{x}}^{2}.
\end{equation}
Therefore, the LLRs can be expressed as
\begin{equation}\label{eq:ch2_LLR2}
\lambda_{n,k}^{\ML} =
\begin{cases}
d^{\ML} - d^{\overline{\ML}}_{n,k}  \ \ \text{if} \  b_{n,k}^{\ML} = 0 \\
d^{\overline{\ML}}_{n,k} - d^{\ML}  \ \ \text{if} \  b_{n,k}^{\ML} = 1
\end{cases}.
\end{equation}
By exploiting the upper triangular structure of $\mbf{R}$, the distance metric $d(\mbf{x})=\norm{\tilde{\mbf{y}} - \mbf{R}\mbf{x}}^{2}$ can be written as
\begin{equation}\label{eq:ch2_EuclideanTree}
d(\mbf{x})=\sum_{m=1}^{M}\abs{\tilde{y}_{n} - \sum_{j=n}^{N} r_{n,j}x_{j}}^{2},
\end{equation}
which in turn can be expressed recursively as
\begin{equation}\label{eq:ch2_recur1}
d_n(x_{n}|x_{n+1}\cdots x_{N})=d_{n+1}(x_{n+1}|x_{n+2}\cdots x_{N}) + e_n(x_{n}|x_{n+1}\cdots x_{N})
\end{equation}
\begin{equation}\label{eq:ch2_recur2}
e_n(x_{n}|x_{n+1}\cdots x_{N})=\abs{\tilde{y}_{n} - \sum_{j=n+1}^{N} r_{n,j}x_{j}-r_{n,n}x_n}^{2},
\end{equation}
for $n\!=\!N,N\!-\!1,\cdots,1$, where $d_n$ is the partial Euclidean distance (PED) of the partial symbol vector (PSV) $[x_n x_{n\!+\!1}\cdots x_N]^{T}$, and $e_n$ is the distance increment (DI) when appending $x_n$ at level $n$ to the PSV $[x_{n\!+\!1} \cdots x_N]^{T}$
\big($d_{N\!+\!1}\!=\!0$ and $d_1\!=\!d(\mbf{x})$\big). Note that accumulated PEDs are reused when exploring lower tree levels.

The recursion in \eqref{eq:ch2_recur1} can be mapped to an $N$-level tree, with a root node at level $N\!+\!1$, leaves at level $1$, and nodes at levels $n\!=\!N,N\!-\!1,\cdots,1$ having $2^{q}$ children. A parent node has a weight $d_n$, and branches to its children have associated weights $e_n$. A path traversed form the root to a leaf corresponds to a lattice point. The first leaf node reached is called the Babai point \cite{babai1986lovasz}, and it gets updated every time a new leaf with a smaller weight is reached. Finding the ML solution corresponds to finding the leaf with the smallest weight.

The counter-ML solution for $b_{n,k}$ can be found by searching for the leaf with the smallest weight that can be reached through paths in the tree whose $\nth{k}$ bit of the $\nth{n}$ symbol is the binary complement of that in the ML solution. This is in effect a traversal of a pruned tree, which gets repeated $N\!\times\!2^q$ times. Consequently, a SO detector requires a total of $N\!\times\!2^q\!+\!1$ tree traversals. However, an alternate solution exists \cite{Studer5571884}, where a single tree traversal is sufficient.

The search space can be limited within a sphere centered at $\tilde{\mbf{y}}$, whose (squared) radius is the minimum distance of any leaf that has already been reached during the current search. In a depth-first (DF) traversal, the children of a node are visited before its siblings, and whenever the PED of an internal node exceeds the radius, that node and its subtree get pruned. However, such pruning is more complicated in a SO detector, where an internal node can only be pruned if it is unable to update any of the $\overline{\ML}$ distances, not only the $\ML$ distance. Moreover, fixed-point implementations require putting constraints on the magnitude of the LLR values, and towards that end, fixed or adaptive radius scaling can be applied, which further reduces the region of search, and hence the node count.

The order in which symbols are enumerated at each level is directly related to the complexity of the detector. The optimal Schnorr-Euchner (SE) \cite{schnorr1994lattice} ordering enumerates symbols in the ascending order of their DIs at each tree level. Moreover, with breadth-first (BRF) traversal, the siblings of a node are visited before its children (the $K$-best algorithm \cite{1010213Wong} for example). As for best-first (BSF) traversal \cite{5955092Shen}, it combines both DF and BRF to reach the shortest path with a reduced search space, however, it is memory-constrained.

%
\section{Nulling-and-Cancellation Detector}
\label{sec:ch2_SIC}

The N/C detector~\cite{choi2006nulling} is used in the widely known vertical Bell Labs layered space time (V-BLAST) architecture \cite{wolniansky1998v}. When combined with QRD, N/C becomes a computationally-efficient procedure which is highly sensitive to layer ordering. Nulling is performed by linearly pre-multiplying the received vector with $\mbf{Q}^\mathcal{H}$, which suppresses the interference from $x_l$, $l >n$, at the $\nth{n}$ layer. This is followed by SIC (back-substitution and slicing) to suppress co-antenna interference; hence, $\hat{\mbf{x}}^{\NC}=[ \hat{x}^{\NC}_1\cdots\hat{x}^{\NC}_n\cdots\hat{x}^{\NC}_N ]$ is computed as
\begin{equation}\label{eq:ch2_SIC2}
\hat{x}^{\NC}_n=\left\lfloor \left(\tilde{y}_{n} - \sum_{l=n+1}^{N} r_{nl}\hat{x}^{\NC}_{l}\right)/r_{nn} \right\rceil_{\mathcal{M}},
\end{equation}
for $n = N,N-1,\cdots, 1$, where $\lfloor \alpha \rceil_{\mathcal{M}} \triangleq \argmin_{x \in \mathcal{M}} \abs{\alpha-x}$ is the slicing operator on the constellation $\mathcal{M}$.

%
\section{Chase Detector}
\label{sec:ch2_CD}

The CD \cite{waters2008chase} mitigates error propagation in SIC by populating a list $\mathcal{S}(\tilde{\mbf{y}},\mbf{R})$ of candidate symbol vectors for final decision. It first partitions $\tilde{\mbf{y}}$, $\mbf{R}$, and $\mbf{x}$ as
\begin{equation}\label{eq:ch2_partitionCD}
    \tilde{\mbf{y}} =
        \begin{bmatrix}
            \tilde{\mbf{y}}_{1} \\
            \tilde{y}_{N}^{}
        \end{bmatrix}, \ \
    \mbf{R} =
        \begin{bmatrix}
            \mbf{A} & \mbf{b} \\
            \mbf{0} & c
        \end{bmatrix}, \ \
    \mbf{x} =
        \begin{bmatrix}
            \mbf{x}_{1} \\
            x_{N}^{}
        \end{bmatrix},
\end{equation}
where $\tilde{\mbf{y}}_{1} \in \mathcal{C}^{(N-1)\times1}$, $\tilde{y}_{N}^{} \in \mathcal{C}^{1\times1}$, $\mbf{A} \in \mathcal{C}^{(N-1)\times(N-1)}$, $\mbf{b}\in \mathcal{C}^{(N-1)\times1}$, $c \in \mathcal{R}^{1\times1}$, $\mbf{x}_{1} \in \mathcal{M}^{N-1}$, $\mbf{0}$ is a $1\times(N-1)$ vector of zero-valued entries, and $x_{N}^{} \in \mathcal{M}$.
Then, for each $x_N$ at the root layer, a candidate vector is calculated as in~\eqref{eq:ch2_SIC2} and added to $\mathcal{S}$. The maximum number of candidate vectors in $\mathcal{S}$ is $\abs{\mathcal{M}}$, and the final HO decision vector is chosen from $\mathcal{S}$ to be
\begin{equation}\label{eq:ch2_sol_CD}
  \hat{\mbf{x}}^{\CD} = \argmin_{\mbf{x} \in \mathcal{S}} \norm{\tilde{\mbf{y}}\!-\!\mbf{R}\mbf{x}}^{2}.
\end{equation}
Note that CD differs from LSD \cite{hochwald2003achieving}. LSD list admission depends on run-time channel conditions, which makes it nondeterministic and more complex. In a SO setting, LSD does not guarantee computing all the required distance metrics.

%
\section{Layered Orthogonal Lattice Detector}
\label{sec:ch2_LORD}

Instead of executing the CD routine once, LORD repeats chase detection with different layer orderings, each time with a different layer as root, by cyclically shifting the columns of $\mbf{H}$. The best output from these trials is the final solution. Each permuted $\mbf{H}$ at step $t$, $t\!=\!1,\!\cdots\!,N$, is QR-decomposed into $\mbf{Q}^{(t)}$ and $\mbf{R}^{(t)}$ according to~\eqref{eq:ch2_partitionCD}. Let $\hat{\mbf{x}}_{(t)}^{\CD}$ denote the output CD solution from step $t$. Then, the final solution $\hat{\mbf{x}}^{\LORD}$ is $\hat{\mbf{x}}_{(t_{min})}^{\CD}$, where
\begin{equation}\label{eq:ch2_LORDout}
    t_{min} = \argmin_{ t \in \{1,\cdots,N\} } \norm{\tilde{\mbf{y}} - \mbf{R}\hat{\mbf{x}}_{(t)}^{\CD}}^{2}.
\end{equation}
Since distances are preserved under different layer orderings with QRD, the accumulated candidate vectors across different partitions form an ``extended'' candidate list, despite the potential overlap of lists from each partition. Therefore, the added gain with LORD compared to CD is significant. Note that optimized layer ordering, using some form of sorted QRD (SQRD), for example, can further enhance the performance.

For dual-layer MIMO systems ($N=2$) LORD achieves exact optimal log-max ML performance, and it only requires $2\times2^q$ instead of $2^{q^2}$ distance computations. By analogy with \eqref{eq:ch2_partitionCD}, the corresponding $2\times2$ modified system model can be represented as:
\begin{equation}\label{eq:ch2_mod_sys_model}
\mbf{y}-\mbf{Hx}\rightarrow \begin{bmatrix} \tilde{y}_{1}\\\tilde{y}_{2}\end{bmatrix} - \begin{bmatrix} a & b \\    0 & c \\ \end{bmatrix}. \begin{bmatrix} x_{1}\\x_{2}\end{bmatrix},
\end{equation}
where $a_{1},c_{1}\in \mathcal{R}^{+}$, and $b_{1}\in\mathcal{C}$. We have, $\min_{\mbf{x}\in\mathcal{X}} \norm{\tilde{\mbf{y}}\!-\!\mbf{R}\mbf{x}}^{2}=\min_{x_{2}\in\mathcal{X}_{2}} d_{2}(x_{2})$:
\begin{align}
\min_{x_{2}\in \mathcal{X}_{2}} d_{2}(x_{2}) &= \min_{\substack{x_{2}\in \mathcal{X}_{2} \\ x_{1}\in \mathcal{X}_{1}}} (\abs{\tilde{y}_{2}\!-\!cx_{2}}^{2}\!+\!\abs{\tilde{y}_{1}\!-\!ax_{1}\!-\!bx_{2}}^{2}) \\ \nonumber
&=\min_{x_{2}\in \mathcal{X}_{2}}(\abs{\tilde{y}_{2}\!-\!cx_{2}}^{2}\!+\!\abs{\tilde{y}_{1}\!-\!a\hat{x}_{1}\!-\!bx_{2}}^{2}),
\end{align}
where $\hat{x}_{1}$ is obtained by slicing $(\tilde{y}_{1}-bx_{2})/a \in \mathcal{C}$ over the constellation $\mathcal{X}_{1}$:
\begin{equation}
 \hat{x}_{1}=\lfloor(\tilde{y}_{1}-bx_{2})/a\rceil_{\mathcal{X}_{1}}\in\mathcal{X}_{1}.
\end{equation}
Note that this implementation requires only $\abs{\mathcal{X}_{2}}=2^q$ distance computations.

In a SO setting, the LLRs of the bits in the symbol $x_{2}$ can be obtained as:
\begin{equation}\label{eq:ch2_LLR22}
\lambda_{2,k}^{\ML} = \min_{x_{2}\in \mathcal{X}_{2,2,k}^{(0)}} d_{2}(x_{2}) - \min_{x_{2}\in \mathcal{X}_{2,2,k}^{(1)}} d_{2}(x_{2}), \ k=1,\ldots,q.
\end{equation}
To obtain the LLRs $\lambda_{1,k}^{\ML}$ for the bits in $x_{1}$, the same operation is repeated in a reversed order, where the $x_{1}$ symbols are exhaustively searched, while the interference over layer 2 is subtracted, followed by simple slicing over $\mathcal{X}_{2}$. Note that to find the hard-decision ML solution only, a 1-sided decomposition is needed on either layer 1 or 2.

%
\section{Subspace Detector}
\label{sec:ch2_subspace}

\begin{figure}[t]
\centering
\includegraphics[width=5in]{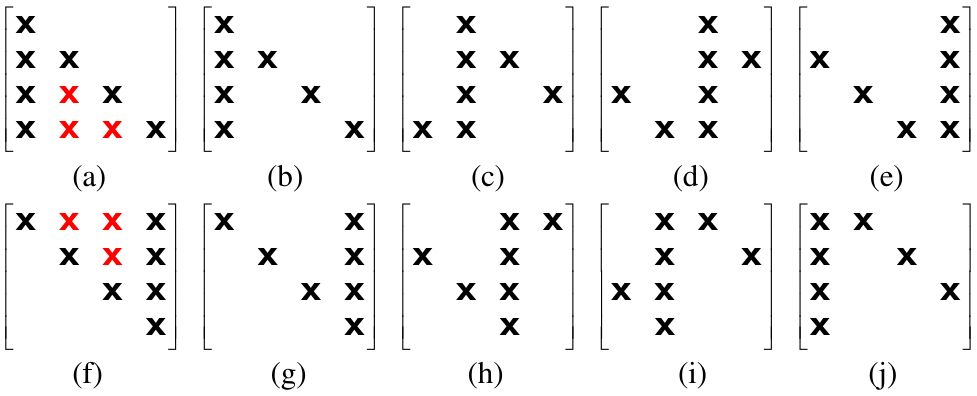}\vspace{-0.1in}
\caption{Channel matrix structures for 4x4 MIMO.}
\label{f:ch2_matrices}
\end{figure}

The aforementioned optimal LORD implementation for $2\!\times\!2$ MIMO cannot scale up for $N\geq3$ without loosing optimality. This is because $\mbf{R}$ would include off-diagonal terms, the red-marked entries in Fig. \ref{f:ch2_matrices}(f), that prevent computing the ML solution by enumerating symbols on one layer and finding the minima through slicing individually on all other layers in parallel. In fact, the ML solution requires enumerating symbols on $N-1$ layers and slicing on the last layer, which results in $\mathcal{O}(2^{q^N})$ complexity.

However, the channel matrix can be punctured to zero-out undesirable entries, as shown in Fig.~\ref{f:ch2_matrices}(g) for a 4-layer MIMO system \cite{2014_mansour_eurasip_WLD}. This configuration allows us to enumerate symbols on layer 4, while finding the minimum distances on layers 3 to 1 in parallel, through slicing only on the corresponding layers. Moreover, to compute the LLRs for bits associated with layers 3 to 1, a similar process is repeated on each layer, after cyclical column shifting followed by channel matrix decomposition. The effective punctured channels are shown in Fig.~\ref{f:ch2_matrices}(h)-(j), respectively. When adopting the complementary QL decomposition (QLD), the corresponding desirable structures are shown in Fig.~\ref{f:ch2_matrices}(b)-(e). In this case, by enumerating symbols on layer 1, the minimum on layers 2 to 4 can be found in parallel through slicing, and a similar process is repeated on other layers.

\subsection{Conventional WR Decomposition}
\label{sec:ch2_WRD}

The first step in SSD is channel matrix decomposition. While LORD only requires QRD, a more powerful WRD scheme is required to puncture the red-marked entries above the diagonal in Fig.~\ref{f:ch2_matrices}(f). WRD transforms $\mbf{H}$ into a punctured UTM $\Rp=[\rp_{ij}]\in\mathcal{C}^{N\times N}$ with $\rp_{ii}\in\mathcal{R}^{+}$, by puncturing entries between the diagonal and the last column through a matrix $\mbf{W}=[\mbf{w}_{1}\cdots \mbf{w}_{n} \cdots \mbf{w}_{N}^{}]\in\mathcal{C}^{M\times N}$, such that $\mbf{W}^\mathcal{H}\mbf{H}=\Rp$. The transformed received symbol vector can be expressed as
\begin{equation}\label{eq:ch2_sysmodel3}
  \bar{\mbf{y}} = \mbf{W}^\mathcal{H}\mbf{y} = \Rp\mbf{x} + \mbf{W}^\mathcal{H}\mbf{n}.
\end{equation}
We assume $\mbf{H}$ to have a full column rank. Setting $\mbf{W}=(\mbf{H}^\mathcal{H}\mbf{H})^{-1}\mbf{H}^\mathcal{H}$ to be the left Moore-Penrose pseudo-inverse of $\mbf{H}$ results in $\Rp=\mbf{I}_{N}$, and choosing $\mbf{W}$ to be an orthonormal basis of the column space of $\mbf{H}$ transforms it into an unpunctured UTM, with $\mbf{W}$ being unitary (QRD). In general, if $\Rp$ is punctured, then $\mbf{W}$ is non-unitary. We impose the condition on the column vectors of $\mbf{W}$ to have unit length, i.e., $\mbf{w}_{n}^\mathcal{H}\mbf{w}_{n}\!=\!1$ for $n\!=\!1,\ldots,N$.

Let $\mbf{P}^{\dag}\!=\!\mbf{H}(\mbf{H}^\mathcal{H}\mbf{H})^{-1}\mbf{H}^\mathcal{H}$ be the orthogonal projection onto the column space of $\mbf{H}$, and $\mbf{P}^{\perp}\!=\!\mbf{I}-\mbf{H}(\mbf{H}^\mathcal{H}\mbf{H})^{-1}\mbf{H}^\mathcal{H}$ be the orthogonal projection onto the left nullspace of $\mbf{H}$. Let $\mbf{H}_{\mathcal{I}}$ be the submatrix formed by the columns of $\mbf{H}$ whose index $n\in{\mathcal{I}}$ (if $\mathcal{I}\!=\!{1,3}, \ \text{then}\  \mbf{H}_{\mathcal{I}}\!=\![\mbf{h}_{1}\mbf{h}_{3}]$). Denote by $\mathcal{I}_{n}$ the column index set of the entries in the $\nth{n}$ row of $\mbf{H}$ to be zeroed out, and define $\tilde{\mbf{w}}_{n}=\mbf{P}_{\mathcal{I}_{n}}^{\perp}\mbf{h}_{n}$, where
\begin{equation}\label{eq:ch2_pperp}
\mbf{P}^{\perp}_{\mathcal{I}_{n}}=\mbf{I}_{N}-\mbf{H}_{\mathcal{I}_{n}}(\mbf{H}^\mathcal{H}_{\mathcal{I}_{n}}\mbf{H}_{\mathcal{I}_{n}})^{-1}\mbf{H}^\mathcal{H}_{\mathcal{I}_{n}},
\end{equation}
and $\mbf{H}_{\mathcal{I}_{n}}\!=\!\{\mbf{h}_{m}\!\mid \!m\!\in\!\mathcal{I}_{n}\}$. The normalized vector is derived as $\mbf{w}\!=\!\tilde{\mbf{w}}_{n}/\norm{\tilde{\mbf{w}}_{n}}$ with $\norm{\tilde{\mbf{w}}_{n}}\!=\!\sqrt{\mbf{h}_{n}^\mathcal{H}\mbf{P}_{\mathcal{I}_{n}}^{\perp}\mbf{h}_{n}}$. Let $\mbf{D}=[d_{n}]\!\in\!\mathcal{R}^{+}$ be a diagonal matrix whose entries are given by $d_{n}\!=\!1/\sqrt{\mbf{h}_{n}^\mathcal{H}\mbf{P}_{\mathcal{I}_{n}}^{\perp}\mbf{h}_{n}},\ n\!=\!1,\ldots,N$. The matrix that zeroes out the entries in the rows of $\mbf{H}$ at column positions given in $\mathcal{I}_{n}$ is

\begin{equation}\label{eq:ch2_w}
\mbf{W}^\mathcal{H}=\mbf{D}\begin{bmatrix}
    \mbf{h}_{1}^\mathcal{H}\mbf{P}_{\mathcal{I}_{1}}^{\perp} \\
    \mbf{h}_{2}^\mathcal{H}\mbf{P}_{\mathcal{I}_{2}}^{\perp} \\
    \vdots \\
    \mbf{h}_{N}^\mathcal{H}\mbf{P}_{\mathcal{I}_{N}}^{\perp}
\end{bmatrix}.
\end{equation}
For example, in a $4\!\times\!4$ MIMO system, we choose the puncturing sets as $\mathcal{I}_{1}\!=\!{2,3}$, $\mathcal{I}_{2}\!=\!{1,3}$, $\mathcal{I}_{3}\!=\!{1,2}$, and $\mathcal{I}_{4}\!=\!{1,2,3}$.

\subsection{Detection Routine}
\label{sec:ch2_det}

To generate SO LLRs for all layers, the $N$ streams are decoupled, one at a time in $N$ steps, by cyclically shifting the columns of $\mbf{H}$ and generating the punctured UTMs, as shown in Fig. \ref{f:ch2_matrices}(g-j). Each permuted $\mbf{H}$ at step $t$ is WR-decomposed into $\mbf{W}^{(t)}$ and $\Rp^{(t)}$. We first partition $\bar{\mbf{y}}^{(t)}$, $\Rp^{(t)}$, and $\mbf{x}$ as in \eqref{eq:ch2_partitionCD}
\begin{equation}\label{eq:partitionCDwrd}
\bar{\mbf{y}}^{(t)} =
    \begin{bmatrix}
        \bar{\mbf{y}}_{1}^{(t)} \\
        \bar{y}_{N}^{(t)}
    \end{bmatrix}, \ \
\Rp =
    \begin{bmatrix}
        \mathring{\mbf{A}}^{(t)} & \mathring{\mbf{b}}^{(t)} \\
        \mbf{0} & \mathring{c}^{(t)}
        \end{bmatrix}, \ \  \mbf{x}=\begin{bmatrix} \mbf{x}_{1} \\ x_{N}^{}  \end{bmatrix},
\end{equation}
where in this case $\mathring{\mbf{A}}^{(t)} \in \mathcal{R}^{(N-1)\times(N-1)}$ is a diagonal matrix. Then, the vector with minimum distance corresponding to a structure $t$ is
\begin{align}\label{eq:ch2_xwr}
\mbf{x}_{(t)}^{\WR} &\!=\!\argmin_{\mbf{x} \in \mathcal{X}}{\norm{\bar{\mbf{y}}^{(t)}\!-\!\Rp^{(t)}\mbf{x}}^{2}} \\
&\!=\!\argmin_{x_{N} \in \mathcal{M}}{\left(\norm{\bar{y}_{N}^{(t)}\!-\!\mathring{c}^{(t)}x_{N}}^{2}\!+\!\norm{\bar{\mbf{y}}_{1}^{(t)}\!-\!\mathring{\mbf{A}}^{(t)}\hat{\mbf{x}}_{1}\!-\!\mathring{\mbf{b}}^{(t)}x_{N}}^{2}\right)},
\end{align}
where $\hat{\mbf{x}}_{1}=\lfloor(\bar{\mbf{y}}_{1}^{(t)}\!-\!\mathring{\mbf{b}}^{(t)}x_{N})/\mathring{\mbf{A}}^{(t)}\rceil_{\mathcal{M}^{N-1}}$ is the sliced output. Since $\mathring{\mbf{A}}^{(t)}$ is diagonal, slicing is applied to individual elements of $\bar{\mbf{y}}^{(t)}_{1}$ over $\mathcal{M}$. To generate soft outputs, we compute two distance metrics defined as
\begin{align}\label{eq:ch2_newLLR1}
  \mbf{u}_{n,k,t}^{\WR} &= \argmin_{x_n \in \mathcal{X}_{n,n,k}^{(0)}}{\norm{\bar{\mbf{y}}^{(t)} - \Rp^{(t)}\mbf{x}}^{2}} \\
  \mbf{v}_{n,k,t}^{\WR} &= \argmin_{x_n \in \mathcal{X}_{n,n,k}^{(1)}}{\norm{\bar{\mbf{y}}^{(t)} - \Rp^{(t)}\mbf{x}}^{2}},
\end{align}
which can be expanded as in \eqref{eq:ch2_xwr}. The LLRs are then calculated as
\begin{equation}\label{eq:ch2_LLRfinal}
  \lambda_{n,k}^{\SSD} = \frac{1}{\sigma^{2}} \left[ \min_t \left( \norm{\bar{\mbf{y}}^{(t)} - \Rp^{(t)}\mbf{u}_{n,k,t}^{\WR}}^{2} \right) - \min_t \left( \norm{\bar{\mbf{y}}^{(t)} - \Rp^{(t)}\mbf{v}_{n,k,t}^{\WR}}^{2} \right) \right],
\end{equation}
for $n\!=\!1,\ldots,N$, $k\!=\!1,\ldots,\log{\abs{\mathcal{M}}}$, and $t\!=\!N\!-\!n\!+\!1$. Note global minimum distances are tracked here, rather than just minimizing over the per stream LLRs.

%% file: Chapter3_TURBO.tex

\chapter{Iterative MIMO Detection with Large Constellations}\label{chapter:turbo}

In this chapter, we build on the LC-LORD \cite{tomasoni2012hardware}, and propose four efficient SO detection schemes. In the first three approaches, we enhance the location of the reduced region of search within a constellation, based on layer ordering, iterative updates of the center of region of search, and HO MMSE detection. In the fourth approach, we propose an enhanced saturation criteria for bit LLRs. The corresponding results are published in \cite{7418324Sarieddeen}.

We limit the discussion to dual-layer MIMO ($N\!=\!M\!=\!2$) and assume a high order MT. Hence, the received signal can be written as $\mbf{y} = \mbf{h}_1 x_1 + \mbf{h}_2 x_2 + \mbf{n}$, where $x_1$ and $x_2$ are drawn from the same Gray-mapped 1024-QAM.

%
\section{Turbo-LORD}
\label{sec:ch3_tlord}

Turbo-LORD (T-LORD) \cite{4411575Tomasoni} \cite{5581209Tomasoni} is a generalization of LORD, that builds on the MAP detector instead of the ML detector, and that is used in the context of iterative detection and decoding. The MAP detector accepts from the decoder, along with the received vector $\mbf{y}$, a priori LLRs $\xi_1$ and $\xi_2$, for bits corresponding to $x_1$ and $x_2$, respectively. The resultant modified distance metric is
\begin{equation}\label{eq:ch3_map}
 \varphi^{\TLORD}(\mbf{x}) = - \frac{1}{\sigma^{2}} \norm{\tilde{\mbf{y}} - \mbf{R}\mbf{x}}^{2} + \mbf{b}_{1}\xi_1 + \mbf{b}_{2}\xi_2,
\end{equation}
and the a-posteriori LLRs after the $\nth{t}$ detection/decoding iteration can be calculated as
\begin{equation}\label{eq:ch3_LLR3}
\lambda_{n,k,t}^{\TLORD} = \max_{\mbf{x}\in \mathcal{S}_{n,k,t}^{(1)}} \varphi^{\TLORD}(\mbf{x}) - \max_{\mbf{x}\in \mathcal{S}_{n,k,t}^{(0)}} \varphi^{\TLORD}(\mbf{x}),
\end{equation}
where $\mathcal{S}_{n,k,t}^{(0)}\!\triangleq\!\{\mbf{x} \in \mathcal{S}_t: b_{n,k}\!=\!0\}$ and $\mathcal{S}_{n,k,t}^{(1)}\!\triangleq\!\{\mbf{x} \in \mathcal{S}_t: b_{n,k}\!=\!1\}$. Note that $\mathcal{S}_t$ is defined at the $\nth{t}$ detection/decoding iteration as $\mathcal{S}$ in Sec. \ref{sec:ch2_CD}.

\subsection{Low-Complexity LORD}
\label{sec:ch3_lclord}

Searching $\abs{\mathcal{M}}=1024$ lattice points is computationally demanding. LC-LORD \cite{tomasoni2012hardware} aims at reducing the number of visited candidate points, by only exploring a subset of the constellation at the root layer, a reduced QAM $\Theta$. For convenience, $\Theta$ is a square subset of $\mathcal{M}$, centered at the equalized output $\tilde{y}_{2}/r_{2,2}$.

LC-LORD does not guarantee the computability of~\eqref{eq:ch3_LLR3}, since one of the two terms will not exist if all points in $\Theta$ have a unique bit value at a specific bit location. This is known as LLR saturation. Note that with Gray coded symbol mapping, the LLRs for low order bits are less likely to get saturated, but this gets more probable as $\abs{\Theta}$ decreases. Moreover, LC-LORD need not be applied to all carriers, in fact, and based on the implementation constraints, the authors in \cite{tomasoni2012hardware} proposed a mechanism in which they isolate the worst carriers and apply full complexity LORD to them. The criteria to identify the worst carriers is to select the smallest of
\begin{equation}\label{ch3_criteria}
  \min_{l=1,2}r_{(2,2)}^{l},
\end{equation}
where $l$ denotes the antenna index at the root layer.

%
\section{Proposed Approaches}
\label{sec:ch3_proposed}

\subsection{Enhancing Search Region Location}
\label{sec:ch3_locality}

The performance of LC-LORD is constrained by the probability of the actual transmitted symbol to lie outside the reduced QAM, causing its failure. The situation is worse with correlated channels, where $\mbf{H}$ tends to be ill-conditioned, and consequently $r_{2,2}$ tends to zero. Towards increasing the likelihood of the actual transmitted symbol to lie inside the reduced QAM, we propose three approaches.

The first approach is based on layer ordering \cite{4224472Siti} followed by N/C, which is also known as ZF with decision-feedback (ZFDF). We first find the equalized output on layer-2 and project its value on layer-1 to obtain $\bar{\mbf{x}}^{1}\!=\![\bar{x}_1^{1}, \bar{x}_2^{1}]$, following the procedure in \eqref{eq:ch2_SIC2}. Then we permute the columns of the channel matrix, find the equalized output on layer-1, and project its value on layer-2 to obtain  $\bar{\mbf{x}}^{2}\!=\![\bar{x}_1^{2}, \bar{x}_2^{2}]$. Finally, the centers of search on both layers would be the components of either $\bar{\mbf{x}}^{1}$ or $\bar{\mbf{x}}^{2}$, with the choice being made on the vector that better minimizes the distance metric $\norm{\mbf{y} - \mbf{H}\mbf{x}}^{2}$. We call the corresponding detector layer-ordered LC-LORD (LO-LC-LORD).

\begin{figure}[!t]
\centering
\includegraphics[width=4.4in]{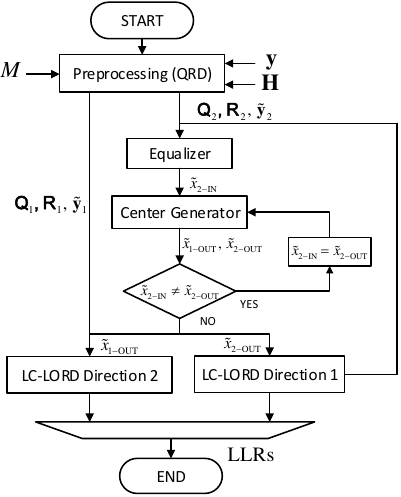}\vspace{-0.1in}
\caption{Iter-LC-LORD algorithm flowchart.}
\label{f:ch3_flow}
\end{figure}

The second approach adds an iterative behavior to the detector, as shown in Fig.~\ref{f:ch3_flow}. It starts by feeding the equalized symbol on the root layer (layer-2 here) to a center generator (CG). The CG accepts a center of reduced search on one layer, and outputs enhanced centers on both layers. The CG functionality is based on HO LC-LORD. This means that CG applies LC-LORD from one direction only, and the components of the hard sub-ML output vector will serve as centers for reduced QAMs in the next iteration. This can iterate as long as the output differs from the input, and in every iteration we get closer to the true ML HO. However, there is no guarantee that the algorithm will reach the true ML value at convergence, since it might get stuck in a local minimum. The algorithm halts after a maximum number of $\bar{J}$ iterations. Once the center is obtained, the algorithm proceeds with LC-LORD as described in Sec. \ref{sec:ch3_lclord}, working in both directions, on reduced constellations $\Theta_{1}$ and $\Theta_{2}$ in layers 1 and 2, respectively. We call the corresponding detector iterative LC-LORD (Iter-LC-LORD).

Moreover, as a third approach, the components of the HO vector of an MMSE detector are used as centers for reduced search regions. We call the latter approach MMSE-LC-LORD. Note that in these approaches the centers are generated on both layers simultaneously, and not independently on each layer, which prevents processing the layers in a fully parallel mode. \\

In the case of T-LORD, the overhead of these approaches can be reduced by only applying them in the first detection/decoding iteration. If the HO of LC-LORD is passed over $\bar{T}$ detection/decoding iterations, the center of reduced search would be updated on every iteration, the same way the CG updates it in Iter-LC-LORD. Note that center updates from one detection/decoding iteration to another can also be driven by the a priori information \cite{tomasoni2012hardware}.

\begin{figure}[t]
\centering
\includegraphics[width=4.2in]{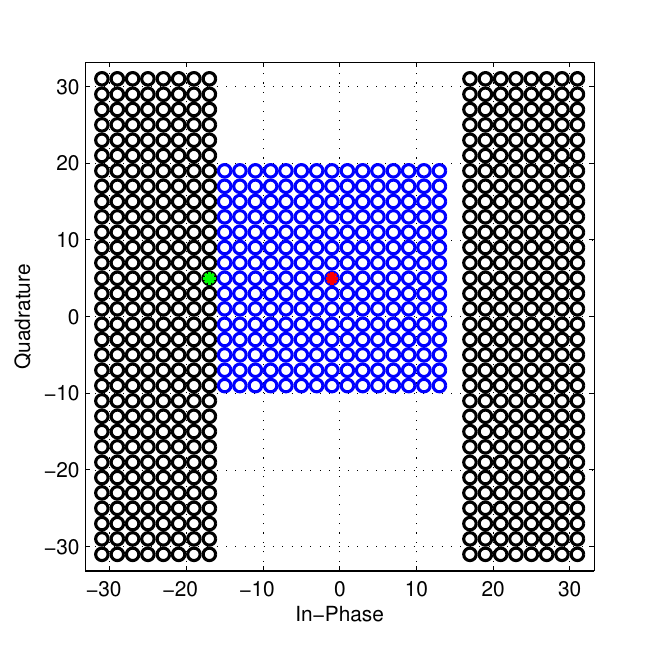}\vspace{-0.1in}
\caption{Constellation schematic - black circles indicate third MSB is 1.}
\label{f:ch3_llr}
\end{figure}

\subsection{Enhancing LLR Saturation}
\label{sec:ch3_saturation}

The authors in \cite{tomasoni2012hardware} suggested either saturating the LLR to a maximum threshold value, or substituting the missing term in \eqref{eq:ch3_LLR3} by the maximum Euclidean norm within $\Theta$. These approaches are easy to implement, but might remarkably degrade performance when $\abs{\Theta}$ is small.

Our proposed approach, region-thresholding LC-LORD (RegTh-LC-LORD), fills the empty component in~\eqref{eq:ch3_LLR3} by an approximate distance metric. We first locate the closest point to the center of $\abs{\Theta}$, having an opposite bit value at the bit location of interest (green point in Fig.~\ref{f:ch3_llr}). Then, we project this point on the other layer (slicing), and find the distance from the resultant vector to the received symbol vector. This mechanism depends on the regions of specific bit values. We augment all our proposed approaches with this thresholding method.

%
\section{Complexity Study}
\label{sec:ch3_complexity}

The computational complexity can be split into two components, complexity of preprocessing stage and complexity of search routine. The preprocessing stage is mainly composed of QRD and equalizations. All LORD-based approaches require two QRDs. However, a solution \cite{6862012Gomaa} exists, in which the equalized outputs on both layers are efficiently computed without QRD, and in \cite{2015_mansour_JSP_2x2QAM}, an optimized scheme for QRD-based distance computations was proposed. All LC-LORD versions have an extra burden of handling the search region boundaries.

On the other hand, the complexity of the search routine is dominated by Euclidian distance computations, that are quantified by the number of visited lattice points. LORD is the most complex with $2\!\times\!\abs{\mathcal{M}}$ computations. After that comes Iter-LC-LORD, which has a variable complexity with a worst case scenario of $(\bar{J}\!+\!2)\!\times\!\abs{\Theta}$ computations. Note that it has a lower complexity on average because the reduced QAMs in subsequent iterations will largely overlap, and hence redundant distance computations can be avoided. Finally, the least complex are LC-LORD and LO-LC-LORD, with each requiring $2\!\times\!\abs{\Theta}$ distance computations. The search cost of SO MMSE is half that of LORD because the computed distances are between points on a single layer. Table \ref{table:ch3_complexity_1} summarizes the approaches and their worst case complexity when applied to a single tone, in terms of distance computations.

\begin{table}[!t]
\caption{\textbf{Detector Complexities in Terms of Distance Computations}}\vspace{0.15in} 
\label{table:ch3_complexity_1} 
\centering 
\begin{tabular}{|c|| c | c |}
  \hline
  Approach & Description & Complexity \\
  \hline\hline
  ML & Full Complexity LORD & $2\times\abs{\mathcal{M}}$ \\
  \hline
  LC-LORD & Low Complexity LORD & $2\times\abs{\Theta}$ \\
  \hline
  LO-LC-LORD & Layer Ordered LC-LORD + Thresholding & $2\times\abs{\Theta}$ \\
  \hline
  Iter-LC-LORD & Iterative LC-LORD   + Thresholding & $(\bar{J}+2)\times\abs{\Theta}$ \\
  \hline
  MMSE-LC-LORD & MMSE-based LC-LORD + Thresholding & $2\times\abs{\Theta}$ \\
  \hline
  RegTh-LC-LORD & LC-LORD + Thresholding & $2\times\abs{\Theta}$ \\
  \hline
  MMSE & SO MMSE & $\abs{\mathcal{M}}$ \\
  \hline
\end{tabular}
\end{table}

\begin{figure}[!t]
\centering
\includegraphics[width=4.7in]{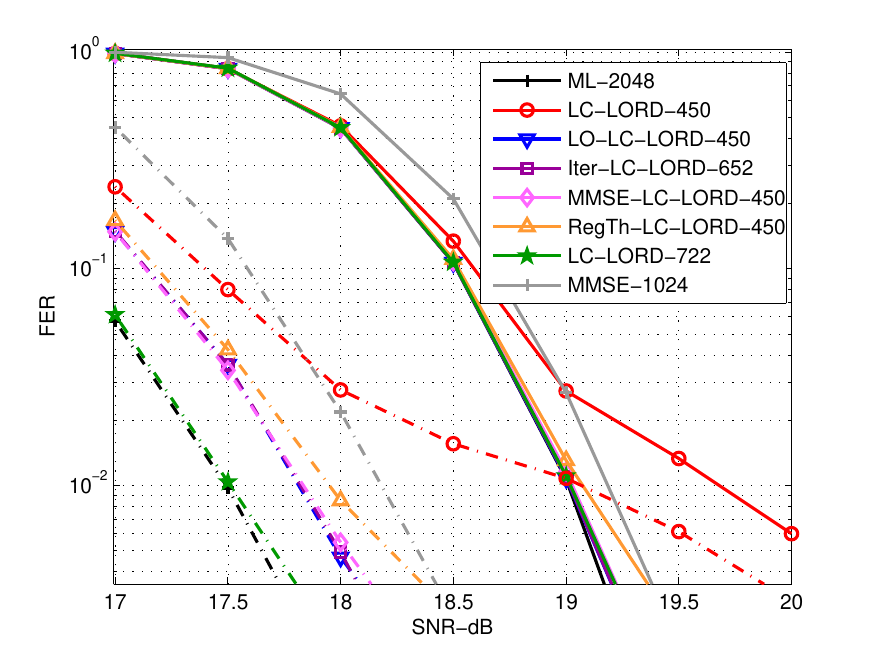}\vspace{-0.1in}
\caption{FER performance - uncorrelated channels - $15\%$ full complexity carriers - $\bar{T} = 1$ (solid) and $\bar{T} = 4$ (dotted).}
\label{f:ch3_feruncorr15}
\end{figure}

\begin{figure}[!t]
\centering
\includegraphics[width=4.7in]{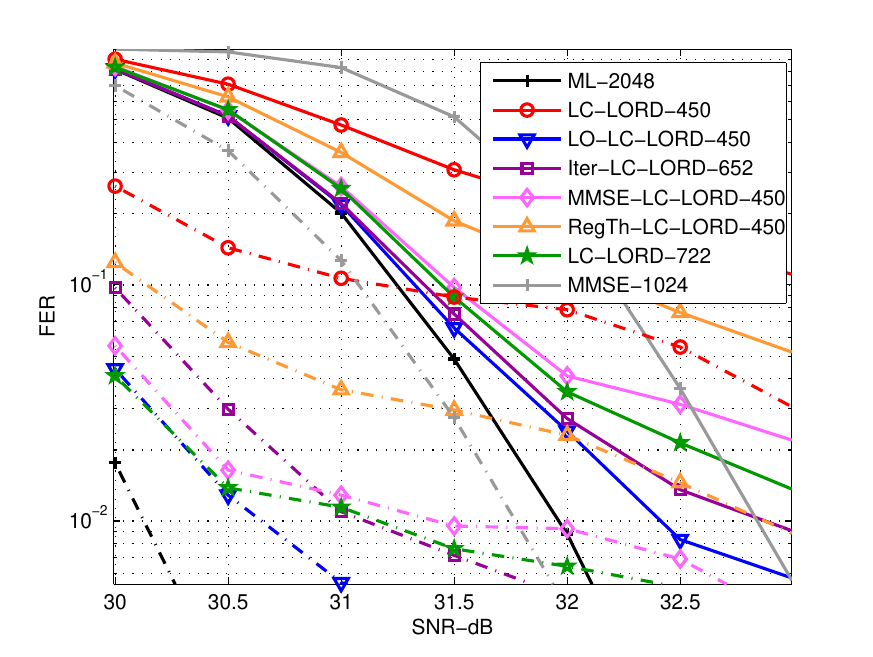}\vspace{-0.1in}
\caption{FER performance - correlated channels - $15\%$ full complexity Carriers - $\bar{T} = 1$ (solid) and $\bar{T} = 4$ (dotted).}
\label{f:ch3_fercorr15}
\end{figure}

\begin{figure}[!t]
\centering
\includegraphics[width=4.7in]{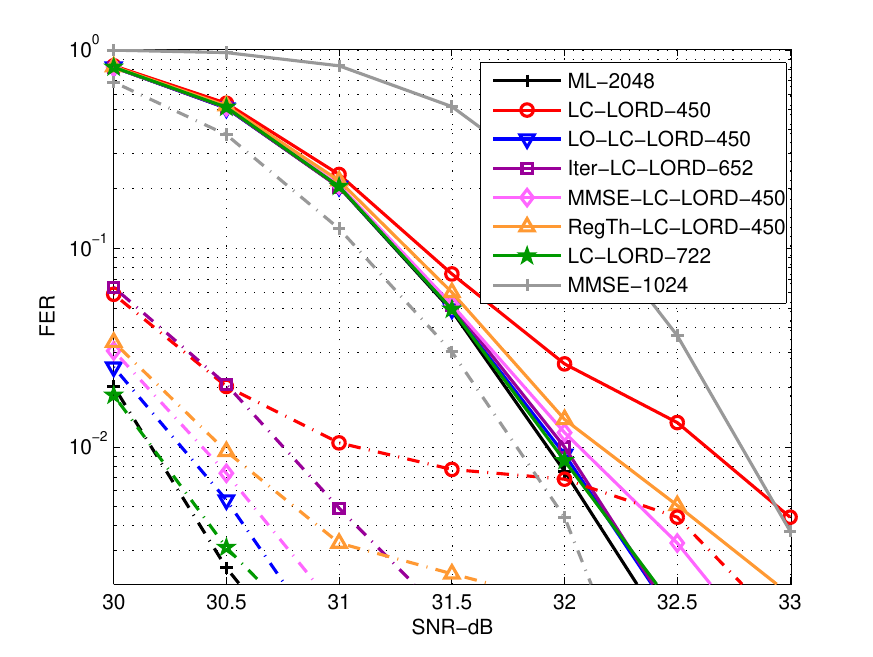}\vspace{-0.1in}
\caption{FER performance - correlated channels - $30\%$ full complexity carriers - $\bar{T} = 1$ (solid) and $\bar{T} = 4$ (dotted).}
\label{f:ch3_fercorr30}
\end{figure}

%
\section{Simulation Results}
\label{sec:ch3_simulation}

The implementation followed the system model of Sec. \ref{sec:ch2_model}. Turbo coding and decoding was used with a code rate of $1/2$. In addition to the zero-mean i.i.d. $\mathcal{CN}(0,1)$ channel (rich scattering), we considered highly correlated channels using the long term evolution (LTE) \cite{LTE_36.211} model, with transmit and receive correlation coefficients of $0.9$.  We assumed $\abs{\Theta}=225$, $\bar{J}=8$ and $\bar{T}=4$.

Figures ~\ref{f:ch3_feruncorr15} to \ref{f:ch3_fercorr30} show the frame error rate (FER) performance. The numbers in the legend correspond to the average complexity in terms of the number of visited points. For Iter-LC-LORD, we avoided redundant computations across iterations, and noted that with $\abs{\Theta}=225$, only $1.8$ out of the $\bar{J}=8$ iterations are required on average to converge. Figure \ref{f:ch3_feruncorr15} corresponds to the case when the worst $15\%$ of carriers were treated with full complexity LORD, and the channels are uncorrelated. All proposed approaches achieved near-optimal performance, restoring the error floors in LC-LORD plots at high SNR. Figure \ref{f:ch3_fercorr15} then shows the respective performance under high channel correlation, where all sub-ML approaches suffer. Compared to LC-LORD, our approaches added a remarkable gain, with the best being LO-LC-LORD, followed by Iter-LC-LORD, then MMSE-LC-LORD, and finally RegTh-LC-LORD. Note that the higher complexity version of LC-LORD ($\abs{\Theta}\!=\!361$) could not beat the less complex Iter-LC-LORD. Finally, despite high channel correlation, the near-optimality of our proposed approaches was restored when the worst $30\%$ of carriers were treated with full complexity, as shown in Fig.~\ref{f:ch3_fercorr30}. \\

\textbf{Conclusion} \\

In this chapter, efficient low-complexity detection in $2\times2$ MIMO systems that use the very high order 1024-QAM has been studied. Several enhancements were proposed to LC-LORD, namely, LO-LC-LORD, Iter-LC-LORD, MMSE-LC-LORD, and RegTh-LC-LORD. The proposed algorithms have been shown to remarkably enhance the performance at a low complexity overhead, both in the context of non-iterative and iterative detection and decoding.

%% file: Chapter4_PRQRD.tex

\chapter{Reduced-Complexity QRD-Based Detection}\label{chapter:prqrd}

In this chapter, we propose computationally efficient detection algorithms, that consist of layer ordering followed by PR-QRD, based on the modified GS (MGS) orthogonalization.

First, the preprocessing complexity of SSD is reduced by using special layer orderings and PR-QRD. A hardware architecture is designed that allows building an 8-layer detector from 4-layer and 2-layer constituent detector blocks. The corresponding results appeared, in parts, in \cite{7471826Sarieddeen} and \cite{7565040Sarieddeen}, and in a more comprehensive manner in \cite{2017Sarieddeen}.

Second, the computational complexity of the SD is reduced by employing an optimized layer-ordering scheme based on the MR criterion. An optimized dataflow architecture employing PR-QRD is proposed, alongside two efficient schedules for channel matrix permutations that optimize its use. A schedule-specific triangular systolic array (TSA) implementation of PR-QRD is also proposed. The corresponding results appeared in \cite{7511059Sarieddeen}.

%
\section{Permutation-Robust QRD}
\label{sec:ch4_QRD}

QRD can be computed using Givens rotation (GR), GS orthogonalization, or Householder transformation (HT) \cite{Golub}. While the hardware implementation of HT is very complex, GR reduces the hardware area, but at the expense of longer clock latency. The classical GS algorithm allows a memory efficient implementation due to its inherent parallelism, resulting in better regularity in data flow and a potential for better hardware-efficiency. However, due to fixed-precision computation and round off errors, it can not guarantee the orthogonality of $\mbf{Q}$. This limitation was overcome by the numerically superior MGS algorithm.

The MGS-based QRD of $\mbf{H}$ is illustrated in Fig.~\ref{f:ch4_alg}. The algorithm consists of two main parts. In the first part, the diagonal elements of $\mbf{R}$ and the columns of $\mbf{Q}$ are computed. In the second part, the non-diagonal elements of $\mbf{R}$ are computed and the columns of $\mbf{H}$ are updated. Considering a $4\times4$ complex matrix, in the first part of the first iteration, the norm of $\mbf{h}_{1}$ is assigned to $r_{11}$, and $\mbf{q}_{1}$ is calculated as $\mbf{q}_{1}=\mbf{h}_{1}/r_{11}$. Then, in the second part $r_{12}$ , $r_{13}$ , and $r_{14}$ are calculated using $\mbf{q}_{1}$, $\mbf{h}_{2}$, $\mbf{h}_{3}$, and $\mbf{h}_{4}$ as
\begin{equation}\label{ch4_vector}
  r_{1j} = \mbf{q}_{1}^{T}\mbf{h}_{j}\,\,\,\,\,\, 2\leq j \leq 4,
\end{equation}
and $\mbf{H}$ gets updated by setting its first column to zero and subtracting from the other columns the length of the projection of $\mbf{q}_{1}$ on them, i-e
\begin{equation}\label{eq:ch4_projection}
\mbf{h}_{j}=\mbf{h}_{j}-\mbf{q}_{1}r_{1j} \,\,\,\,\,\, 2\leq j \leq 4.
\end{equation}
This procedure is repeated with one less column every new iteration.

\begin{algorithm}[t]
\centering
\begin{algorithmic}[1]
\Procedure{MGS-QRD}{$\mbf{H}$}
\State $k\gets 1$;
\For{$k= 1:N$}
\State $r_{kk} \gets \sqrt{ \mbf{h}_{k}^\mathcal{H}\mbf{h}_{k} } $
\State $\mbf{q}_{k} \gets \mbf{h}_{k} / r_{kk} $
\State $j \gets k+1$
\For{$j=1:N$}
\State $r_{kj} \gets \mbf{q}_{k}^{T}\mbf{h}_{j} $
\State $\mbf{h}_{j} \gets \mbf{h}_{j} -  \mbf{q}_{k} r_{kj} $
\EndFor
\EndFor
\EndProcedure
\end{algorithmic}
\caption{MGS QRD algorithm.}\label{algorithm}
\label{f:ch4_alg}
\end{algorithm}

\begin{figure}[!t]
\centering
\includegraphics[width=5.2in]{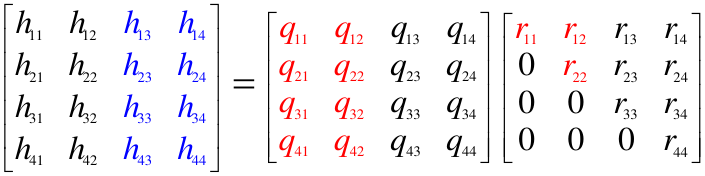}\vspace{-0.1in}
\caption{QRD savings under column permutations.}
\label{f:ch4_savings}
\end{figure}

When computing the QRD of a matrix, which is derived from another matrix, of known decomposition, by some column permutations, computational savings can be made. Part of the decomposition result remains unaltered under specific permutations. For example, assume as shown in Fig.~\ref{f:ch4_savings}, columns $3$ and $4$ in $\mbf{H}$ (in blue) were permuted. The first two columns of $\mbf{Q}$ and $\mbf{R}$ (in red) depend only on the first two columns of $\mbf{H}$, and hence there is no need to recompute them. We propose a PR-QRD that saves these redundant computations.

Furthermore, note that for detection the product $\mbf{W}^\mathcal{H}\mbf{y}$ must also be formed. This can be efficiently computed by first right-augmenting $\mbf{y}$ to $\mbf{H}$, and then performing QRD on the augmented matrix to form $\tilde{\mbf{Q}}\tilde{\mbf{R}}=[\mbf{H}|\mbf{y}]$. When carrying out the orthogonalization procedure, the same operations applied to the columns of $\mbf{H}$ are applied to the augmented column. This results in $\tilde{\mbf{Q}}=[\mbf{Q}|\mbf{0}_{N\times1}]$, with $\tilde{\mbf{R}}=[\mbf{R}|\tilde{\mbf{y}}]$, where $\mbf{H}=\mbf{QR}$ and $\tilde{\mbf{y}}=\mbf{Q}^\mathcal{H}\mbf{y}$. Consequently, $\tilde{\mbf{y}}$ is generated as a by-product. Then, carrying out the operations to puncture a given entry, these operations are also applied on the rightmost column of $\tilde{\mbf{R}}$.

%
\section{Application to Subspace Detector}
\label{sec:ch4_app_SSD}

Denote by the reference SSD algorithm of Sec. \ref{sec:ch2_det} the cyclic SSD (CYSD), that cyclically shifts the columns of $\mbf{H}$ before each decomposition $t$. Since the $\mbf{H}^{(t)}$ matrices differ by one swap operation, simplifications can be introduced.

\subsection{Single-Permutation Subspace Detector}
\label{sec:ch4_SPSD}
When cyclically shifting the columns of $\mbf{H}$, the number of WRD operations required is equal to the number of layers to be processed, which is a significant computational burden that forms a bottleneck in high order MIMO. An alternative minimal swapping operation can reduce this computational overhead. For example, in the case of $4\times4$ MIMO, if we want to compute the LLRs of the bits on layer 2, we can swap $\mbf{h}_{2}$ with $\mbf{h}_{4}$, and use the matrix decomposition of Fig.~\ref{f:ch2_matrices}(g). We represent this swapping operation by a permutation:
\begin{equation}\label{eq:ch4_permutation}
\pi^{(t)}(i)=\left\{
                \begin{array}{ll}
                  N \ \ \text{if} \  i=t\\
                  t \ \ \text{if} \  i=N\\
                  i \ \ \text{otherwise}
                \end{array}
              \right.
\end{equation}
for $t\!=\!1,\ldots,N$ and $i\!=\!1,\ldots,N$. The remainder of the SSD derivation, up to~\eqref{eq:ch2_LLRfinal}, remains intact. We call this algorithm single-permutation SSD (SPSD).

\subsection{Pairwise Subspace Detector}
\label{sec:ch4_PWSD}
Another approach, which we will later argue to be of a practical interest, is what we call pairwise SSD (PWSD). This approach consists of lumping the channel columns in pairs (assuming $N$ even), and handling each pair of layers at a time. First, the pair of interest is swapped with the rightmost two columns. Then, the columns of the pair get swapped so that each can be at position $i\!=\!N$. For example, in the case of $4\!\times\!4$ MIMO, the $4$ permuted channel matrices can be $\mbf{H}_{1}\!=\![\mbf{h}_{3}\mbf{h}_{4}\mbf{h}_{1}\mbf{h}_{2}]$, $\mbf{H}_{2}\!=\![\mbf{h}_{3}\mbf{h}_{4}\mbf{h}_{2}\mbf{h}_{1}]$, $\mbf{H}_{3}\!=\![\mbf{h}_{1}\mbf{h}_{2}\mbf{h}_{3}\mbf{h}_{4}]$, and $\mbf{H}_{4}\!=\![\mbf{h}_{1}\mbf{h}_{2}\mbf{h}_{4}\mbf{h}_{3}]$.  After each of the $N$ permutations, the permuted channel matrix is decomposed, and the LLRs for the corresponding layer are computed as in~\eqref{eq:ch2_LLRfinal}.

\begin{figure}[t]
\centering
\includegraphics[width=4.5in]{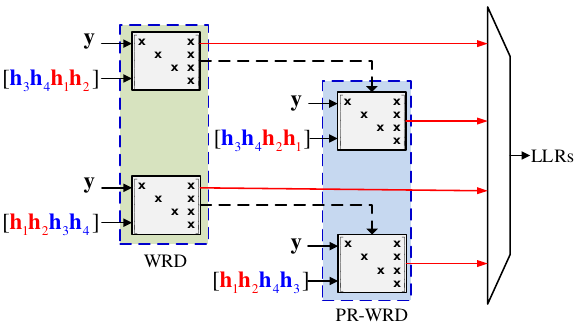}\vspace{-0.1in}
\caption{A $2$-stage $4\times4$ MIMO PWSD architecture.}
\label{f:ch4_architecture}
\end{figure}

\begin{figure}[h!]
\centerline{
\includegraphics[width=8in,angle=90]{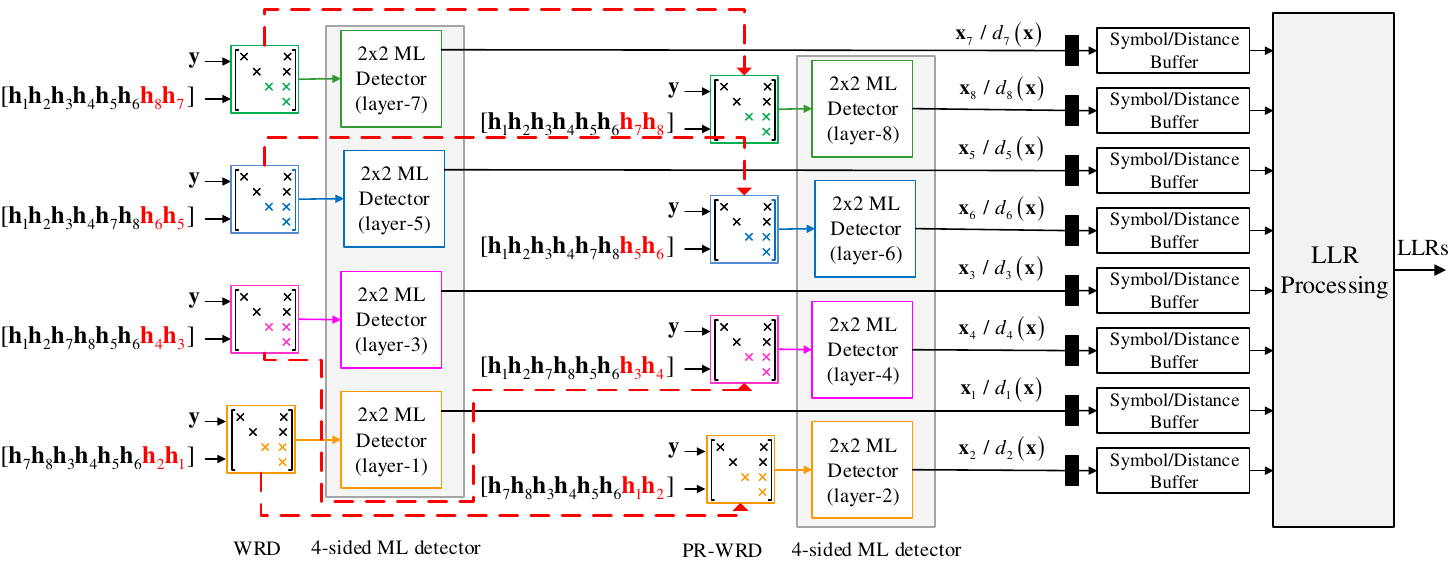}\vspace{-0.1in}
}
\caption{A $2$-stage $8\times8$ MIMO PWSD architecture.}
\label{f:ch4_architecture1}
\end{figure}

\subsection{2-Stage Subspace Detector}
\label{sec:ch4_implementation2stage}

The reference CYSD with cyclic permutations does not allow further savings because all column positions are altered from one permutation to another. However, parallelism is an inherent feature in it, where the process on each layer can run on a separate core. If we discard this parallelism, and use a pipelined architecture, the decomposition output from one layer can be fed to the subsequent layer, allowing computational savings.

A 2-stage architecture for PWSD is shown in Figs.~\ref{f:ch4_architecture} and \ref{f:ch4_architecture1}, for $4\times4$ and $8\times8$ MIMO, respectively. The odd channel permutations can execute in parallel, but with no redundant computations to save. The LLRs of their corresponding layers are sent to a buffer, and the WRD output is passed to the next stage, to assist the WRD of even permutations. A PR-WRD is thus applied in the second stage, making use of previous decompositions. Finally, the collected LLRs are processed as previously described. To implement SPSD in $8\times8$ MIMO, for example, an 8-stage architecture is required, in which the decompositions are carried out serially, and each stage can make use of computations in all previous stages. Such an architecture, if used with PWSD, results in more savings than a 2-stage architecture. However, adding more stages complicates the architecture, and increases its size and latency.

\begin{table}[!t]
\caption{\textbf{Computational Savings in Permutation-Robust SSDs}}\vspace{0.15in} 
\label{table:ch4_complexity1} 
\centering 

\begin{tabular}{| c || c | c |}
\hline
\multirow{1}{*}{} & Permutations & Saved Computations \Tstrut \Bstrut \\
\cline{1-3}

\multirow{3}{*}{\vtop{\hbox{\strut SPSD}\hbox{\strut $4\times4$ MIMO}}}  & 1: $\mbf{h}_{1}\mbf{h}_{2}\mbf{h}_{3}\mbf{h}_{4}$  & \text{none} \Tstrut \\
& 2: $\mbf{h}_{1}\mbf{h}_{2}\mbf{h}_{4}\mbf{h}_{3}$ & $88\  \RML\!+\!40\  \RAD$\\
& 3: $\mbf{h}_{1}\mbf{h}_{4}\mbf{h}_{3}\mbf{h}_{2}$ & $28\ \RML\!+\!8\ \RAD$ \Tstrut  \\
& 4: $\mbf{h}_{4}\mbf{h}_{2}\mbf{h}_{3}\mbf{h}_{1}$ & \text{none} \Tstrut \\
\cline{1-3}

\multirow{3}{*}{\vtop{\hbox{\strut PWSD}\hbox{\strut $4\times4$ MIMO}}}  & 1: $\mbf{h}_{3}\mbf{h}_{4}\mbf{h}_{1}\mbf{h}_{2}$ & \text{none} \Tstrut \\
& 2: $\mbf{h}_{3}\mbf{h}_{4}\mbf{h}_{2}\mbf{h}_{1}$ & $88\  \RML\!+\!40\  \RAD$ \Tstrut \\
& 3: $\mbf{h}_{1}\mbf{h}_{2}\mbf{h}_{3}\mbf{h}_{4}$ & \text{none} \Tstrut \\
& 4: $\mbf{h}_{1}\mbf{h}_{2}\mbf{h}_{4}\mbf{h}_{3}$ & $88\  \RML\!+\!40\  \RAD$ \Tstrut \\
\cline{1-3}

\multirow{4}{*}{\vtop{\hbox{\strut SPSD}\hbox{\strut $8\times8$ MIMO}}}  &  1: $\mbf{h}_{1}\mbf{h}_{2}\mbf{h}_{3}\mbf{h}_{4}\mbf{h}_{5}\mbf{h}_{6}\mbf{h}_{7}$\red{$\mbf{\underline{h}}_{8}$} & \text{none} \Tstrut \\
& 2: $\mbf{h}_{1}\mbf{h}_{2}\mbf{h}_{3}\mbf{h}_{4}\mbf{h}_{5}\mbf{h}_{6}$\blue{$\mbf{h}_{8}$}$$\red{$\mbf{\underline{h}}_{7}$} & $1296\ \RML\!+\!816\ \RAD$ \Tstrut\\
& 3: $\mbf{h}_{1}\mbf{h}_{2}\mbf{h}_{3}\mbf{h}_{4}\mbf{h}_{5}$\blue{$\mbf{h}_{8}$}$\mbf{h}_{7}$\red{$\mbf{\underline{h}}_{6}$} & $920\ \RML\!+\!560\ \RAD$ \Tstrut  \\
& 4: $\mbf{h}_{1}\mbf{h}_{2}\mbf{h}_{3}\mbf{h}_{4}$\blue{$\mbf{h}_{8}$}$\mbf{h}_{6}\mbf{h}_{7}$\red{$\mbf{\underline{h}}_{5}$} & $608\ \RML\!+\!352\ \RAD$ \Tstrut \\
& 5: $\mbf{h}_{1}\mbf{h}_{2}\mbf{h}_{3}$\blue{$\mbf{h}_{8}$}$\mbf{h}_{5}\mbf{h}_{6}\mbf{h}_{7}$\red{$\mbf{\underline{h}}_{4}$} & $360\ \RML\!+\!192\ \RAD$ \Tstrut \\
& 6: $\mbf{h}_{1}\mbf{h}_{2}$\blue{$\mbf{h}_{8}$}$\mbf{h}_{4}\mbf{h}_{5}\mbf{h}_{6}\mbf{h}_{7}$\red{$\mbf{\underline{h}}_{3}$} & $176\ \RML\!+\!80\ \RAD$ \Tstrut\\
& 7: $\mbf{h}_{1}$\blue{$\mbf{h}_{8}$}$\mbf{h}_{3}\mbf{h}_{4}\mbf{h}_{5}\mbf{h}_{6}\mbf{h}_{7}$\red{$\mbf{\underline{h}}_{2}$} & $56\ \RML\!+\!16\ \RAD$ \Tstrut  \\
& 8: \blue{$\mbf{h}_{8}$}$\mbf{h}_{2}\mbf{h}_{3}\mbf{h}_{4}\mbf{h}_{5}\mbf{h}_{6}\mbf{h}_{7}$\red{$\mbf{\underline{h}}_{1}$} & \text{none} \Tstrut \\
\cline{1-3}

\multirow{3}{*}{\vtop{\hbox{\strut 8-Stage PWSD}\hbox{\strut $8\times8$ MIMO}}}  &  1: $\mbf{h}_{1}\mbf{h}_{2}\mbf{h}_{3}\mbf{h}_{4}\mbf{h}_{5}\mbf{h}_{6}$\red{$\mbf{h}_{8}\mbf{\underline{h}}_{7}$} & \text{none} \Tstrut  \\
& 2: $\mbf{h}_{1}\mbf{h}_{2}\mbf{h}_{3}\mbf{h}_{4}\mbf{h}_{5}\mbf{h}_{6}$\red{$\mbf{h}_{7}\mbf{\underline{h}}_{8}$} & $1296\ \RML\!+\!816\ \RAD$ \Tstrut \\
& 3: $\mbf{h}_{1}\mbf{h}_{2}\mbf{h}_{3}\mbf{h}_{4}$\blue{$\mbf{h}_{7}\mbf{h}_{8}$}\red{$\mbf{h}_{6}\mbf{\underline{h}}_{5}$} & $608\ \RML\!+\!352\ \RAD$ \Tstrut \\
& 4: $\mbf{h}_{1}\mbf{h}_{2}\mbf{h}_{3}\mbf{h}_{4}$\blue{$\mbf{h}_{7}\mbf{h}_{8}$}\red{$\mbf{h}_{6}\mbf{\underline{h}}_{6}$} & $1296\ \RML\!+\!816\ \RAD$ \Tstrut \\
& 5: $\mbf{h}_{1}\mbf{h}_{2}$\blue{$\mbf{h}_{7}\mbf{h}_{8}$}$\mbf{h}_{5}\mbf{h}_{6}$\red{$\mbf{h}_{4}\mbf{\underline{h}}_{3}$} & $176\ \RML\!+\!80\ \RAD$ \Tstrut  \\
& 6: $\mbf{h}_{1}\mbf{h}_{2}$\blue{$\mbf{h}_{7}\mbf{h}_{8}$}$\mbf{h}_{5}\mbf{h}_{6}$\red{$\mbf{h}_{3}\mbf{\underline{h}}_{4}$} & $1296\ \RML\!+\!816\ \RAD$ \Tstrut \\
& 7: \blue{$\mbf{h}_{7}\mbf{h}_{8}$}$\mbf{h}_{3}\mbf{h}_{4}\mbf{h}_{5}\mbf{h}_{6}$\red{$\mbf{h}_{2}\mbf{\underline{h}}_{1}$} & \text{none} \Tstrut \\
& 8: \blue{$\mbf{h}_{7}\mbf{h}_{8}$}$\mbf{h}_{3}\mbf{h}_{4}\mbf{h}_{5}\mbf{h}_{6}$\red{$\mbf{h}_{1}\mbf{\underline{h}}_{2}$} & $1296\ \RML\!+\!816\ \RAD$ \Tstrut \\
\cline{1-3}

\multirow{3}{*}{\vtop{\hbox{\strut 2-Stage PWSD}\hbox{\strut $8\times8$ MIMO}}}  &  1: $\mbf{h}_{1}\mbf{h}_{2}\mbf{h}_{3}\mbf{h}_{4}\mbf{h}_{5}\mbf{h}_{6}$\red{$\mbf{h}_{8}\mbf{\underline{h}}_{7}$} & \text{none} \Tstrut  \\
& 2: $\mbf{h}_{1}\mbf{h}_{2}\mbf{h}_{3}\mbf{h}_{4}\mbf{h}_{5}\mbf{h}_{6}$\red{$\mbf{h}_{7}\mbf{\underline{h}}_{8}$} & $1296\ \RML\!+\!816\ \RAD$ \Tstrut \\
& 3: $\mbf{h}_{1}\mbf{h}_{2}\mbf{h}_{3}\mbf{h}_{4}$\blue{$\mbf{h}_{7}\mbf{h}_{8}$}\red{$\mbf{h}_{6}\mbf{\underline{h}}_{5}$} & \text{none} \Tstrut \\
& 4: $\mbf{h}_{1}\mbf{h}_{2}\mbf{h}_{3}\mbf{h}_{4}$\blue{$\mbf{h}_{7}\mbf{h}_{8}$}\red{$\mbf{h}_{6}\mbf{\underline{h}}_{6}$} & $1296\ \RML\!+\!816\ \RAD$ \Tstrut \\
& 5: $\mbf{h}_{1}\mbf{h}_{2}$\blue{$\mbf{h}_{7}\mbf{h}_{8}$}$\mbf{h}_{5}\mbf{h}_{6}$\red{$\mbf{h}_{4}\mbf{\underline{h}}_{3}$} & \text{none} \Tstrut  \\
& 6: $\mbf{h}_{1}\mbf{h}_{2}$\blue{$\mbf{h}_{7}\mbf{h}_{8}$}$\mbf{h}_{5}\mbf{h}_{6}$\red{$\mbf{h}_{3}\mbf{\underline{h}}_{4}$} & $1296\ \RML\!+\!816\ \RAD$ \Tstrut \\
& 7: \blue{$\mbf{h}_{7}\mbf{h}_{8}$}$\mbf{h}_{3}\mbf{h}_{4}\mbf{h}_{5}\mbf{h}_{6}$\red{$\mbf{h}_{2}\mbf{\underline{h}}_{1}$} & \text{none} \Tstrut \\
& 8: \blue{$\mbf{h}_{7}\mbf{h}_{8}$}$\mbf{h}_{3}\mbf{h}_{4}\mbf{h}_{5}\mbf{h}_{6}$\red{$\mbf{h}_{1}\mbf{\underline{h}}_{2}$} & $1296\ \RML\!+\!816\ \RAD$ \Tstrut \\

\hline 
\end{tabular}
\end{table}

We analyze the complexity in terms of floating-point operations (FLOPs) based on real multiplication (RML) and addition (RAD). Real division and square-root operations are equivalent to a RML. Also, complex multiplication requires $4$ RMLs and $2$ RADs, while complex addition requires $2$ RADs.

Table \ref{table:ch4_complexity1} summarizes the redundant QRD computations that can be saved in the efficient implementations, depending on the permutations and their order, for $4\times4$ and $8\times8$ MIMO systems (the setup of permutations is not unique). The complete QRD in $4\times4$ MIMO requires a total of $304$ RML and $176$ RAD, and the savings are $88$ RML and $40$ RAD. The complete QRD in $8\times8$ MIMO requires a total of $2240$ RML and $1472$ RAD, and the savings in the PR-QRD reach $1296$ RML and $816$ RAD. This means that the overhead is reduced by around $30\%$ with a 2-stage PWSD. The impact of the proposed approaches is more pronounced in higher order systems, $32\times32$ MIMO for example, but worse with lower order systems such as $4\!\times\!4$ MIMO, where the rightmost two columns constitute the majority of required computations. When the PR-WRD does not include matrix puncturing, CYSD, SPSD, and PWSD reduce to cyclic LORD (CYLD), single-permutation LORD (SPLD), and pairwise LORD (PWLD), respectively. The savings are more visible with LORD detectors where preprocessing is solely constituted of QRDs.

\begin{figure}[!t]
\centering
\includegraphics[width=5in]{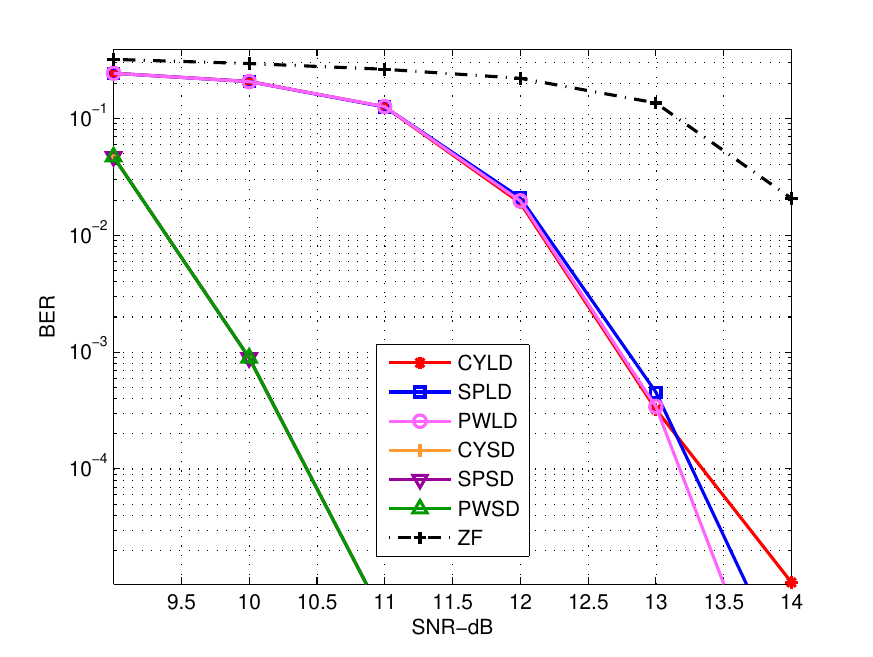}\vspace{-0.1in}
\caption{BER performance - $8\times8$ MIMO - 16-QAM - uncorrelated channels.}
\label{f:ch4_ber8x8}
\end{figure}

Figure \ref{f:ch4_ber8x8} shows the BER performance of the proposed MIMO approaches, compared to that of CYSD/CYLD, and the linear ZF detector, for $8\!\times\!8$ MIMO with 16-QAM. The PWSD and SPSD curves coincided with the CYSD curve, and so did PWLD and SPLD with CYLD. This means that the savings came at no performance degradation cost. \\

Note that in these simulations only per-layer LLRs where computed, hence the SSD and LORD schemes were symbol-based, which explains why the SSD schemes performed better (more on that in Sec \ref{sec:ch5_propSSsb}).

%
\section{Application to Sphere Decoder}
\label{sec:ch4_app_SD}

The number of tree nodes that get visited in a SD is highly nondeterministic, and depends on several factors such as the SNR and the degree of orthogonality of $\mbf{H}$. In particular, the order of the columns of $\mbf{H}$ can be adjusted to reduce the tree search complexity without compromising performance. Adjusting the detection order of the spatial streams according to the channel realization is achieved by performing QRD on a permuted channel matrix $\mbf{H}\bar{\mbf{P}}$, rather than $\mbf{H}$, where $\bar{\mbf{P}}=[\bar{\mbf{p}}_1 \bar{\mbf{p}}_2 \cdots \bar{\mbf{p}}_N]$ is a permutation matrix, and $\bar{\mbf{p}}_i$ is a unit vector having a value of $1$ in the $\nth{i}$ position. The modified system model is thus represented as

\begin{equation}\label{eq:ch4_sysmodel2}
  \mbf{y} = \mbf{H}\mbf{z} + \mbf{n} = (\mbf{H}\bar{\mbf{P}})(\bar{\mbf{P}}^{-1}\mbf{z}) + \mbf{n} = \mbf{Q}\mbf{R}\mbf{x} + \mbf{n}
\end{equation}
\begin{equation}\label{eq:ch4_sysmodel3}
  \mbf{Q}^\mathcal{H}\mbf{y}=\tilde{\mbf{y}}=\mbf{R}\mbf{x} + \mbf{Q}^\mathcal{H}\mbf{n}.
\end{equation}

Studies \cite{4444760Studer,964288Wubben,waters2008chase,5370660Dai} show that more efficient pruning of the search tree is obtained when streams with higher effective SNR are mapped to tree levels closer to the root, which translates into the main diagonal entries of $\mbf{R}$ in $\mbf{H}\bar{\mbf{P}}\!=\!\mbf{QR}$ being sorted in an ascending order. While solving the precise solution to this problem has a prohibitive complexity, SQRD\cite{964288Wubben,5370660Dai} is a popular heuristic algorithm, that achieves a good complexity/performance trade-off. The SQRD is an extension of the MGS orthogonalization algorithm \cite{Golub} for QRD computation, that orders the columns of $\mbf{H}$ in each orthogonalization step. Although this scheme is effective for HO detection at high SNR, its performance degrades when applied to SO detection at low SNR. Nevertheless, other schemes that are more effective at low SNR are substantially more complex, such as the one in \cite{1494671Su}, which is based on orthogonal projections.

\subsection{Improved Layer Ordering Using MRQRD}
\label{sec:ch4_MRQRD}

A more effective layer ordering scheme was proposed in \cite{2014_sphereP1_mansour}, in which layers are ordered such that the corresponding Babai solution has MR among all possible orderings. The resulting ordered QRD is thus called minimum cumulative residual QRD (MRQRD). Starting with the LS solution of the unconstrained system \cite{Golub}
\begin{align}\label{eq:ch4_LS}
\mbf{z}^{\text{LS}} &= \argmin _{\mbf{z}\in\mathcal{C}^{N}} \norm{\mbf{y} - \mbf{H}\mbf{z}}^{2}  \\
& = \bar{\mbf{P}}\cdot\argmin _{\mbf{x}\in\mathcal{C}^{N}} \norm{\tilde{\mbf{y}} - \mbf{R}\mbf{x}}^{2} = \bar{\mbf{P}}\cdot\mbf{x}^{\text{LS}},
\end{align}
and assuming that $\mbf{H}$ has a full column rank, the LS solution is found to be unique, with a residual defined as

\begin{equation}\label{eq:ch4_Res}
\text{Res}^{\text{LS}} = \norm{\mbf{y} - \mbf{H}\mbf{z}^{\text{LS}}}^{2} = \norm{\tilde{\mbf{y}} - \mbf{R}\mbf{x}^{\text{LS}}}^{2},
\end{equation}
that is minimal and independent of the column order. Moreover, the smaller the residual is, the better we can predict $\mbf{y}$ with the columns of $\mbf{H}$ \cite{Golub}.

However, for a given subset of $\mbf{H}$, $\{\mbf{h}_{p_1},\cdots,\mbf{h}_{p_k}\}$ ($p_i$, composed of $k$ permuted columns, is the location of 1 in $\bar{\mbf{p}}_i$), the partial LS solution has a corresponding residual that is not unique, which is expressed as
\begin{align}\label{eq:ch4_parRes}
\text{Res}^{\text{LS}}[\mbf{h}_{p_1},\cdots,\mbf{h}_{p_k}] &= \min _{\mbf{z}\in\mathcal{C}^{k}} \norm{\mbf{y} - [\mbf{h}_{p_1},\cdots,\mbf{h}_{p_k}]\mbf{z}}^{2} \\
& \neq \text{Res}^{\text{LS}}[\mbf{h}_{1},\cdots,\mbf{h}_{k}].
\end{align}
Casting this in the context of the tree-search scheme, the Babai solution and its residual both depend on the permutation $\bar{\mbf{P}}$. We choose $\bar{\mbf{P}}$, from all possible permutations $\Pi$, such that the cumulative residual of the corresponding partial Babai solutions, when derived from layer $N$ back to layer $1$, is minimal:
\begin{equation}\label{eq:ch4_cumRes}
\text{CRes}^{\text{Bab}} [\mbf{h}_{p_1},\cdots,\mbf{h}_{p_N}] = \min_{\text{all}\  \Pi} \sum_{k=1}^{N} \text{Res}^{\text{Bab}} [\mbf{h}_{p_k},\cdots,\mbf{h}_{p_N}].
\end{equation}
The Babai solution and its residual are defined as
\begin{equation}\label{eq:ch4_BabaiSol}
x_k^{\text{Bab}} = \argmin_{x\in \mathcal{M}} \abs{\tilde{y}_k - \sum_{j=k+1}^{N} r_{k,j}x_j - r_{k,k}x}^{2}
\end{equation}
\begin{equation}\label{eq:ch4_BabaiRes}
\text{Res}^{\text{Bab}} [\mbf{r}_{p_k},\cdots,\mbf{r}_{p_N}] = \norm{[\mbf{r}_{p_k},\cdots,\mbf{r}_{p_N}]\mbf{x}_k^{\text{Bab}}-\tilde{\mbf{y}}}^{2},
\end{equation}
for $k\!=\!N,N\!-\!1,\cdots,1$, where $\mbf{x}_k^{\text{Bab}}\!=\![x_k^{\text{Bab}},\cdots,x_N^{\text{Bab}}]^T$.

Reordering according to the MR criterion of~\eqref{eq:ch4_cumRes} is a pre-detection stage that is capable of reducing the node count. The price to pay is a moderate increase in the number of computations and memory locations to determine the MR. We propose optimized architectures based on PR-QRD to decrease this computational overhead.

\subsection{MRQRD Dataflow Architecture}
\label{sec:ch4_proposed}

\begin{figure}[!t]
\centering
\includegraphics[width=4.2in]{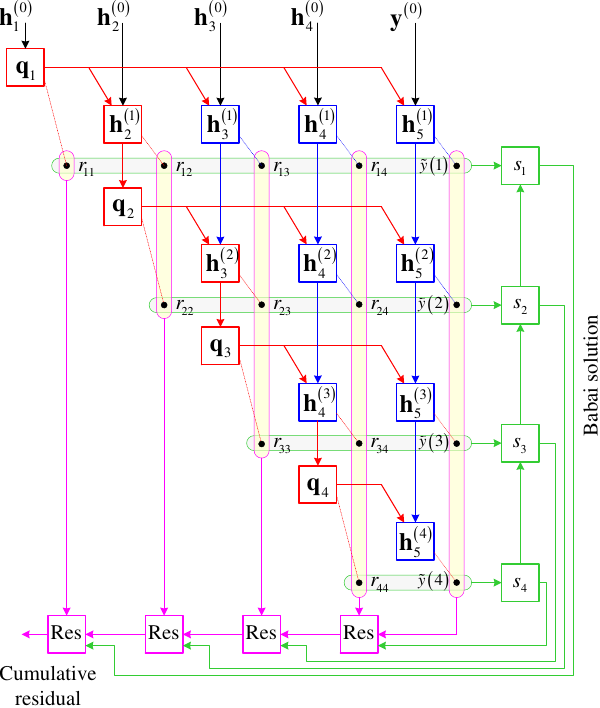}\vspace{-0.1in}
\caption{Optimized dataflow graph for $4\!\times\!4$ MRQRD.}
\label{f:ch4_architecture2}
\end{figure}

For a relatively small number of layers, the desired permutation can be efficiently determined. In what follows, we consider a $4\!\times\!4$ MIMO system. An efficient dataflow architecture that simultaneously performs QRD and finds the Babai solution and its residual is shown in Fig.~\ref{f:ch4_architecture2}. First, the elements of $\mbf{R}$ are derived row-wise from top to bottom. Then, the Babai solution and the residuals are computed simultaneously from bottom to top and right to left, respectively. In order to compute the residuals for all $4!\!=\!24$ possible permutations and identify the minimum, the block should repeat the computations according to a specific schedule. In what follows, we propose two efficient schedules.

\begin{figure}[!t]
\centering
\includegraphics[width=3.8in]{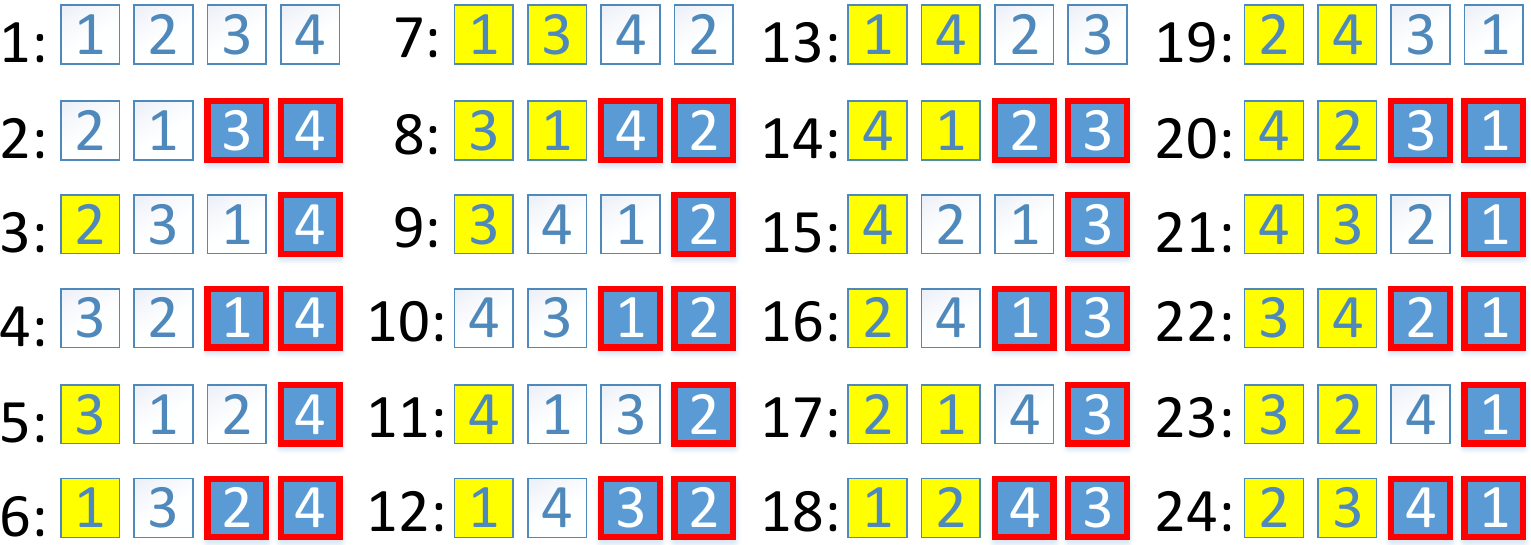}\vspace{-0.1in}
\caption{Permutation schedule 1.}
\label{f:ch4_flow1}
\end{figure}

\begin{figure}[!t]
\centering
\includegraphics[width=5.6in]{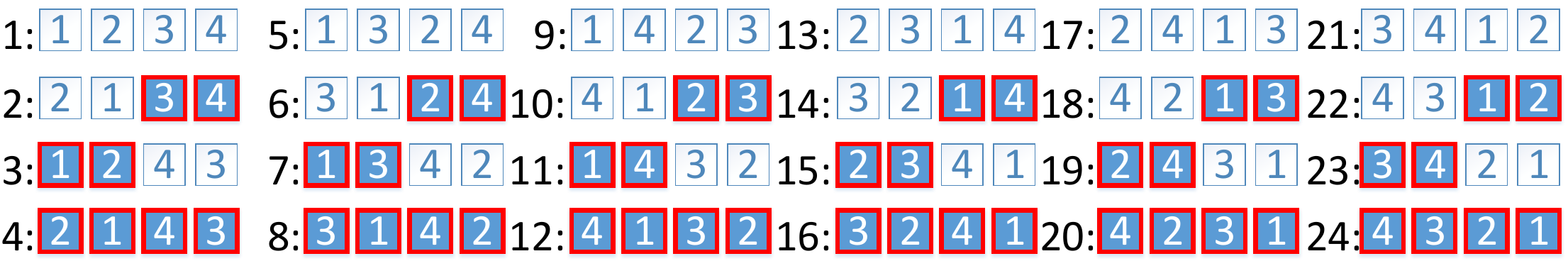}\vspace{-0.1in}
\caption{Permutation schedule 2.}
\label{f:ch4_flow2}
\end{figure}

The first schedule is shown in Fig.~\ref{f:ch4_flow1}. Note that the square numbers correspond to the indices of the channel matrix columns, where the highlighted indices correspond to locations of QRD redundant computations. In a straightforward implementation that does not require additional memory, and that only considers savings when the leftmost columns get permuted, only blue-highlighted positions are saved. This is an intuitive design, since it saves computations in finding the Babai solution as well as QRD. For example, if the first two layers are swapped, the block only recomputes the first two rows of $\mbf{R}$, and then finds the remaining two Babai components and computes the residuals.

Assuming a more advanced circuitry, that allows the storage of several decomposition outputs in memory and supports savings when column permutations take place at either side of $\mbf{H}$, enhanced schedules can be designed. This is feasible since the MRQRD computations are parallelizable, and are not on the critical path. In an extreme case where all $24$ decompositions are stored in memory, additional computational savings occur at yellow locations in Fig.~\ref{f:ch4_flow1}. The second proposed schedule allows a good tradeoff between space and computational complexity, as shown in Fig.~\ref{f:ch4_flow2}. Here, the hardware implementation is assumed to store the outputs of only four consecutive decomposition stages in memory.

\subsection{TSA Architecture}
\label{sec:ch4_implementation}

\begin{figure*}[!t]
\centering
\includegraphics[width=5.55in]{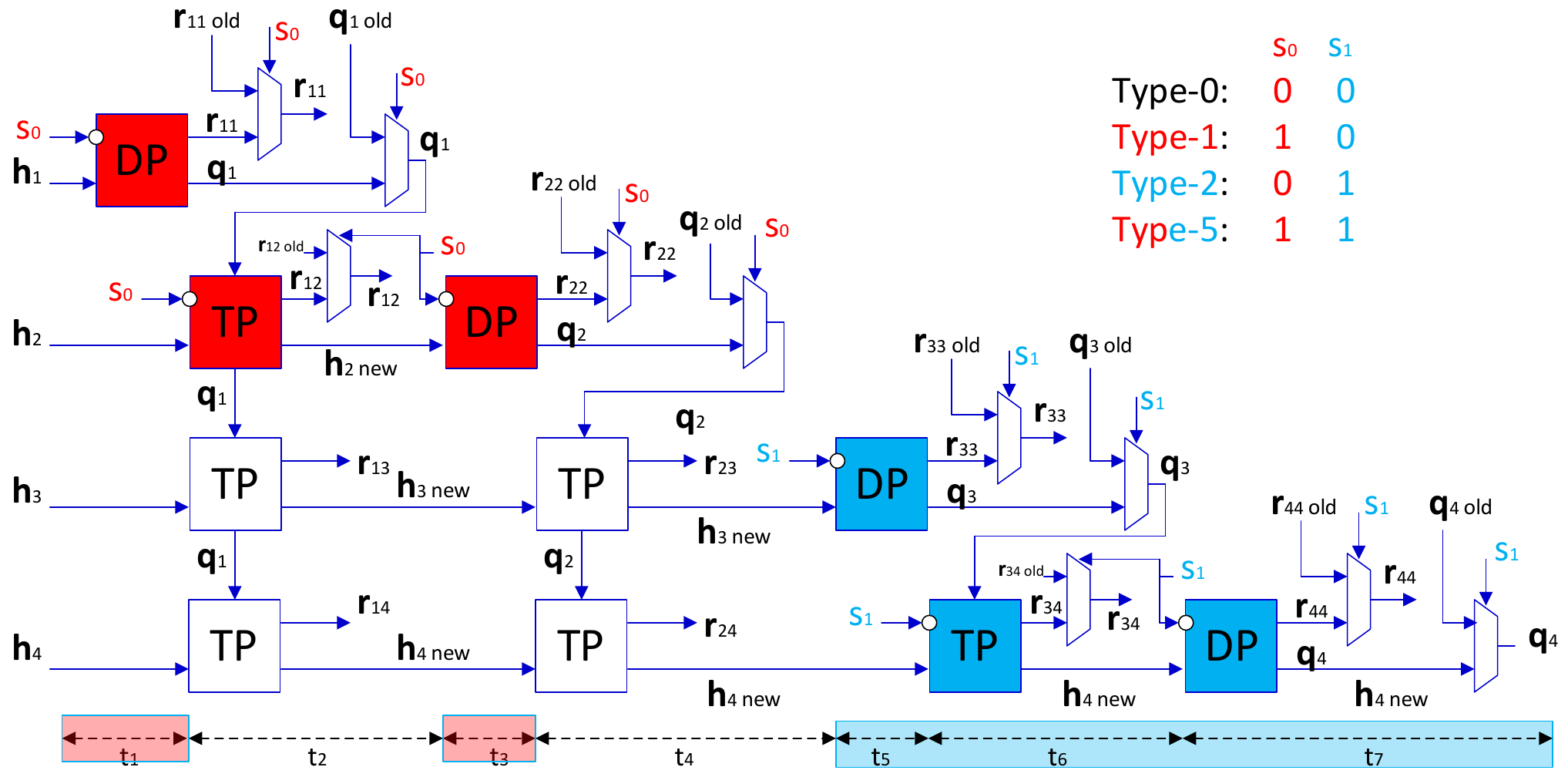}\vspace{-0.1in}
\caption{Modified TSA-QRD for a $4\!\times\!4$ Matrix.}
\label{f:ch4_fig4}
\end{figure*}

A computationally efficient and numerically stable TSA QRD architecture for a $4\!\times\!4$ matrix using MGS was presented in \cite{5463050Chang}. The first stage of QRD is executed by a diagonal-process (DP) unit, which computes the diagonal elements of $\mbf{R}$ and the columns of $\mbf{Q}$. In the second stage, a triangular-process (TP) unit computes the non-diagonal elements of $\mbf{R}$, and updates the remaining columns of $\mbf{H}$. The TSA QRD architecture can thus be formed by repetitive DP and TP operations. Figure \ref{f:ch4_fig4} shows a TSA architecture modified to cope with PR-QRD. \\

Because savings are not always possible, condition signals depending on specific permutations are added, that decide whether a block should execute or not. Note that the colored DPs and TPs are active low on these signals, and that a multiplexer exists at their output, that either selects a newly computed value or a value stored in a buffer. The operation requires seven time slots, with DPs executing on odd time slots and TPs on even ones. For example, in time slot $1$, $\mbf{h}_{1}$ is fed to DP and $r_{11}$ and $\mbf{q}_{1}$ are obtained. The remaining columns of $\mbf{H}$ are delayed in a buffer, waiting for $\mbf{q}_{1}$ to be available at all TPs of time slot $2$ in parallel. The TPs then pass their output to the subsequent stages, and so on.

\begin{table}[!t]
\caption{\textbf{Computational Savings in PR-QRD}}\vspace{0.15in} 
\label{table:ch4_complexity2} 
\centering 
\begin{tabular}{| c || c | c |}
\hline
Permutations & Redundant & Saved \Tstrut \Bstrut \\
\hline\hline
Type-0 & \text{none} & \text{none} \Tstrut \\
Type-1 & $\mbf{q}_{1}$, $\mbf{q}_{2}$, $r_{11}$, $r_{12}$, $r_{22}$ & $88\  \text{RML}\!+\!40\  \text{RAD}$  \Tstrut \\
Type-2 & $\mbf{q}_{3}$, $\mbf{q}_{4}$, $r_{33}$, $r_{34}$, $r_{44}$ & $88\  \text{RML}\!+\!40\  \text{RAD}$ \Tstrut \\
Type-3 & $\mbf{q}_{4}$, $r_{44}$ & $28\ \text{RML}\!+\!8\ \text{RAD}$  \Tstrut \\
Type-4 & $\mbf{q}_{1}$, $r_{11}$ & $28\ \text{RML}\!+\!8\ \text{RAD}$  \Tstrut \\
Type-5 & \text{Type1} $\bigcup$ \text{Type2} & $176\  \text{RML}\!+\!80\  \text{RAD}$ \Tstrut \\
\hline
\end{tabular}
\end{table}

There exist five types of permutations. Some do not make use of any redundant computations, and others make full use of them. We refer to these permutations as Type-0 and Type-5, respectively. Type-1 corresponds to the case when the last two columns are swapped, and Type-2 to the case when the first two columns are swapped instead. Note that Type-1 does not allow for savings in residual computations. Finally, Type-3 and Type-4 correspond to the cases when only the last and first columns remain intact, respectively.

Table \ref{table:ch4_complexity2} summarizes the permutation types, assuming $[\mbf{h}_{1}\mbf{h}_{2}\mbf{h}_{3}\mbf{h}_{4}]$ is initially decomposed, and shows the redundant computations in QRD, as well as the computational savings that can be achieved using a PR-QRD. Note that one complete QRD requires a total of $304$ RMLs and $176$ RADs, and the savings in the PR-QRD reach $176$ RMLs and $80$ RADs. In total, the schedule of Fig.~\ref{f:ch4_flow1} allows for savings in QRD equal to $1280$ RMLs and $544$ RADs, while the schedule of Fig.~\ref{f:ch4_flow2} allows for savings equal to $2112$ RMLs and $960$ RADs. Noting that the total 24 QRDs require $7296$ RMLs and $4224$ RADs, the schedule of Fig.~\ref{f:ch4_flow2} reduces the QRD overhead by more than $25\%$. The savings in residual computations are less significant, since the Babai solution of~\eqref{eq:ch4_BabaiSol} can be found by a simple slicing operation, and~\eqref{eq:ch4_BabaiRes} needs to be recomputed.

Moreover, in addition to computational savings, time savings are also achievable. Figure \ref{f:ch4_fig4} highlights the time slots that can be saved with different permutation types. Two from the seven time slots are saved with Type-1 permutations, three with Type-2, and five with Type-4 (Fig.~\ref{f:ch4_fig4} corresponds to schedule 2, and hence Type-3 and Type-4 are not highlighted). Thus, $60$ time slots from the total $168$ time slots that are required by all $24$ permutations are saved in the second proposed schedule, which accounts for a time saving percentage of $36\%$ in QRD computations. \\

\begin{figure}[!t]
\centering
\includegraphics[width=4.8in]{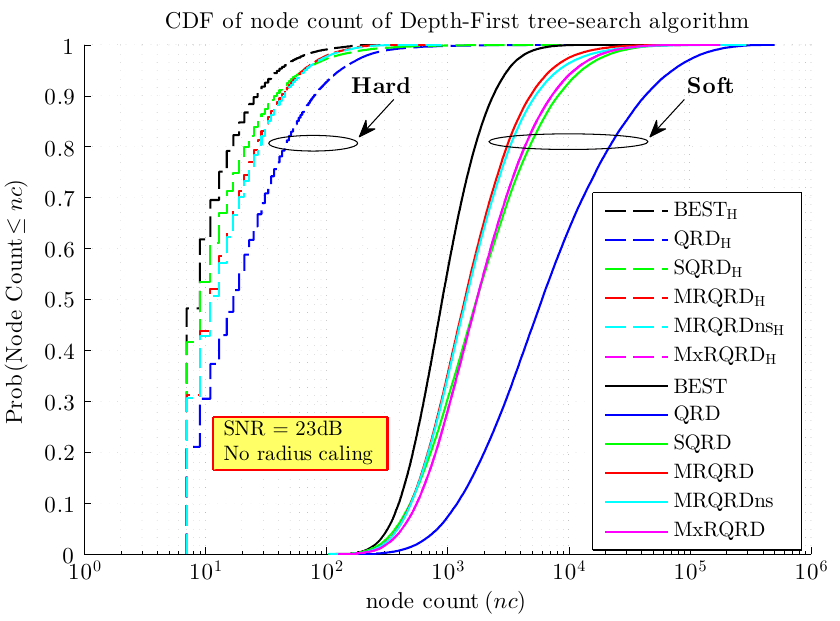}\vspace{-0.1in}
\caption{Node count CDF for various layer ordering}
\label{f:ch4_plot1}
\end{figure}

Figure \ref{f:ch4_plot1} from \cite{2014_sphereP1_mansour} shows the cumulative distribution function (CDF) of the node count for various QRD schemes, with HO and SO detection, when DF tree-search was employed. Six different ordering schemes were studied, including the best ordering among all possible permutations (Best). The MRQRDns is the same as MRQRD, but with no symbol slicing when propagating values in the recursion. MxRQRD, on the other hand, orders the layers based on maximum forward residuals (orders layers in ascending order of residuals). The gap in the median node count between hard and soft ML detection ranged between 2 to 3 orders of magnitude. Moreover, SO ML detection is found to be more sensitive to the ordering scheme than HO detection, where the gap between the best ordering scheme and the case when no ordering is applied is one order of magnitude. The figure illustrates the advantage of the proposed ordering schemes based on MRs in reducing the node count compared to SQRD, both with or without slicing in the recursion. \\~\\~\\~\\~\\

\textbf{Conclusion} \\

Enhanced layer-ordering schemes have been proposed for SO SSDs and SDs, and low-complexity hardware architectures have been proposed, with corresponding execution schedules. The implementations employed a PR-QRD, based on the MGS orthogonalization algorithm. It has been shown through simulations that using the proposed scheme with SSD, the QRD overhead can be reduced by $30\%$ for $8\!\times\!8$ MIMO with no performance degradation cost, and it has been argued that the savings are more profound with higher order MIMO. With SD, the QRD overhead has been reduced by $25\%$ in computations and $36\%$ in time, when reducing the node count by one order of magnitude, for $4\!\times\!4$ MIMO.

%% file: Chapter5_PUNDET.tex

\chapter{Large MIMO Detection via Channel Puncturing }\label{chapter:ch5_pundet}

In this chapter we present a family of WRD-based detectors that build on popular QRD-based detectors (Table~\ref{table:ch5_summary}). In particular, we propose a punctured ML (PML) detector, a punctured N/C (PN/C) detector, a punctured CD (PCD), as well as a HO SSD. We then propose efficient architectures and analyze the computational complexity of the proposed detectors. We show that the computational savings are much more pronounced with large MIMO dimensions. Finally, we study the performance of the proposed detectors in the context of large MIMO with high order MTs, and in the presence of spatial channel correlation. We show that the performance of these schemes scales up efficiently with high orders, and that they are superior to their QRD-based counterparts in the presence of channel correlation. The results of this section appeared, in parts, in \cite{8292405Sarieddeen} and \cite{8186206Sarieddeen}.

\begin{table}[!t]
\centering
\caption{\textbf{Summary of Proposed and Studied Detectors}}\vspace{0.15in} 
\label{table:ch5_summary} 
\centering 
\begin{tabular}{| c || c |} 
\hline
WRD-Based Detectors & QRD-Based Detectors \\
\hline\hline
ML & PML = PCD  \\\hline %
N/C & PN/C  \\\hline %
CD & PCD \\\hline
LORD & SSD  \\\hline
SLORD & SSSD \\
\hline 
\end{tabular}
\end{table}

%
\section{Punctured QR Decomposition (WRD)}
\label{sec:ch5_WRD}

A brute force approach for computing $\mbf{W}$ \cite{ojard2008method} (Sec. \ref{sec:ch2_WRD}) involves matrix inversions, which is complex and prone to roundoff error. However, an alternative approach exists that employs QRD followed by elementary matrix operations~\cite{2014_mansour_eurasip_WLD}.

Let $\mbf{H}$ be QR-decomposed such that $\mbf{Q}^\mathcal{H}\mbf{H}=\mbf{R}$. Obviously, $\mbf{q}_{N}^\mathcal{H}\mbf{q}_{N}^{}=1$ and $\mbf{q}^\mathcal{H}_{N}\mbf{h}_{n}=0$ for all $n=1,\ldots,N\!-\!1$, hence, $\mbf{w}_{N}^{}=\mbf{q}_{N}^{}$. Now assume the $\nth{n}$ entry $r_{mn}$ in row $m$ of $\mbf{R}$ is to be nulled, for $m\!=\!1,\cdots,N\!-\!2$ and $n\!=\!m\!+\!1,\cdots,N\!-\!1$. We have $\mbf{q}_{m}^\mathcal{H}\mbf{h}_{n}=r_{mn}\in\mathcal{C}$ and $\mbf{q}_{n}^\mathcal{H}\mbf{h}_{n}=r_{nn}\in\mathcal{R}^{+}$, from which it follows that $\left(\mbf{q}^\mathcal{H}_{m}-\mbf{q}^\mathcal{H}_{n}\frac{r_{mn}}{r_{nn}}\right)\mbf{h}_{n}=0$. Hence, with $\rho_{mn}^{}\!\triangleq\!\tfrac{r_{mn}}{r_{nn}} \in\mathcal{C}$, the equations
\begin{align}\label{eq:ch5_punc1}
    \mbf{q}_{m}^{} &= \mbf{q}_{m}^{}  -   \mbf{q}_{n}^{} \rho_{mn}^\mathcal{H}, \\
    r_{mn}^{}      &= r_{mn}^{}       -   r_{nn}^{} \rho_{mn}^{},~~\text{and}     \\
    r_{mN}^{}      &= r_{mN}^{}       -   r_{nN}^{} \rho_{mn}^{},
\end{align}
when repeated for $n=N-1,N-2,\cdots,m+1$, would puncture the required $\nth{n}$ entry and update the $\nth{N}$ entry in row $m$ of $\mbf{R}$, as well as update the $\nth{m}$ column of $\mbf{Q}$ accordingly, while
\begin{align}\label{eq:ch5_punc2}
  r_{mm}^{} &= r_{mm}/\norm{\mbf{q}_{m}}, \\
  r_{mN}^{} &= r_{mN}/\norm{\mbf{q}_{m}},~~\text{and} \\
  \mbf{q}_{m} &= \mbf{q}_{m}/\norm{\mbf{q}_{m}},
\end{align}
would normalize $\mbf{q}_{m}$ in $\mbf{Q}$ and update the non-zero entries in row $m$ of $\mbf{R}$ accordingly. All these operations are to be carried for $m\!=\!N\!-\!2,N\!-\!3,\cdots,1$. The resultant $\mbf{Q}$ is $\mbf{W}$, and the resultant $\mbf{R}$ is $\Rp$.

In matrix form, we can write \eqref{eq:ch5_punc1}-\eqref{eq:ch5_punc2} using elementary matrices $\mbf{E}_{m} = [e_{nj}], 1\leq m\leq N$, which differ from $\mbf{I}_{N}$ by a single elementary row operation, defined as follows:
\begin{equation}\label{eq:ch5_elementary}
e_{nj}=\left\{
                \begin{array}{ll}
                  1 \ \ \text{if} \  j=n\\
                  -r_{nm}/r_{mm} \ \ \text{if} \  j\!=\!m, \ j\in\mathcal{I}_{n}\\
                  0 \ \ \text{otherwise}
                \end{array}
              \right.
\end{equation}
The product of these matrices forms the unscaled matrices $\mbf{R}_{2}\!=\!(\mbf{E}_{n}\ldots\mbf{E}_{1})\mbf{R}$ and $\mbf{Q}_{2}^\mathcal{H}\!=\!(\mbf{E}^\mathcal{H}_{n}\ldots\mbf{E}^\mathcal{H}_{1})\mbf{Q}^\mathcal{H}$. The scaling operations can be written using the diagonal matrix $\bar{\mbf{D}}\!=\![d_{n}] \in \mathcal{R}^{+}$, where $d_{n}\!=\!1/\sqrt{[\mbf{Q}_{1}^\mathcal{H}\mbf{Q}_{2}]_{nn}}$ and $[\cdot]_{nn}$ denotes the \nth{n} diagonal element. The desired (scaled) matrices are given by $\mbf{W}^\mathcal{H}=\bar{\mbf{D}}\mbf{Q}_{2}^\mathcal{H}$ and $\Rp=\bar{\mbf{D}}\mbf{R}_{2}$. Note That unlike QRD, there is no permutation-robust implementation for puncturing. The punctured elements are in the upper rows, affecting the leftmost columns of $\mbf{Q}$. Furthermore, if the system had more receive antennas ($M\!>\!N$), the ``thin'' form of the QRD for tall matrices would have been used, and other modifications would have immediately followed.

The transformed received symbol vector after applying $\mbf{W}^\mathcal{H}$ can then be expressed as
\begin{equation}\label{eq:ch5_sysmodel3}
  \bar{\mbf{y}} = \mbf{W}^\mathcal{H}\mbf{y} = \Rp\mbf{x} + \mbf{W}^\mathcal{H}\mbf{n},
\end{equation}
where by analogy with \ref{eq:ch2_partitionCD} we have
\begin{equation}\label{eq:ch5_partitionCDwrd}
\bar{\mbf{y}} =
    \begin{bmatrix}
        \bar{\mbf{y}}_{1} \\
        \bar{y}_{N}^{}
    \end{bmatrix}, \ \
\Rp =
    \begin{bmatrix}
        \mathring{\mbf{A}} & \mathring{\mbf{b}} \\
        \mbf{0} & \mathring{c}
        \end{bmatrix}, \ \  \mbf{x}=\begin{bmatrix} \mbf{x}_{1} \\ x_{N}^{}  \end{bmatrix},
\end{equation}
and in this case $\mathring{\mbf{A}} \in \mathcal{R}^{(N-1)\times(N-1)}$ is a diagonal matrix. For example, in the special case of $4\!\times\!4$ MIMO, $\Rp$ is obtained from $\mbf{R}$ by puncturing entries $r_{23}^{}$, $r_{12}^{}$, and $r_{13}^{}$, respectively:
\begin{equation}\label{eq:ch5_SICmatrices}
  \mbf{R} =
    \begin{bsmallmatrix}
        r_{11}  & r_{12}     & r_{13}     & r_{14} \\
        0       & r_{22}     & r_{23}     & r_{24} \\
        0       & 0          & r_{33}     & r_{34} \\
        0       & 0          & 0          & r_{44}
    \end{bsmallmatrix},\quad
  \Rp =
    \begin{bsmallmatrix}
        \rp_{11} & 0 & 0 & \rp_{14} \\
        0 & \rp_{22} & 0 & \rp_{24} \\
        0 & 0 & \rp_{33} & \rp_{34} \\
        0 & 0 & 0 & \rp_{44}
    \end{bsmallmatrix}.
\end{equation}

%
\section{Punctured ML Detector (PML)}
\label{sec:ch5_propML}

Similar to the ML detector, an ``exhaustive'' PML detector searches $\mathcal{X}$ to find
\begin{equation}\label{eq:ch5_PML_dist}
   \hat{\mbf{x}}^{\PML} = \min_{\mbf{x} \in \mathcal{X}}\norm{\mbf{W}^\mathcal{H}(\mbf{y} - \mbf{H}\mbf{x})}^{2}.
\end{equation}
Pre-multiplying by $\mbf{W}$, unlike $\mbf{Q}$, modifies Euclidean distances, hence we have
\begin{align}\label{eq:ch5_mod_dist}
d(\mbf{x}) = \norm{\mbf{y} - \mbf{H}\mbf{x}}^{2} &= \norm{\mbf{Q}^\mathcal{H}(\mbf{y} - \mbf{H}\mbf{x})}^{2} \\ &\neq \norm{\mbf{W}^\mathcal{H}(\mbf{y} - \mbf{H}\mbf{x})}^{2} = \norm{\bar{\mbf{y}}\!-\!\Rp\mbf{x}}^{2} = \bar{d}(\mbf{x}).
\end{align}
Due to colored noise, this minimum distance detector is not optimal.

%
\section{Punctured N/C Detector (PN/C)}
\label{sec:ch5_propSIC}

With PN/C, we null by pre-multiplying by $\mbf{W}^\mathcal{H}$ instead of $\mbf{Q}^\mathcal{H}$, and perform SIC as
\begin{equation}\label{eq:ch5_SICpunctured}
    \hat{x}^{\PNC}_{n} = \left\lfloor \left(\bar{y}_{n} - \rp_{nN}^{}\hat{x}_{N}^{\PNC}\right)/\rp_{nn} \right\rceil_{\mathcal{M}},
\end{equation}
for $n\!=\!N\!-\!1,\cdots, 1$, where $\hat{\mbf{x}}^{\PNC}\!=\![ \hat{x}^{\PNC}_1\!\cdots\!\hat{x}^{\PNC}_n\cdots\hat{x}^{\PNC}_N ]$, and $\hat{x}^{\PNC}_{N} \!=\! \left\lfloor \bar{y}_{N}/\rp_{NN}\right\rceil_{\mathcal{M}}$. Note that slicing on layers $n \!=\! N\!-\!1,\cdots, 1$ can be done in parallel since $\mathring{\mbf{A}}$ is diagonal.

%
\section{Punctured Chase Detector (PCD)}
\label{sec:ch5_propCD}

The PCD builds on the partition in~\eqref{eq:ch5_partitionCDwrd}, and performs the operations of a CD (Sec. \ref{sec:ch2_CD}). A modified list of candidate symbol vectors $\mathcal{P}(\bar{\mbf{y}},\Rp)$ is thus created. The distance of a vector $\mbf{x}=[\mbf{x}_1,x_N]^T$ is given by
\begin{equation}\label{eq:ch5_xwr}
    \bar{d}(\mbf{x})    \!=\!
        \norm{\bar{\mbf{y}}\!-\!\Rp\mbf{x}}^{2} \!=\! \abs{\bar{y}_{N}^{}\!-\!\mathring{c} x_{N}^{}}^{2}\!+\!\norm{\bar{\mbf{y}}_{1}\!-\!\mathring{\mbf{A}}\mbf{{x}}_{1}\!-\!\mathring{\mbf{b}}x_{N}^{}}^{2}.
\end{equation}
For a given $x_N^{}\in\mathcal{M}$, the distance in~\eqref{eq:ch5_xwr} is minimized as
\begin{align}\label{eq:ch5_xwr2}
    \min_{\mbf{x}_1\in \mathcal{M}^{N-1}} \bar{d}(\mbf{x})
        &= \abs{\bar{y}_{N}^{}\!-\!\mathring{c} x_{N}^{}}^{2}\!+\!
              \min_{\mbf{x}_1\in \mathcal{M}^{N-1}} \norm{\bar{\mbf{y}}_{1}\!-\!\mathring{\mbf{A}}\mbf{{x}}_{1}\!-\!\mathring{\mbf{b}}x_{N}^{}}^{2}\\
        &= \abs{\bar{y}_{N}^{}\!-\!\mathring{c} x_{N}^{}}^{2} +
              \norm{\bar{\mbf{y}}_{1}\!-\!\mathring{\mbf{A}}\hat{\mbf{x}}_{1}(x_N)\!-\!\mathring{\mbf{b}}x_{N}^{}}^{2}\\
        &\triangleq \bar{d}^\mathcal{H}\left(\mbf{x}(x_N)\right),
\end{align}
where $\hat{\mbf{x}}_{1}(x_N)=\lfloor(\bar{\mbf{y}}_{1}\!-\!\mathring{\mbf{b}} x_{N}^{})/\mathring{\mbf{A}}\rceil_{\mathcal{M}^{N-1}}$, which is a vectorized slicing operation, and $\mbf{x}(x_N)=[\hat{\mbf{x}}_1(x_N),x_N]^T$. The symbol vector $\mbf{x}(x_N)$ is then added to $\mathcal{P}$, together with its distance $\bar{d}^\mathcal{H}\left(\mbf{x}(x_N)\right)$. The final HO symbol vector $\hat{\mbf{x}}^{\PCD}$ is found from $\mathcal{P}$ as the one with smallest distance.

While the PCD computes distances only to $\abs{\mathcal{P}(\tilde{\mbf{y}},\Rp)}=\abs{\mathcal{M}}$ candidate symbol vectors, for a given layer ordering and channel partition, it is clear from~\eqref{eq:ch5_xwr2} that it achieves the exact performance as that of the PML detector. In other words, there is no vector in the lattice $\mathcal{X}$, outside the set $\mathcal{P}(\tilde{\mbf{y}},\mbf{R})$, that can have a smaller distance metric than that of the PCD solution. The proof goes as follows:
\begin{align}\label{eq:ch5_xwr3}
    \min_{\mbf{x}\in \mathcal{X}} \bar{d}(\mbf{x})
        &=  \min_{x_N\in\mathcal{M},\mbf{x}_1\in \mathcal{M}^{N-1}}
            \left\{
            \abs{\bar{y}_{N}^{}\!-\!\mathring{c} x_{N}^{}}^{2}\!+\!
              \norm{\bar{\mbf{y}}_{1}\!-\!\mathring{\mbf{A}}\mbf{{x}}_{1}\!-\!\mathring{\mbf{b}}x_{N}^{}}^{2}
            \right\}\\
        &= \min_{x_N\in\mathcal{M}}
            \left\{
                \abs{\bar{y}_{N}^{}\!-\!\mathring{c} x_{N}^{}}^{2}\!+\!
                \min_{\mbf{x}_1\in \mathcal{M}^{N-1}} \norm{\bar{\mbf{y}}_{1}\!-\!\mathring{\mbf{A}}\mbf{{x}}_{1}\!-\!\mathring{\mbf{b}}x_{N}^{}}^{2}
            \right\}\\
        &= \min_{x_N\in\mathcal{M}}
            \left\{
                \abs{\bar{y}_{N}^{}\!-\!\mathring{c} x_{N}^{}}^{2} +
                \norm{\bar{\mbf{y}}_{1}\!-\!\mathring{\mbf{A}}\hat{\mbf{x}}_{1}(x_N)\!-\!\mathring{\mbf{b}}x_{N}^{}}^{2}
            \right\}   \\
        &= \min_{\mbf{x}(x_N)\in\mathcal{P}} \bar{d}^\mathcal{H}\left(\mbf{x}(x_N)\right).
\end{align}

%
\section{Vector-Based Subspace Detector (VSSD)}
\label{sec:ch5_propSSvb}

The VSSD is an extension to PCD, the same way LORD is an extension to CD. The columns of $\mbf{H}$ are cyclically shifted, and punctured UTMs are generated. Each permuted $\mbf{H}$ at step $t$, $t = 1,\cdots,N$, is WR-decomposed into $\mbf{W}^{(t)}$ and $\Rp^{(t)}$ according to~\eqref{eq:ch5_partitionCDwrd}. Let $\hat{\mbf{x}}_{(t)}^{\PCD}$ denote the PCD solution from step $t$. The final solution $\hat{\mbf{x}}^{\VSSD}$ is $\hat{\mbf{x}}_{(t_{min})}^{\PCD}$, where $t_{min}$ is defined as:
\begin{equation}\label{eq:ch5_SSDout}
t_{\min} = \argmin_{ t \in \{1,\cdots,N\} } \norm{\mbf{y} - \mbf{H}\hat{\mbf{x}}_{(t)}^{\PCD}}^{2}.
\end{equation}
Note that we revert back to the original space of $\mbf{H}$ to compute the true Euclidean distance metrics in~\eqref{eq:ch5_SSDout}. The gain achieved by VSSD compared to PCD is limited, since each $\Rp^{(t)}$ generates an independent space, and hence we end up taking the best output from $N$ independent trials. The VSSD is in effect the HO version of the reference SO SSD \cite{2014_mansour_SPL_WLD} (Sec. \ref{sec:ch2_det}), and we refer to it by simply SSD in the remainder of this chapter.

%
\section{Symbol-Based Subspace Detector (SSSD)}
\label{sec:ch5_propSSsb}

As a variation of SSD, the SSSD selects at each step $t$, only the root symbol of the output vector as a component of the final output vector. Thus, the output vector $\hat{\mbf{x}}^{\SSSD}=[ \hat{x}^{\SSSD}_1\cdots\hat{x}^{\SSSD}_n\cdots\hat{x}^{\SSSD}_N ]$ gets assembled one symbol at a time over $N$ executions of PCD, where
\begin{equation}\label{eq:ch5_SSSDout}
\hat{x}^{\SSSD}_n = \hat{x}_{N-n+1(t=n)}^{\PCD}.
\end{equation}
For example, in a $4\!\times\!4$ MIMO system, we have $\hat{x}^{\SSSD}_1 = \hat{x}_{4(t=1)}^{\PCD}$, where $\hat{\mbf{x}}_{(t=1)}^{\PCD}$ is the HO solution of a PCD following the partition in Fig.~\ref{f:ch2_matrices}(g). Similarly $\hat{x}^{\SSSD}_2 = \hat{x}_{3(t=2)}^{\PCD}$, $\hat{x}^{\SSSD}_3 = \hat{x}_{2(t=3)}^{\PCD}$, and $\hat{x}^{\SSSD}_4 = \hat{x}_{2(t=4)}^{\PCD}$, are obtained following the partitions (h), (i), and (j), respectively. Note that we can define symbol-based LORD (SLORD) in a similar manner:
\begin{align}\label{eq:ch5_SLORD}
\hat{\mbf{x}}^{\SLORD}&=[\hat{x}^{\SLORD}_1\! \cdots\! \hat{x}^{\SLORD}_n\! \cdots\hat{x}^{\SLORD}_N] \\ \hat{x}^{\SLORD}_n &= \hat{x}_{N\!-\!n\!+\!1(t=n)}^{\CD}.
\end{align}

%
\section{Soft-Output Detection}
\label{sec:ch5_soft_output}

To generate the LLRs with SSSD, the $N$ streams should be decoupled in $N$ steps, where in each step $t \in \{1,\cdots,N\}$ the LLRs for the bits corresponding to symbol $x_n$ ($n=t$) are calculated. Hence, for each bit, we compute
\begin{equation}\label{eq:ch5_LLR_SSSD}
  \lambda_{n,k,t}^{\SSSD}\!=\!\frac{1}{\sigma^{2}} \left(\min_{\mbf{x} \in \mathcal{P}^{(0)}_{n,k,t}}{\norm{\bar{\mbf{y}}^{(t)}\!-\!\Rp^{(t)}\mbf{x}}^{2}}\!-\!\min_{\mbf{x} \in \mathcal{P}^{(1)}_{n,k,t}}{\norm{\bar{\mbf{y}}^{(t)}\!-\!\Rp^{(t)}\mbf{x}}^{2}} \right)
\end{equation}
for $t\!=\!1,\!\cdots\!,N$ and $k\!=\!1,\!\cdots\!,\log(\abs{\mathcal{M}})$, where the sets $\mathcal{P}^{(0)}_{n,k,t}\!\triangleq\!\{\mbf{x} \in \mathcal{P}(\bar{\mbf{y}^{(t)}},\Rp^{(t)}): b_{n,k}\!=\!0\}$ and $\mathcal{P}^{(1)}_{n,k,t}\!\triangleq\!\{\mbf{x} \in \mathcal{P}(\bar{\mbf{y}^{(t)}},\Rp^{(t)}): b_{n,k}\!=\!1\}$ correspond to subsets of symbol vectors in $\mathcal{P}(\bar{\mbf{y}^{(t)}},\Rp^{(t)})$, having in the corresponding $\nth{k}$ bit of the $\nth{n}$ symbol a value of $0$ and $1$, respectively. Note that these distance metrics can be expanded as in~\eqref{eq:ch5_xwr}. Similarly, we can define the LLRs for SLORD as
\begin{equation}\label{eq:ch5_LLRSLORD}
  \lambda_{n,k,t}^{\SLORD}\!=\!\frac{1}{\sigma^{2}} \left(\min_{\mbf{x} \in \mathcal{S}^{(0)}_{n,k,t}}{\norm{\tilde{\mbf{y}}^{(t)}\!-\!\mbf{R}^{(t)}\mbf{x}}^{2}}\!-\!\min_{\mbf{x} \in \mathcal{S}^{(1)}_{n,k,t}}{\norm{\tilde{\mbf{y}}^{(t)}\!-\!\mbf{R}^{(t)}\mbf{x}}^{2}} \right),
\end{equation}
where $\mathcal{S}^{(0)}_{n,k,t}\!\triangleq\!\{\mbf{x} \in \mathcal{S}(\tilde{\mbf{y}^{(t)}},\mbf{R}^{(t)}): b_{n,k}\!=\!0\}$ and $\mathcal{S}^{(1)}_{n,k,t}\!\triangleq\!\{\mbf{x} \in \mathcal{S}(\tilde{\mbf{y}^{(t)}},\mbf{R}^{(t)}): b_{n,k}\!=\!1\}$.

With SO SSD and LORD, tighter LLRs can be computed with an extra processing overhead by tracking global distances rather than per stream distances:
\begin{equation}\label{eq:ch5_LLRSSD}
  \lambda_{n,k}^{\SSD} = \frac{1}{\sigma^{2}} \left[ \min_t \left( \min_{\mbf{x} \in \mathcal{P}^{(0)}_{n,k,t}}{\norm{\bar{\mbf{y}}^{(t)} - \Rp^{(t)}\mbf{x}}^{2}} \right) - \min_t \left( \min_{\mbf{x} \in \mathcal{P}^{(1)}_{n,k,t}}{\norm{\bar{\mbf{y}}^{(t)} - \Rp^{(t)}\mbf{x}}^{2}} \right) \right]
\end{equation}
\begin{equation}\label{eq:ch5_LLRLORD}
  \lambda_{n,k}^{\LORD} = \frac{1}{\sigma^{2}} \left[ \min_t \left( \min_{\mbf{x} \in \mathcal{S}^{(0)}_{n,k,t}}{\norm{\tilde{\mbf{y}}^{(t)} - \mbf{R}^{(t)}\mbf{x}}^{2}} \right) - \min_t \left( \min_{\mbf{x} \in \mathcal{S}^{(1)}_{n,k,t}}{\norm{\tilde{\mbf{y}}^{(t)} - \mbf{R}^{(t)}\mbf{x}}^{2}} \right) \right].
\end{equation}

\begin{figure}[!t]
\centering
\includegraphics[width=4.6in]{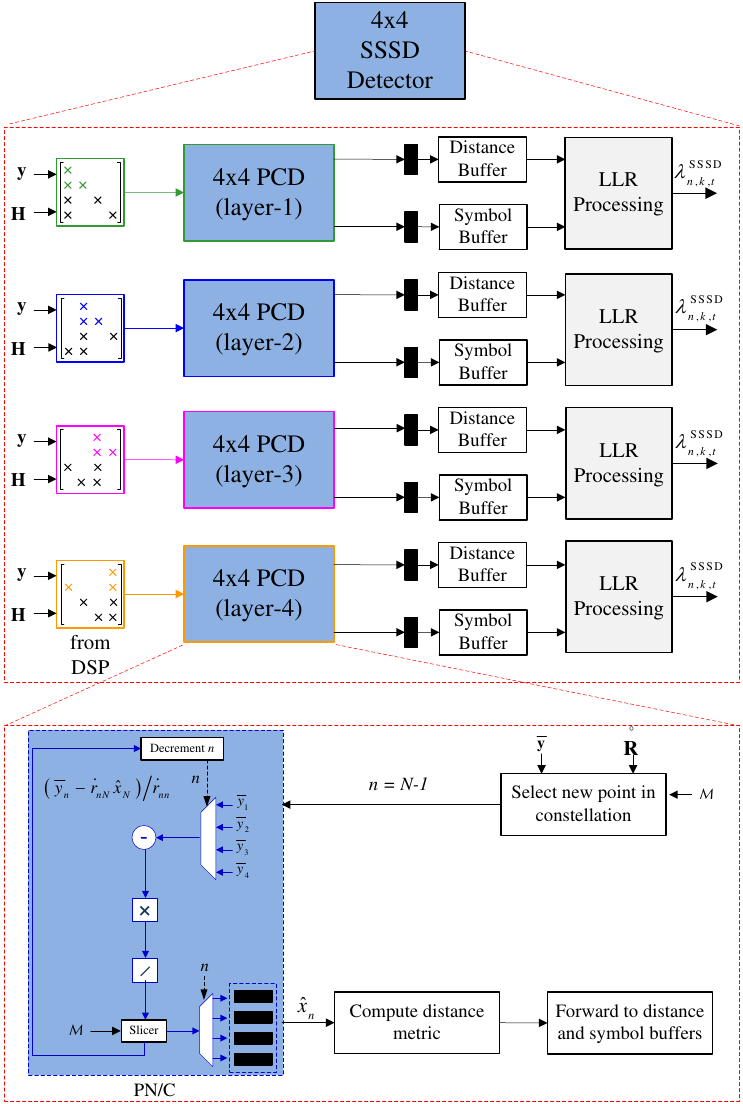}\vspace{-0.1in}
\caption{Hierarchical architectural design for SO SSSD.}
\label{archfig:ch5_a}
\end{figure}
\begin{figure}[!t]
\centering
\includegraphics[width=4.6in]{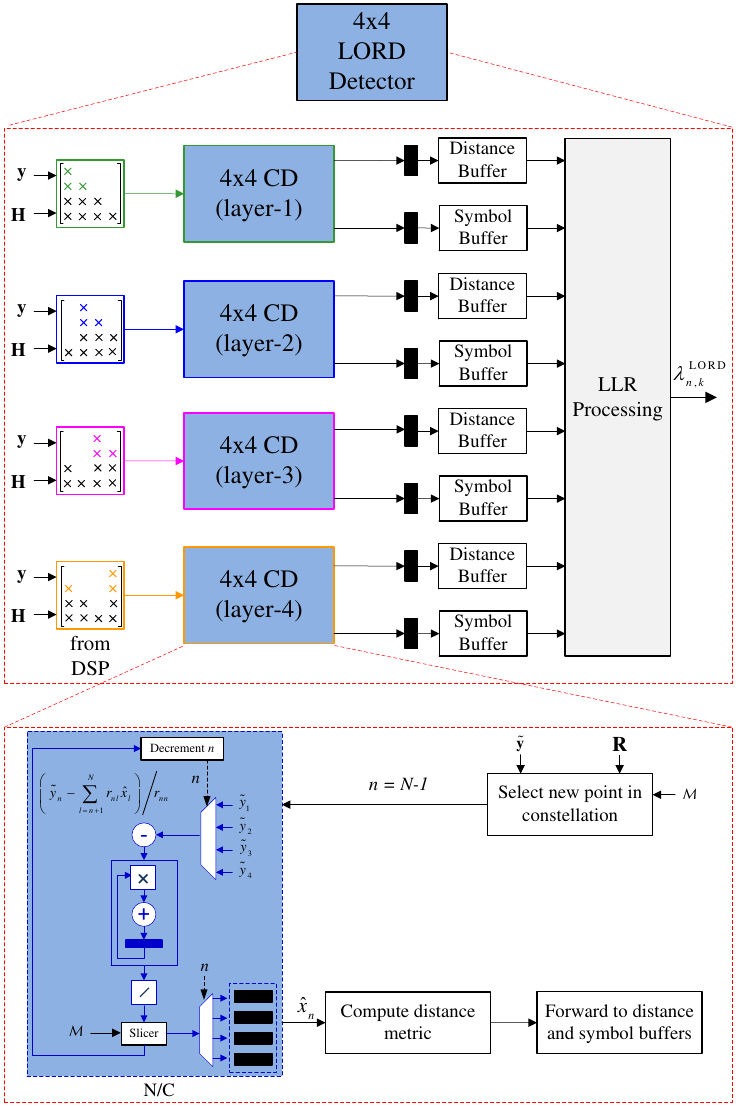}\vspace{-0.1in}
\caption{Hierarchical architectural design for SO LORD.}
\label{archfig:ch5_b}
\end{figure}

%
\section{Architectures and Complexity Analysis}
\label{sec:ch5_complexity}

The main motive behind channel puncturing is reducing complexity. To support this fact, a cost efficient architecture that implements the SSSD algorithm is shown in Fig.~\ref{archfig:ch5_a}, together with a counterpart architecture that implements LORD in Fig.~\ref{archfig:ch5_b}. The designs are hierarchical, showing SSSD using PCD building blocks, that themselves use PN/C, while LORD uses CD and N/C blocks.

With SSSD, the distances computed in PCD and their symbol vectors are directly forwarded to an LLR processing unit at the corresponding layer of interest. The PCD processes on all other layers can run in parallel, and the complete LLR vector will be available at the output after the processing delay of one layer. However, the LORD architecture is not fully parallelizable. Moreover, the PCD routine itself is much less complex than the CD routine because it performs fewer computations due to punctured entries. 

Every time the product $\Rp\mbf{x}$ is computed instead of $\mbf{R}\mbf{x}$, $(N-2)(N-1)/2$ complex multiplications are saved, which amounts to $\theta_1=(N^2-3N+2)\,\RAD+(2N^2-6N+4)\,\RML$ FLOPs. For example, in a $16\!\times\!16$ MIMO system, $77\%$ of the multiplications required in N/C are saved with puncturing, and the savings increase to $94\%$ in a $64\!\times\!64$ MIMO system. Therefore, the SO SSD and SSSD can save $N\times\abs{\mathcal{M}}\times\theta_1$ FLOPs compared to the SO LORD and SLORD, respectively.

\begin{table}[!t]
\centering
\caption{\textbf{Preprocessing Cost of Studied Detectors}}\vspace{0.15in} 
\label{table:ch5_detectors1} 
\centering 
\begin{tabular}{| c || c | c | c |} 
\hline
Detection Scheme & QRD Cost & Puncturing Cost \\
\hline\hline
ML $\rightarrow$ PML = PCD & $\theta_2/J$ & $\theta_3/J$ \\\hline %
N/C $\rightarrow$ PN/C & $\theta_2/J$ & $\theta_3/J$ \\\hline %
CD $\rightarrow$ PCD & $\theta_2/J$ & $\theta_3/J$ \\\hline
LORD $\rightarrow$ SSD & $N\times\theta_2/J$ & $N\times\theta_3/J$  \\\hline
SLORD $\rightarrow$ SSSD & $N\times\theta_2/J$ & $N\times\theta_3/J$  \\
\hline 
\end{tabular}
\end{table}

\begin{table}[!t]
\centering
\caption{\textbf{Savings in Studied HO Detectors}}\vspace{0.15in} 
\label{table:ch5_detectors2} 
\centering 
\begin{tabular}{| c || c | c |} 
\hline
Detection Scheme & Savings in Computations (FLOPs) \\
\hline\hline
ML $\rightarrow$ PML = PCD  & $\left(\abs{\mathcal{M}}^N-\abs{\mathcal{M}}\right)\times\theta_1$\\\hline %
N/C $\rightarrow$ PN/C & $\theta_1$ \\\hline %
CD $\rightarrow$ PCD & $\abs{\mathcal{M}}\times\theta_1$ \\\hline
LORD $\rightarrow$ SSD  & $N\!\times\!(\abs{\mathcal{M}}\!\times\!\theta_1\! -\! (4N^2\!+\!4N\!-\!2)\,\RAD\!-\!(4N^2\!+\!4N)\,\RML)$ \\\hline
SLORD $\rightarrow$ SSSD & $N\times\abs{\mathcal{M}}\times\theta_1$ \\
\hline 
\end{tabular}
\end{table}

The only computational drawback in subspace schemes is in channel decomposition. As shown in ~\cite{2014_mansour_SPL_WLD,2014_mansour_eurasip_WLD}, regular QRD requires $\theta_2=(4N^3-N^2-N)\,\RAD+(4N^3+3N^2)\,\RML$ FLOPs, and puncturing alone requires $\theta_3=\frac{2}{3}(8N^3-15N^2+4N-12)\,\RAD+(\frac{16}{3}N^3-7N^2+\frac{8}{3}N-20)\,\RML$ FLOPs (this overhead was reduced in \cite{7565040Sarieddeen} for SSD). However, channel matrix decompositions are only performed in the preprocessing stage of detection, and with slow fading channels, the decomposition outputs can be retained for a very large number of frames $J$.

For HO computations, PN/C saves $\theta_1$ FLOPs compared to N/C, PCD saves $\abs{\mathcal{M}}\times\theta_1$ FLOPs compared to CD, SSSD saves $N\times\abs{\mathcal{M}}\times\theta_1$ FLOPs compared to SLORD, and SSD saves $N\times(\abs{\mathcal{M}}\times\theta_1 - (4N^2+4N-2)\,\RAD-(4N^2+4N)\,\RML)$ FLOPs compared to LORD, where the subtracted terms in the latter account for distance computations in~\eqref{eq:ch5_SSDout}. These results are summarized in Tables \ref{table:ch5_detectors1} and \ref{table:ch5_detectors2}. Note that the substantial savings with PML are based on the fact that PML and PCD are identical. PML has no practical significance, and it is only included as a reference. Furthermore, extra memory is needed with LORD to compute the global minimum distance for every bit after layer processing \cite{2015_mansour_JSP_2x2QAM}, where it is required to store not only distances, but also their corresponding symbol vectors. The WRD-based approaches are thus computationally efficient, especially with slow fading channels and high order modulation constellations.

%
\section{Channel Shortening vs Channel Puncturing}
\label{sec:ch5_air}

Channel puncturing can be conceived as an alternative efficient implementation of channel shortening. However, efficient implementation does not imply best performance. By maximizing the mutual information that the transceiver system can achieve with a mismatched channel model, a framework for constructing optimal channel shortening for subspace detection was proposed in \cite{rusek2012optimal}, building on an original framework in \cite{rusek2012bounds}. The achievable information rate (AIR) metric was employed, which is a generalized mutual information that the transceiver system can achieve with a mismatched channel model at the receiver.

Building on the work in \cite{rusek2012optimal}, a SO MIMO detector was proposed in \cite{hu2017soft} by utilizing AIR-based partial marginalization (AIR-PM). Partial marginalization (PM) \cite{larsson2008fixed,persson2011partial} is a method of calculating LLRs without spanning entire lattices. The AIR-PM detector exploits a tree-based representation, where parent layers are exhaustively searched, and least-square estimates are used for marginalization on child layers. As in the case of SSD, since connections among all child layers are broken, PMs can be executed independently and in parallel. For AIR-based detection, the distance metric can be expressed as
\begin{equation}\label{eq:ch5_AIR_1}
 d(\mbf{y}|\mbf{x}) = \frac{1}{\sigma^{2}} \left( 2\Re\{\mbf{x}^\mathcal{H}\mbf{H}^\mathcal{H}\mbf{y}\} - \mbf{x}^\mathcal{H}\mbf{H}^\mathcal{H}\mbf{H}\mbf{x} - \mbf{y}^\mathcal{H}\mbf{y} \right).
\end{equation}
Neglecting the last term and absorbing the noise variance into $\mbf{H}_r$ and $\mbf{G}_r$, the probability function of the resultant detection model would be
\begin{equation}\label{eq:ch5_AIR_2}
 \tilde{\Prb}(\mbf{y}|\mbf{x}) = \exp \left( 2\Re\{\mbf{x}^\mathcal{H}\mbf{H}_r^\mathcal{H}\mbf{y}\} - \mbf{x}^\mathcal{H}\mbf{G}_r\mbf{x} \right).
\end{equation}
Note that all lattice processing is contained in $\mbf{G}_r$. The AIR is defined as
\begin{equation}\label{eq:ch5_AIR_3}
 I_{\AIR}(\mbf{y};\mbf{x}) = \mathsf{E}_{\mbf{x},\mbf{y}}[\ln \tilde{\Prb}(\mbf{y}|\mbf{x})] - \mathsf{E}_{\mbf{y}}[\ln \tilde{\Prb}(\mbf{y})],
\end{equation}
where $\tilde{\Prb}(\mbf{y})\!=\!\sum_{\mbf{x}\in\mathcal{X}} \tilde{\Prb}(\mbf{y}|\mbf{x})\Prb(\mbf{x})$. Assuming complex Gaussian inputs, a closed form $I_{\AIR}$ expression was reached by optimizing \eqref{eq:ch5_AIR_3} over a prefilter matrix $\mbf{H}_r$:
\begin{align}\label{eq:ch5_AIR_4}
 I_{\AIR}(\mbf{y};\mbf{x}) &= N + \ln \det(\mbf{I}_N + \mbf{G}_r) - \Tr( \bar{\mbf{B}}(\mbf{I}_N + \mbf{G}_r) ) \\
 \mbf{H}_r &= \bar{\mbf{W}}^\mathcal{H}(\mbf{I}_N + \mbf{G}_r)\\
 \bar{\mbf{W}} &= \mbf{H}^\mathcal{H}\left(\mbf{H}\mbf{H}^\mathcal{H} + \sigma^{2}\mbf{I}_N \right) \\
 \bar{\mbf{B}} &= \mbf{I}_N - \bar{\mbf{W}}\mbf{H} \\
 \mbf{G}_r &= \mbf{U}\mbf{U}^\mathcal{H} - \mbf{I}_N,
\end{align}
The matrix $ \mbf{G}_r$, and hence $\mbf{U}$, is then chosen to maximize \eqref{eq:ch5_AIR_4} under the decomposition constraints. The resultant $\mbf{U}$ matrix is shown in equations (26) and (27) in \cite{hu2017soft}, and the resultant $I_{\AIR}$ reads:
\begin{equation}\label{eq:ch5_AIR_5}
 I_{\AIR}(\mbf{y};\mbf{x}) = 2 \sum_{n=1}^{N} \ln u_{n,n}.
\end{equation}
The matrix $\mbf{U}$ has the same structure as the matrix $\mbf{R}$ in WRD, and its punctured shape is responsible for decreasing the complexity of lattice processing. It can be noted that computing $\mbf{U}$ with multiple parent layers is computationally intensive due to matrix inversions. However, for a single parent layer this overhead is graceful, where we have the $\nth{n}$ diagonal element of $\mbf{U}$ computed from the elements of $\bar{\mbf{B}}$ as follows:
\begin{equation}\label{eq:ch5_AIR_6}
 u_{n,n} = \sqrt{\left( \bar{b}_{n,n} - \frac{\abs{\bar{b}_{n,N}}^2}{\bar{b}_{N,N}} \right)^{-1} }.
\end{equation}

Despite the fact that WRD in SSD is not optimal in the sense of maximizing the AIR, it can be shown to have a much lower complexity compared to AIR-PM when more than one layer is a parent layer, as well as a near-optimal performance. AIR-based detectors need to compute $\mbf{H}_r$ and $\mbf{G}_r$. Computing $\mbf{H}_r$ requires a matrix inversion, which is a difficult task in the context of large MIMO. Moreover, computing $\mbf{G}_r$ (equation (26) and in \cite{hu2017soft}) also requires multiple matrix inversions when the number of parent layers is large. On the contrary, QRD/WRD-based schemes only require executing a QRD (followed by elementary matrix operations in case of WRD), which is a much simpler task, especially if a dedicated decomposition engine is used.

\begin{figure}[!t]
\centering
\includegraphics[width=5.8in]{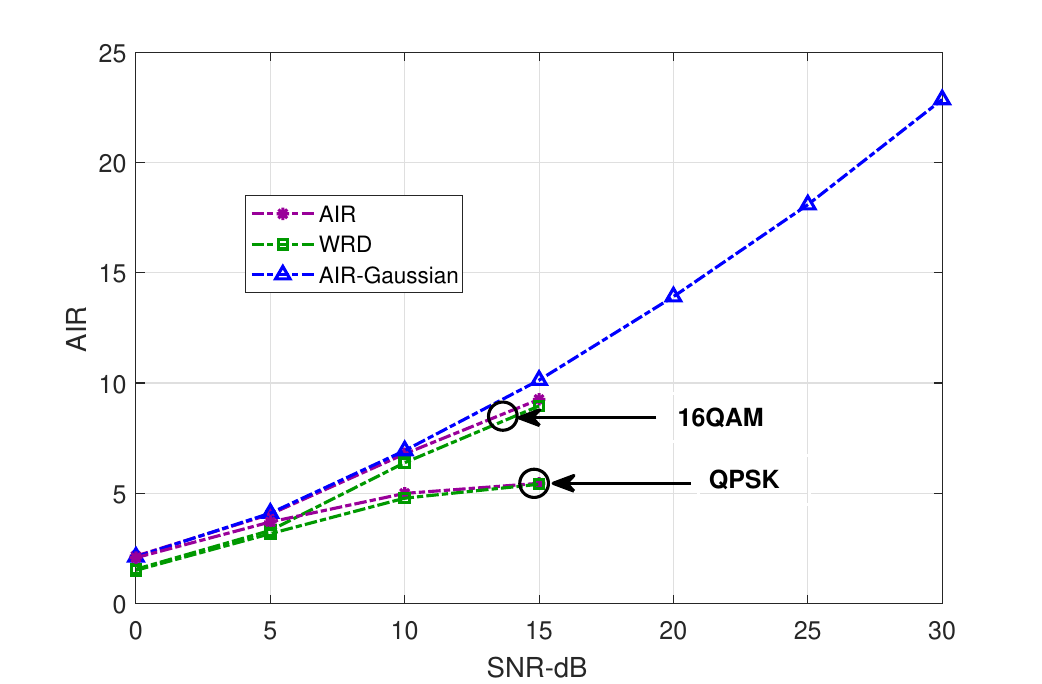}\vspace{-0.1in}
\caption{Achievable information rates - $4\times4$ MIMO.}
\label{capfig:ch5_a}
\end{figure}

The AIRs of WRD-based and AIR-based schemes were obtained empirically via a Monte Carlo simulation by computing \eqref{eq:ch5_AIR_3} for different MTs. These rates were compared to the theoretical AIR of the AIR-based detectors assuming Gaussian inputs (equation \eqref{eq:ch5_AIR_5}). As shown in Fig.~\ref{capfig:ch5_a}, AIRs with $16$-QAM are much closer to the theoretical bound than those with QPSK. Furthermore, the gap in the AIR between WRD-based and AIR-based detectors is shown to be small at low SNR and negligible at high SNR. 

%
\section{Simulation Results}
\label{sec:ch5_results}

The proposed detectors were simulated following the system model of Sec.~\ref{sec:ch2_model}. Both HO and SO scenarios were considered, where in the latter turbo coding was used, with a code rate of $1/2$ and $8$ decoding iterations. In addition to $\mbf{H}$, we considered $\mbf{H}_{c}\!=\!\mbf{C}_{r}^{1/2}\mbf{H}\mbf{C}_{t}^{1/2}$ that accounts for antenna correlation, where $\mbf{C}_{t}$ and $\mbf{C}_{r}$ are the transmit and receive antenna correlation matrices, respectively, with correlation factors $\alpha=\bar{\alpha}=0.9$. We assume, for convenience, the generic exponential model~\cite{loyka2001channel}. Hence, in the case of $4\!\times\!4$ MIMO, we have

\begin{equation}\label{ch5_matrix}
 \mbf{C}_{t} = \begin{bmatrix}
    1 & \alpha & \alpha^2 & \alpha^3  \\
    \alpha & 1 & \alpha & \alpha^2  \\
    \alpha^2 & \alpha & 1 & \alpha \\
    \alpha^3 & \alpha^2 & \alpha & 1
\end{bmatrix},\ \ \ \  \mbf{C}_{r} = \begin{bmatrix}
    1 & \bar{\alpha} & \bar{\alpha}^2 & \bar{\alpha}^3  \\
    \bar{\alpha} & 1 & \bar{\alpha} & \bar{\alpha}^2  \\
    \bar{\alpha}^2 & \bar{\alpha} & 1 & \bar{\alpha} \\
    \bar{\alpha}^3 & \bar{\alpha}^2 & \bar{\alpha} & 1
\end{bmatrix}.
\end{equation}

\begin{figure}[!t]
\centering
\includegraphics[width=5.95in]{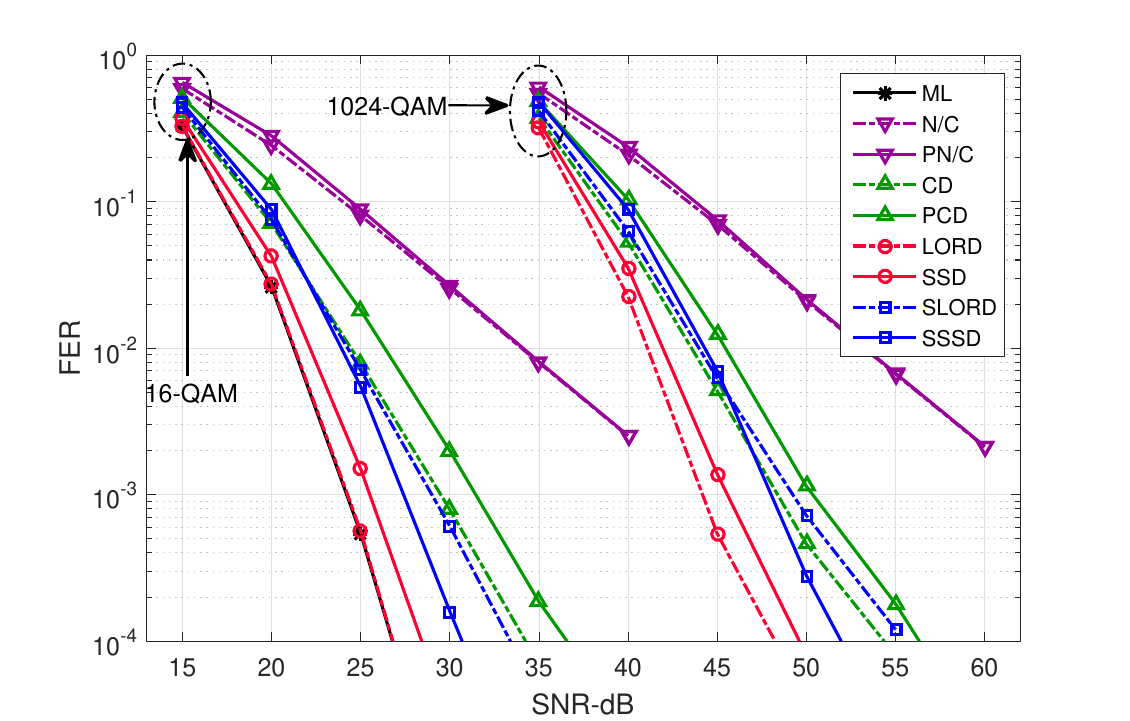}\vspace{-0.1in}
\caption{HO FER performance - $4\times4$ MIMO.}
\label{hardfig:ch5_a}
\end{figure}
\begin{figure}[!t]
\centering
\includegraphics[width=5.95in]{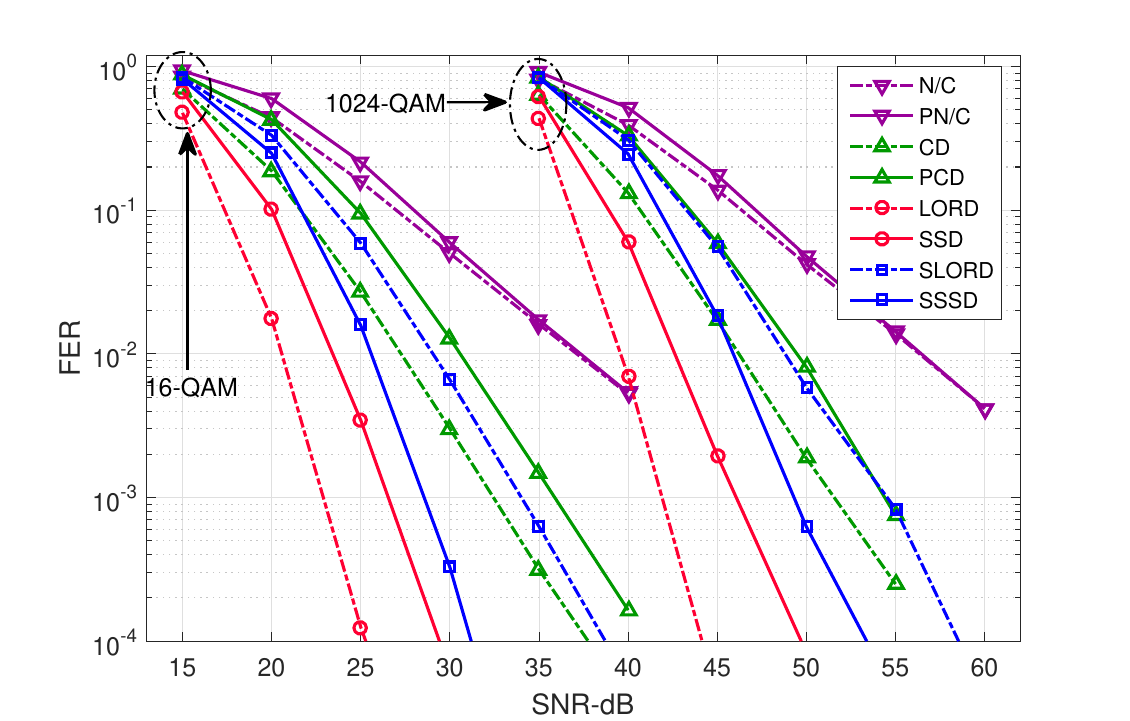}\vspace{-0.1in}
\caption{HO FER performance - $8\times8$ MIMO.}
\label{hardfig:ch5_b}
\end{figure}
\begin{figure}[!h]
\centering
\includegraphics[width=5.95in]{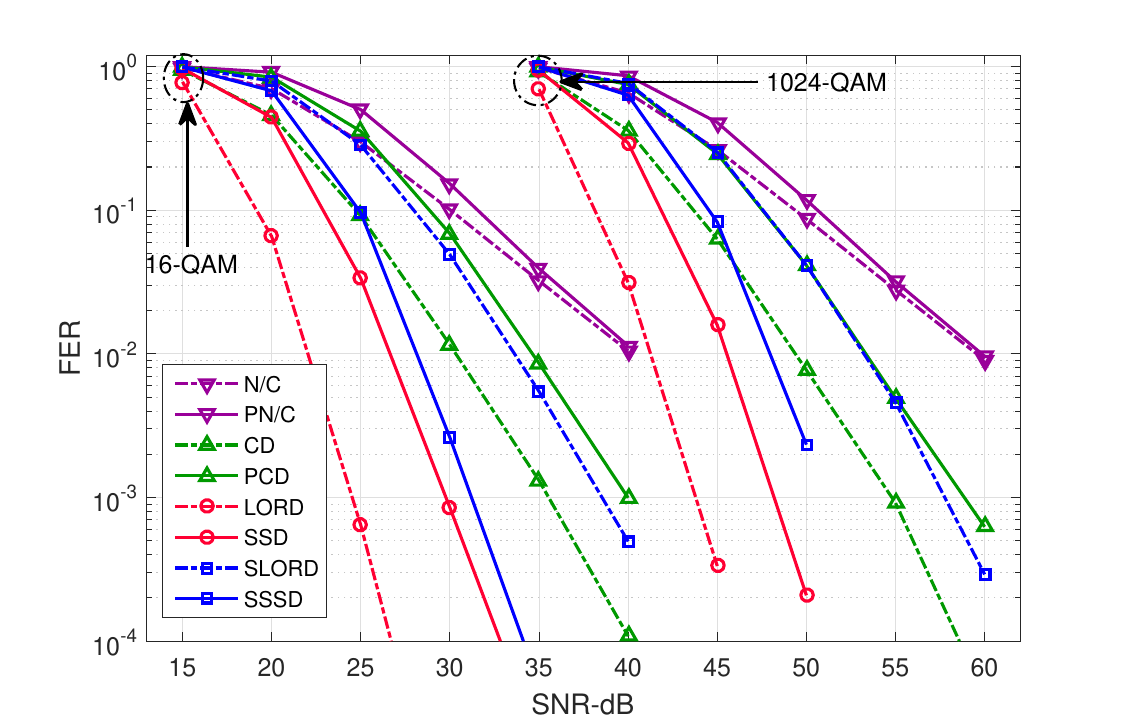}\vspace{-0.1in}
\caption{HO FER performance - $16\times16$ MIMO.}
\label{hardfig:ch5_c}
\end{figure}
\begin{figure}[!h]
\centering
\includegraphics[width=5.95in]{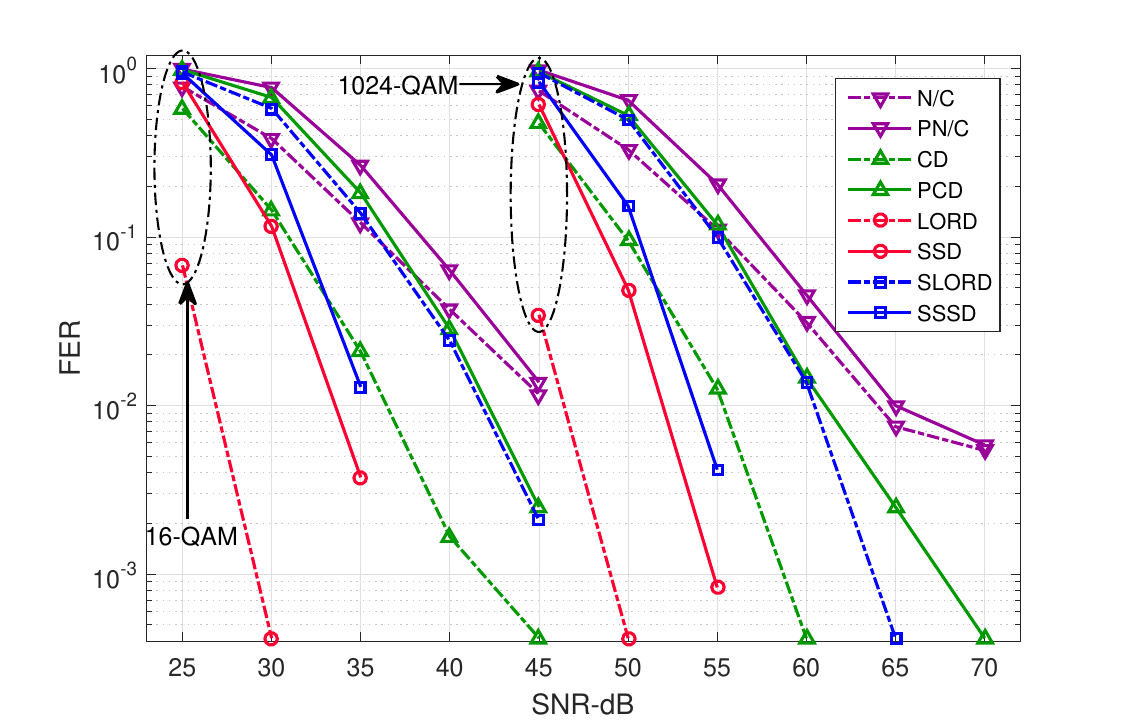}\vspace{-0.1in}
\caption{HO FER performance - $64\times64$ MIMO.}
\label{hardfig:ch5_d}
\end{figure}

Figures \ref{hardfig:ch5_a} to \ref{hardfig:ch5_d} show the HO FER performance with various MIMO configurations, when $\mathcal{M}$ is $16$-QAM and $1024$-QAM \cite{7418324Sarieddeen}. In the context of $4\!\times\!4$ MIMO (Fig.~\ref{hardfig:ch5_a}), the performance degradation in PN/C compared to N/C is negligible, the chase detectors cut the gap between N/C and ML in half, and the PCD introduces a $\unit[2]{dB}$ loss compared to the CD. Moreover, while LORD achieves exact ML performance, SSD lags behind by also $\unit[2]{dB}$. The relative performances of the detectors are maintained with very large constellations. Figures \ref{hardfig:ch5_b}, \ref{hardfig:ch5_c}, and \ref{hardfig:ch5_d} then show the performance of the proposed schemes in the context of $8\!\times\!8$ MIMO, $16\!\times\!16$ MIMO, and $64\!\times\!64$ MIMO, respectively. The relative performances are maintained, but the gap between WRD and QRD-based schemes increases from $\unit[2]{dB}$, to $\unit[4]{dB}$, $\unit[5]{dB}$, and $\unit[7]{dB}$, respectively. The SSSD was the only WRD-based detector to achieve a performance gain, compared to SLORD. Note that large MIMO systems do not achieve the gains of massive MIMO systems with very large receive-to-transmit antenna ratios, which, in addition to our definition of $\mathsf{SNR}$, explain the high SNR range.

\begin{figure}[!t]
\centering
\includegraphics[width=5.95in]{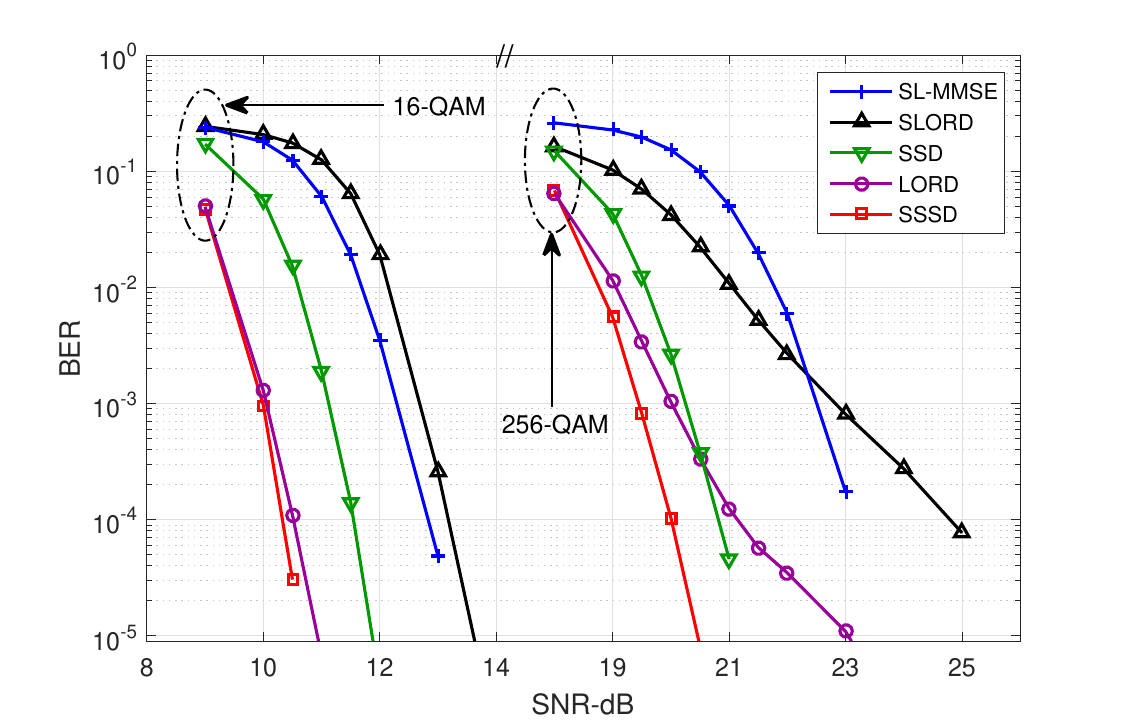}\vspace{-0.1in}
\caption{SO FER performance - $8\times8$ MIMO.}
\label{softfig:ch5_a}
\end{figure}
\begin{figure}[!t]
\centering
\includegraphics[width=5.95in]{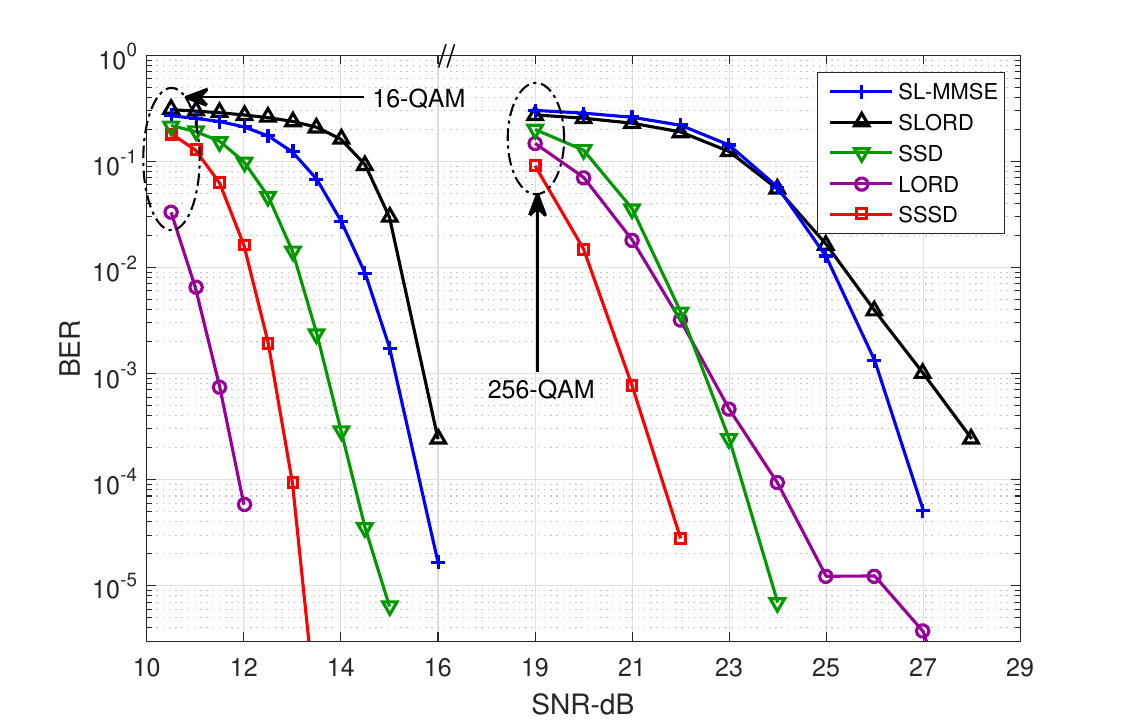}\vspace{-0.1in}
\caption{SO FER performance - $16\times16$ MIMO.}
\label{softfig:ch5_b}
\end{figure}
\begin{figure}[!h]
\centering
\includegraphics[width=5.95in]{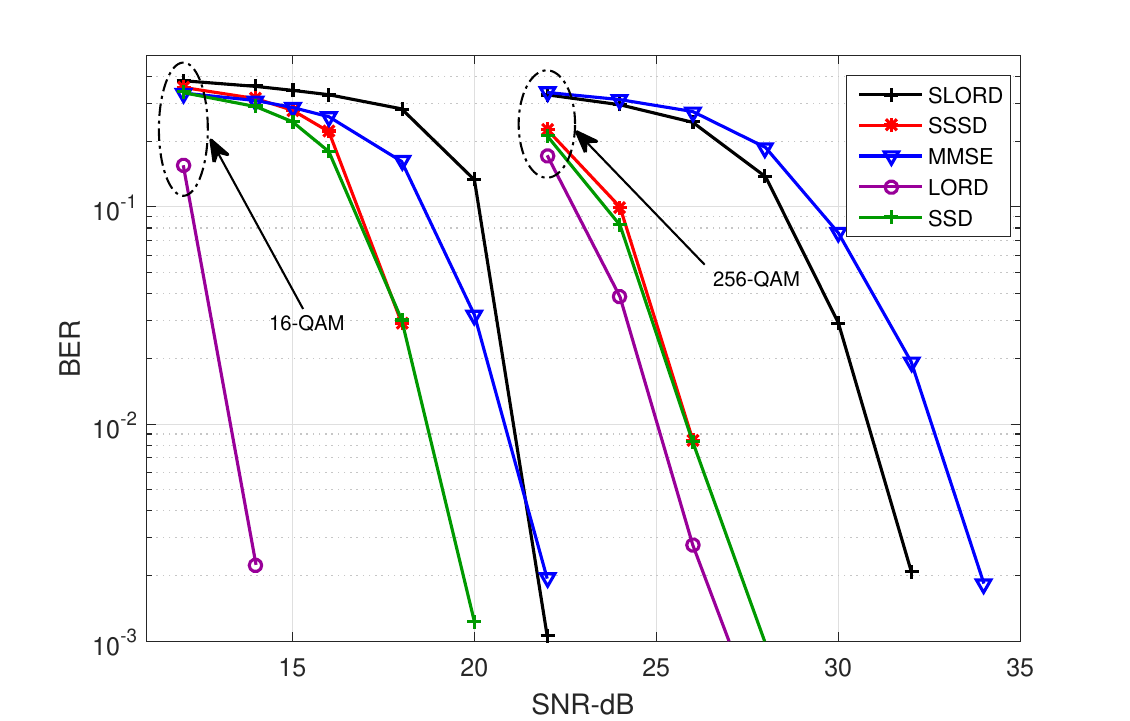}\vspace{-0.1in}
\caption{SO FER performance - $64\times64$ MIMO.}
\label{softfig:ch5_c}
\end{figure}
\begin{figure}[!h]
\centering
\includegraphics[width=5.95in]{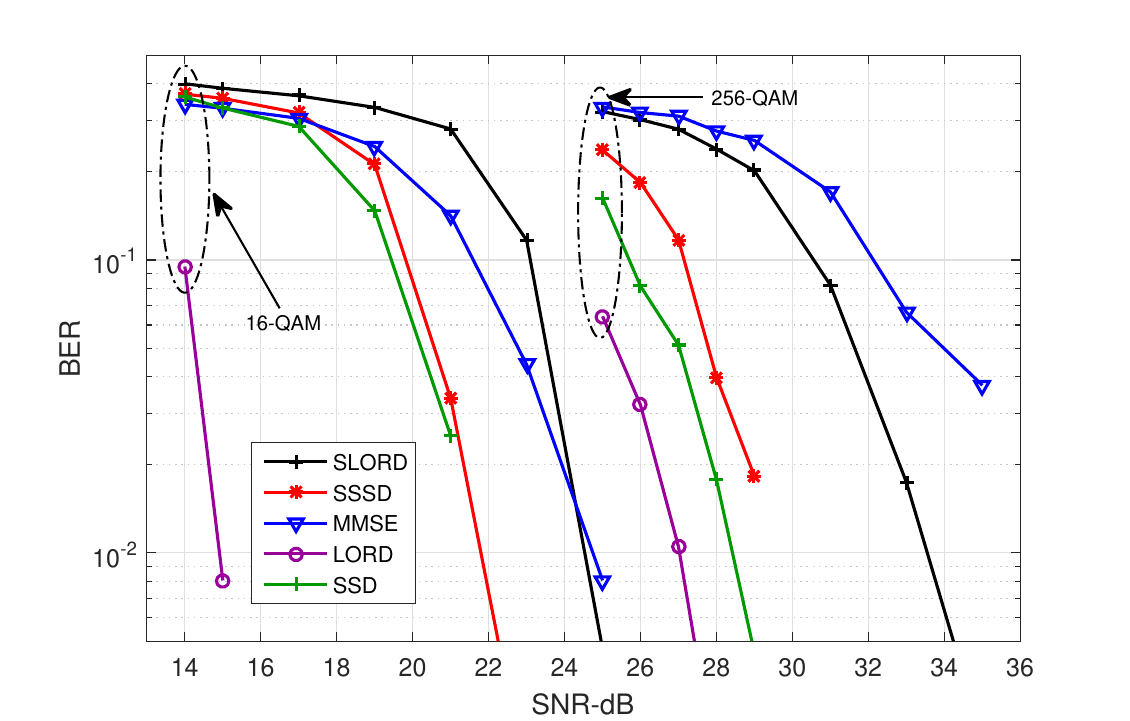}\vspace{-0.1in}
\caption{SO FER performance - $128\times128$ MIMO.}
\label{softfig:ch5_d}
\end{figure}

Figures \ref{softfig:ch5_a} to \ref{softfig:ch5_d} show the BER performance of the studied SO detectors, compared to a reference low-complexity MMSE detector \cite{Studer_MMSE}, with various MIMO configurations, when $\mathcal{M}$ is $16$-QAM and $256$-QAM. With relatively low MIMO orders, the SSSD (or SO PCD) outperforms LORD, and so does the SSD with high order MTs, while SLORD and MMSE lag behind. For example, the SO PCD achieves a $\unit[2.5]{dB}$ gain compared to LORD, at a BER of $10^{-4}$ in $16\!\times\!16$ MIMO with $256$-QAM. However, with high order MIMO, SSSD can not beat LORD ($\unit[5]{dB}$ and $\unit[7]{dB}$ gaps are noticed with $16$-QAM). Nevertheless, at very high MIMO orders, the reduction in complexity with WRD-based detectors is particularly large, and the gap in performance can be as low as $\unit[1]{dB}$ or $\unit[2]{dB}$ with $256$-QAM, where the effect of interference is reduced with larger MTs.

\begin{figure}[!h]
\centering
\includegraphics[width=5.95in]{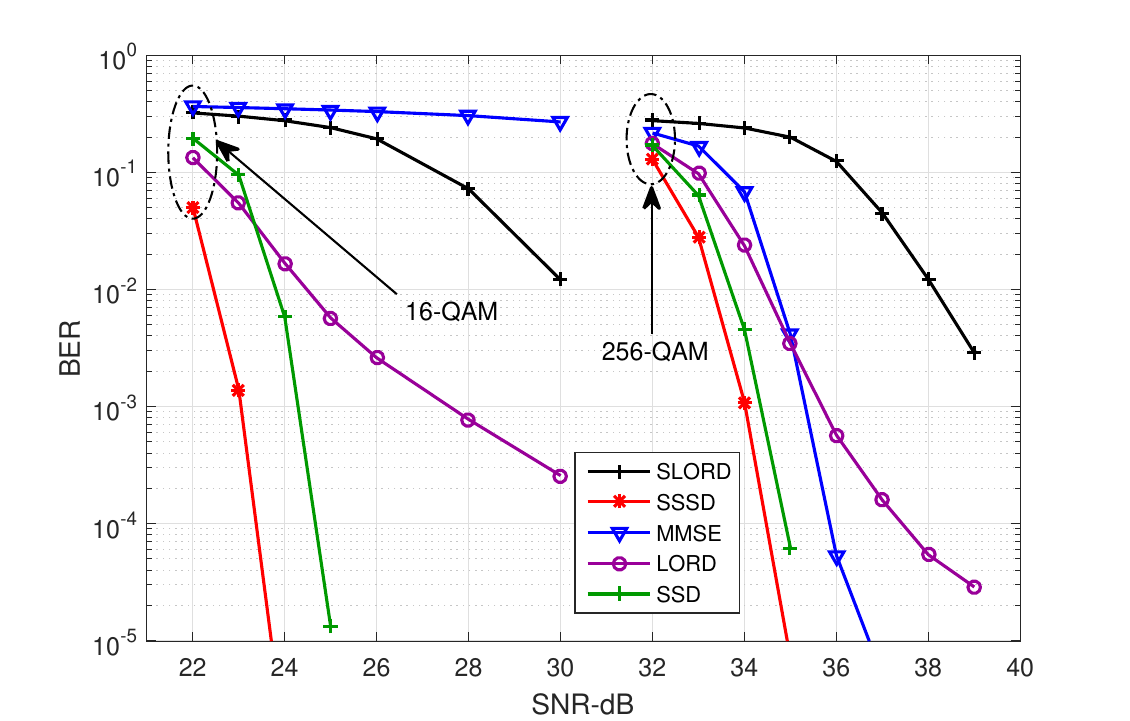}\vspace{-0.1in}
\caption{SO FER performance - $4\times4$ MIMO - correlated channels.}
\label{corrsoftfig:ch5_a}
\end{figure}
\begin{figure}[!h]
\centering
\includegraphics[width=5.95in]{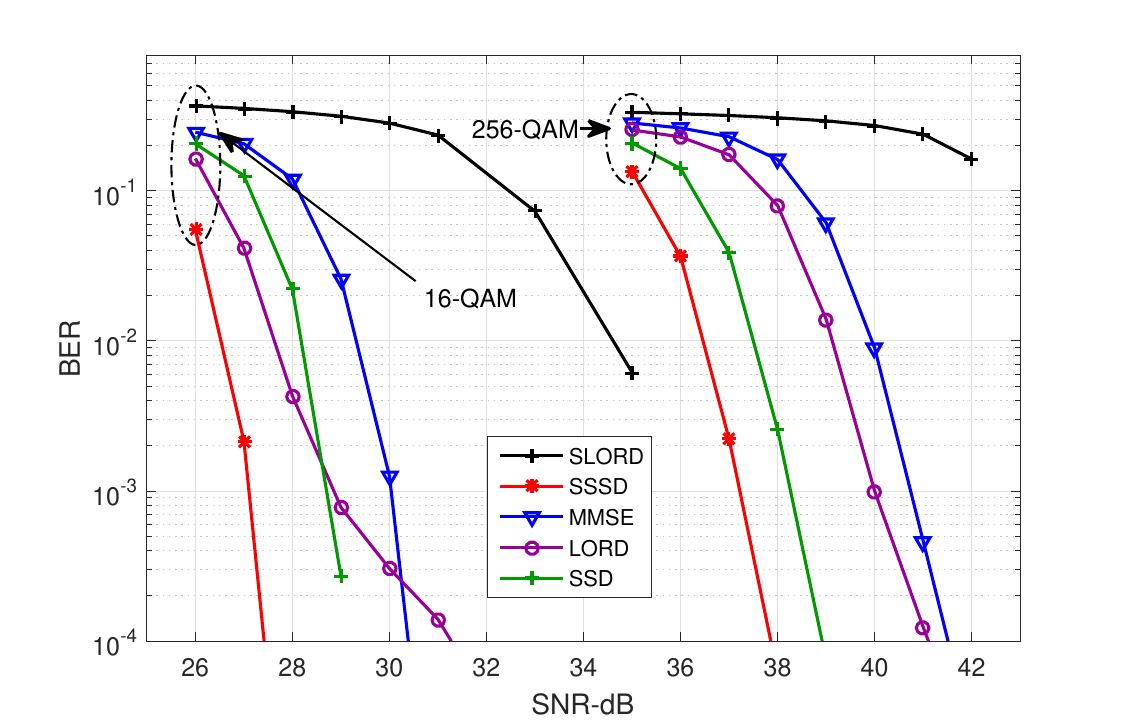}\vspace{-0.1in}
\caption{SO FER performance - $8\times8$ MIMO - correlated channels.}
\label{corrsoftfig:ch5_b}
\end{figure}
\begin{figure}[!h]
\centering
\includegraphics[width=5.95in]{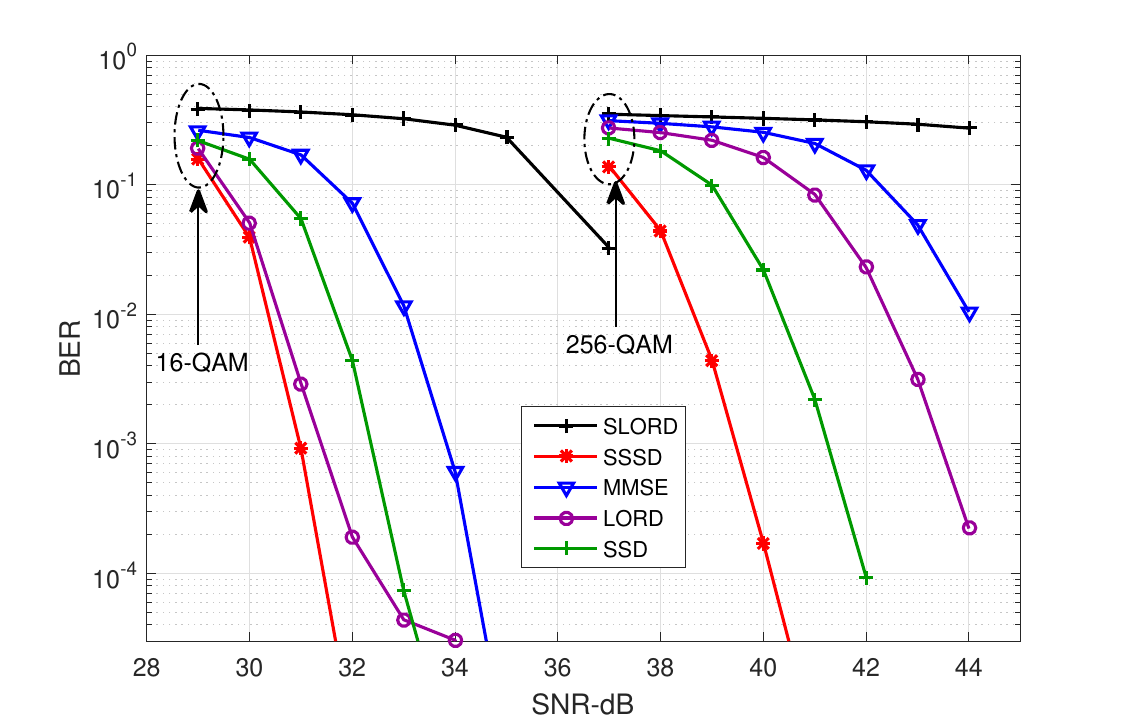}\vspace{-0.1in}
\caption{SO FER performance - $16\times16$ MIMO - correlated channels.}
\label{corrsoftfig:ch5_c}
\end{figure}
\begin{figure}[!h]
\centering
\includegraphics[width=5.95in]{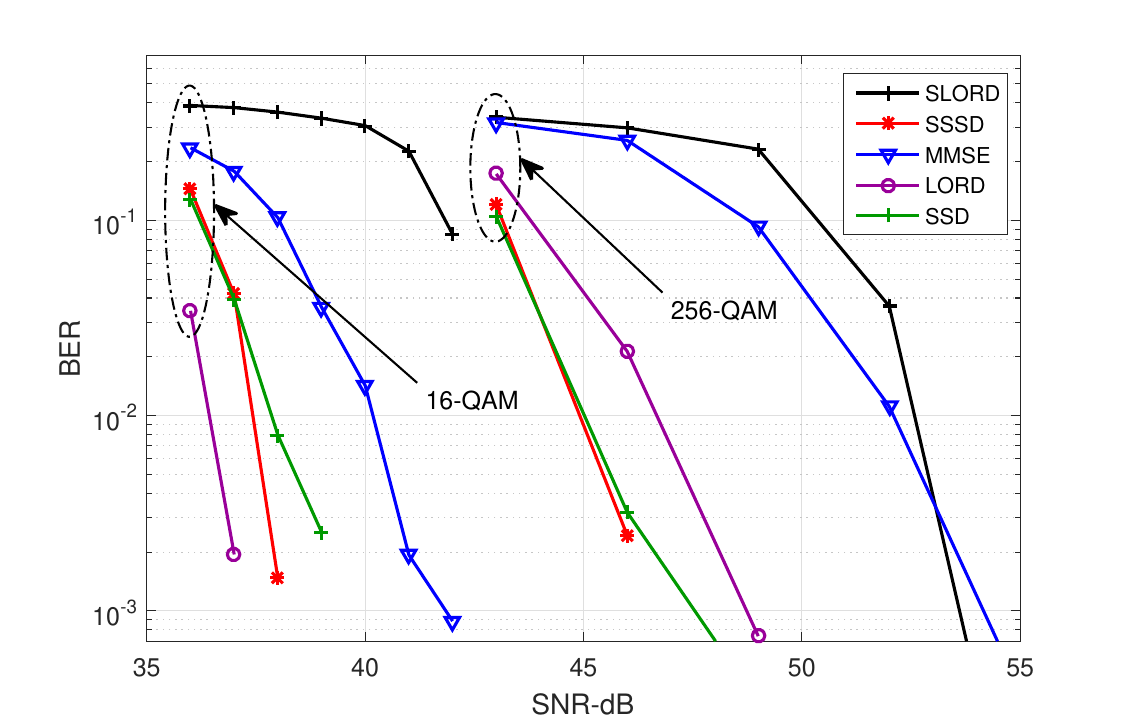}\vspace{-0.1in}
\caption{SO FER performance - $64\times64$ MIMO - correlated channels.}
\label{corrsoftfig:ch5_d}
\end{figure}

Figures \ref{corrsoftfig:ch5_a} to \ref{corrsoftfig:ch5_d} show the SO BER performance of the detectors under high channel correlation. Subspace detectors, SSD and SSSD, outperform the much more complex LORD. It is only at very high MIMO orders with low order MTs that LORD slightly outperforms SSSD. This declares the SSSD the winning detector in the presence of channel correlation. \\

\textbf{Conclusion} \\

A family of low-complexity MIMO detectors that employ punctured QRD in lieu of regular QRD has been proposed. The proposed detectors have been shown to achieve significant computational savings in the context of large MIMO systems. Furthermore, significant performance gains have been observed with highly correlated channels. An architectural design has been proposed, by using the detectors of lower complexity as building blocks in their more complex extensions, and it has been established that the proposed schemes scale up efficiently both in the number of antennas and the constellation size. In particular, SO per-layer subspace detection has been shown to achieve a $\unit[2.5]{dB}$ SNR gain in $256$-QAM $16\!\times\!16$ MIMO, while saving $77\%$ of N/C computations.

%% file: Chapter6_ANALYSIS.tex

\chapter{Capacity, Diversity, and BER Analysis}\label{chapter:ch6_Analysis}

We analyze mathematically the capacity and  BER performance of the proposed HO detectors. First, capacity bounds under puncturing are derived. Second, the diversity gain is characterized and used to show that channel matrix puncturing does not negatively affect the diversity gain in HO detection. Third, the performance of these detectors is studied via a probabilistic BER characterization. We extend the study for several variations of SO detection schemes, and show that significant performance gains can be achieved with channel puncturing. The results of this section appeared, in parts, in \cite{8292405Sarieddeen}, \cite{ISIT18_Sarieddeen}, and \cite{8186206Sarieddeen}.

%
\section{Statistical Properties of Punctured Channels}
\label{sec:ch6_distributions}

Note that after puncturing, the column at the root layer in $\mbf{W}$ (layer $N$ here), remains orthogonal to all other columns. Hence, taking the expectation of $\mbf{W}^\mathcal{H}\mbf{nn}^\mathcal{H}\mbf{W}$ over $\mbf{n}$, we have:
\begin{equation}\label{eq:ch6_matrix}
    \mathsf{E}_{\mbf{n}}[\mbf{W}^\mathcal{H}\mbf{nn}^\mathcal{H}\mbf{W}] =
        \begin{bsmallmatrix}
            \sigma^2    & e_{12}    & e_{13}    & 0 \\
            e_{12}^\mathcal{H}    & \sigma^2  & e_{23}    & 0 \\
            e_{13}^\mathcal{H}    & e_{23}^\mathcal{H}  & \sigma^2  & 0 \\
            0           & 0         & 0         & \sigma^2
        \end{bsmallmatrix}.
\end{equation}
Therefore, although the resultant noise is colored, WRD preserves the noise variance at the layer of interest. However, the statistical properties of the elements of $\Rp$ get distorted under puncturing.
\begin{figure}[!t]
\centering
\includegraphics[width=5.2in]{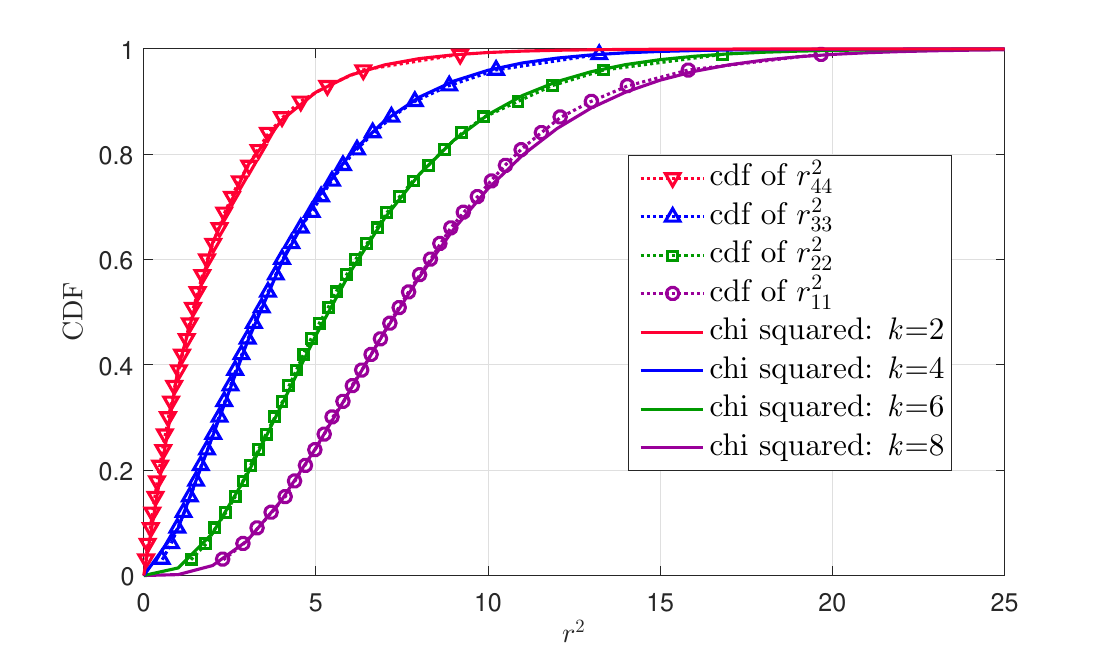}\vspace{-0.1in}
\caption{Empirical vs theoretical CDFs of the diagonal elements of $\mbf{R}$.}
\label{distfig:ch6_a}
\end{figure}
\begin{figure}[!t]
\centering
\includegraphics[width=5.2in]{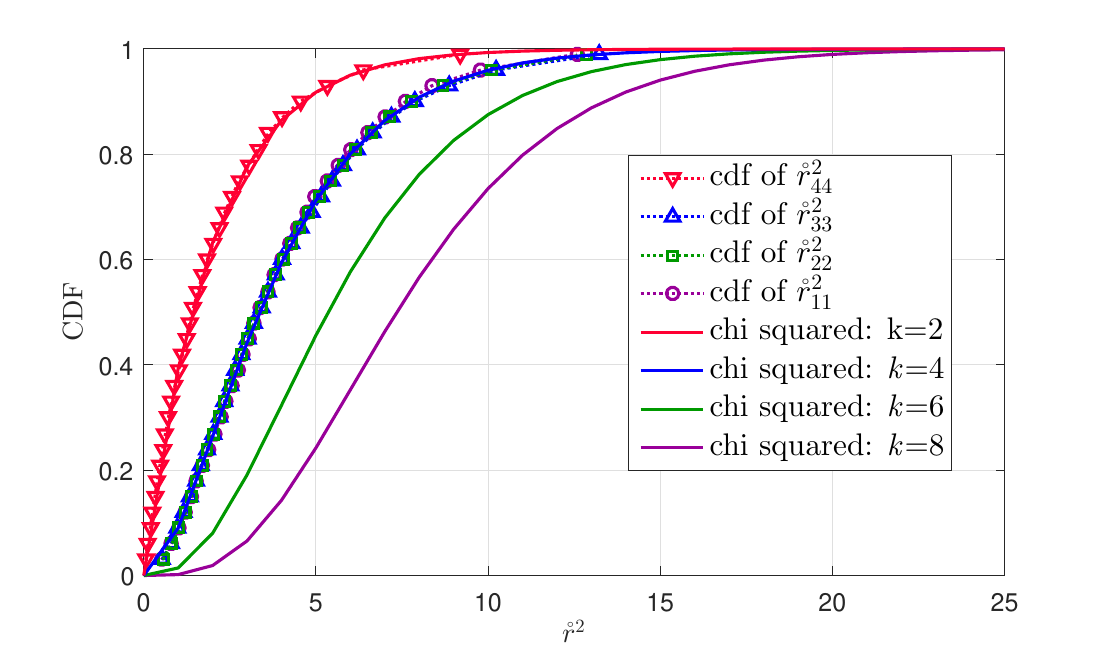}\vspace{-0.1in}
\caption{Empirical vs theoretical CDFs of the diagonal elements of $\Rp$.}
\label{distfig:ch6_b}
\end{figure}
The non-zero elements of $\mbf{R}$ (given i.i.d. Rayleigh fading) are known to be independent random variables with the following distributions~\cite{edelman1988eigenvalues,choi2006nulling}:

\begin{itemize}
  \item The off-diagonal elements are circular symmetric complex Gaussian with unit variance.
  \item The square of the $\nth{n}$ diagonal element is chi-squared distributed with $2(N-n+1)$ degrees of freedom, with a probability density function \\
      \begin{equation}
      f(g=r_{nn}^2) = \frac{1}{(N-n)!} g^{N-n}e^{-g}, \ \ g \geq 0,
      \end{equation}
\end{itemize}
where chi-squared comes from the sum of squares of Rayleigh distributed random variables. This can be verified by analyzing the Householder matrix construction, as shown in Appendix B of \cite{marzetta2016fundamentals}.

While the distributions of non-zero off-diagonal elements remain intact, the distributions of diagonal elements at upper layers $n\!=\!1,\cdots,N\!-\!3$, lose degrees of freedom from $2(N\!-\!n\!+\!1)$ down to $4$, as depicted in Figs.~\ref{distfig:ch6_a} and \ref{distfig:ch6_b} for a $4\times4$ channel matrix, where empirical CDFs of the diagonal elements are compared to theoretical chi-squared CDFs in solid lines. This is caused by the fact that each puncturing operation at layer $n$ renders the $\nth{n}$ column of $\mbf{W}$ dependent on one of the remaining columns, thus eliminating two degrees of freedom from the corresponding distribution of $\rp_{nn}^2$.

%
\section{Analysis of Capacity}
\label{sec:ch6_capacity}

In this section, we analyse the impact of channel puncturing on the capacity of MIMO systems. Recall from \cite{telatar1999capacity,foschini1998limits} that the capacity of a channel from our system model ($\mathsf{E}[\mbf{n}\mbf{n}^\mathcal{H}]=\sigma^{2}\mbf{I}_M$ and $\mathsf{E}[\mbf{x}\mbf{x}^\mathcal{H}]=\mbf{I}_N$), expressed in bits per seconds per Herts (bps/Hz) is
\begin{equation}\label{eq:ch6_cap_1}
C_{\mbf{H}} = \log \det \left( \mbf{I}_M + \frac{1}{\sigma^{2}}\mbf{H}\mbf{H}^\mathcal{H} \right),
\end{equation}
which can be achieved when the transmitter uses Gaussian codebooks. Furthermore, a more generic equation for capacity is
\begin{equation}\label{eq:ch6_cap_2}
C = \log \frac{\det \left( \mbf{Z} + \mbf{H}\Sigma\mbf{H}^\mathcal{H} \right) }{\det{\mbf{Z}}},
\end{equation}
where $\mbf{Z}$ is the covariance matrix of the altered noise and $\Sigma\!=\!\mathsf{E}[\mbf{x}\mbf{x}^\mathcal{H}]$ is the transmit covariance matrix. This equation can be used to account for colored noise \cite{jorswieck2004performance} resulting from any source of interference. Since $\mbf{n}$ is circular symmetric complex Gaussian, then the colored noise $\mbf{W}^\mathcal{H}\mbf{n}$ is also circular symmetric complex Gaussian, with a covariance matrix $\mathsf{E}_{\mbf{n}}[\mbf{W}^\mathcal{H}\mbf{nn}^\mathcal{H}\mbf{W}]$. Therefore, the capacity of the modified system model, with puncturing as a source of interference, can be expressed as
\begin{equation}\label{eq:ch6_cap_3}
C_{\Rp,\Opt} = \log \frac{\det \left(  \mathsf{E}_{\mbf{n}}[\mbf{W}^\mathcal{H}\mbf{nn}^\mathcal{H}\mbf{W}] + \Rp\Rp^\mathcal{H} \right) }{\det{\mathsf{E}_{\mbf{n}}[\mbf{W}^\mathcal{H}\mbf{nn}^\mathcal{H}\mbf{W}]}}.
\end{equation}
Under perfect knowledge of colored noise, $C_{\mbf{H}}$ and $C_{\Rp,\Opt}$ are identical. Hence, in principle, if an optimal wightening filter is used in the proposed detectors to account for colored noise, there should be no loss in performance. However, the computational complexity of such a filter is arguably larger than the computational savings that are caused by puncturing, which is why we neglected the effect of colored noise in our proposed algorithms. The resultant achievable rate is thus expressed as
\begin{equation}\label{eq:ch6_cap_4}
C_{\Rp} = \log \det \left( \mbf{I}_M  + \frac{1}{\sigma^{2}}\Rp\Rp^\mathcal{H} \right).
\end{equation}

In what follows, we seek a lower bound on the achievable rates, assuming symmetric $N\times N$ MIMO, for simplicity. As described in \cite{foschini1998limits} for regular MIMO, using results on random matrices \cite{edelman1988eigenvalues} and the notion of unitarily equivalent rectangular matrices, we say that $\mbf{H}$ is unitarily equivalent to the bidiagonal matrix

 \begin{equation}\label{eq:ch6_matrix1}
    \begin{bmatrix}
        \chi_{2N}     &        &       &    \\
        \chi_{2(N-1)} &   \chi_{2(N-1)}    &      &  \\
                &    \ddots       & \ddots     &  \\
                &            &      \chi_2     & \chi_2
    \end{bmatrix},
\end{equation}
where $\chi_j^2$ is a chi-squared distributed random variable with $j$ degrees of freedom. Note that this can be proved by performing the HT on $\mbf{H}$. Building on the distributions of the punctured matrix, and applying the HT once on the full column of $\Rp$, we can infer by analogy that $\Rp$ is unitarily equivalent to the matrix
\begin{equation}\label{eq:ch6_matrix2}
    \begin{bmatrix}
        \chi_{4}      &        &       &    \\
        \chi_{2(N-1)}  &   \chi_{4}    &      &  \\
                &         & \ddots     &  \\
                &            &           & \chi_2
    \end{bmatrix},
\end{equation}
with the diagonal entries being all $\chi_{4}$ except for the last entry, and where all the entries that are not shown are understood to be zeros. We next use the representation in \eqref{eq:ch6_matrix2} to derive a lower bound on \eqref{eq:ch6_cap_4}. Define $\Upsilon_{j} = \frac{1}{\sigma} \chi_{j}$, the matrix $\mbf{I}_M  + \frac{1}{\sigma^{2}}\Rp\Rp^\mathcal{H}$ has the form
\begin{equation}\label{eq:ch6_matrix3}
    \begin{bmatrix}
        1+\Upsilon_{4}^2              &   \Upsilon_{4}\Upsilon_{2(N-1)}           &                    &           &                      &  \\
        \Upsilon_{4}\Upsilon_{2(N-1)} &   1+\Upsilon_{4}^2+\Upsilon_{2(N-1)}^2    &                    &           &                      &  \\
                                      &                                           & 1+\Upsilon_{4}^2   &           &                      &  \\
                                      &                                           &                    & \ddots    &                      &  \\
                                      &                                           &                    &           &   1+\Upsilon_{4}^2   &        \\
                                      &                                           &                    &           &                      &   1+\Upsilon_{2}^2
    \end{bmatrix}.
\end{equation}
In evaluating $\det \left( \mbf{I}_M  + \frac{1}{\sigma^{2}}\Rp\Rp^\mathcal{H} \right)$ we get, from the product of the $N$ main diagonal terms in \eqref{eq:ch6_matrix3}, a contribution of the form $L_{\Rp} + \varpi$ where
\begin{equation}\label{eq:ch6_bound_1}
L_{\Rp} = \left( 1+\Upsilon_{4}^2  \right)^{N-1}(1+\Upsilon_{2}^2),
\end{equation}
and $\varpi$ is a positive number. It can be easily noted that every negative term in the remainder of determinant computations is cancelled by a distinct positive contribution to $\varpi$, and that $\varpi$ contains more terms than needed. Therefore $C_{\Rp}>L_{\Rp}$ with probability one, and we have our lower bound. Comparing this bound with the lower bound on $C_{\mbf{H}}$ \cite{foschini1998limits}
\begin{equation}\label{eq:ch6_bound_2}
L_{\mbf{H}} = \prod_{j=1}^N\left(1+\Upsilon_{2j}^2  \right),
\end{equation}
we note that despite the loss in degrees of freedom, both bounds are dominated by the term $(1+\Upsilon_{2}^2)$, and hence the gap in achievable rates will appear as a shift that grows wider with larger number of antennas, and not a change in the slope. \\

\begin{figure}[!t]
\centering
\includegraphics[width=5.5in]{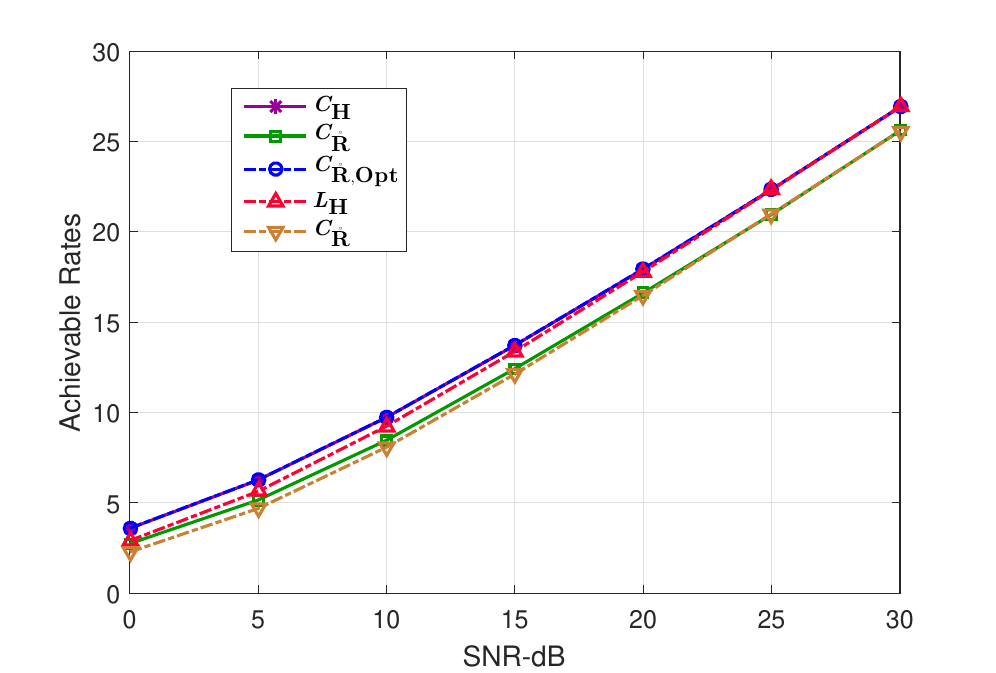}\vspace{-0.1in}
\caption{Achievable rates (bps/Hz) with $4\times4$ MIMO.}
\label{capfig:ch6_a}
\end{figure}
\begin{figure}[!t]
\centering
\includegraphics[width=5.5in]{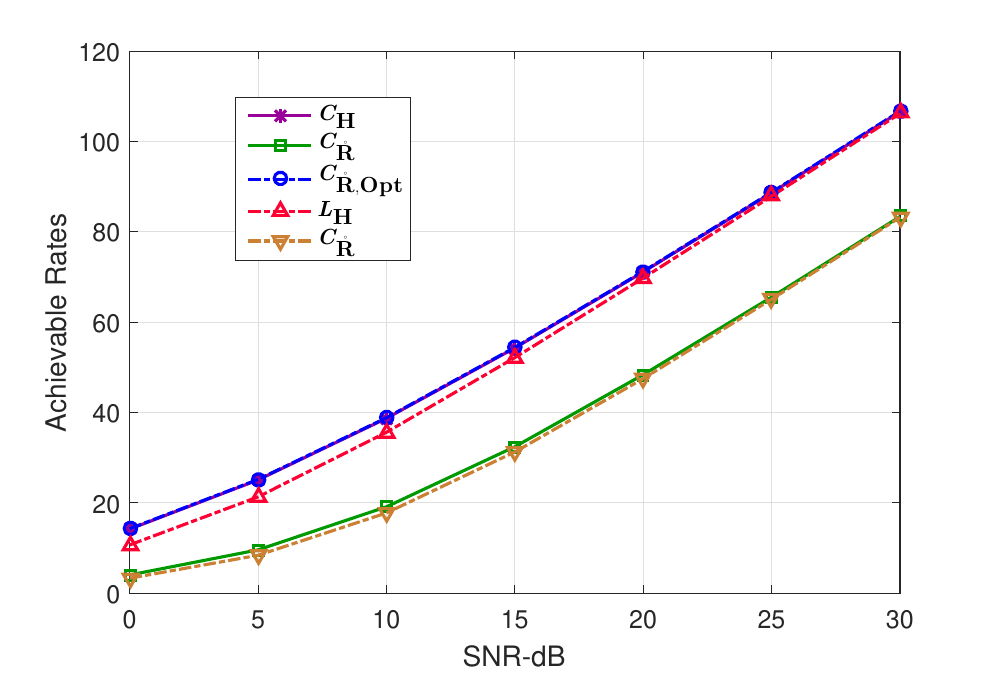}\vspace{-0.1in}
\caption{Achievable rates (bps/Hz) with $16\times16$ MIMO.}
\label{capfig:ch6_b}
\end{figure}
\begin{figure}[!t]
\centering
\includegraphics[width=5.5in]{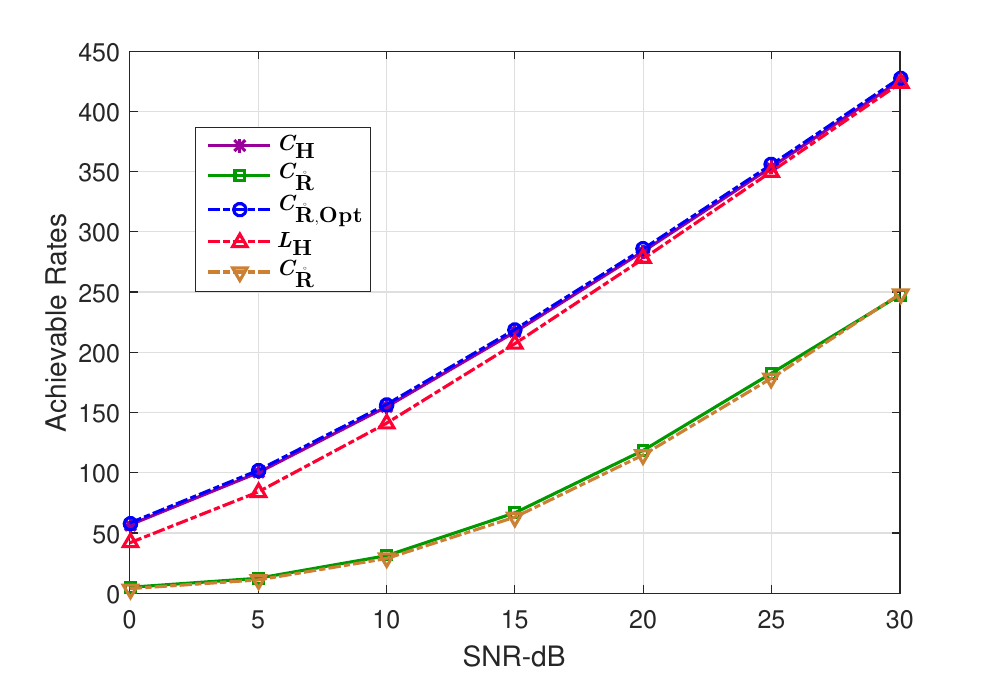}\vspace{-0.1in}
\caption{Achievable rates (bps/Hz) with $64\times64$ MIMO.}
\label{capfig:ch6_c}
\end{figure}

Figures \ref{capfig:ch6_a}, \ref{capfig:ch6_b}, and \ref{capfig:ch6_c} show the capacity plots alongside the bounds for $4\times4$, $16\times16$, and $64\times64$ MIMO, respectively. The plots verify the analysis: $C_{\mbf{H}}$ and $C_{\Rp,\Opt}$ are identical, the lower bounds are tight, and the gap is a shift that grows with MIMO order. Note that the studied capacities correspond to making use of the entire channel under a specific decomposition. It can be argued that with special layer ordering and layer-of-interest selection, such as with SSSD, or with optimized power allocation schemes, these gaps can be significantly reduced.

%
\section{Analysis of Achievable Diversity Gain}
\label{sec:ch6_diversity}

It is known that ML detection achieves full receive diversity $M$, and it can be shown that the N/C and PN/C detectors, being special cases of ZF with decision feedback, can only achieve a receive diversity gain of $1$. Moreover, it can be argued that both SSD (VSSD) and LORD also achieve full diversity, since they exploit the full channel matrix $\mbf{H}$ to compute distance metrics. In what follows, we study the achievable diversity gains of PML (PCD), SSSD, and SLORD.

%
\subsection{Punctured ML Detector / Punctured Chase Detector}
\label{sec:ch6_diversityCD}

To capture the diversity order of PML, we derive the pairwise error probability (PEP). Suppose that $\mbf{x}^{(1)}$ is transmitted, while $\mbf{x}^{(2)}$ is erroneously detected, the PEP can be expressed as
\begin{align}\label{eq:ch6_app2_1}
    \Prb(\mbf{x}^{(1)} \rightarrow \mbf{x}^{(2)}) & = \Prb\left(\norm{\mbf{W}^\mathcal{H}(\mbf{y} - \mbf{H}\mbf{x}^{(2)})}^{2} \leq \norm{\mbf{W}^\mathcal{H}(\mbf{y} - \mbf{H}\mbf{x}^{(1)})}^{2} \right)\\
    & = \Prb\left(\norm{\mbf{W}^\mathcal{H}\mbf{H}(\mbf{x}^{(1)} - \mbf{x}^{(2)}) + \mbf{W}^\mathcal{H}\mbf{n}}^{2} \leq \norm{\mbf{W}^\mathcal{H}\mbf{n}}^{2} \right) \\
    & = \Prb\left(\Re(\mbf{n}^\mathcal{H}\mbf{W}\Rp\mbf{d}) \geq \tfrac{1}{2}\norm{\Rp\mbf{d}}^{2} \right),
\end{align}
where $\mbf{d} \triangleq \mbf{x}^{(1)} - \mbf{x}^{(2)}$. Since $\mbf{n}$ consists of circular symmetric complex Gaussian random variables, then so is $\mbf{n}^\mathcal{H}\mbf{W}\Rp\mbf{d}$. It is easy to show that
\begin{align}\label{eq:ch6_app2_2}
    \mathsf{E}\left[\mbf{n}^\mathcal{H}\mbf{W}\Rp\mbf{d}\right] & = 0 \\
    \mathsf{E}\left[(\mbf{n}^\mathcal{H}\mbf{W}\Rp\mbf{d})(\mbf{n}^\mathcal{H}\mbf{W}\Rp\mbf{d})^\mathcal{H}\right] & = \mathsf{E}\left[\Tr(\Rp^\mathcal{H}\mbf{W}^\mathcal{H}\mbf{n}\mbf{n}^\mathcal{H}\mbf{W}\Rp\mbf{d}\mbf{d}^\mathcal{H})\right] \\
    &= \sigma^{2} \Tr\left( (\mbf{W}\Rp)^\mathcal{H}(\mbf{W}\Rp)\mbf{d}\mbf{d}^\mathcal{H} \right),
\end{align}
where $\Tr(\cdot)$ is introduced since $(\mbf{n}^\mathcal{H}\mbf{W}\Rp\mbf{d})(\mbf{n}^\mathcal{H}\mbf{W}\Rp\mbf{d})^\mathcal{H}$ is a scalar. Hence, we have
\begin{align}
    \Re(\mbf{n}^\mathcal{H}\mbf{W}\Rp\mbf{d}) &\sim \mathcal{N} \left( 0, \frac{\sigma^{2}}{2} \Tr \left( (\mbf{W}\Rp)^\mathcal{H}(\mbf{W}\Rp)\mbf{d}\mbf{d}^\mathcal{H} \right) \right) \\
    &= \mathcal{N} \left( 0, \frac{\sigma^{2}}{2} \norm{\mbf{W}\Rp\mbf{d}}^2 \right),
\end{align}
and therefore,
\begin{align}\label{eq:ch6_app2_5}
\Prb(\mbf{x}^{(1)} \rightarrow \mbf{x}^{(2)}) &= Q\left( \frac{\norm{\Rp\mbf{d}}^2}{\sqrt{2\sigma^{2}}\norm{\mbf{W}\Rp\mbf{d}} }  \right)\\
& \leq Q\left( \sqrt{\frac{\norm{\Rp\mbf{d}}^2}{2\sigma^{2}\norm{\mbf{W}}^2_{\F}}}  \right),
\end{align}
where the inequality holds since $\norm{\mbf{W}\Rp\mbf{d}}\!\leq\!\norm{\mbf{W}}_{\F}\norm{\Rp\mbf{d}}$ (Sec. 5.2 in \cite{meyer2000matrix}). Moreover, using union bound, we have
\begin{align}\label{eq:ch6_app2_6}
\Prb(\mbf{x}^{(1)}\!\rightarrow\!\mbf{x}^{(2)}) & \leq Q\left( \sqrt{\frac{\norm{\Rp\mbf{d}_{\min}}^2}{2\sigma^{2}\norm{\mbf{W}}^2_{\F}}}  \right) \\
&\leq \sum_{\mbf{d}\in \Omega,\mbf{d}\neq0}\!Q\left( \sqrt{\frac{\norm{\Rp\mbf{d}}^2}{2\sigma^{2}\!\norm{\mbf{W}}^2_{\F}}}  \right),
\end{align}
where $\Omega\triangleq\{\mbf{d}\!=\!\mbf{x}\!-\!\acute{\mbf{x}} \ |\  \mbf{x},\acute{\mbf{x}}\!\in\!\mathcal{X} \}$, and $\mbf{d}_{\min}\!=\!\argmin_{\mbf{d}\in \Omega,\mbf{d}\neq0} \norm{\Rp\mbf{d}}^2$. Finally, using the Chernoff bound, the average PEP is upper bounded as
\begin{align}\label{eq:ch6_app2_7}
\mathsf{E}\left[\Prb(\mbf{x}^{(1)} \rightarrow \mbf{x}^{(2)})\right] & \leq \sum_{\mbf{d}\in \Omega,\mbf{d}\neq0} \mathsf{E}\left[ \exp \left(-\frac{\norm{\Rp\mbf{d}}^2}{4\sigma^{2}\norm{\mbf{W}}^2_{\F}}\right)  \right] \\ & = \sum_{\mbf{d}\in \Omega,\mbf{d}\neq0} \mathsf{E}\left[ \exp \left(-\frac{\norm{\Rp\mbf{d}}^2}{4N\sigma^{2}}\right)  \right],
\end{align}
where $\norm{\mbf{W}}^2_{\F} = N$ since the columns of $\mbf{W}$ were normalized in~\eqref{eq:ch5_punc2}.

For regular ML detection \cite{Biglieri_2002,bai2014low,larsson2008space}, we have
\begin{align}\label{eq:ch6_app2_8}
\mathsf{E}\left[\Prb(\mbf{x}^{(1)} \rightarrow \mbf{x}^{(2)})\right] & \leq \sum_{\mbf{d}\in \Omega,\mbf{d}\neq0} \mathsf{E}\left[ \exp \left(-\frac{\norm{\mbf{H}\mbf{d}}^2}{4\sigma^{2}}\right)  \right] \\
& \leq \sum_{\mbf{d}\in \Omega,\mbf{d}\neq0} \det \left( \mbf{I}_N + \frac{\mbf{d}\mbf{d}^\mathcal{H}}{4\sigma^{2}} \right)^{-M},
\end{align}
where the expected value over the elements of $\mbf{H}$ results in full receive diversity $M$, because each column of $\mbf{H}$ contains $M$ independent Rayleigh distributed random variables, whose square is exponentially distributed. However, with $\Rp$ instead of $\mbf{H}$ in PML detection, the first $N\!-\!1$ columns have single diagonal elements, whose squares are chi-squared distributed with $4$ degrees of freedom, which corresponds to two exponentially distributed complex random variables, and hence a receive diversity order equal to $2$. Only column $N$ of $\Rp$ provides a diversity equal to $M$. Therefore, by analogy with \eqref{eq:ch6_app2_8}, the average PEP for the PML detector is
\begin{equation}\label{eq:ch6_app2_9}
\mathsf{E}\left[\Prb(\mbf{x}^{(1)} \rightarrow \mbf{x}^{(2)})\right] \leq \sum_{\mbf{d}\in \Omega,\mbf{d}\neq0} \det \left( \mbf{I}_N + \frac{\mbf{d}\mbf{d}^\mathcal{H}}{4N\sigma^{2}} \right)^{-2},
\end{equation}
and hence PML detection can not achieve a receive diversity gain of order greater than 2. However, noting that PML and PCD are identical (Sec.~\ref{sec:ch5_propCD}), and knowing that the regular CD achieves a receive diversity order of 2 (more on that in Sec.~\ref{sec:ch6_berCD}), we conclude that channel puncturing does not reduce the diversity gain of the CD.

%
\subsection{Symbol-Based Subspace Detector}
\label{sec:ch6_diversitySSSD}

To capture the diversity order of SSSD, we derive a modified PEP. Without loss of generality, we assume that layer $N$ is the root layer of interest. Hence, an error occurs when $x_N^{(1)}$ is transmitted and $x_N^{(2)}$ is erroneously detected, with probability
\begin{align}\label{eq:ch6_app3_1}
\Prb(x_N^{(1)} \rightarrow x_N^{(2)}) & \!=\!\Prb\left(\!\norm{\mbf{W}^\mathcal{H}\!\left(\mbf{y} \!-\! \mbf{H}_1\mbf{x}_1^{(2)}\! - \!\mbf{h}_N x_N^{(2)}\!\right)}^{2} \!\leq\! \norm{\mbf{W}^\mathcal{H}\!\left(\!\mbf{y}\! -\! \mbf{H}_1\mbf{x}_1^{(1)}\! -\! \mbf{h}_N x_N^{(1)}\!\right)}^{2} \right)\\
 & \!= \!\Prb\left(\!\norm{\!\mbf{W}^\mathcal{H}\mbf{H}_1\left(\!\mbf{x}_1^{(1)} \!-\! \mbf{x}_1^{(2)}\!\right) \!+ \!\mbf{W}^\mathcal{H}\mbf{h}_N\left(\!x_N^{(1)} \!-\! x_N^{(2)}\!\right) \!+ \!\mbf{W}^\mathcal{H}\mbf{n}\!}^{2} \!\leq\! \norm{\!\mbf{W}^\mathcal{H}\mbf{n}\!}^{2} \!\right) \\
 & \!=\! \Prb\left(\norm{\Rp_{1}\left(\mbf{x}_1^{(1)} \!-\! \mbf{x}_1^{(2)}\right) \!+\! \rrp_{N}\left(x_N^{(1)} \!-\! x_N^{(2)}\right)\! + \!\mbf{W}^\mathcal{H}\mbf{n}}^{2}\! \leq \!\norm{\mbf{W}^\mathcal{H}\mbf{n}}^{2} \right),
\end{align}
where $\mbf{y} = \mbf{H}_1\mbf{x}_1^{(1)} + \mbf{h}_N x_N^{(1)} + \mbf{n}$, $\mbf{x}_1^{(2)} = \hat{\mbf{x}}_{1}(x_N^{(2)})$ is computed as in Sec.~\ref{sec:ch5_propCD}, $\mbf{h}_N$ and $\rrp_{N}$ are the \nth{\emph{N}} column of $\mbf{H}$ and $\Rp$, and $\mbf{H}_1\in\mathcal{C}^{M\times (N-1)}$ and $\Rp_{1}\in\mathcal{C}^{M\times (N-1)}$ are the first $N-1$ columns of $\mbf{H}$ and $\Rp$, respectively. Let $\mathring{\mbf{\Delta}} = \Rp_{1} \left(\mbf{x}_1^{(1)} - \mbf{x}_1^{(2)}\right)$, and let $d = x_N^{(1)} - x_N^{(2)}$; we have
\begin{align}\label{eq:ch6_app3_2}
 \Prb\left(x_N^{(1)} \rightarrow x_N^{(2)}\right)
& \!= \Prb\left(\norm{\rrp_{N} d + \left(\mbf{W}^\mathcal{H}\mbf{n} + \mathring{\mbf{\Delta}}\right)}^{2} \leq \norm{\mbf{W}^\mathcal{H}\mbf{n}}^{2} \right)\notag \\
&\! =\! \Prb\left(\!\norm{\rrp_{N} d}^{2} \!\leq\! -2\Re\left(\!\left(\!\mbf{W}^\mathcal{H}\mbf{n} \!+\! \mathring{\mbf{\Delta}}\!\right)^\mathcal{H}\!\rrp_{N} d\!\right) \!-\! \norm{\!\mbf{W}^\mathcal{H}\mbf{n} \!+\! \mathring{\mbf{\Delta}}\!}^{2} + \norm{\mbf{W}^\mathcal{H}\mbf{n}}^{2} \!\right) \notag\\
 & \!\leq \Prb\left( -2\Re\left(\left(\mbf{W}^\mathcal{H}\mbf{n} + \mathring{\mbf{\Delta}}\right)^\mathcal{H}\rrp_{N} d\right) \geq  \norm{\rrp_{N} d}^{2} - \norm{\mbf{W}^\mathcal{H}\mbf{n}}^{2} \right).
\end{align}
Since $\mbf{n}$ and the columns of $\mbf{H}$ are circular symmetric complex Gaussian, then so is $\mbf{n}^\mathcal{H}\mbf{W}\rrp_{N} d + \mathring{\mbf{\Delta}}^\mathcal{H}\rrp_{N} d$. Thus, it can be shown that
\begin{equation}\label{eq:ch6_app3_3_1}
    \mathsf{E}\left[(\mbf{W}^\mathcal{H}\mbf{n} + \mathring{\mbf{\Delta}})^\mathcal{H}\rrp_{N} d\right] = 0
\end{equation}
\begin{equation}\label{eq:ch6_app3_3_2}
    \mathsf{E}\left[\left(\left(\mbf{W}^\mathcal{H}\mbf{n} + \mathring{\mbf{\Delta}}\right)^\mathcal{H}\rrp_{N} d\right)\left(\left(\mbf{W}^\mathcal{H}\mbf{n} + \mathring{\mbf{\Delta}}\right)^\mathcal{H}\rrp_{N} d\right)^\mathcal{H}\right] = \sigma^{2}\norm{\mbf{W}\rrp_{N} d}^2 + \norm{\mathring{\mbf{\Delta}}^\mathcal{H}\rrp_{N} d}^2
\end{equation}
\begin{equation}\label{eq:ch6_app3_3_3}
        \Re\left(\left(\mbf{W}^\mathcal{H}\mbf{n} + \mathring{\mbf{\Delta}}\right)^\mathcal{H}\rrp_{N} d\right) \sim \mathcal{N} \left( 0, \frac{\sigma^{2}}{2}\norm{\mbf{W}\rrp_{N}d}^2 + \frac{1}{2}\norm{\mathring{\mbf{\Delta}}^\mathcal{H}\rrp_{N}d}^2   \right).
\end{equation}
Hence, continuing from~\eqref{eq:ch6_app3_2}, we have
\begin{align}\label{eq:ch6_app3_6}
 \Prb\left(x_N^{(1)} \rightarrow x_N^{(2)}\right)
 & \leq Q\left( \frac{\norm{\rrp_{N} d}^{2} - \norm{\mbf{W}^\mathcal{H}\mbf{n}}^{2}} {\sqrt{2\sigma^{2}\norm{\mbf{W}\rrp_{N}d}^2 + 2\norm{\mathring{\mbf{\Delta}}^\mathcal{H}\rrp_{N}d}^2} }  \right) \\
 & = Q\left( \sqrt{\frac{ \norm{\rrp_{N} d}^{4} - 2\norm{\rrp_{N} d}^{2}\norm{\mbf{W}^\mathcal{H}\mbf{n}}^{2} +\norm{\mbf{W}^\mathcal{H}\mbf{n}}^{4} } {2\sigma^{2}\norm{\mbf{W}\rrp_{N}d}^2 + 2\norm{\mathring{\mbf{\Delta}}^\mathcal{H}\rrp_{N}d}^2 }}  \right) \\
 & \leq Q\left( \sqrt{\frac{ \norm{\rrp_{N} d}^{4} - 2\norm{\rrp_{N} d}^{2}\norm{\mbf{W}^\mathcal{H}\mbf{n}}^{2} } {\norm{\rrp_{N}d}^2 \left( 2\sigma^{2}\norm{\mbf{W}}^2_{\F} + 2\norm{\mathring{\mbf{\Delta}}}^2 \right)}}  \right) \\
 & = Q\left( \sqrt{\frac{ \norm{\rrp_{N} d}^{2} - 2\norm{\mbf{W}^\mathcal{H}\mbf{n}}^{2} } {2\sigma^{2}\norm{\mbf{W}}^2_{\F} + 2\norm{\mathring{\mbf{\Delta}}}^2}}  \right) \\
 & = Q\left( \sqrt{\frac{ d^2\norm{\rrp_{N}}^{2} - 2N\sigma^{2} } {2N\sigma^{2} + 2\norm{\mathring{\mbf{\Delta}}}^2}}  \right).
\end{align}
Then, using union and Chernoff bounds, with $\Phi\triangleq\{d\!=\!x\!-\!\acute{x} \ |\  x,\acute{x}\!\in\!\mathcal{M} \}$ ($\abs{d}^2\!=\!dd^\mathcal{H}$), the average PEP can be upper bounded as
\begin{align}\label{eq:ch6_app3_7}
\mathsf{E}\left[\Prb \left(x_N^{(1)} \rightarrow x_N^{(2)}\right)\right]
& \leq \sum_{d\in \Phi,d\neq0} \mathsf{E}\left[ \exp \left(-\frac{ \abs{d}^2\norm{\rrp_{N}}^{2} - 2N\sigma^{2} } {4N\sigma^{2} + 4\norm{\mathring{\mbf{\Delta}}}^2}\right)  \right] \\
& = \sum_{d\in \Phi,d\neq0} \mathsf{E}\left[ \exp \left(-\frac{ \abs{d}^2\norm{\rrp_{N}}^{2} } {4N\sigma^{2} + 4\norm{\mathring{\mbf{\Delta}}}^2}\right) \exp \left(\frac{ 2N\sigma^{2} } {4N\sigma^{2} + 4\norm{\mathring{\mbf{\Delta}}}^2}\right)  \right] \\
& \approx \sum_{d\in \Phi,d\neq0} \mathsf{E}\left[ \exp \left(-\frac{ \abs{d}^2\norm{\rrp_{N}}^{2} } {4N\sigma^{2} + 4\norm{\mathring{\mbf{\Delta}}}^2}\right) \right],
\end{align}
where the last approximation holds since the second exponential term is less that $\exp(1)$, with equality at high SNR ($\sigma^{2}\!=\!0)$. Finally, taking the expectation over all squared elements of $\rrp_{N}$, which are exponentially distributed, we obtain
\begin{align}\label{eq:ch6_app3_8}
\mathsf{E}\left[\Prb \left(x_N^{(1)} \rightarrow x_N^{(2)}\right)\right] & \leq \sum_{d\in \Phi,d\neq0} \prod_{l=1}^{M} \left( \frac{\abs{d}^2}{4N\sigma^{2} + 4\norm{\mathring{\mbf{\Delta}}}^2} +1 \right)^{-1} \\
& \leq \sum_{d\in \Phi,d\neq0} \left( \frac{\abs{d}^2}{4N\sigma^{2} + 4\norm{\mathring{\mbf{\Delta}}}^2} \right)^{-M}.
\end{align}
The denominator $4N\sigma^{2} \!+\! 4\norm{\mathring{\mbf{\Delta}}}^2$ represents noise plus interference, hence, SSSD appears to achieve a full receive diversity gain at the layer of interest when BERs are plotted in terms of signal-to-interference-plus-noise ratio (SINR). In the case of SLORD, following a similar derivation, the average PEP can be expressed as

\begin{equation}\label{eq:ch6_app3_9}
\mathsf{E}\left[\Prb \left(x_N^{(1)} \rightarrow x_N^{(2)}\right)\right] \leq \sum_{d\in \Phi,d\neq0}  \left( \frac{\abs{d}^2}{4N\sigma^{2} + 4\norm{\mbf{\Delta}}^2} \right)^{-M},
\end{equation}
where $\mbf{\Delta} = \mbf{R}_{1} \left(\mbf{x}_1^{(1)} - \mbf{x}_1^{(2)}\right)$, and $\mbf{R}_1\in\mathcal{C}^{M\times (N-1)}$ consists of the first $N-1$ columns of $\mbf{R}$. Note that $\mathring{\mbf{\Delta}}$ can be expressed as
\begin{equation}\label{eq:ch6_app3_10}
\mathring{\mbf{\Delta}} = \left[ \rp_{11}\left(x_1^{(1)}-x_1^{(2)}\right),\ \ldots \ , \rp_{(N\!-\!1)(N\!-\!1)}\left(x_{N\!-\!1}^{(1)}-x_{N\!-\!1}^{(2)}\right), 0\right]^T,
\end{equation}
and consequently, $\norm{\mathring{\mbf{\Delta}}}^2$ is upper bounded by
\begin{equation}\label{eq:ch6_app3_11}
\norm{\mathring{\mbf{\Delta}}}^2_{\max} = \rp_{11}^2\varrho^2 + \ \cdots \ + \rp_{(N\!-\!1)(N\!-\!1)}^2\varrho^2,
\end{equation}
when a one-bit slicing error (assuming Gray mapping) occurs on all upper layers, with $\varrho = 2/\log(2^q)$ for a $2^q$-QAM constellation. Since the expected values of the chi-squared-distributed square of diagonal elements in $\mbf{R}$ are greater than those in $\Rp$ (expected value equals degrees of freedom), we have $\norm{\mbf{\Delta}}^2 > \norm{\mathring{\mbf{\Delta}}}^2$, and hence SSSD outperforms SLORD. Furthermore, with higher order constellations, $\norm{\mathring{\mbf{\Delta}}}^2_{\max}$ is significantly reduced, boosting the performance of SSSD.

%
\section{Characterization and Analysis of BER}
\label{sec:ch6_probBER}

%
\subsection{Punctured N/C Detector}
\label{sec:ch6_berSIC}
Let $P_n^{}(r_{nn}^{})$ and $\acute{P}_n^{}(\rp_{nn}^{})$ be the probabilities of bit error conditioned on $r_{nn}$ and $\rp_{nn}$, when detecting $x_n$ ($1\!\leq\!n\!\leq\!N$), for N/C and PN/C, respectively. Consider the PN/C detector, and assume as in \cite{choi2006nulling} normalized binary phase shift keying (BPSK), where $\mathcal{M}=\{-1,1\}$, we have
\begin{equation}\label{eq:ch6_qfun1}
    \acute{P}_N^{}(\rp_{NN}^{}) = Q \left( \sqrt{ \frac{ 2\rp_{NN}^2 }{ \sigma^{2} } } \right)
\end{equation}
at layer $N$. At the remaining layers, if the cancellation at layer $N$ is correct, we get $\acute{P}_n(\rp_{nn}) = Q \left( \sqrt{ \frac{ 2\rp_{nn}^2 }{ \sigma^{2} } } \right)$. Otherwise, we have
\begin{equation}\label{eq:ch6_cancellation}
\bar{y}_{n}\!-\! \rp_{nN}^{}\hat{x}_{N}^{\PNC} = \rp_{nn}x_{n} + \rp_{nN}^{}(x_{N}^{} - \hat{x}_{N}^{\PNC}) + w_n.
\end{equation}
Noting that $x_{N}^{}\!-\!\hat{x}_{N}^{\PNC}\!=\!\pm 2$, the variance of interference plus noise is $\sigma^{2}\!+\!4$, and hence the BER becomes $\acute{P}_n(\rp_{nn}) = Q \left(  \sqrt{ \frac{ 2\rp_{nn}^2 }{ \sigma^{2} + 4 } } \right) $. Thus, the resultant BER for detecting $x_n$ can be written as
\begin{equation}\label{eq:ch6_berPSIC}
     \acute{P}_n(\rp_{nn}) \!=\!  Q \left( \!\sqrt{ \frac{ 2\rp_{nn}^2 }{ \sigma^{2} } } \right) \left(1\!-\!\acute{P}_N^{}(\rp_{NN}^{})\right) \!+\! Q \left(\! \sqrt{ \frac{ 2\rp_{nn}^2 }{ \sigma^{2} \!+\! 4 } } \right) \acute{P}_N^{}(\rp_{NN}^{}).
\end{equation}
The BER for N/C with regular QRD at layer $n$ ($n\!=\!N\!-\!1,\cdots,1$) is \cite{choi2006nulling}
\begin{align}\label{eq:ch6_berSIC}
& P_n(r_{nn}) = \sum_{\psi_{n+1}\in D_{n+1}} P_n(\err|r_{nn},\psi_{n+1})P_{n+1}(\psi_{n+1}) \\
& P_n(\err|r_{nn},\psi_{n+1}) = Q \left( \sqrt{ \frac{ 2r_{nn}^2 }{ \sigma^{2} + 4\psi_{n+1}\psi_{n+1}^{T} } } \right),
\end{align}
where $P_{n+1}(\psi_{n+1})$ can be computed recursively, and $\psi_n$ is an instance of $D_n$, the set of all possible error patterns leading to layer $n$, which are represented as binary vectors with $1$ in the place of incorrect layer detection:
\begin{equation}\label{eq:ch6_Dn}
     D_n = \{[\underbrace{0 0 \cdots 0}_{N-n+1}],[\underbrace{0 0 \cdots 1}_{N-n+1}],\cdots[\underbrace{1 1 \cdots 1}_{N-n+1}] \}.
\end{equation}
Note that with WRD, we have a smaller set $\acute{D}_n = \{[0 0 \cdots 0],[0 0 \cdots 1] \} \subset D_n$, where $\abs{\acute{D}_n} = 2 < \abs{D_n} = 2^{N-n+1}$. Therefore, error propagation is largely reduced.

However, having fewer terms in the BER formula does not mean a better BER performance. The average BER at layer $n$ is obtained by taking the expectation over $r_{nn}^2$ and $\rp_{nn}^2$. Since $\rp_{nn}^2$ has smaller values than $r_{nn}^2$ at layers $1\!\leq\!n\!\leq\!N\!-\!2$, and since $Q(\cdot)$ is a monotonically decreasing function, PN/C will result in performance degradation. But layer $N$ has the worst performance, despite not being affected by noise coloring, since more errors occur at this layer due to low array gain, and they get propagated to higher layers. Therefore, the performance of both N/C and PN/C detectors will be dominated by $\acute{P}_N^{}(\rp_{NN}^{})=P_N^{}(r_{NN}^{})$, and all computational savings in the proposed PN/C detector come at a negligible cost.

In equation form, the function $G(d,\gamma)$ provides the average BER over a $d$-fold diversity Rayleigh fading channel with mean branch SNR $\gamma$:
\begin{equation}\label{eq:ch6_ber_avg_2}
    G \left( d, \gamma\right) = \left[ \frac{1}{2} \left( 1-\mu\right)\right]^d \sum_{k=0}^{d-1} \left( \begin{array}{c} d-1+k \\ k \end{array} \right)\left[ \frac{1}{2} \left( 1+\mu\right) \right]^k,
\end{equation}
where $\mu=\sqrt{\gamma/(1+\gamma)}$. The average BER with PN/C at layers $1\!\leq\!n\!\leq\!N\!-\!1$ is thus expressed as
\begin{equation}\label{eq:ch6_ber_avg_1}
     \acute{P}_n =  G \left( 2, \frac{1}{\sigma^{2}} \right) (1-\acute{P}_N) + G \left( 2, \frac{1}{\sigma^{2} + 4} \right) \acute{P}_N,
\end{equation}
where layers $1\!\leq\!n\!\leq\!N\!-\!1$ only provide a $2$-fold diversity due to puncturing. Similarly, the average BERs at layer $n<N$ for N/C can be obtained by replacing $P_n(\err|r_{nn},\psi_{n+1})$ in~\eqref{eq:ch6_berSIC} by its average over $r_{nn}$, $P_n(\err|\psi_{n+1})$, where
\begin{align}\label{eq:ch6_ber_avg_3}
    P_n(\err|\psi_{n+1}) &= \mathsf{E}[P_n(\err|r_{nn},\psi_{n+1})] \\
    & = G \left( N-n+1, \frac{1}{\sigma^{2} + 4\psi_{n+1}\psi_{n+1}^{T}} \right),
\end{align}
with the numerator in $\gamma$ being 1 because we assume normalization. We have
\begin{align}\label{eq:ch6_ber_avg_4}
    \acute{P}_N^{}(\rp_{NN}^{}) = P_N^{}(r_{NN}^{}) &=  G(1,1/\sigma^{2}) \\
    & = \frac{1}{2} \left( 1 - \sqrt{\frac{1/\sigma^{2}}{1+1/\sigma^{2}}} \right).
\end{align}
Note that these equations can be extended to an arbitrary constellation size by expressing the function $G(d,\gamma)$ as $G(d,\gamma,\abs{\mathcal{M}})$. The function $G(d,\gamma,2^q)$ provides the average BER of $2^q$-QAM over a $d$-fold diversity Rayleigh fading channel with mean branch SNR $\gamma$ \cite{Kim_1997,Kim_2008}:
\begin{align}\label{eq:ch6_app1_1}
    & G(d,\gamma,2^q) = \frac{\sqrt{2^q}-1}{2^q} \left[ \left(\sqrt{2^q}-1\right) + 4I_1 - \left(\sqrt{2^q}-1\right)I_2 \right]\\
    & I_1 = \left[ \frac{1}{2} \left( 1-\mu\right)\right]^d \sum_{k=0}^{d-1} \left( \begin{array}{c} d-1+k \\ k \end{array} \right)\left[ \frac{1}{2} \left( 1+\mu\right) \right]^k\\
& I_2=\begin{cases}
                \displaystyle  \ \frac{4}{\pi} \ \mu \tan^{-1}\mu  \ \ \ \ \text{for} \ \ d = 1 \\~\\
                \vspace{+0.05in}
                \displaystyle  \ \frac{4}{\pi}  \sum_{k=0}^{d-1} \frac{(2k)!}{2^{2k}(k!)^2} \bigg[ \left(\frac{1}{1+\bar{\varrho}\gamma}\right)^k \mu \tan^{-1}\mu \bigg] \\ \vspace{+0.05in}
                \displaystyle \ + \frac{2}{\pi}  \sum_{k=1}^{d-1} \frac{(2k)!}{2^{2k}(k!)^2} \bigg[ \sum_{v=1}^{k} \frac{2^{2v}v!}{(2v)!} \left(\frac{1}{1+\bar{\varrho}\gamma}\right)^{k-v+1} \\
                \displaystyle \ \times \left(\frac{\bar{\varrho}\gamma}{1+2\bar{\varrho}\gamma}\right) (v-1)! \left(\frac{1}{1+2\bar{\varrho}\gamma}\right)^{v-1} \bigg]  \ \ \ \ \text{for} \ \ d \geq 2
            \end{cases} \\
& \mu=\sqrt{\frac{\bar{\varrho}\gamma}{1+\bar{\varrho}\gamma}}, \ \ \ \ \ \ \bar{\varrho} = \frac{3\log(2^q)}{2(2^q-1)}
\end{align}
When using $G(d,\gamma,2^q)$ in \eqref{eq:ch6_berSIC} and \eqref{eq:ch6_berPSIC}, the constant term 4, which represents the variance of the error caused by wrong decisions on lower layers, should be modified. With $2^q$-QAM, and assuming normalized constellations, an error at a lower layer will more likely incur a noise of variance $\varrho^2=(2/\log(2^q))^2$ at upper layers.

%
\subsection{Punctured Chase Detector}\label{sec:ch6_berCD}
To capture the performance of the PCD, we follow a probabilistic approach similar to that in \cite{radji2009interference}. Denote by $\hat{\mbf{x}}^{\ML}$, $\hat{\mbf{x}}^{\CD}$, and $\hat{\mbf{x}}^{\PCD}$, the vector outputs, and by $P^{\ML}$, $P^{\CD}$, and $P^{\PCD}$, the vector error rates, of the ML detector, the CD, and the PCD, respectively (vector error rates and BERs are related by a scaling factor). We start by the case of a regular CD; we have
\begin{equation}\label{eq:ch6_prob1}
    P^{\CD} = \mathsf{E}_{\mbf{R},\mbf{x}} \bigg[\Prb\left(\hat{\mbf{x}}^{\CD}\neq\mbf{x}~|~\mbf{R},\mbf{x}\right)\bigg].
\end{equation}
For clarity of presentation, we drop the expectation operator in what follows. Clearly, $P^{\ML}<P^{\CD}$ is a lower bound; we seek a tight upper bound. We further have
\begin{align}\label{eq:ch6_prob2}
  P^{\CD} & = \Prb(\hat{\mbf{x}}^{\CD}\neq\mbf{x}~|~\mbf{R},\mbf{x}) \notag\\
  & = \Prb(\hat{\mbf{x}}^{\CD}\neq\mbf{x}~|~\mbf{R},\mbf{x},\hat{\mbf{x}}^{\ML}\neq\mbf{x})\Prb(\hat{\mbf{x}}^{\ML}\neq\mbf{x}~|~\mbf{R},\mbf{x}) \notag\\
  & + \Prb(\hat{\mbf{x}}^{\CD}\neq\mbf{x}~|~\mbf{R},\mbf{x},\hat{\mbf{x}}^{\ML}=\mbf{x})\Prb(\hat{\mbf{x}}^{\ML}=\mbf{x}~|~\mbf{R},\mbf{x}).
\end{align}
The first term in~\eqref{eq:ch6_prob2} is upper bounded by $P^{\ML}=\Prb(\hat{\mbf{x}}^{\ML}\neq\mbf{x}~|~\mbf{R},\mbf{x})$. To simplify the second term, we expand
\begin{align}\label{eq:ch6_prob3}
  & \Prb(\hat{\mbf{x}}^{\CD}\neq\mbf{x}~|~\mbf{R},\mbf{x},\hat{\mbf{x}}^{\ML}=\mbf{x}) \\
  & = \Prb(~\hat{\mbf{x}}^{\CD}\neq\mbf{x}~|~\mbf{R},\mbf{x},\hat{\mbf{x}}^{\ML}=\mbf{x},\mbf{x}\in \mathcal{S}(\tilde{\mbf{y}},\mbf{R})~) \times \Prb(\mbf{x}\in \mathcal{S}(\tilde{\mbf{y}},\mbf{R})~|~\mbf{R},\mbf{x},\hat{\mbf{x}}^{\ML}=\mbf{x}) \\
  & + \Prb(~\hat{\mbf{x}}^{\CD}\neq\mbf{x}~|~\mbf{R},\mbf{x},\hat{\mbf{x}}^{\ML}=\mbf{x},\mbf{x}\not\in \mathcal{S}(\tilde{\mbf{y}},\mbf{R})~) \times \Prb(\mbf{x}\not\in \mathcal{S}(\tilde{\mbf{y}},\mbf{R})~|~\mbf{R},\mbf{x},\hat{\mbf{x}}^{\ML}=\mbf{x}).
\end{align}
We can note that
\begin{align}\label{eq:ch6_prob4}
& \Prb(~\hat{\mbf{x}}^{\CD}\neq\mbf{x}~|~\mbf{R},\mbf{x},\hat{\mbf{x}}^{\ML}=\mbf{x},\mbf{x} \in \mathcal{S}(\tilde{\mbf{y}},\mbf{R})~) \triangleq P^{\A} = 0,~\text{and} \\
& \Prb(~\hat{\mbf{x}}^{\CD}\neq\mbf{x}~|~\mbf{R},\mbf{x},\hat{\mbf{x}}^{\ML}=\mbf{x},\mbf{x} \not\in \mathcal{S}(\tilde{\mbf{y}},\mbf{R})~) = 1.
\end{align}
Hence, substituting back in~\eqref{eq:ch6_prob3}, we get
\begin{equation}\label{eq:ch6_prob5}
  \Prb(\hat{\mbf{x}}^{\CD}\!\neq\!\mbf{x} |~\mbf{R},\mbf{x},\hat{\mbf{x}}^{\ML}\!=\!\mbf{x})
  = \Prb(\mbf{x}\!\not\in\! \mathcal{S}(\tilde{\mbf{y}},\!\mbf{R})|~\mbf{R},\mbf{x},\hat{\mbf{x}}^{\ML}\!=\!\mbf{x}) \!\triangleq\! P^{\B}.
\end{equation}
Substituting back in~\eqref{eq:ch6_prob2}, the second term is upper bounded by $P^{\B}$, which effectively is equivalent to the probability that the generated list does not contain the true vector. But since we exhaustively search over all possible values of $\hat{x}_N^{}$ in $\mathcal{M}$, this probability will in effect be the probability of error in SIC on upper layers. Therefore, we have
\begin{equation}\label{eq:ch6_boundsCD}
P^{\ML}<P^{\CD}<P^{\ML}+P^{\B}.
\end{equation}

\begin{figure}[t]
\centering
\includegraphics[width=6in]{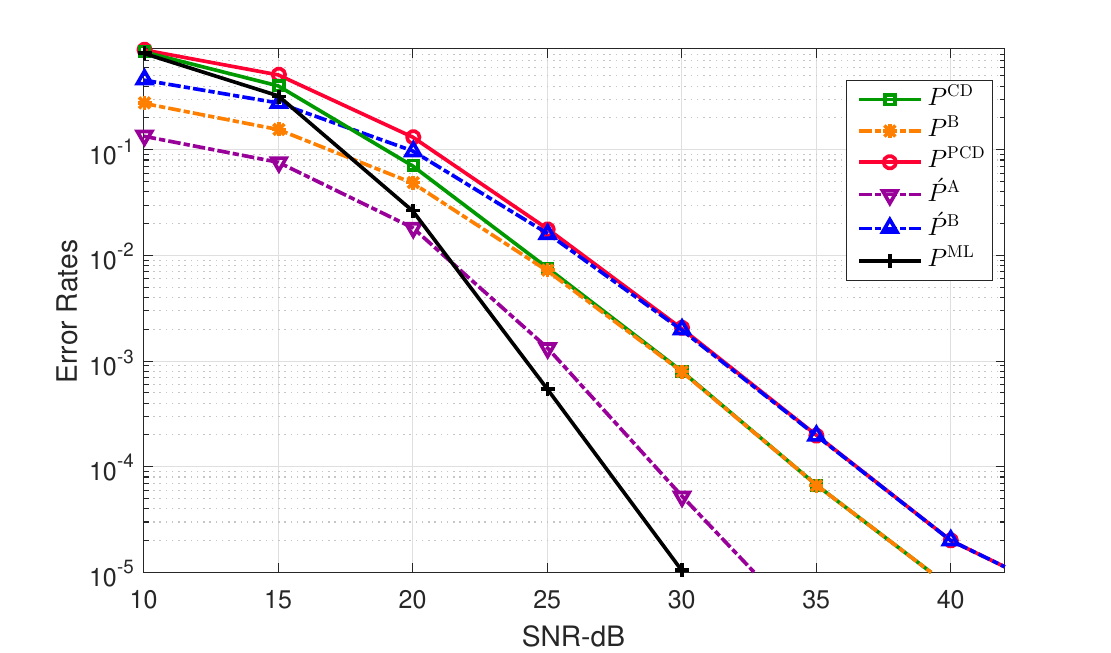}\vspace{-0.1in}
\caption{Empirical error probabilities with CD - $4\times4$ MIMO - $16$-QAM.}
\label{perffig:ch6_a}
\end{figure}

Similarly, we can derive the bounds for $P^{\PCD}$ by expanding $\Prb(\hat{\mbf{x}}^{\PCD}\neq\mbf{x}~|~\Rp,\mbf{x})$ and substituting $\mathcal{S}(\tilde{\mbf{y}},\mbf{R})$ with $\mathcal{P}(\bar{\mbf{y}},\Rp)$. However, the modified $P^{\A}$ and $P^{\B}$, $\acute{P}^{\A}$ and $\acute{P}^{\B}$, will evaluate differently:
\begin{align}\label{eq:ch6_prob6}
& \acute{P}^{\A}\triangleq\Prb(~\hat{\mbf{x}}^{\PCD}\!\neq\!\mbf{x}~|~\Rp,\mbf{x},\hat{\mbf{x}}^{\ML}\!=\!\mbf{x},\mbf{x} \!\in \!\mathcal{P}(\bar{\mbf{y}},\Rp)~) \neq 0 \\
& \acute{P}^{\B}\triangleq \Prb(\mbf{x}\not\in \mathcal{P}(\bar{\mbf{y}},\Rp)~|~\Rp,\mbf{x},\hat{\mbf{x}}^{\ML}=\mbf{x}) > P^{\B}.
\end{align}
Note that $\acute{P}^{\A}$ is not zero, because even if $\hat{\mbf{x}}^{\ML}$ is within the generated list, the modified distance metric might not select it as the HO vector. Therefore, the bounds for the PCD are
\begin{equation}\label{eq:ch6_boundsPCD}
P^{\ML}<P^{\PCD}<P^{\ML}+\acute{P}^{\B}+\xi,
\end{equation}
where $\xi=\acute{P}^{\A}(1-\acute{P}^{\B})(1-P^{\ML})$. As shown in Fig.~\ref{perffig:ch6_a}, the dominant factors that affect the performances of $P^{\CD}$ and $P^{\PCD}$ at high SNR are $P^{\B}$ and $\acute{P}^{\B}$, respectively. Thus, $\acute{P}^{\A}$, and hence $\xi$, can be safely neglected. Nevertheless, since SIC over a punctured channel matrix is more prone to errors, we have $\acute{P}^{\B} > P^{\B}$, and accordingly $P^{\PCD} > P^{\CD}$. But layers with smallest degrees of freedom (equal to 4) dominate the BER performance, and they exist in both the CD and the PCD. Therefore, the performance gap will be in the form of a shift, and there is no loss in diversity gain.

An alternative approximate approach to study the BER performance of the CD starting from N/C exists \cite{waters2008chase}, where the gain of generating a list by considering more candidate symbols at layer $N$ can be seen as an effective SNR gain at that layer. The gain factor is given by
\begin{figure}[!t]
\centering
\includegraphics[width=6in]{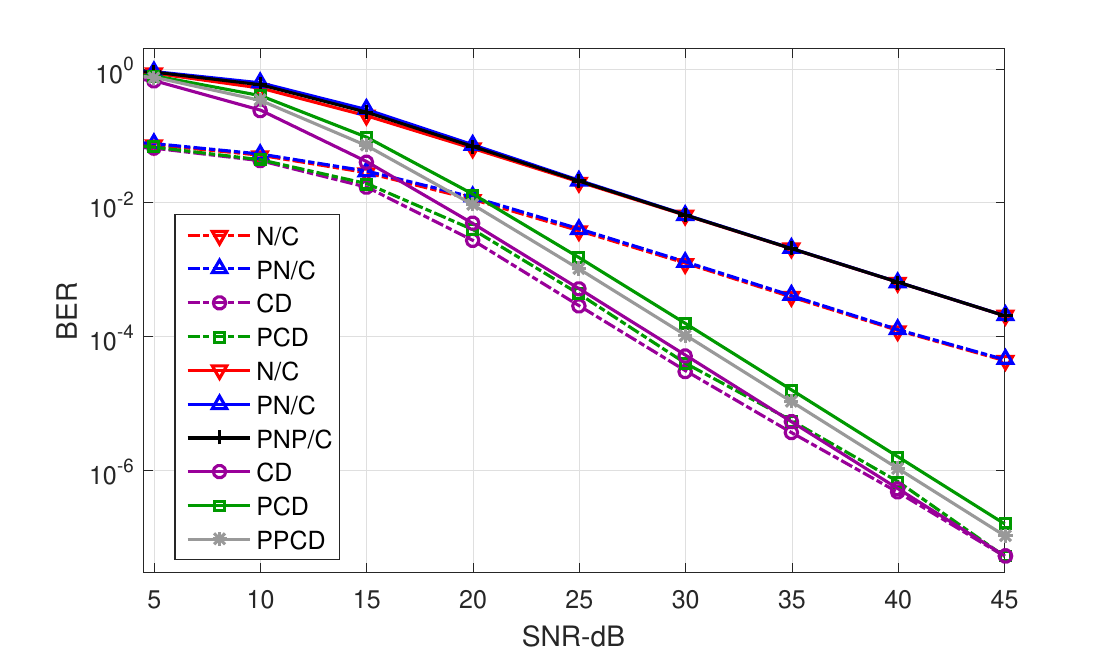}\vspace{-0.1in}
\caption{Theoretical (solid lines) vs simulated BERs - $4\!\times\!4$ MIMO - $16$-QAM.}
\label{theofig:ch6_a}
\end{figure}
\begin{figure}[!t]
\centering
\includegraphics[width=6in]{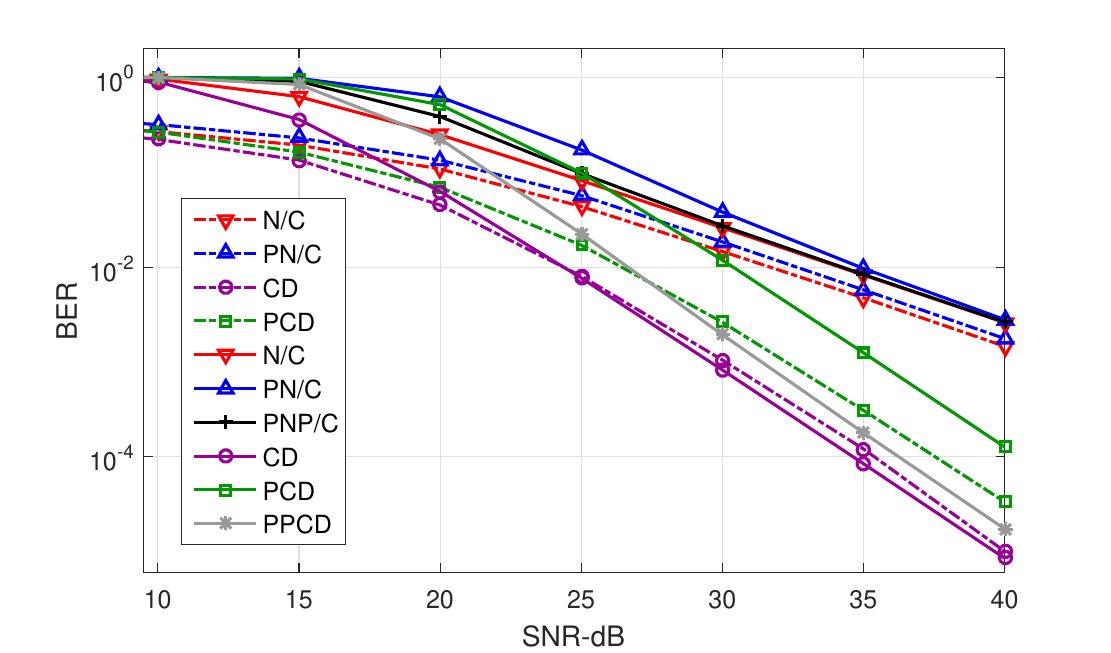}\vspace{-0.1in}
\caption{Theoretical (solid lines) vs simulated BERs - $16\!\times\!16$ MIMO - $16$-QAM.}
\label{theofig:ch6_b}
\end{figure}
\begin{equation}\label{eq:ch6_SNRgain}
\Gamma = \frac{\delta_{Ls}}{\delta_N^{}},
\end{equation}
where $\delta_N^{}$ is the distance between the transmitted symbol and the nearest decision boundary in $\mathcal{M}$, and $\delta_{Ls}$ is the distance to the nearest decision boundary of the list (an error occurs when $x_N^{}$ is not in the list). Since in our case the list is the entirety of $\mathcal{M}$, the SNR gain is $\Gamma=\infty$. In other words, no error is propagated from layer $N$, which was the limiting layer in N/C. Therefore, the BER of CD and PCD can be obtained by summing the combinations of BERs on all layers $n<N$, which are computed using~\eqref{eq:ch6_berSIC} and \eqref{eq:ch6_berPSIC}, respectively, while setting $\acute{P}_N^{}(\rp_{NN}^{}) = P_N^{}(r_{NN}^{}) = 0$ in these equations. Hence, we have
\begin{equation}\label{eq:ch6_ber_PCD}
     P^{\PCD} =  (N-1) \left[ G \left( 2, \frac{1}{\sigma^{2}} \right) \right].
\end{equation}
Note here that $P^{\PCD}$ stands for BER instead of vector error rate.

Figures \ref{theofig:ch6_a} and \ref{theofig:ch6_b} show the theoretical and simulated BERs with $16$-QAM, for $4\times4$ and $16\times16$ MIMO systems. Note that the CD and PCD theoretical curves are not true upper bounds on error performance, but rather close approximations \cite{waters2008chase}. The gap between punctured and unpunctured schemes increases with the number of antennas. However, this gap can be reduced by employing partial puncturing, at the expense of less computational savings. For example, the partially-punctured N/C (PPN/C) detector and the partially-punctured CD (PPCD) correspond to the case where the entries above the diagonal in the third column are not zeroed-out during puncturing. In the remainder of this thesis, we always consider full puncturing that results in maximal complexity reduction.

%
\subsection{Symbol-Based Subspace Detector}
\label{sec:ch6_berSSsb}

\begin{figure}[t]
\centering
\includegraphics[width=6in]{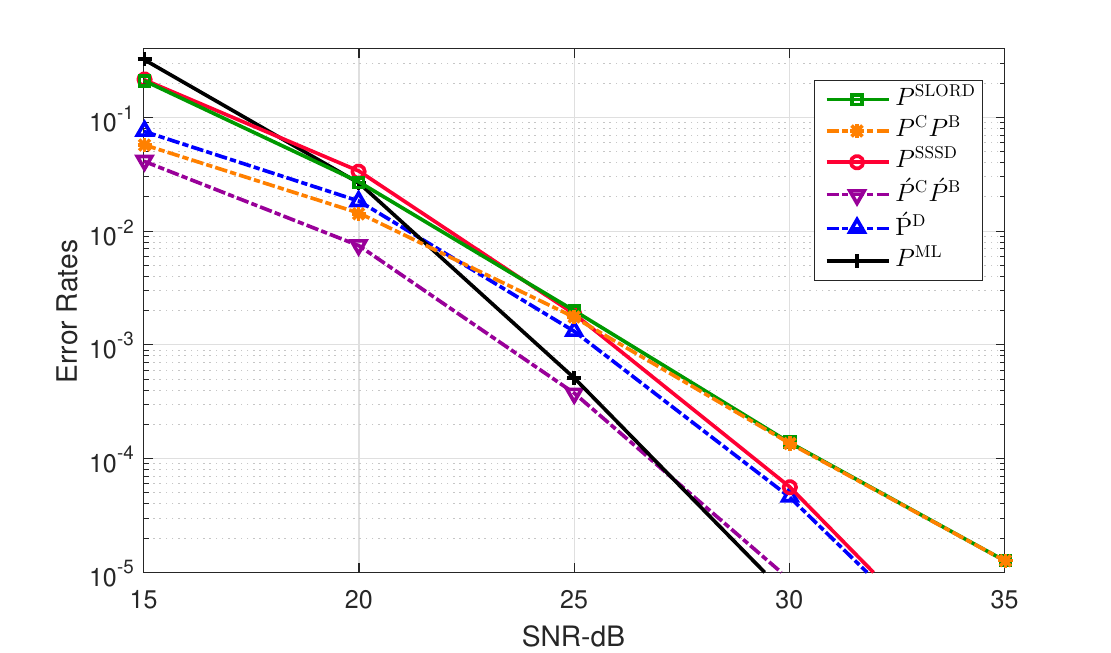}\vspace{-0.1in}
\caption{ Empirical error probabilities with SSD - $4\times4$ MIMO - $16$-QAM.}
\label{perffig:ch6_b}
\end{figure}

Denote by $P^{\SLORD}$ and $P^{\SSSD}$ the vector error rates, of SSSD and SLORD, respectively. We start by the case of SLORD. Noting that the performance can be captured from one partition, say $t\!=\!1$, we investigate:
\begin{align}\label{eq:ch6_prob22}
  & \Prb(\hat{x}^{\SLORD}_N\neq x_N^{}~|~\mbf{R},\mbf{x}) \\
  & = \Prb(\hat{x}^{\SLORD}_N\neq x_N^{}~|~\mbf{R},\mbf{x},\hat{\mbf{x}}^{\ML}\neq\mbf{x})\Prb(\hat{\mbf{x}}^{\ML}\neq\mbf{x}~|~\mbf{R},\mbf{x}) \\
  & + \Prb(\hat{x}^{\SLORD}_N\neq x_N^{}~|~\mbf{R},\mbf{x},\hat{\mbf{x}}^{\ML}=\mbf{x})\Prb(\hat{\mbf{x}}^{\ML}=\mbf{x}~|~\mbf{R},\mbf{x})
\end{align}
\begin{align}\label{eq:ch6_prob32}
  & \Prb(\hat{x}^{\SLORD}_N\neq x_N^{}~|~\mbf{R},\mbf{x},\hat{\mbf{x}}^{\ML}=\mbf{x}) \\
  &  = \Prb(~\hat{x}^{\SLORD}_N\neq x_N^{}~|~\mbf{R},\mbf{x},\hat{\mbf{x}}^{\ML}=\mbf{x},\mbf{x}\in \mathcal{S}(\tilde{\mbf{y}},\mbf{R})~) & \\
  & \times \Prb(\mbf{x}\in \mathcal{S}(\tilde{\mbf{y}},\mbf{R})~|~\mbf{R},\mbf{x},\hat{\mbf{x}}^{\ML}=\mbf{x}) \\
  & + \Prb(~\hat{x}^{\SLORD}_N\neq x_N^{}~|~\mbf{R},\mbf{x},\hat{\mbf{x}}^{\ML}=\mbf{x},\mbf{x}\not\in \mathcal{S}(\tilde{\mbf{y}},\mbf{R})~) \\
  & \times \Prb(\mbf{x}\not\in \mathcal{S}(\tilde{\mbf{y}},\mbf{R})~|~\mbf{R},\mbf{x},\hat{\mbf{x}}^{\ML}=\mbf{x}).
\end{align}
We define:
\begin{align}\label{eq:ch6_prob42}
& \Prb(~\hat{x}^{\SLORD}_N\neq x_N^{}~|~\mbf{R},\mbf{x},\hat{\mbf{x}}^{\ML}=\mbf{x},\mbf{x} \in \mathcal{S}(\tilde{\mbf{y}},\mbf{R})~) ~\triangleq~ P^{\D} = 0 \\
& \Prb(~\hat{x}^{\SLORD}_N\neq x_N^{}~|~\mbf{R},\mbf{x},\hat{\mbf{x}}^{\ML}=\mbf{x},\mbf{x} \not\in \mathcal{S}(\tilde{\mbf{y}},\mbf{R})~) ~\triangleq~ P^{\C} \neq 1\\
& \Prb(\mbf{x}\not\in \mathcal{S}(\tilde{\mbf{y}},\mbf{R})~|~\mbf{R},\mbf{x},\hat{\mbf{x}}^{\ML}=\mbf{x}) = P^{\B},
\end{align}
and substitute back in~\eqref{eq:ch6_prob32} to get
\begin{equation}\label{eq:ch6_prob52}
  \Prb(\hat{x}^{\SLORD}_N\neq x_N^{}~|~\mbf{R},\mbf{x},\hat{\mbf{x}}^{\ML}=\mbf{x}) = P^{\C}P^{\B}.
\end{equation}
Following the same argument as in Sec.\ref{sec:ch6_berCD} we have.
\begin{align}\label{eq:ch6_boundsSLORD}
& P^{\ML}<P^{\SLORD}<P^{\ML}+P^{\C}P^{\B}(1-P^{\ML})\\
& P^{\ML}<P^{\SLORD}<P^{\ML}+P^{\C}P^{\B}
\end{align}
Similarly, we can derive the bounds for $P^{\SSSD}$ by expanding $\Prb(\hat{x}^{\SSSD}_N\neq x_N^{}~|~\Rp,\mbf{x})$. We define
\begin{align}\label{eq:ch6_prob43}
& \Prb(~\hat{x}^{\SSSD}_N\neq x_N^{}~|~\Rp,\mbf{x},\hat{\mbf{x}}^{\ML}=\mbf{x},\mbf{x}\!\in \mathcal{P}(\bar{\mbf{y}},\Rp)~) = \acute{P}^{\D} \neq 0 \\
& \Prb(~\hat{x}^{\SSSD}_N\neq x_N^{}~|~\Rp,\mbf{x},\hat{\mbf{x}}^{\ML}=\mbf{x},\mbf{x} \not\in \mathcal{P}(\bar{\mbf{y}},\Rp)~) = \acute{P}^{\C} \neq 1 \\
& \Prb(\mbf{x}\not\in \mathcal{P}(\bar{\mbf{y}},\Rp)~|~\Rp,\mbf{x},\hat{\mbf{x}}^{\ML}=\mbf{x}) = \acute{P}^{\B},
\end{align}
where following the same procedure we get
\begin{align}\label{eq:ch6_boundsSLORD2}
& P^{\ML}<P^{\SSSD}<P^{\ML}+(~\acute{P}^{\C}\acute{P}^{\B}+\acute{P}^{\D}(1-\acute{P}^{\B})~)(1-P^{\ML}) \\
& P^{\ML}<P^{\SSSD}<P^{\ML}+\acute{P}^{\C}\acute{P}^{\B}+\acute{P}^{\D}.
\end{align}
Therefore, the dominant factor that affects $P^{\SLORD}$ at high SNR is $P^{\C}P^{\B}$, and the factors that affect $P^{\SSSD}$ are $\acute{P}^{\C}\acute{P}^{\B}$ and $\acute{P}^{\D}$ (we can show that $\acute{P}^{\D}$ is identical to $\acute{P}^{\A}$ from~\eqref{eq:ch6_prob6}). Note that $P^{\C}$ can be expressed as the probability that an error occurs at layer $N$ in the CD, given that an error occurred at the upper layers, and $\acute{P}^{\C}$ is similarly defined in the case of a PCD (recall also that noise remains uncolored at layer N under puncturing). Since the PCD does not propagate errors at upper layers, the distance metric \eqref{eq:ch5_xwr} is not severely distorted, and $\hat{x}_N^{}$ can still be recovered. We thus have $\acute{P}^{\C}\ll P^{\C}$. Moreover, as Fig.~\ref{perffig:ch6_b} shows, $\acute{P}^{\D}$ is the limiting term for $P^{\SSSD}$ at high SNR because $\acute{P}^{\C}\acute{P}^{\B}\ll \acute{P}^{\D}$. Hence, although $P^{\B} < \acute{P}^{\B}$, we still have $P^{\SSSD} < P^{\SLORD}$, which is a gain caused by puncturing. This is in accordance with the conclusion in Sec.~\ref{sec:ch6_diversitySSSD}.

\begin{figure}[t]
\centering
\includegraphics[width=6.1in]{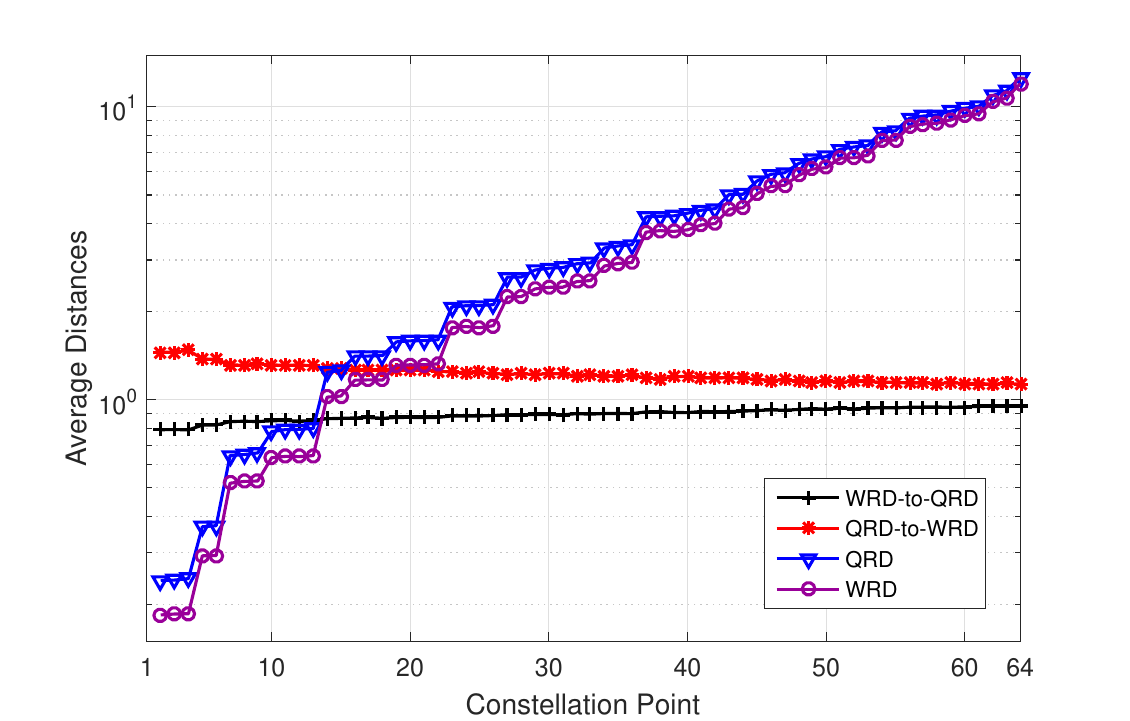}\vspace{-0.1in}
\caption{QRD-based vs WRD-based distance metrics.}
\label{f:ch6_scalings}
\end{figure}

%
\section{Comments on Soft-Output Detection}
\label{sec:ch6_soft_output}

Using the results of Sec.~\ref{sec:ch6_berSSsb}, we deduce that SO SSSD outperforms SO SLORD. Moreover, note that with CD and PCD, only the layer of interest is exhaustively searched, which provides the required distance metrics to compute the LLRs on this layer only. Repeating the same process on all layers, the resultant detectors would be SO SLORD and SO SSSD, respectively:
\begin{align}\label{eq:ch6_equality}
  \lambda_{n,k,t}^{\PCD} &= \lambda_{n,k,t}^{\SSSD}\\
   \lambda_{n,k,t}^{\CD} &= \lambda_{n,k,t}^{\SLORD}.
\end{align}
But since distance metrics from different channel decompositions in SSD are independent, SO SSD will not achieve a better performance than SO SSSD.

Finally, the performance of SO detectors is sensitive to factors that are not captured in the HO analysis. For example, computing distances via WRD as $\norm{\mbf{W}^\mathcal{H}(\mbf{y} - \mbf{H}\mbf{x})}^{2}$, instead of $\norm{\mbf{Q}^\mathcal{H}(\mbf{y} - \mbf{H}\mbf{x})}^{2}$, is subject to downscaling, as Fig.~\ref{f:ch6_scalings} shows. The scaling effect is higher with smaller distances, that will end up as ML or counter-ML distances to be used in LLR computations. Hence, the LLRs get scaled accordingly. This results in less confidence in LLR outputs, which is beneficial since LORD schemes produce overconfident LLRs. Furthermore, Fig.~\ref{f:ch6_scalings} shows a comparison of QRD-based and WRD-based distance metrics versus symbols from the constellation at the root layer. A noiseless $4\times4$ MIMO system with $64$-QAM is assumed. The symbols on the x-axis are sorted in increasing order of distance metrics (constellation point 1 with a zero distance corresponds to the true transmitted symbol vector). Both absolute distances as well as distance ratios are shown. On average, the order of distance metrics is retained under puncturing. However, at specific instances, the order gets distorted with a probability greater than $\acute{P}^{\A}$. \\

\textbf{Conclusion} \\

In this chapter,  the performance of the proposed detectors has been characterized and analyzed mathematically, by deriving bounds on the achievable rates, diversity gains, and error probability. It has been shown that puncturing results in a graceful loss in system capacity, and it does not negatively impact the receive diversity gain in HO detectors. The analysis has been extended to SO detection, and it has been shown that significant performance gains are attainable by ordering the layer of interest to be at the root when puncturing the channel.

%% file: Chapter7_MUMIMO2x2.tex

\chapter{Dual-Layer Multiuser MIMO Detection}\label{chapter:mumimo2x2}

In this chapter, we consider near optimal detection methods for $2\times2$ MU-MIMO systems by treating the interfering signal as a constrained unknown to be estimated. We propose joint likelihood-based MC of the co-scheduled user and data detection of the user of interest. By adjusting the Max-Log-MAP MC approach to the structure of the ML detector or LC-LORD, and expanding it to include distances of counter maximum likelihood hypothesis symbols, the decision metric for MC is shown to be an accumulation over a set of tones of Euclidean distance computations also used by the detectors for bit LLR generation. Hence, we show that an enhanced near IA MU-MIMO detector can be \emph{efficiently} implemented with a slight modification to the SO versions of the detectors. An efficient hardware implementation scheme is presented. The results of this section appeared, in parts, in \cite{7418322Sarieddeen}, \cite{7472369Sarieddeen}, and \cite{7564725Sarieddeen}.

%
\section{System Model}
\label{sec:ch7_sysmodel}

We follow the system model of Sec. \ref{sec:ch2_model}, but limit the discussion to the special case of dual-layer MIMO ($N\!=\!M\!=\!2$). We also define a new notation for the MTs. We assume $x_1$ to be drawn from the arbitrary, but known, constellation $\bar{\Lambda}$ (MT of the user of interest), that could be QPSK, 16-QAM or 64-QAM, and $x_2$ to be drawn from an unknown constellation $\Lambda_{j}$, $j\in\{0,1,2,3\}$, where $\Lambda_{0}$, $\Lambda_{1}$, $\Lambda_{2}$ and $\Lambda_{3}$ correspond to the constellations $\Phi$, QPSK, 16-QAM and 64-QAM, respectively, with $\Phi$ representing a constellation having one entry of zero power, corresponding to the case when there is no interferer.

\begin{figure}[!t]
\centering
\includegraphics[width=5in]{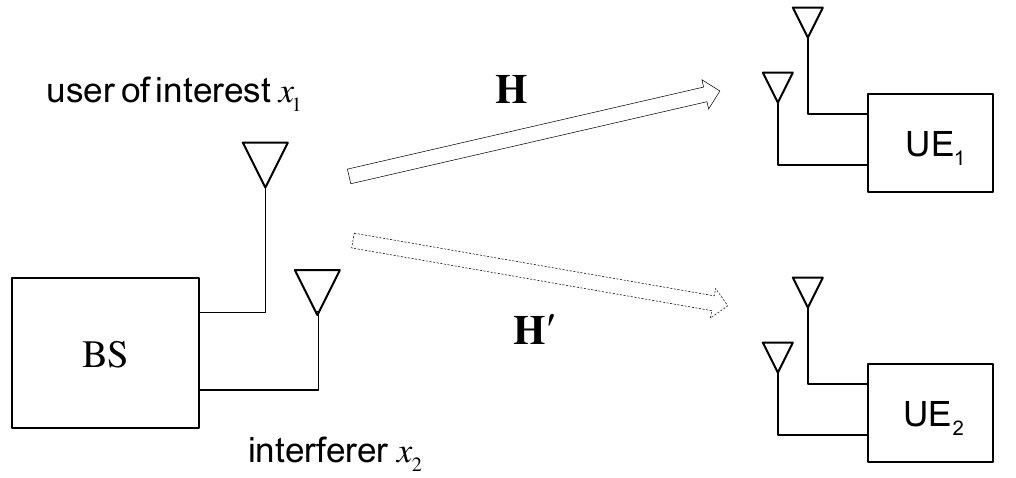}\vspace{-0.1in}
\caption{$2\times2$ MU-MIMO system model}
\label{f:ch7_system_model}
\end{figure}

%
\section{Interference Rejection Combining}
\label{sec:ch7_IRC}

Linear IRC detection is employed when estimates of both the desired and co-scheduled users' channels are available at the receiver, but knowledge of the MT of the co-scheduled user is \emph{not}. IRC works as a linear MMSE receiver \cite{2011_Bai}, performing whitening followed by MRC: \\

\begin{equation}\label{eq:ch7_IRC}
\mbf{h}_1^\mathcal{H}\mbf{L}^{-1}\mbf{y} = \mbf{h}_1^\mathcal{H}\mbf{L}^{-1}\mbf{h}_1 x_1 + \mbf{h}_1^\mathcal{H}\mbf{L}^{-1}(\mbf{h}_2 x_2 + \mbf{n}),
\end{equation}
with $\mbf{L} = \mbf{h}_2 \mbf{h}_2^\mathcal{H} + \sigma^{2} \mbf{I}_{2}$ being the covariance matrix of the sum of interference and noise components. The resultant distance metric to be used in LLR computation is generated as
\begin{equation}\label{eq:ch7_dist_IRC}
\varphi^{\IRC}(x_{1}) = \frac{1}{\sigma_{\mbf{n},\IRC}^{2}}\abs{\mbf{h}_1^\mathcal{H}\mbf{L}^{-1}\mbf{y} - \mbf{h}_1^\mathcal{H} \mbf{L}^{-1}  \mbf{h}_1 x_{1}}^{2},
\end{equation}
where unlike in \cite{Bae}, we have accounted for the variability of the variance from tone to tone by the scaling factor $\frac{1}{\sigma_{\mbf{n},\IRC}^{2}}$, with $\sigma_{\mbf{n},\IRC}^{2} = \mbf{h}_1^\mathcal{H}\mbf{L}^{-1}\mbf{h}_1$. Then the LLRs are computed as

\begin{equation}\label{eq:ch7_LLR_IRC}
\lambda^{\IRC}_i = \min_{x_{1}\in \bar{\Lambda}_i^{(1)}} \varphi^{\IRC}(x_{1}) - \min_{x_{1}\in \bar{\Lambda}_i^{(0)}} \varphi^{\IRC}(x_{1}),
\end{equation}
where $\bar{\Lambda}_i^{(1)}$ and $\bar{\Lambda}_i^{(0)}$ correspond to points in $\bar{\Lambda}$ having in the bit position $i$ of the symbol of interest the values of 1 and 0, respectively. Note that since the interference is discrete and not Gaussian, IRC is not optimal.

%
\section{ML MC for $2\times2$ MU-MIMO Systems}
\label{sec:ch7_mumimo}

The optimal likelihood-based MC scheme decides on the modulation format that has the maximum likelihood within multiple hypotheses. Following the Bayesian formulation, hypothesis testing is performed on the possible modulation formats to estimate the constellation of the interferer. We consider four hypotheses: $\mbf{y}\sim \Prb(\mbf{y};x_{1}\in \bar{\Lambda},x_{2}\in \Lambda_{j}),\ j\in\{0,1,2,3\}$, with likelihoods
\begin{equation}\label{eq:ch7_gen}
\Prb(\mbf{y};\Lambda_{j})=\sum_{x_{1}\in \tilde{\Lambda},\,x_{2}\in \Lambda_{j}} \Prb(\mbf{y}|x_{1},x_{2})\Prb(x_{1},x_{2}).
\end{equation}
Under statistical independence between $x_{1}$ and $x_{2}$, and assuming uniform priors, $\Prb(x_{1})=1/\abs{\bar{\Lambda}}$ and $\Prb(x_{2})=1/\abs{\Lambda_{j}}$, where $\abs{\cdot}$ denotes the cardinality of the constellation ($\Prb(x_{1})$ is fixed over hypotheses and thus can be dropped), the ML MC decision metric can be derived as
\begin{equation}\label{eq:ch7_gen2}
\hat{j}=\argmax_{j\in\{0,1,2,3\}}\sum_{x_{1}\in \bar{\Lambda},\,x_{2}\in \Lambda_{j}} \Prb(\mbf{y}|x_{1},x_{2})\frac{1}{\abs{\Lambda_{j}}}.
\end{equation}
Noting that $\Prb(\mbf{y}|x_{1},x_{2}) = \frac{1}{({\pi\sigma^{2}})^{N_{r}}} \exp(-\frac{1}{\sigma^{2}} \norm{\mbf{y} - \mbf{Hx}}^{2})$, and neglecting the term $\frac{1}{({\pi\sigma^{2}})^{N_{r}}}=\frac{1}{({\pi\sigma^{2}})^{2}}$ which is assumed fixed over hypotheses, the resultant Log-MAP decision metric is
\begin{equation}\label{eq:ch7_erl}
\hat{j}_{\LogMAP}=\argmax_{j\in\{0,1,2,3\}} \left(\log\frac{1}{\abs{\Lambda_{j}}} + \log\sum_{x_{1}\in \bar{\Lambda},\,x_{2}\in \Lambda_{j}} \exp\left(-\frac{1}{\sigma^{2}} \norm{\mbf{y} - \mbf{Hx}}^{2}\right) \right),
\end{equation}
which is the optimal ALRT solution. Note that neglecting the correction term $\log(1/\abs{\Lambda_{j}})$ results in the GLRT solution \cite{1311278Yucek}.

Solving~\eqref{eq:ch7_erl} is computationally intensive, because for each $j$ we have to calculate $\abs{\bar{\Lambda}}\times\abs{\Lambda_{j}}$ exponential terms. However, one of these terms is dominant and corresponds to the ML distance
\begin{equation}
d^{\ML}_{j} = \min_{x_{1}\in \tilde{\Lambda},\,x_{2}\in \Lambda_{j}} \varphi^{\ML}(\mbf{x})
\end{equation}
\begin{equation}
\varphi^{\ML}(\mbf{x}) = \frac{1}{\sigma^{2}} \norm{\mbf{y} - \mbf{Hx}}^{2}.
\end{equation}
Hence, following the approximation ($\log\sum_{r}{\exp(a_{r})}\approx\max_{r}a_{r}$), we obtain
\begin{equation}\label{eq:ch7_er2}
\hat{j}_{\MaxLogMAP}=\argmax_{j\in\{0,1,2,3\}} \left(\log\frac{1}{\abs{\Lambda_{j}}} - d^{\ML}_{j}\right),
\end{equation}
which is the sub-optimal Max-Log-MAP classifier \cite{7450603Gomaa}\cite{Bae}.

%
\section{Proposed MU-MIMO Receivers }
\label{sec:ch7_proposed}

\begin{figure}[!t]
\centering
\includegraphics[width=5in]{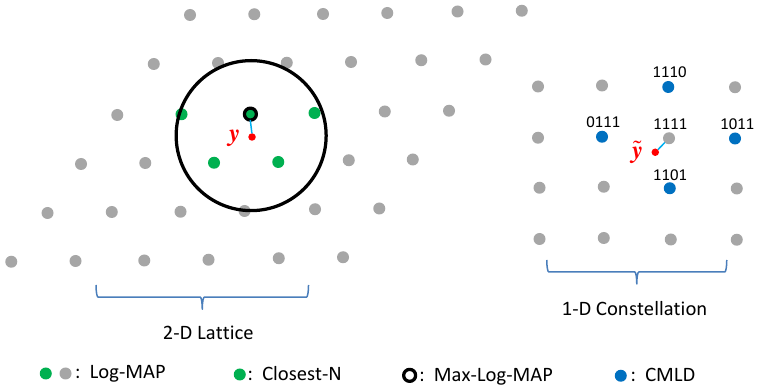}\vspace{-0.1in}
\caption{Lattice points used in MC schemes}
\label{f:ch7_points}
\end{figure}

Since the more distance metrics that get accumulated in~\eqref{eq:ch7_erl}, the better the approximation is, we can enhance the classifier by considering the most influential \emph{N} distances that best minimize $\varphi^{\ML}(\mbf{x})$. We call this approach the Closest-\emph{N} classifier, and we will use it as a reference to compare our second proposed approach to.

We next consider a special subset of distances, that consists of the counter ML distances corresponding to bits of the symbol of interest in addition to the ML distance. We call the corresponding scheme counter-ML-distance-based MC (CMLD). Note that with CMLD, the distances considered are not the smallest, and hence not the most influential. The counter ML distance corresponding to a specific bit is defined as
\begin{equation}
d_{\cML,j,i} =
\begin{cases}
\displaystyle\min_{x_{1}\in \bar{\Lambda},\,x_{2}\in \Lambda_{j}|b_{i}=0} \varphi^{\ML}(\mbf{x}) & \ \  b_{i}^{(\ML)}=1 \\
\displaystyle\min_{x_{1}\in \bar{\Lambda},\,x_{2}\in \Lambda_{j}|b_{i}=1} \varphi^{\ML}(\mbf{x}) & \ \ b_{i}^{(\ML)}=0
\end{cases}
\end{equation}
with $b_{i}^{(\ML)}$ being the value of the $\nth{i}$ bit in the bit vector of the ML solution.

The lattice points (symbol vectors) to which the distances are taken in each approach are shown in Fig.~\ref{f:ch7_points}. For Log-MAP, Max-Log-MAP and Closest-\emph{N}, the points are drawn directly from the 2-D MIMO lattice structure. However, only the component of interest is shown for the CMLD approaches, which is drawn from its corresponding constellation,

After the classifier decides on $\hat{j}$, an ML SO detector generates the bit LLRs as follows:
\begin{equation}\label{eq:ch7_det}
\lambda^{\ML}_i = \min_{x_{1}\in \bar{\Lambda}_i^{(1)},\,x_{2}\in \Lambda_{\hat{j}}} \varphi^{\ML}(\mbf{x}) - \min_{x_{1}\in \bar{\Lambda}_i^{(0)},\,x_{2}\in \Lambda_{\hat{j}}} \varphi^{\ML}(\mbf{x}).
\end{equation}
Hence, the main component of the decision metric for MC is found to be an accumulation over a set of tones of Euclidean distance computations, which are also used by the ML detector for bit LLR soft decision generation. Combining MC and detection routines is thus computationally efficient.

Furthermore, if we consider LC-LORD instead of ML detection, the proposed MC schemes should be modified. First,~\eqref{eq:ch7_erl} does not hold with LORD, since only layer-1 is exhaustively searched. Moreover, while~\eqref{eq:ch7_er2} still holds with LORD, $d^{\ML}_{j}$ is not guaranteed to be within the reduced QAM $\Theta$, and thus it does not hold with LC-LORD. Consequently, we introduce the Quasi-Log-MAP and Quasi-Max-Log-MAP MC schemes, that best approximate the original schemes. With Quasi-Log-MAP, the summation in~\eqref{eq:ch7_erl} is over the $\abs{\Theta}$ lattice points searched by LC-LORD, with corresponding distance metrics $d_{1}(x_{1})$, while with Quasi-Max-Log-MAP, the modified ML distance metric considered is $\min_{x_{1}\in\Theta} d_{1}(x_{1})$. Similarly, we can define the Closest-\emph{N} and CMLD classifiers starting from the LC-LORD solution.

Equation \eqref{eq:ch7_er3} generalizes the likelihood function assuming $T$ observations (tones) are accumulated under a constant interfering MT before deciding on a winning hypothesis, where $\bar{S}$ corresponds to the subset of lattice points to consider in each approach:
\begin{equation} \label{eq:ch7_er3}
\hat{j}=\argmax_{j\in\{0,1,2,3\}} \sum_{t=1}^{T} \left(\log\frac{1}{|\Lambda_{j}|} + \log\sum_{\mbf{x}\in \bar{S}} \exp\left(-\varphi^{\ML}(\mbf{x})\right) \right).
\end{equation}

The joint MC and detection setup is described as follows. After observing $T$ vectors, and for each of the four possible hypotheses, the detection routine is called $T$ times and the outputs are stored in memory. Concurrently, the likelihood for each hypothesis gets computed. Eventually, the hypothesis that gets the maximum likelihood is declared a winner and the corresponding output is retrieved. The price to pay is an increase in space complexity.

\begin{figure}[t]
\centering
\includegraphics[width=5.7in]{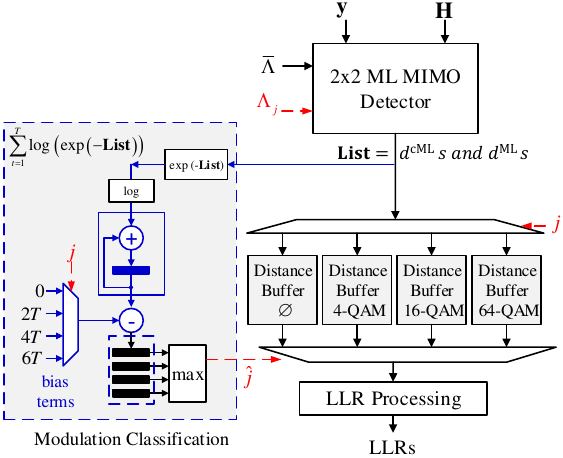}\vspace{-0.1in}
\caption{Architecture for a $2\times 2$ MU-MIMO detector.}
\label{f:ch7_architecture}
\end{figure}

\section{Joint MC and Detection Architecture}
\label{sec:ch7_implementation}

An optimized architecture for a $2\!\times\! 2$ MU-MIMO detector following the CMLD approach is shown in Fig.~\ref{f:ch7_architecture}. At the core of this architecture is a ML MIMO detector (or LC-LORD), that detects $\mbf{x}$ assuming all possible choices of the interferer's MT, and generates the corresponding lists of $d^{\cML}$s and $d^{\ML}$s for all $T$ vectors. These distances and symbols are stored in buffers of size $T\times(2^q+1)$. The sum of the logarithm of the exponential of the distance metrics are passed to an adder that accumulates them over a span of $T$ tones, during which the interferer modulation is assumed to be static. The resulting accumulated distances for each interferer hypothesis are stored in a buffer, and after deciding on a winning hypothesis, the corresponding stored distances are forwarded for LLR processing.

\begin{table}[!t]
\caption{\textbf{Computational Complexity of MC Schemes with ML Detection}}\vspace{0.15in} 
\label{table:ch7_complexity1} 
\centering 
\begin{tabular}{| c || c | c |} 
\hline
Approach & Exponentials & Distance Computations \Tstrut \Bstrut \\
\hline\hline

Log-MAP  & $ \abs{\bar{\Lambda}}(\abs{\Lambda_{0}}\!+\!\abs{\Lambda_{1}}\!+\!\abs{\Lambda_{2}}\!+\!\abs{\Lambda_{3}})$ & $\abs{\bar{\Lambda}} (\abs{\Lambda_{0}}\!+\!\abs{\Lambda_{1}}\!+\!\abs{\Lambda_{2}}\!+\!\abs{\Lambda_{3}})$\Tstrut \\ %
Closest-\emph{N} & $ 4   N $ & $\abs{\bar{\Lambda}}(\abs{\Lambda_{0}}\!+\!\abs{\Lambda_{1}}\!+\!\abs{\Lambda_{2}}\!+\!\abs{\Lambda_{3}})$\Tstrut\\
CMLD  & $ 4    (2^q\!+\!1)$ & $\abs{\bar{\Lambda}} (\abs{\Lambda_{0}}\!+\!\abs{\Lambda_{1}}\!+\!\abs{\Lambda_{2}}\!+\!\abs{\Lambda_{3}})$\Tstrut \\
Max-Log-MAP &  $ 4   $ & $\abs{\bar{\Lambda}} (\abs{\Lambda_{0}}\!+\!\abs{\Lambda_{1}}\!+\!\abs{\Lambda_{2}}\!+\!\abs{\Lambda_{3}})$\Tstrut \\
\hline

Joint Log-MAP  & $ \abs{\bar{\Lambda}}(\abs{\Lambda_{0}}\!+\!\abs{\Lambda_{1}}\!+\!\abs{\Lambda_{2}}\!+\!\abs{\Lambda_{3}})$ & $0$ \Tstrut \\ %
Joint Closest-\emph{N} &  $ 4    N $ & $0$ \Tstrut \\
Joint CMLD  & $ 4  (2^q\!+\!1)$ & $0$ \Tstrut \\
Joint Max-Log-MAP & $ 4  $ & $0$  \Tstrut\\ 
\hline 
\end{tabular}
\end{table}

\begin{table}[!h]
\caption{\textbf{Computational Complexity of MC Schemes with LC-LORD}}\vspace{0.15in} 
\label{table:ch7_complexity2} 
\centering 

\begin{tabular}{| c || c | c |} 
\hline
Approach & Exponentials & Distance Computations \Tstrut \Bstrut \\
\hline\hline

Log-MAP  & $ \abs{\bar{\Lambda}}$ & $\abs{\bar{\Lambda}} $\Tstrut \\ %
Quasi-Log-MAP  & $ \abs{\bar{\Theta}}$ & $\abs{\bar{\Theta}} $\Tstrut \\ %
Closest-\emph{N} & $ 4  N $ & $\abs{\bar{\Lambda}}$\Tstrut\\
CMLD  & $ 4  (2^q\!+\!1)$ & $\abs{\bar{\Lambda}} $\Tstrut \\
Max-Log-MAP &  $ 4  $ & $\abs{\bar{\Lambda}}$\Tstrut \\
Quasi-Max-Log-MAP &  $ 4  $ & $\abs{\bar{\Theta}}$\Tstrut \\
\hline

Joint Log-MAP  & $ \abs{\bar{\Lambda}}$ & $0$ \Tstrut \\ %
Joint Quasi-Log-MAP  & $ \abs{\bar{\Theta}}$ & $0$ \Tstrut \\ %
Joint Closest-\emph{N} &  $ 4  N $ & $0$ \Tstrut \\
Joint CMLD  & $ 4  (2^q\!+\!1)$ & $0$ \Tstrut \\
Joint Max-Log-MAP & $ 4 $ & $0$  \Tstrut\\
Joint Quasi-Max-Log-MAP & $ 4  $ & $0$  \Tstrut\\ 
\hline 
\end{tabular}
\end{table}

This algorithm can be used in 802.11ac (WiFi) \cite{802.11}, which supports $80$MHz of bandwidth with $242$ usable tones, $8$ of which are reserved for pilots and $234$ data tones (worst case is $20$MHz, $4$ pilots, and $52$ data tones). The length of the data field in a WiFi frame can be a very large number of OFDM symbols ($L$). Since the interferer's modulation constellation remains static over $T$ tones and $L$ symbols, the particular choice of $T\!=\!234$ results in substantial savings in computations. The detector only needs to run in this mode to identify the interferer's constellation for one OFDM symbol in the frame. It can then switch back to normal ML detection mode (without MC) to generate the LLRs for the remaining ODFM symbols. Moreover, the algorithm can be used in LTE \cite{LTE_36.211} transmission modes 7, 8, and 9, where estimates of desired and co-scheduled users channels are available at the receiver.

The total number of distance computations needed to generate the LLRs from the $234\!\times\!L$ data tones is $234\!\times\!L\!\times\!\abs{\bar{\Lambda}}\!\times\!\abs{\Lambda_{j}}$, and the average overhead of MC is $234\!\times\!3\!\times\!\abs{\bar{\Lambda}}\!\times\!\abs{\Lambda_{j}}$. This corresponds to an increase of only $3/L\%$, compared to distances computed by an ML detector with perfect knowledge of the interferer. The size $L$ of the data field can take a range of values from 8 to more than 1024, hence, the increase in distance computations ranges between $37.5\%$ and less than $0.3\%$. However, an additional burden of MC is the number of exponential and logarithmic operations it requires (blue box in Fig.~\ref{f:ch7_architecture}). Tables \ref{table:ch7_complexity1} and \ref{table:ch7_complexity2} compare the complexity of different MC routines, per tone, when applied solely or in a joint setup, in terms of Euclidean distance computations and number of  exponential operations, in the context of ML detection and LC-LORD, respectively.

%
\section{Simulation Results}
\label{sec:ch7_simulation}

The joint MC and detection schemes were simulated. The decision on the winning hypothesis was made after receiving $T\!=\!52$ or $T\!=\!12$ tones, with the first corresponding to WiFi's worst case, and the second to an LTE scenario. Turbo coding was used, with a code rate of $1/3$ and number of decoding iterations equal to $4$. We considered the scenario where the user of interest uses 16-QAM with ML detection and 64-QAM with LC-LORD (since LC-LORD is of less interest with smaller constellations), while the interferer hops over the four hypotheses with equal probability on every new frame. Furthermore, we considered both rich scattering and correlated channels, with transmit and receive correlation coefficients of $0.9$ and $0.6$ as defined for LTE.

\begin{figure}[!t]
\centering
\includegraphics[width=4.6in]{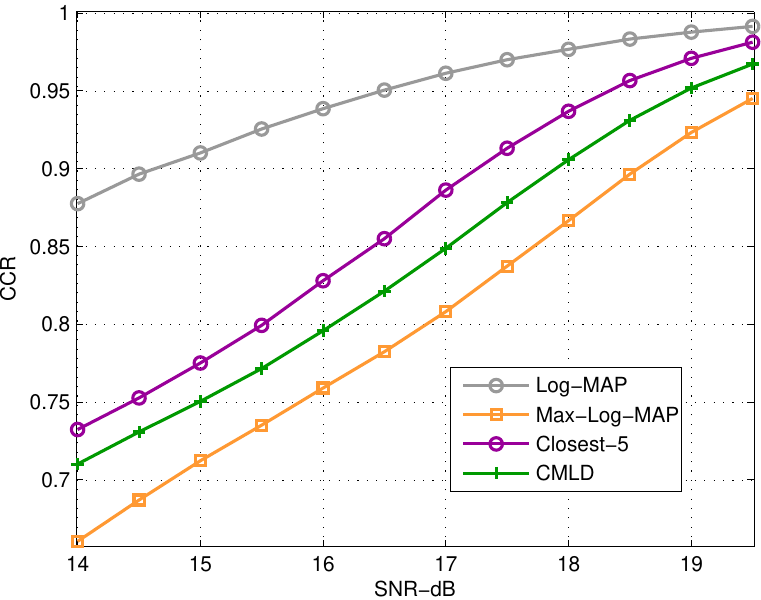}\vspace{-0.1in}
\caption{CCR performance - $\bar{\Lambda}$ is 16-QAM - $T=12$ - $0.9$ correlation.}
\label{f:ch7_ccrcorr12}
\end{figure}
\begin{figure}[!t]
\centering
\includegraphics[width=4.6in]{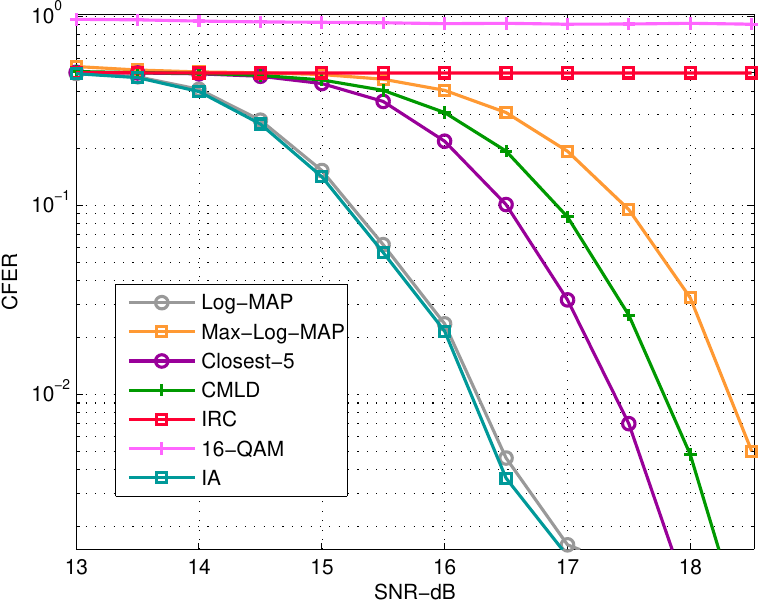}\vspace{-0.1in}
\caption{CFER performance - ML Detection - $\bar{\Lambda}$ is 16-QAM -  $T=52$ - $0.9$ correlation.}
\label{f:ch7_fercorr}
\end{figure}
\begin{figure}[!t]
\centering
\includegraphics[width=4.6in]{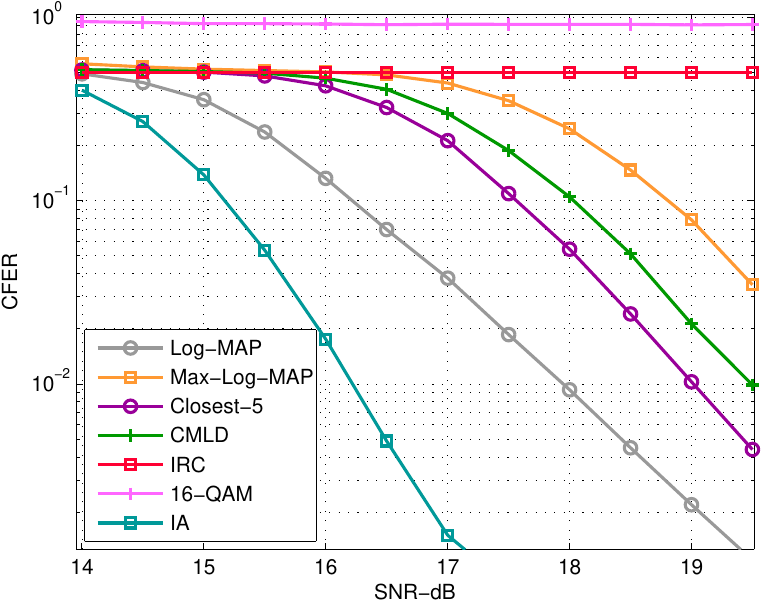}\vspace{-0.1in}
\caption{CFER performance - ML Detection - $\bar{\Lambda}$ is 16-QAM -  $T=12$ - $0.9$ correlation.}
\label{f:ch7_fercorr12}
\end{figure}
\begin{figure}[!t]
\centering
\includegraphics[width=4.6in]{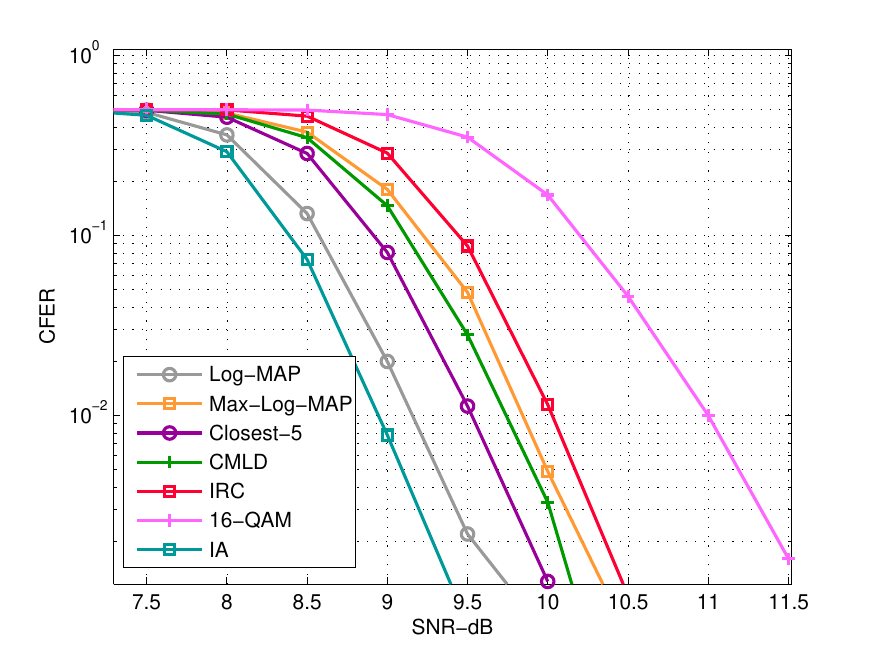}\vspace{-0.1in}
\caption{CFER performance - ML Detection - $\bar{\Lambda}$ is 16-QAM -  $T=12$ - uncorrelated channels.}
\label{f:ch7_feruncorr12}
\end{figure}

For ML detection, we simulated four receivers that are assisted by the studied MC schemes: Log-MAP, Max-Log-Map, Closest-$5$ (Closest-\emph{N} with \emph{N} $\!=\!5$) and CMLD. In addition, we included the receiver that always assumes the interferer to be 16-QAM, as well as the ideal IA receiver. Figure~\ref{f:ch7_ccrcorr12} shows, for highly correlated channels, the correct classification ratio (CCR) for the MC approaches with $T\!=\!12$. The CCR gaps are significant, but all approaches converge to unity at high SNR. Figures ~\ref{f:ch7_fercorr} and ~\ref{f:ch7_fercorr12} show the coded frame error rate (CFER) plots with high channel correlation, for $T\!=\!52$ and $T\!=\!12$, respectively. The choice $T=52$ made the Log-MAP MC-based receiver approach the performance of the IA receiver. On average, compared to Max-Log-MAP, CMLD resulted in a CFER SNR gain of $\unit[0.6]{dB}$, Closest-$5$ a gain of $\unit[1.1]{dB}$, and Log-MAP a gain of $\unit[2.2]{dB}$. Moreover, the IRC receiver and the receiver that assumes the interferer to be 16-QAM performed badly under high channel correlation.

\begin{figure}[t!]
\centering
\includegraphics[width=4.6in]{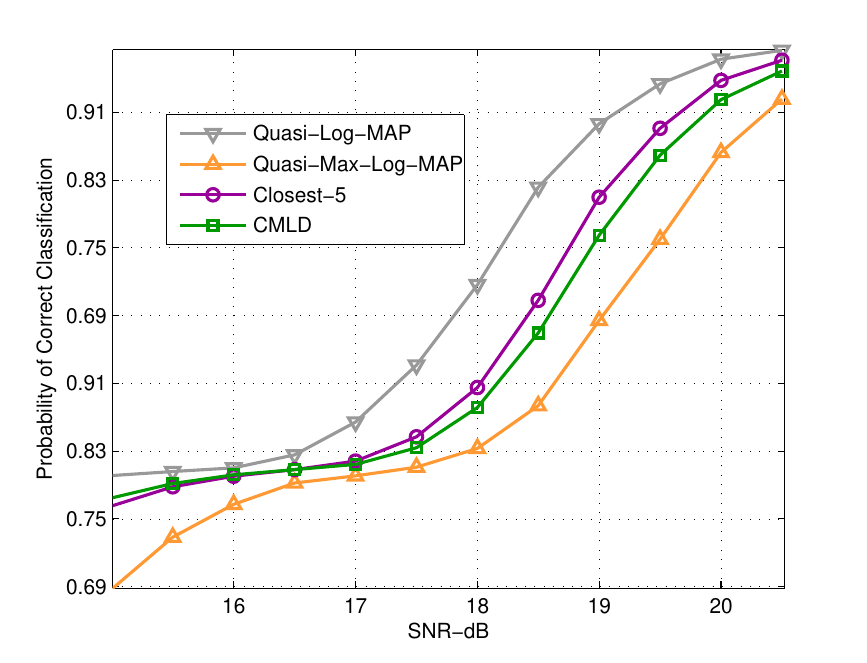}\vspace{-0.1in}
\caption{CCR Performance - LC-LORD - $\bar{\Lambda}$ is 64-QAM - $\abs{\Theta}\!=\!49$ - $0.6$ correlation.}
\label{f:ch7_ccrcorr3}
\end{figure}
\begin{figure}[t!]
\centering
\includegraphics[width=4.6in]{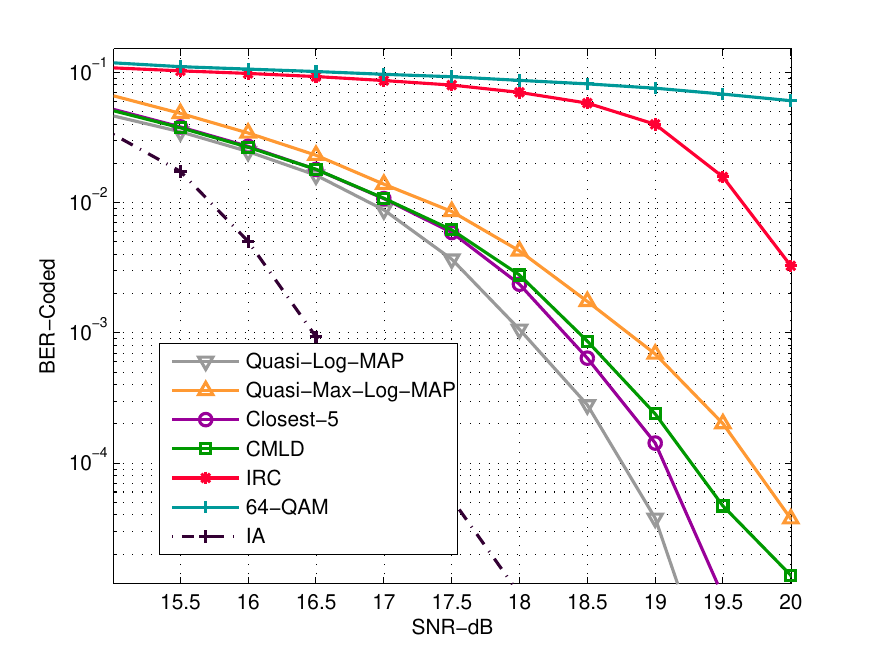}\vspace{-0.1in}
\caption{BER Performance - LC-LORD - $\bar{\Lambda}$ is 64-QAM - $\abs{\Theta}\!=\!49$ - $0.6$ correlation.}
\label{f:ch7_corr3}
\end{figure}
\begin{figure}[t!]
\centering
\includegraphics[width=4.6in]{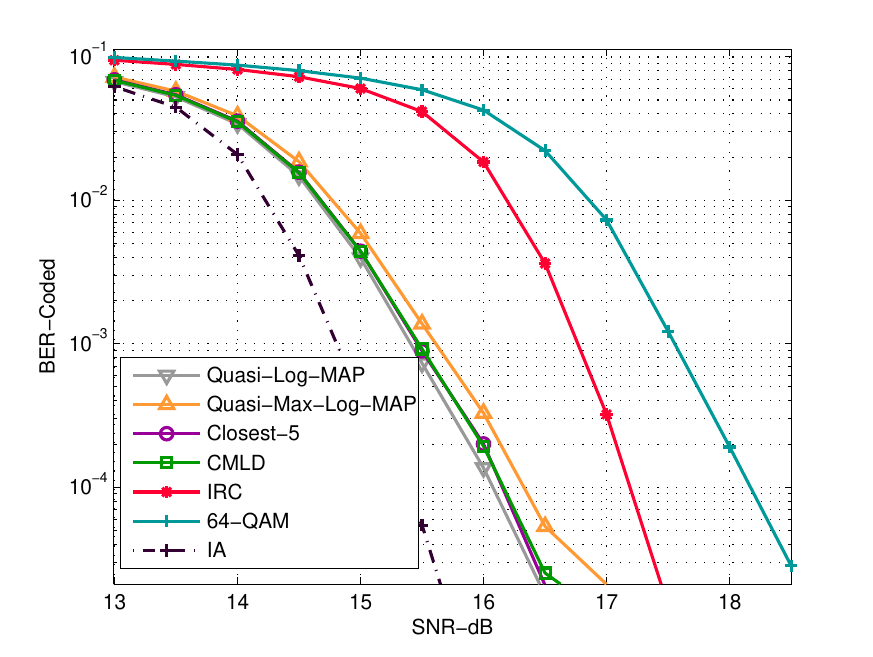}\vspace{-0.1in}
\caption{BER Performance - LC-LORD - $\bar{\Lambda}$ is 64-QAM - $\abs{\Theta}\!=\!49$ - ucorrelated channels.}
\label{f:ch7_uncorr3}
\end{figure}
\begin{figure}[t!]
\centering
\includegraphics[width=4.6in]{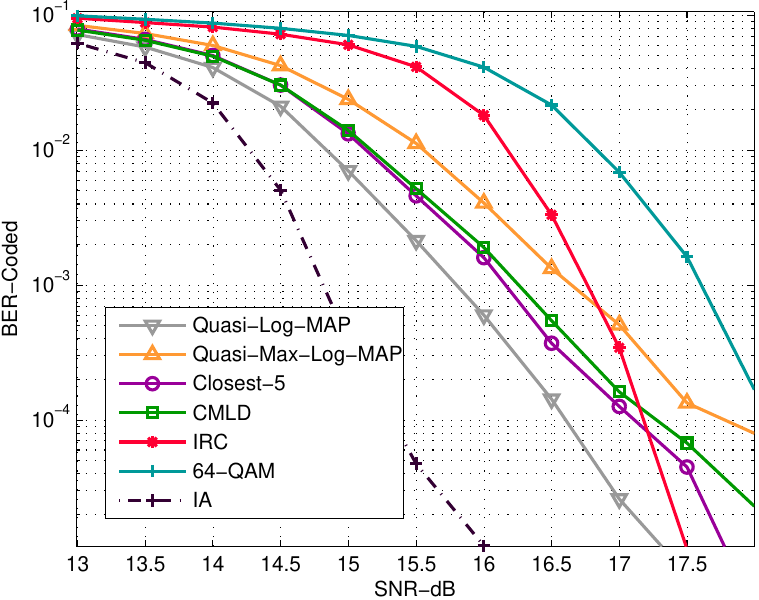}\vspace{-0.1in}
\caption{BER Performance - LC-LORD - $\bar{\Lambda}$ is 64-QAM - $\abs{\Theta}\!=\!25$ - uncorrelated channels.}
\label{f:ch7_uncorr2}
\end{figure}
\begin{figure}[t]
\centering
\includegraphics[width=4.6in]{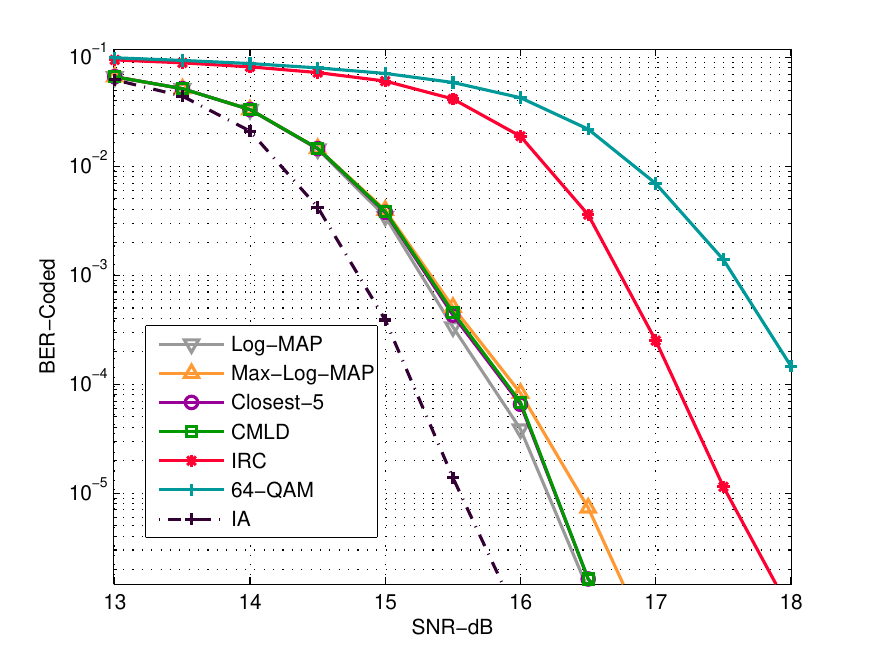}\vspace{-0.1in}
\caption{BER Performance - LC-LORD - $\bar{\Lambda}$ is 64-QAM - $\Theta\!=\!\bar{\Lambda}$ - uncorrelated channels.}
\label{f:ch7_uncorrfull}
\end{figure}

With ideal channel conditions, the total gap between the Log-MAP and Max-Log-MAP MC-based receivers does not exceed $\unit[0.7]{dB}$, as shown in Fig.~\ref{f:ch7_feruncorr12} for $T\!=\!12$. Compared to Max-Log-MAP, CMLD resulted in a CFER SNR gain of $\unit[0.15]{dB}$ and Closest-$5$ a gain of $\unit[0.3]{dB}$. Note that the Log-Map MC-based detector did not approach the optimal ML IA receiver here, because the CCR values are far from unity over low SNR range.

For LC-LORD, we simulated the receivers that are assisted by the MC schemes studied in Sec.~\ref{sec:ch7_proposed}: Quasi-Log-MAP, Quasi-Max-Log-Map, Closest-$5$ (Closest-\emph{N} with \emph{N} $\!=\!5$) and CMLD (note that with LC-LORD, counter ML distances in case of CMLD might get saturated). In addition, we included the receiver that always assumes the interferer to be 64-QAM, as well as the ideal IA receiver.

We considered two scenarios of reduced search in LC-LORD, one with a worst case search region of $\abs{\Theta}\!=\!49$ and another with a worst case of $\abs{\Theta}=25$. However, even the first case results in remarkable complexity savings when $\abs{\bar{\Lambda}}\!=\!64$, because most of times the center of $\Theta$ is close to the boundaries of ${\bar{\Lambda}}$, and thus $\Theta$ gets truncated, which further reduces its size. Moreover, we applied LC-LORD to $70\%$ of the streams, keeping the less reliable $30\%$ in full complexity mode.

Figure ~\ref{f:ch7_ccrcorr3} shows, for high correlation and $\abs{\Theta}\!=\!49$, the CCR plots for the MC approaches. All approaches reach unity CCR at high SNR. The gap between Quasi-Log-MAP and Quasi-Max-Log-MAP classifiers is around $\unit[1.2]{dB}$, and Closest-$5$ and CMLD cut this gap in half, with Closest-$5$ being slightly superior.

The coded BER performance of the proposed approaches, compared to the reference detectors, is shown in Fig.~\ref{f:ch7_corr3}. Assuming the interferer to always be 64-QAM performs badly with high channel correlation, and so does IRC to a less extent. The gap between IRC and the IA receiver is $\unit[4]{dB}$, and the MC-based approaches cut this gap in half. Quasi-Log-MAP and Quasi-Max-Log-MAP are $\unit[1]{dB}$ apart, and Closest-$5$ and CMLD beat Quasi-Max-Log-MAP by $\unit[0.5]{dB}$.

The rest of the plots show the BER performance with uncorrelated channels. For $\abs{\Theta}\!=\!49$ (Fig.~\ref{f:ch7_uncorr3}), the gap between IRC and IA reduces to $\unit[2]{dB}$, and assuming the interferer 64-QAM is only $\unit[0.8]{dB}$ away from IRC. Moreover, MC-based detectors almost coincide, both being less than $\unit[1]{dB}$ away form IA. However, if the search region is further reduced to $\abs{\Theta}\!=\!25$, IRC will beat the MC-based detectors at high SNR, even when the channels are uncorrelated, as shown in Fig.~\ref{f:ch7_uncorr2}. Finally, Fig.~\ref{f:ch7_uncorrfull} shows the BER performance, when $\Theta\!=\!\bar{\Lambda}$, which is in effect the optimal joint MC and LORD detection case. Comparing its plots with those of Fig.~\ref{f:ch7_uncorr3}, we conclude that setting $\abs{\Theta}=49$ and running full complexity LORD at the worst $30\%$ of carriers is near optimal.

From these results, we can see that both Closet-$5$ and CMLD offer enhanced low complexity realizations for MC-based MU-MIMO receivers. From a practical perspective, CMLD has the advantage as shown in Sec.~\ref{sec:ch7_implementation}, where the proposed joint MC and detection setup does not apply to Closest-$5$ (extra processing is required to sort the closest 5 distances as well as extra space).

The benefits of CMLD become clearer in the context of joint MC and SO SD. A SD reduces the number of visited lattice points, however, points leading to counter ML distances are never omitted and efficient joint CMLD MC and detection is maintained. On the other hand, the Closest-\emph{N} classifier is more useful if applied jointly with LSD, which keeps track of distances to closest neighbouring symbols, or with sphere decoding with adaptive radius pruning, both of which do not guarantee the inclusion of counter ML symbols. Moreover, this work can be combined with \cite{classon2000turbo,valenti2001umts}, using constant and linear Max-Log-MAP to enhance the approximation performance. Finally, we could have considered the distances to counter ML hypothesis symbols that correspond to the bits of the interfering symbol $x_2$ as well, but soft detectors do not keep track of the LLRs of the interferer, and hence less savings can be made with the joint MC and detection setup. \\

\textbf{Conclusion} \\

In this chapter, low-complexity receivers have been proposed for $2\times2$ MU-MIMO systems, based on joint MC and ML detection or LC-LORD, where the ML MC scheme has been adapted to cope with both detectors. These receivers have been compared to other state-of-the-art receivers. The performance of the proposed schemes has been shown to lie between the MC-based schemes that use Log-MAP and Max-Log-MAP classifiers, remarkably beating the latter, especially with high channel correlation. An efficient implementation has been proposed by making use of a hardware architecture that enables joint MC and detection. Finally, it has been shown that the proposed approaches can be used in various communication standards, where in the special case of WiFi they result in a negligible complexity overhead.

%% file: Chapter8_MUMIMOLARGE.tex

\chapter{High-Order Multiuser MIMO Detection}\label{chapter:mumimolarge}

In this chapter, we extend the work on $2\times2$ MU-MIMO and propose an efficient SSSD for larger antenna configurations, in which the MT of the interferer is estimated, while multiple decoupled streams are being individually detected. We propose low-complexity versions of the optimal log-MAP and Max-Log-MAP modulation classifiers, that adapt to the limitations of the proposed SSSD scheme, and employ them in a hierarchical fashion with feature-based classifiers. We show that the proposed algorithms can be \emph{efficiently} implemented, by proposing a corresponding low-complexity architecture and studying its complexity in the context of 802.11ac. The corresponding results appeared in \cite{2017Sarieddeen}.

%
\section{System Model}
\label{sec:ch8_sysmodel}

\begin{figure}[t]
\centering
\includegraphics[width=5in]{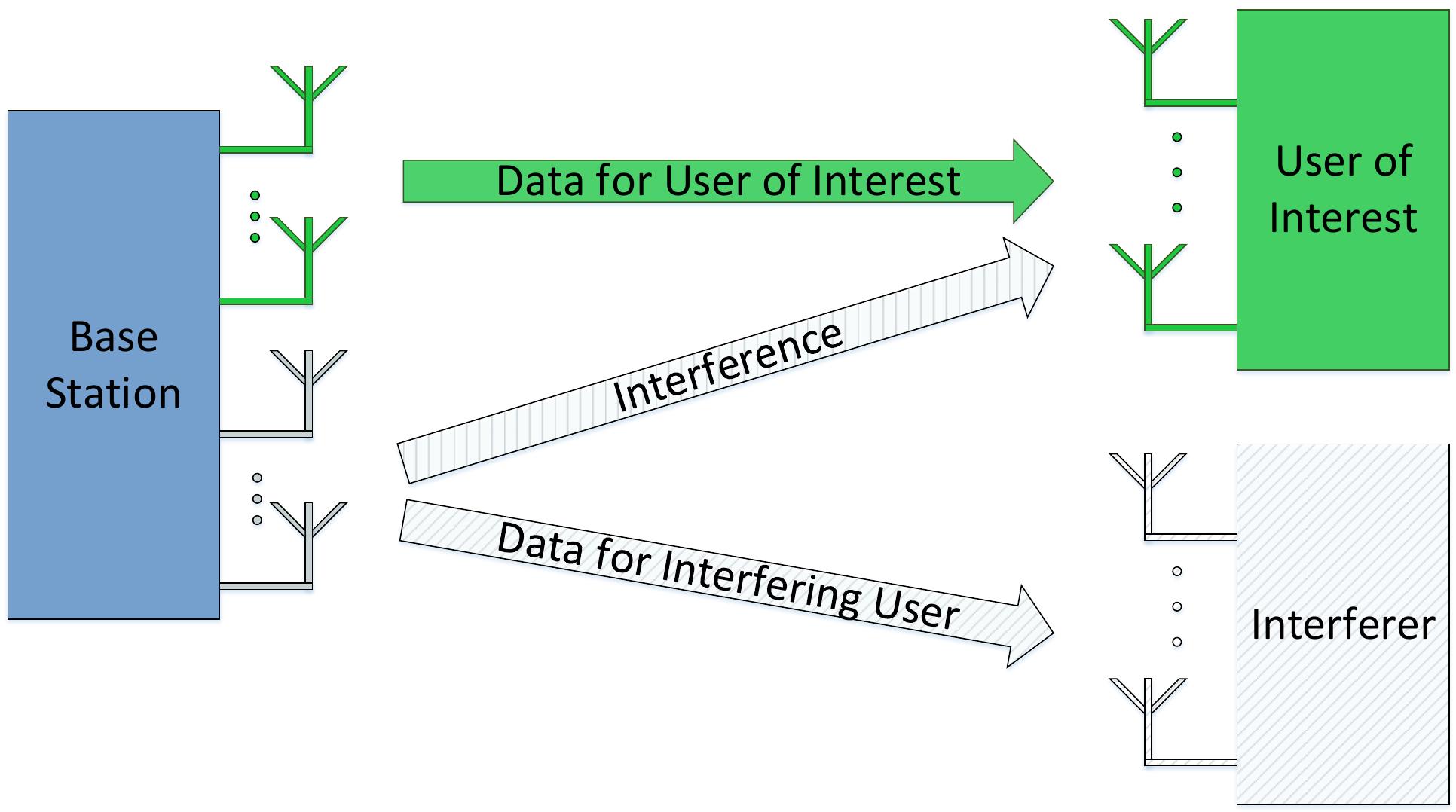}\vspace{-0.1in}
\caption{MU-MIMO system model.}
\label{f:ch8_sysmodel}
\end{figure}

We modify the system model of Sec.~\ref{sec:ch2_model} by considering a scenario in which $N_{\user}\!\leq\!N$ antennas transmit useful data to the user of interest, while the remaining $N_{\inter}\!=\!N\!-\!N_{\user}$ antennas send interfering data (Fig.~\ref{f:ch8_sysmodel}). Note that the entries of $\mbf{H}$ are still considered i.i.d $\mathcal{CN}(0,1)$, and no weighting is applied.

We assume that the $N_{\user}$ symbols of the user of interest that form $\mbf{x}_{\user}\!=\![x_1 \cdots x_{N_{\user}}]^{T}$ are drawn from the arbitrary, but known, constellation $\mathcal{M}$. We also assume, without loss of generality, that the $N_{\inter}$ symbols of the interferer that form $\mbf{x}_{\inter}\!=\![x_{N_{\user}\!+\!1}\cdots x_N]^{T}$ are drawn from the same unknown constellation $\mathcal{U}_j$, $j\in\{0,1,2,3,4\}$, where $\mathcal{U}_0$, $\mathcal{U}_1$, $\mathcal{U}_2$, $\mathcal{U}_3$ and $\mathcal{U}_4$ correspond to the constellations ${\Phi}$, QPSK, 16-QAM, 64-QAM and 256-QAM, respectively.

%
\section{Reference MU-MIMO MC Schemes}
\label{sec:ch8_mumimo}

\subsection{Likelihood-Based MC}
\label{sec:ch8_likelihoodbased}

In a derivation similar to that of Sec. \ref{sec:ch7_mumimo}, we consider five hypotheses: $\mbf{y}\!\sim\! \Prb(\mbf{y};\mbf{x}_{\user}\in\mathcal{M}^{N_{\user}},\ \mbf{x}_{\inter}\in\mathcal{U}_j^{N\inter}),\ j\in\{0,1,2,3,4\}$, with likelihoods
\begin{equation}\label{eq:ch8_gen}
\Prb(\mbf{y};\mathcal{U}_j)=\sum_{\mbf{x}_{\user}\in\mathcal{M}^{N_{\user}},\,\mbf{x}_{\inter}\in\mathcal{U}_j^{N\inter}} \Prb(\mbf{y}|\mbf{x})\Prb(\mbf{x}).
\end{equation}
Under statistical independence between the components of $\mbf{x}$, and assuming uniform priors, $\Prb(x_{1})\!=\!\cdots\!=\!\Prb(N_{\user})\!=\!1/\abs{\mathcal{M}}$ and $\Prb(N_{\user\!+\!1})\!=\!\cdots\!=\!\Prb(x_N)\!=\!1/\abs{\mathcal{U}_j}$ (the probabilities of user symbols are independent of the interferer and thus can be dropped), the ML MC decision metric can be expressed as
\begin{equation}\label{eq:ch8_gen2}
\hat{j}=\argmax_{j\in\{0,1,2,3,4\}}\sum_{\mbf{x}_{\user}\in\mathcal{M}^{N_{\user}},\,\mbf{x}_{\inter}\in\mathcal{U}_j^{N\inter}} \Prb(\mbf{y}|\mbf{x})\frac{1}{\abs{\mathcal{U}_j}^{N_{\inter}}}.
\end{equation}
Noting that $\Prb(\mbf{y}|\mbf{x})\!=\!\frac{1}{({\pi\sigma^{2}})^{M}} \exp(-\frac{1}{\sigma^{2}} \norm{\mbf{y} - \mbf{Hx}}^{2})$, and neglecting $\frac{1}{({\pi\sigma^{2}})^{M}}$ which is fixed over hypotheses, the resultant Log-MAP metric (ALRT solution) is
\begin{equation}\label{eq:ch8_logmap}
\hat{j}_{\text{Log-MAP}}=\argmax_{j\in\{0,1,2,3,4\}} \left(N_{\inter}\log\frac{1}{\abs{\mathcal{U}_j}} +  \log\sum_{\substack{ \mbf{x}_{\user}\in\mathcal{M}^{N_{\user}}  \\ \mbf{x}_{\inter}\in\mathcal{U}_j^{N\inter}} } \exp\left(-\frac{1}{\sigma^{2}} \norm{\mbf{y} - \mbf{Hx}}^{2}\right) \right).
\end{equation}
Solving equation \eqref{eq:ch8_logmap} is computationally intensive, because for each $j$ we have to calculate $\abs{\mathcal{M}}^{N_{\user}}\!\times\!\abs{\mathcal{U}_j}^{N_{\inter}}$ exponential terms. However, one of these terms is dominant and corresponds to the scaled ML distance
\begin{equation}
d^{\ML}_{j} = \min_{\mbf{x}_{\user}\in\mathcal{M}^{N_{\user}},\,\mbf{x}_{\inter}\in\mathcal{U}_j^{N\inter}} \frac{1}{\sigma^{2}} \norm{\mbf{y} - \mbf{Hx}}^{2}.
\end{equation}
Hence, following the Jacobian-logarithm approximation we obtain
\begin{equation}\label{eq:ch8_maxlogmap}
\hat{j}_{\text{Max-Log-MAP}}=\argmax_{j\in\{0,1,2,3,4\}} \left(N_{\inter}\log\frac{1}{\abs{\mathcal{U}_j}} - d^{\ML}_{j}\right),
\end{equation}
which is the sub-optimal Max-Log-MAP classifier \cite{7450603Gomaa}\cite{Bae}.

Therefore, as in the case of $2\times2$ MIMO, the main component of the decision metric for MC is found to be an accumulation over a set of tones of Euclidean distance computations, which are also used by the ML detector for bit LLR generation. Combining MC and detection routines is thus computationally efficient.

\subsection{MC Using Higher-Order Cumulants}
\label{sec:ch8_featurebased}

For feature-based classification, feature vectors containing higher-order CCs are used. These features cannot be directly extracted from the components of $\mbf{y}$, since they consist of linear mixtures of the components of the transmitted signal vector and additive noise. First, the channel is compensated, using ZF for example, where the received vector is multiplied by the pseudo-inverse of the channel matrix. Then, the MT-specific features are estimated from the noisy recovered symbol streams, the components of the vector $\hat{\mbf{y}}_{\ZF}$.

The general expression of a cumulant of order $u$, $v$-times conjugated, for a complex random variable $s$ is given as \cite{William1994}:
\begin{equation}\label{eq:ch8_cumulant}
\kappa_{s}^{u,v} = \sum_{\beth_u} \bigg[ k(p) \prod _{j=1}^{p} \mathsf{E} \{ s^{u_j-v_j} s^{*v_j} \} \bigg]
\end{equation}
where $\beth_u$ is the set of the partitions of the elements $\{1,2,\cdots,u\}$. A partition $\rho$ consists of $p$ sets $\nu_j:\rho\!=\!\{\nu_j\}_{j=1}^{p}$, where $u_j$ is the size of the set $\nu_i$, $v_j$ is the number of conjugated terms, and $k(p)\!=\!(-1)^{p-1}(p-1)!$.

Assume that we require estimating the MT from which the $\nth{i}$ symbol $\hat{y}_{\ZF,i}$ of $\hat{\mbf{y}}_{\ZF}$ was drawn, we replace $s$ by $\hat{y}_{\ZF,i}$, and compute~\eqref{eq:ch8_cumulant} with the required $u$ and $v$. Equation \eqref{eq:ch8_cumulant} can be simplified for specific values of $u$ and $v$. For example, we have:

\begin{equation}\label{eq:ch8_cumulant-20}
\kappa_{s}^{2,0} = \mathsf{E}\{s^{2}\}
\end{equation}
\begin{equation}\label{eq:ch8_cumulant-21}
\kappa_{s}^{2,1} = \mathsf{E}\{\abs{s}^{2}\}
\end{equation}
\begin{equation}\label{eq:ch8_cumulant-40}
\kappa_{s}^{4,0} = \mathsf{E}\{s^{4}\} - 3\mathsf{E}\{s^{2}\}^2
\end{equation}
\begin{equation}\label{eq:ch8_cumulant-42}
\kappa_{s}^{4,2} = \mathsf{E}\{s^{2} s^{*2}\} - \abs{\mathsf{E}\{s^{2}\}}^2 - 2\mathsf{E}\{s s^\mathcal{H}\}^2.
\end{equation}

Due to symmetry in constellations, only cumulants of even order are non-zero for linearly modulated signals, and hence are useful for MC. The theoretical values for various cumulants for QAM modulations are shown in table \ref{table:ch8_theoreticalcumulants}. Note that when only discriminating between QAMs, $\kappa_{s}^{4,1}$, $\kappa_{s}^{6,0}$, and $\kappa_{s}^{6,2}$ are also all zeros, and hence can not be used. Moreover, $\kappa_{s}^{2,0}$ and $\kappa_{s}^{2,1}$ can only be used to know whether interference exists or not.

\begin{table}[!t]
\caption{\textbf{Theoretical Cumulants for Different Constellations}}\vspace{0.15in} 
\label{table:ch8_theoreticalcumulants} 
\centering 
\begin{tabular}{| c || c | c | c | c | c | } 
\hline
Cumulant & ${\Phi}$ & QPSK & 16-QAM & 64-QAM & 256-QAM\\
\hline
$\kappa_{s}^{2,0}$ & 0 & 1 & 1 & 1 & 1  \Tstrut \\ %
$\kappa_{s}^{2,1}$ & 0 & 1 & 1 & 1 & 1  \Tstrut \\ %
$\kappa_{s}^{4,0}$ & 0 & 1 & -0.68 & -0.619 & -0.6047  \Tstrut \\ %
$\kappa_{s}^{4,1}$ & 0 & 0 & 0 & 0 & 0  \Tstrut \\ %
$\kappa_{s}^{4,2}$ & 0 & -1 & -0.68 & -0.619 & -0.6047  \Tstrut \\ %
$\kappa_{s}^{6,0}$ & 0 & 0 & 0 & 0 & 0  \Tstrut \\ %
$\kappa_{s}^{6,1}$ & 0 & -4 & 2.08 & 1.7972 & 1.7345  \Tstrut \\ %
$\kappa_{s}^{6,2}$ & 0 & 0 & 0 & 0 & 0  \Tstrut \\ %
$\kappa_{s}^{6,3}$ & 0 & 4 & 2.08 & 1.7972 & 1.7345  \Tstrut \\ %
\hline 
\end{tabular}
\end{table}

Eventually, the decision on a specific modulation scheme is made by choosing the MT that minimizes the Euclidean distance between the feature vector estimate and the theoretical feature vector. When multiple branches transmit symbols from the same MT, selection combining can be applied to select the feature estimate from the branch that has the highest SNR, or a more sophisticated MRC mechanism.

%
\section{Proposed MU-MIMO MC Schemes}
\label{sec:ch8_proposedMC}

\subsection{Modified Likelihood-Based MC}
\label{sec:ch8_proposedMC-1}

Since in this study we use SSSD instead of ML detection, major modifications should be made to likelihood-based MC. Equation \eqref{eq:ch8_logmap} does not hold with subspace decomposition, since the entire lattice is not exhaustively searched, neither does~\eqref{eq:ch8_maxlogmap}, since $d^{\ML}_{j}$ is not guaranteed to be within the search region. Moreover, aiming at efficiently combining MC and SSSD, and since SSSD only makes use of the layers of interest, we propose carrying the summation over the desired signal constellations while MC is for the interfering user.

We thus introduce the Quasi-Log-MAP and Quasi-Max-Log-MAP MC schemes, that best approximate the original schemes. With Quasi-Log-MAP, the summation in \ref{eq:ch8_logmap} is over the $\abs{\mathcal{M}}$ lattice points searched by SSSD, on one of the detected streams, say $l$, and hence the modified likelihood function can be represented as
\begin{equation}\label{eq:ch8_quasilogmap}
\hat{j}_{\text{Quasi-Log-MAP}}=\argmax_{j\in\{0,1,2,3,4\}} \left(\log\frac{1}{\abs{\mathcal{U}_j}} + \log\sum_{x_l\in\abs{\mathcal{M}}} \exp\left(-\frac{1}{\sigma^{2}} \norm{\bar{\mbf{y}}^{(t)}\!-\!\Rp^{(t)}\mbf{x}}^{2}\right) \right),
\end{equation}
where the Euclidean distance is expanded as in~\eqref{eq:ch2_xwr}.

With Quasi-Max-Log-MAP, the modified ML distance metric ($d^{{\ML}^{'}}$) is considered to be the minimum of the spanned scaled distances in Quasi-Log-MAP. Moreover, better (average) performance can be achieved when accumulating the minimum distances from all layers of interest. Equation \eqref{eq:ch8_er3} generalizes the proposed likelihood function assuming $T$ observations (tones) are accumulated under a constant interfering MT before deciding on a winning hypothesis.
\begin{equation} \label{eq:ch8_er3}
\hat{j}_{\text{Quasi-Max-Log-MAP}}=\argmax_{j\in\{0,1,2,3,4\}} \sum_{t=1}^{T} \left(\log\frac{1}{|\mathcal{U}_j|} - \sum_{l=1}^{N_{\user}} d^{{\ML}^{'}}_{j,l,t} \right)
\end{equation}
Since distances from different layers of interest that undergo different decompositions are independent, combining them is equivalent to repeating the observation and taking the average, as opposed to having a more powerful observation. Hence, dropping the inner summation over $N_{\user}$ layers in \eqref{eq:ch8_er3} is efficient.

\subsection{Hierarchical MC}
\label{sec:ch8_proposedMC-2}

While likelihood-base MC applies subspace decomposition a number of times equal to the number of hypotheses, feature-based MC requires only one subspace decomposition routine following MC. However, we can have a combination of both, that reduces the entailed complexity with minimum effect on performance.

Note that the theoretical CC values for higher order QAM constellations are very close. Also, despite the fact that higher order cumulants have more distant theoretical values, their corresponding variance is high. Thus, higher order cumulants do not necessarily result in better performance. Nevertheless, at least second order cumulants can be used to eliminate the hypothesis of no interferer, before proceeding with likelihood based MC to estimate other hypotheses.

We propose hierarchical MC as follows: First, the ZF solution is computed, and $\kappa_{s}^{2,0}$ is calculated. If the result is closer to $0$, we assume no interference, and the entire likelihood-based MC routine is skipped. Otherwise, if the result is closer to $1$, likelihood-based MC follows, but with one less hypothesis to check.

\subsection{Assuming High-Order Interfering Modulation Types}
\label{sec:ch8_proposedMC-3}

Instead of adding a MC routine, an attractive solution is to assume the interfering MT to be a high order QAM, without attempting to estimate it. A similar solution was presented in~\cite{2011_Ghaffar_a,2011_Ghaffar_b}, where the interfering MT was assumed to be 16-QAM, and an ML detector followed. In our case, assuming very high order constellations is feasible, because the number of distances computed in SSSD is not affected by the size of the interfering constellation. The only increase in complexity by assuming higher order interfering constellations is in the slicing operation they undergo, which is negligible. Therefore, we propose to assume the interfering MT to be 64-QAM, 256-QAM, or 1024-QAM, where the latter is not even one of the possible hypotheses.

\begin{figure}[t!]
\centering
\includegraphics[width=5.5in]{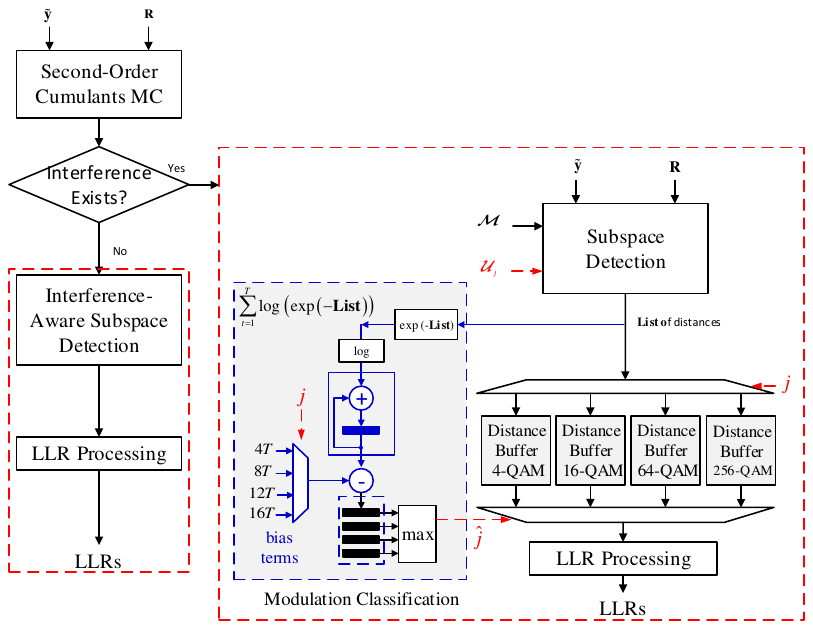}\vspace{-0.1in}
\caption{Architecture for joint hierarchical MC and SSSD.}
\label{f:ch8_architecture-2}
\end{figure}
\begin{figure}[t!]
\centering
\includegraphics[width=5in]{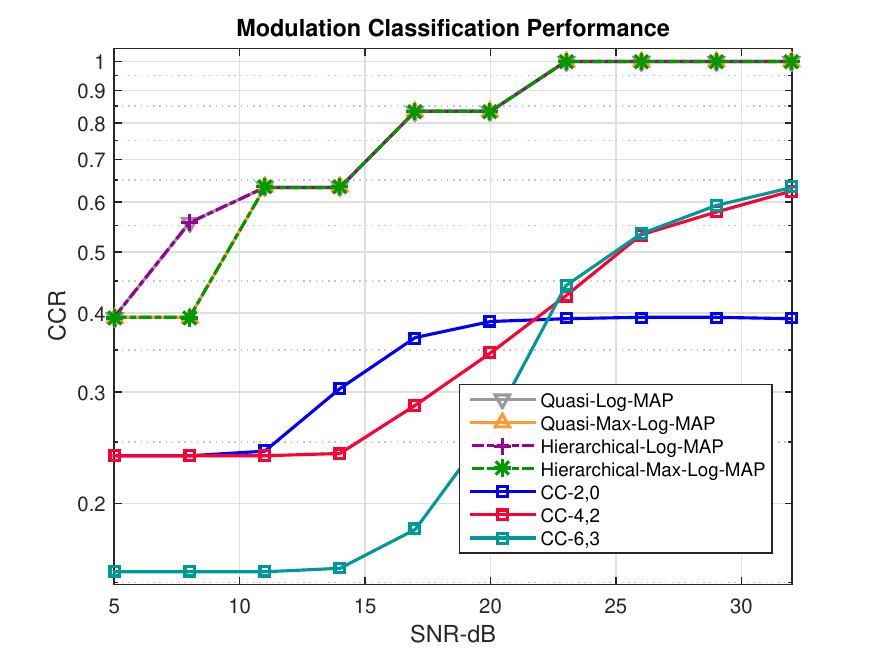}\vspace{-0.1in}
\caption{CCR performance - $4\times4$ MU-MIMO - $N_{\user}=2$ - $\mathcal{M}$ is 64-QAM - uncorrelated.}
\label{f:ch8_ccr4x4uncorr}
\end{figure}
\begin{figure}[t!]
\centering
\includegraphics[width=5in]{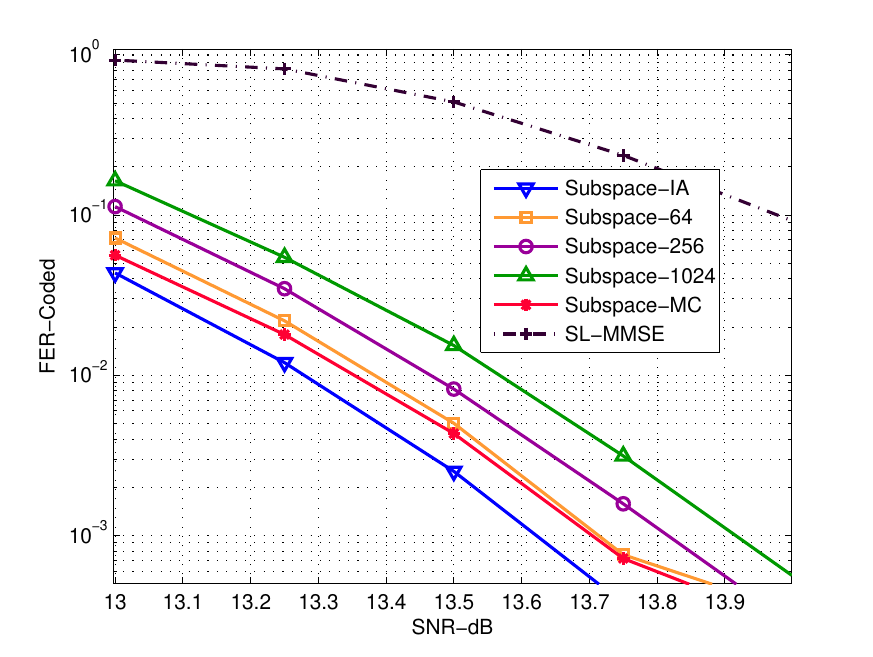}\vspace{-0.1in}
\caption{FER performance - $4\times4$ MU-MIMO - $N_{\user}=2$ - $\mathcal{M}$ is 64-QAM - uncorrelated.}
\label{f:ch8_fer4x4uncorr}
\end{figure}
\begin{figure}[t!]
\centering
\includegraphics[width=5in]{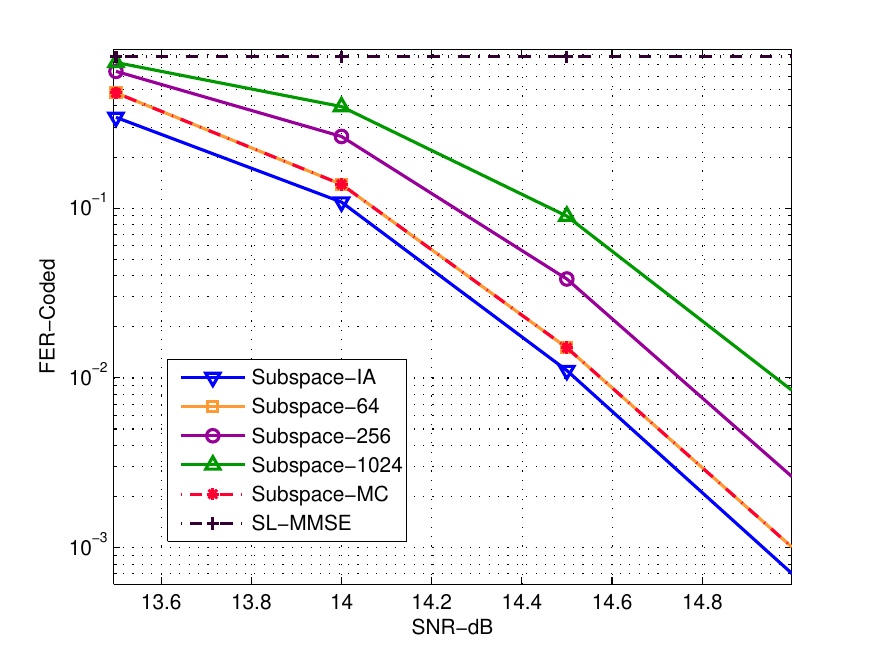}\vspace{-0.1in}
\caption{FER performance - $8\times8$ MU-MIMO - $N_{\user}=2$ - $\mathcal{M}$ is 64-QAM - uncorrelated.}
\label{f:ch8_fer8x8uncorr}
\end{figure}
\begin{figure}[t!]
\centering
\includegraphics[width=4.83in]{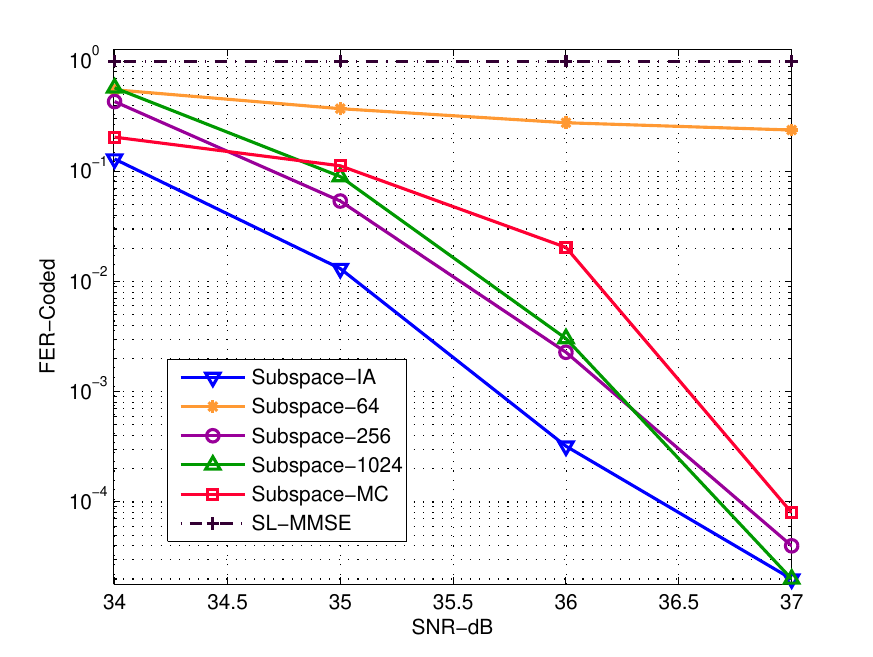}\vspace{-0.1in}
\caption{FER performance - $4\times4$ MU-MIMO - $N_{\user}=2$ - $\mathcal{M}$ is 64-QAM - $0.3$ correlation.}
\label{f:ch8_fer4x4corr0.3}
\end{figure}
\begin{figure}[t!]
\centering
\includegraphics[width=4.83in]{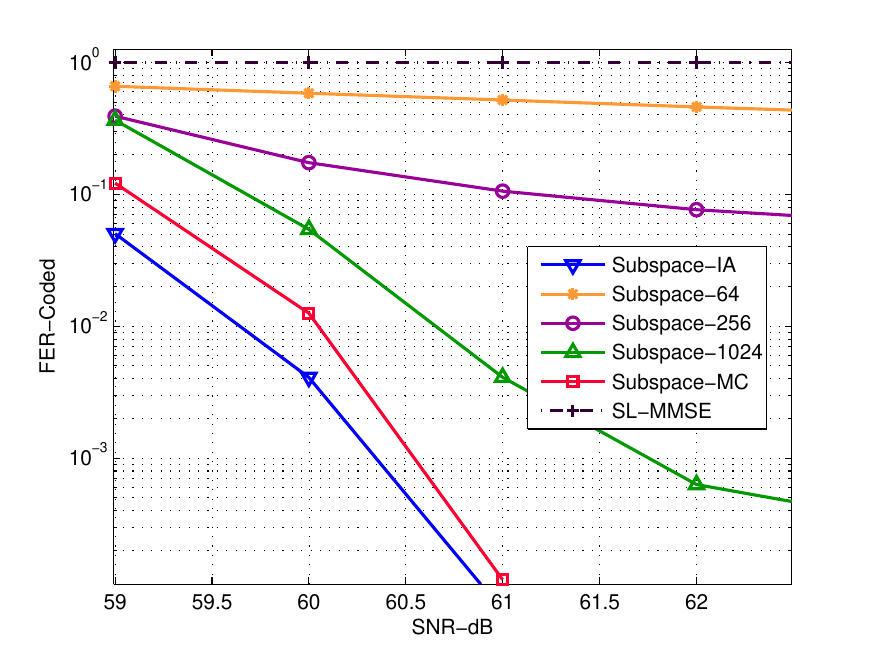}\vspace{-0.1in}
\caption{FER performance - $4\times4$ MU-MIMO - $N_{\user}=2$ - $\mathcal{M}$ is 64-QAM - $0.9$ correlation.}
\label{f:ch8_fer4x4corr0.9}
\end{figure}

%
\section{Joint MC and Detection Architecture}
\label{sec:ch8_implementationjoint}

After the classifier decides on $\hat{j}$, SO SSSD generates the bit LLRs. These two tasks can be realized jointly as previously described. The optimized architecture for the proposed MU-MIMO detector with hierarchical likelihood and feature-based MC is shown in Fig.~\ref{f:ch8_architecture-2}. At the core of this architecture is a SSSD, that in the first stage detects the first received symbol assuming all possible choices of the interferer's MT (except the one corresponding to no interference), and generates the corresponding list of Euclidian distance metrics for all $T$ vectors. These distances and symbols are stored in buffers (increased space complexity). The sum of the logarithm of the exponential of the distance metrics (no logarithms and exponentials with Max-Log-MAP and Quasi-Max-Log-MAP) are passed to an adder that accumulates them over a span of $T$ tones, during which the interferer modulation is assumed to be static. The resulting accumulated distances for each interference hypothesis are saved, and after deciding on a winning hypothesis, the corresponding distances are forwarded for LLR processing. This block is only activated when the feature-based classifier decides that an interferer exists. Otherwise, SSSD is applied once, assuming the interfering MT is ${\Phi}$, and the distances are forwarded for LLR processing.

As described in Sec.\ref{sec:ch7_implementation}, the proposed algorithms can be used in many communication standards. Since the interferer's modulation constellation remains static over $T$ tones and $L$ symbols, the particular choice of $T\!=\!234$ results in substantial computational savings in the context of 802.11ac. The detector only needs to run in the above mode to identify the interferer's constellation for one OFDM symbol in the frame. It can then switch back to normal SSSD. The total number of distance computations needed to generate the LLRs from the $234\!\times\!L$ data tones is $234\!\times\!L\!\times\!\abs{\mathcal{M}}\!\times\!N_{\user}$. With likelihood-based MC, the average overhead of MC is $234\!\times\!4\!\times\!\abs{\mathcal{M}}\!\times\!N_{\user}$. This corresponds to a maximum increase of only $4/L\%$, compared to the distances computed by a SSSD with perfect knowledge about the interferer. The size $L$ of the data field can take values between 8 and 1024, hence, the increase in distance computations ranges between $50\%$ and less than $0.4\%$. With hierarchical MC, the overhead in distance computations is $3/L\%$, which only occurs $80\%$ of times, and it thus ranges between $30\%$ and $0.23\%$. However, we have to add the complexity of ZF and computing second order CCs.

%
\section{Simulation Results}
\label{sec:ch8_simulation}

Joint MC and SSSD was implemented following the studied system model. The decision on the hypothesis is done after receiving $T\!=\!234$ tones. Turbo coding is used, with a code rate of $1/2$ and $8$ decoding iterations. The interferer was assumed to hop over the five hypotheses with equal probability on every new frame. Moreover, in addition to the regular channel $\mbf{H}$, we considered another channel $\mbf{H}_{c}$, that accounts for antenna correlation as defined for LTE.

The CCR of the MC schemes is shown in Fig.~\ref{f:ch8_ccr4x4uncorr}. Classifiers based on CC did not perform well since all hypotheses are QAMs, and the Quasi-Log-MAP and Quasi-Max-Log-MAP classifiers had a very similar performance. Hierarchical versions of the classifiers had exact likelihood-based classification performance, which means that the reduction in complexity came at no performance cost.

In the remaining figures, we illustrate the FER performance of five SSSD schemes: the IA SSSD, the MC-based SSSD with a hierarchical classifier, and the SSSD schemes that assume the interfering MT to be 64-QAM, 256-QAM, and 1024-QAM. For a $4\!\times\!4$ MU-MIMO system with $N_{\user}\!=\!2$, Fig.~\ref{f:ch8_fer4x4uncorr} shows the corresponding FER performance with 64-QAM and uncorrelated channels. Adding a MC routine and assuming the 64-QAM hypothesis are found to achieve near-IA performance. Moreover, less than $\unit[0.5]{dB}$ apart are the schemes that assume 256-QAM and 1024-QAM, respectively. Similar performance is noted for $8\!\times\!8$ MU-MIMO, with $N_{\user}\!=\!2$ and six interfering layers, as shown in Fig.~\ref{f:ch8_fer8x8uncorr}.

Upon adding channel correlation, some of the SSSD schemes that assume the interferer without MC will exhibit an error floor. Figure \ref{f:ch8_fer4x4corr0.3} shows the case when medium correlation is added (transmit and receive correlation factors of 0.3). The SSSD that assumes the interferer to be 16-QAM saturated, while the remaining schemes maintained near-IA performance. Pushing this further, Fig.~\ref{f:ch8_fer4x4corr0.9} shows the case when the channel is highly correlated (transmit correlation factor of 0.6 and receive correlation factor of 0.9). Here, all interference-assuming detectors saturated, each at a different FER level, with the best of them being the SSSD that assumes the 1024-QAM hypothesis.

To better understand these results we note the following: SSSD is less sensitive to interference than other detection schemes, which explains why assuming an interfering MT without estimation works fine. Moreover, 64-QAM is closer to the median of the hypotheses, and hence slicing over it is more likely to result in a similar output to slicing over the correct hypothesis. However, with high channel correlation, the slicing operation will cause larger errors. Thus, assuming the interfering MT to be of high order reduces this error while maintaining the structure of a QAM. Finally, correlation shifts the plots to a higher SNR range, where the CCR of MC is near 1, and therefore near-IA performance is maintained with MC-based SSSD. \\

\textbf{Conclusion} \\

Several low-complexity SSSDs for MU-MIMO systems have been proposed, alongside an architectural implementation. It has been concluded that while assuming the MT of the interferer without estimation is sufficient with good channel conditions, MC is required at high SNR with correlated channels. The MC complexity overhead has been shown to reduce to only $0.23\%$ in WiFi.

%% file: Chapter9_MC_MIMO.tex

\chapter{Per-Layer MC for Adaptive MIMO Systems}
\label{chapter:mumimoMC}

In this chapter, the problem of efficient MC in regular single-user MIMO systems is considered. Per-layer likelihood-based MC is proposed by employing subspace decomposition to partially decouple the transmitted streams. When detecting the MT of the stream of interest, a dense constellation is assumed on all remaining streams. The proposed classifier outperforms existing MC schemes at a lower complexity cost, and can be efficiently implemented in the context of joint MC and subspace data detection. The corresponding results appeared in \cite{7588171Sarieddeen}.

We follow the system model of Sec. \ref{sec:ch2_model}. In a MIMO system that supports non-uniform MTs, each of the $N$ transmitted symbols is assumed to be drawn from one of $S$ possible MTs, with equal probability. We develop MC schemes to estimate the MT per-layer, using the received signal $\mbf{y}$ and assuming perfect CSI.

%
\section{Likelihood-Based MIMO MC}
\label{sec:ch9_likelihoodbased}

With regular single-user MIMO, Bayesian hypothesis testing is performed on all the $S^N$ possible hypotheses, corresponding to $\mathcal{\bar{X}}_j\!=\!\mathcal{X}_{j,1}\times \ldots \times\mathcal{X}_{j,N}$ finite lattices ($j\in\{1,\ldots,S^N\}$), with likelihoods
\begin{equation}\label{eq:ch9_gen}
\Prb(\mbf{y};\mathcal{\bar{X}}_j)=\sum_{\mbf{x}\in\mathcal{\bar{X}}_j} \Prb(\mbf{y}|\mbf{x})\Prb(\mbf{x}).
\end{equation}
Under statistical independence between the components of $\mbf{x}$, and assuming uniform priors, $\Prb(x_{n})\!=\!1/\abs{\mathcal{X}_n}$, the decision metric can be expressed as
\begin{equation}\label{eq:ch9_gen2}
\eta = \argmax_{j\in\{1,\ldots,S^N\}}\sum_{\mbf{x}\in\mathcal{\bar{X}}_j} \Prb(\mbf{y}|\mbf{x})\frac{1}{\abs{\mathcal{X}_{j,1}}}\times\cdots\times\frac{1}{\abs{\mathcal{X}_{j,N}}}.
\end{equation}
Consequently, the Log-MAP decision metric (ALRT solution) is
\begin{equation}\label{eq:ch9_logmap}
\eta_{\text{L}} = \argmax_{j\in\{1,\ldots,S^N\}} \left(\log\frac{1}{\abs{\mathcal{X}_{j,1}}} + \cdots + \log\frac{1}{\abs{\mathcal{X}_{j,N}}}\\ + \log\sum_{\mbf{x}\in\mathcal{\bar{X}}_j}  \exp\left(-\frac{1}{\sigma^{2}} \norm{\mbf{y} - \mbf{Hx}}^{2}\right) \right).
\end{equation}
Solving~\eqref{eq:ch9_logmap} is computationally intensive, because for each $j$ we have to calculate $\abs{\mathcal{X}_{j,1}}\times\cdots\times\abs{\mathcal{X}_{j,N}}$ exponential terms. The Max-Log-MAP solution is:
\begin{equation}\label{eq:ch9_maxlogmap}
\eta_{\text{M}} = \argmax_{j\in\{1,\ldots,S^N\}} \left(\log\frac{1}{\abs{\mathcal{X}_{j,1}}} + \cdots + \log\frac{1}{\abs{\mathcal{X}_{j,N}}} - d^{\ML}_{j} \right)
\end{equation}
\begin{equation}
d^{\ML}_{j} = \min_{\mbf{x}\in\mathcal{\bar{X}}_j} \frac{1}{\sigma^{2}} \norm{\mbf{y} - \mbf{Hx}}^{2}.
\end{equation}
While Max-Log-MAP eliminates exponential operations, the number of Euclidean distance computations per hypothesis remains exponential, and computing the likelihood functions $S^N$ times is exhaustive. An alternative approach is required, that separates the transmitted signals for individual treatment, which results in only $\abs{\mathcal{X}_{j,n}}$ distance computations per layer $n$ and hypothesis $j\in\{1,\ldots,S\}$. This is achieved by the per-layer sub-optimal ALRT solution. With perfect CSI at the receiver, the sub-optimal ALRT classifier finds the ZF equalized output $\hat{\mbf{y}}^{\ZF}$, computes the scaled noise variance $\sigma^{2}_{\ZF}\!=\!(\mbf{h}_n^\mathcal{H}\mbf{h}_n)^{-1}\sigma^{2}$, and generates the likelihood function per layer $n$ as follows:
\begin{equation}\label{eq:ch9_zflogmap}
\eta_{\text{S}} =\argmax_{j\in\{1,\ldots,S\}} \left(\log\frac{1}{\abs{\mathcal{X}_{j,n}}} + \log\sum_{x_n\in\mathcal{X}_{j,n}}  \exp\left(-\frac{1}{\sigma^{2}_{\ZF}} \abs{\hat{y}^{ZF}_{n} - x_n}^{2}\right) \right).
\end{equation}
We seek a classifier that decouples the layers while maintaining distance metrics that are close to that of Log-MAP.

%
\section{Proposed MIMO MC}
\label{sec:ch9_proposed_MC}

We build on the WRD decomposition of Sec. \ref{eq:partitionCDwrd}. To generate the likelihood functions on all layers, the $N$ streams are decoupled, one at a time, by cyclically shifting the columns of $\mbf{H}$ and generating the punctured UTMs. Alternatively, a minimal swapping operation can put the layer of interest $n$ at the rightmost column location as described in Sec. \ref{sec:ch4_SPSD}. Each permuted $\mbf{H}^{(n)}$ is then WR-decomposed into $\mbf{W}^{(n)}$ and $\Rp^{(n)}$. By accumulating $T$ observations before deciding on a winning hypothesis, the proposed likelihood functions at layer $n$ can be expressed as:
\begin{equation}\label{eq:ch9_sublogmap}
\hat{\eta}_{\text{L}}=\argmax_{j\in\{1,\ldots,S\}} \sum_{t=1}^{T} \left(\log\frac{1}{\abs{\mathcal{X}_{j,n}}} \\ + \log\sum_{\tilde{x}_2\in\mathcal{X}_{j,n}} \exp\left(-\frac{1}{\sigma^{2}} \norm{\tilde{\mbf{y}}^{(n)} - \Rp^{(n)}\mbf{x}}^{2}\right) \right)
\end{equation}
\begin{equation}\label{eq:ch9_submaxlogmap}
\hat{\eta}_{\text{M}}=\argmax_{j\in\{1,\ldots,S\}} \sum_{t=1}^{T} \left(\log\frac{1}{\abs{\mathcal{X}_{j,n}}} -  \hat{d}^{\ML}_{j} \right)
\end{equation}
\begin{equation}
\hat{d}^{\ML}_{j} = \min_{\tilde{x}_2\in\mathcal{X}_{j,n}} \frac{1}{\sigma^{2}} \norm{\bar{\mbf{y}}^{(n)}\!-\!\Rp^{(n)}\mbf{x}}^{2}.
\end{equation}

Note that the knowledge of MTs on all remaining layers is required, which is infeasible in an independent per-layer scheme. Therefore, we propose to do slicing assuming dense constellations, 1024-QAM for example. The idea of slicing over a dense constellation comes from the work on MU-MIMO in Sec.\ref{sec:ch8_proposedMC-3}, where we have shown that near-optimal data detection can be achieved while assuming interferers to have high order MTs, which captures the geometry of constellations while minimizing errors. Dense constellations are not an issue in our case, since subspace decomposition only employs these constellations in slicing operations.

Table \ref{table:ch9_complexity} compares the upper bounds on computational complexity of studied classifiers, in terms of the number of Euclidean distance computations, as well as exponential and logarithmic operations, where $\mathcal{X}_{\text{max}}$ is the largest possible MT. Note that the table does not account for the less significant preprocessing computations (ZF equalization, QRD/WRD) that can be computed once for a large number of observations when the channel variation is slow. While computations in optimal ALRT are exponential in the number of transmit antennas, they are linear in the proposed subspace-decomposition-based classifiers and sub-optimal ALRT solution (the latter is less complex since its distance computations are one dimensional).

Note that had we used QRD instead of WLD, the distance metrics in~\eqref{eq:ch9_sublogmap} and \eqref{eq:ch9_submaxlogmap} would have been executed via SIC as done in LORD, resulting in LORD-Log-MAP and LORD-Max-Log-MAP classifiers, respectively.

\begin{table}[!t]
\caption{\textbf{Computational Complexity of MIMO MC Schemes}}\vspace{0.15in}
\label{table:ch9_complexity}
\centering
\begin{tabular}{| c || c | c | c | }
\hline
Approach & Euc. Dist. & Exp. & Log. \\
\hline
Log-MAP (ALRT) & $S^N\times\abs{\mathcal{X_{\text{max}}}}^N$ & $S^N\times\abs{\mathcal{X_{\text{max}}}}^N$ & $S^N$ \Tstrut \\ %
Max-Log-MAP & $S^N\times\abs{\mathcal{X_{\text{max}}}}^N$ & $0$ & $0$ \Tstrut \\ %
Sub-optimal ALRT & $N\times S\times\abs{\mathcal{X_{\text{max}}}}$ & $N\times S\times\abs{\mathcal{X_{\text{max}}}}$ & $S$ \Tstrut \\ %
Subspace-Log-MAP & $N\times S\times\abs{\mathcal{X_{\text{max}}}}$ & $N\times S\times\abs{\mathcal{X_{\text{max}}}}$ & $S$ \Tstrut \\
Subspace-Max-Log-MAP & $N\times S\times\abs{\mathcal{X_{\text{max}}}}$ & $0$ & $0$ \Tstrut \\
\hline
\end{tabular}
\end{table}

\begin{algorithm}[t]
\caption{Proposed per-layer joint MC and detection}\label{ch9_propalg}
\begin{algorithmic}[1]
\State Swap the column of interest $n$ with column $N$ in $\mbf{H}$
\State Decompose the channel matrix
\State Calculate the distance metrics for all hypotheses while assuming the MTs on the remaining layers to be 1024-QAM
\State Calculate the classifier likelihood function
\State Repeat steps 3 and 4 for $T$ observations, accumulate likelihoods, and decide on the winning hypothesis
\State Forward the distance metrics that correspond to the winning hypothesis for bit LLR generation
\end{algorithmic}
\end{algorithm}

%
\section{Joint MC and Detection Architecture}
\label{sec:ch9_joint}

Be it SSSD or MC, while independently processing a layer of interest, the MTs on the remaining layers are unknown, and parallel slicing or SIC is conducted assuming 1024-QAM. This means that the distance metrics computed for data detection are identical to those computed in~\eqref{eq:ch9_sublogmap} and \eqref{eq:ch9_submaxlogmap}, and thus combining MC and detection results in a minimal MC overhead.

\begin{figure}[!t]
\centering
\includegraphics[width=4.4in]{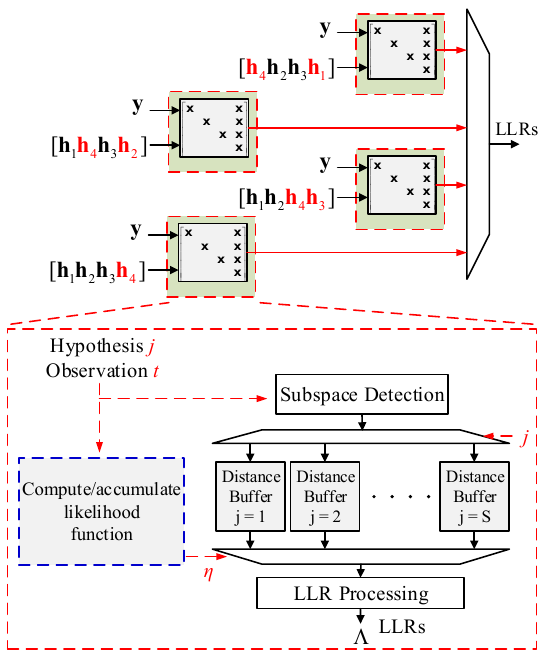}\vspace{-0.1in}
\caption{Joint MC and subspace detection architecture.}
\label{f:ch9_architecture}
\end{figure}

The joint MC and detection setup is summarized in algorithm \ref{ch9_propalg} and architecturally illustrated in Fig.~\ref{f:ch9_architecture}. For $T$ observations, the detection routine is executed $T$ times for all hypotheses, and the resulting distance metrics are stored in memory. Concurrently, the likelihood of each hypothesis is computed. Eventually, the metrics corresponding to the winning hypothesis are retrieved for LLR processing. The receiver can run in this joint mode for a sufficient number of observations and then switch back to regular data detection. Moreover, since the operations on different layers are independent, the proposed algorithm can be parallelized on multiple processing units.

%
\section{Simulation Results}
\label{sec:ch9_simulation}

Several MC and detection schemes were simulated in the context of $4\!\times\!4$ MIMO. We considered five hypotheses of MTs per layer, varying with equal probability on every new frame, which are ${\Phi}$, QPSK, 16-QAM, 64-QAM and 256-QAM, with ${\Phi}$ here corresponding to the case when the transmitting antenna is silent, as opposed to no interference in MU-MIMO. Note that an all-QAM set of hypotheses that only differs by modulation order is hard to classify, but is more likely to occur in future standards. The winning hypothesis was decided after accumulating $T\!=\!1000$ observations. Turbo coding was used, with a code rate of $1/2$ and $8$ decoding iterations. Moreover, in addition to the regular channel $\mbf{H}$, we considered a correlated channel with a correlation factor of 0.3 as defined for LTE.
\begin{figure}[t]
\centering
\includegraphics[width=5.83in]{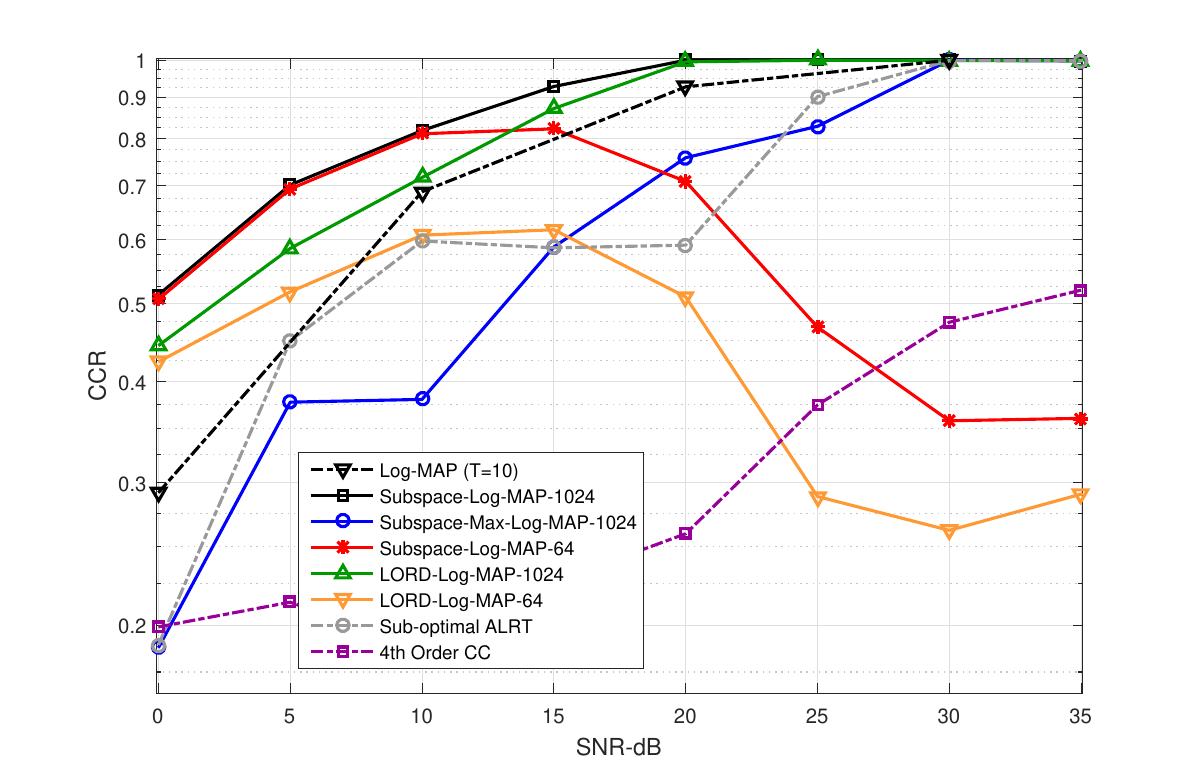}\vspace{-0.1in}
\caption{CCR performance - uncorrelated channels.}
\label{f:ch9_plot1}
\end{figure}
\begin{figure}[h!]
\centering
\includegraphics[width=5.83in]{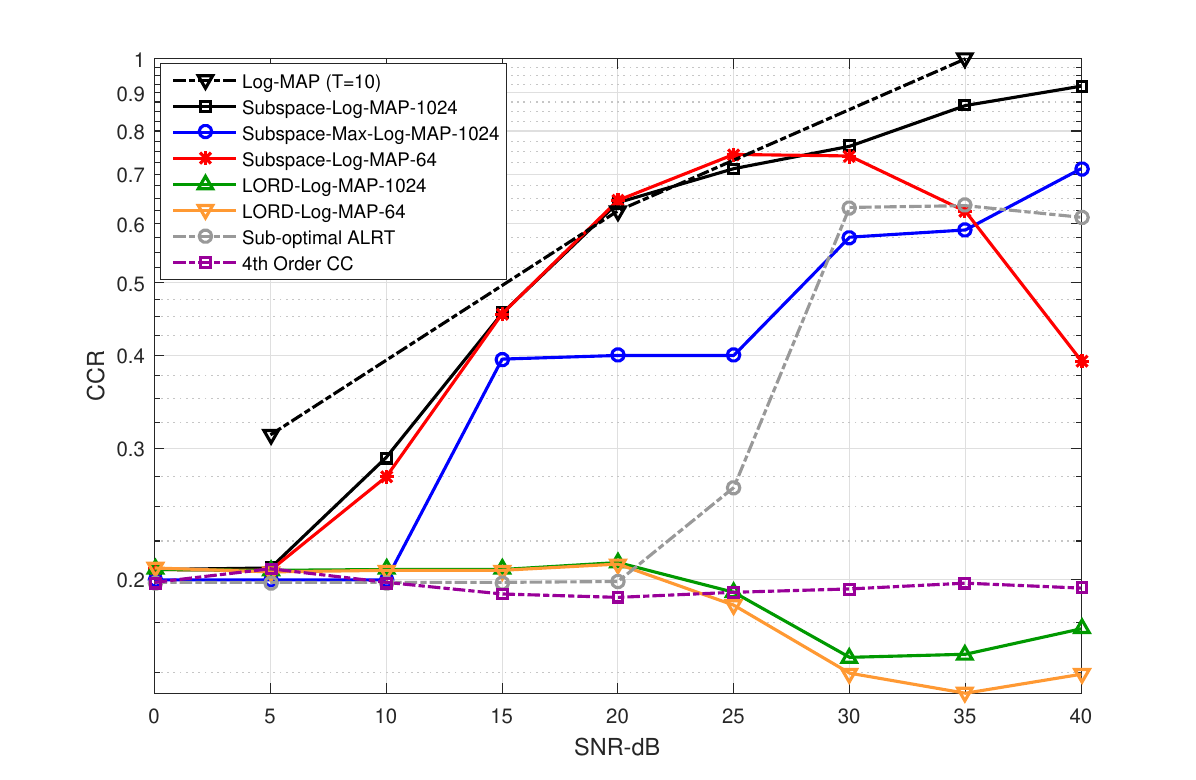}\vspace{-0.1in}
\caption{CCR performance - correlated channels.}
\label{f:ch9_plot2}
\end{figure}
\begin{figure}[h!]
\centering
\includegraphics[width=5.83in]{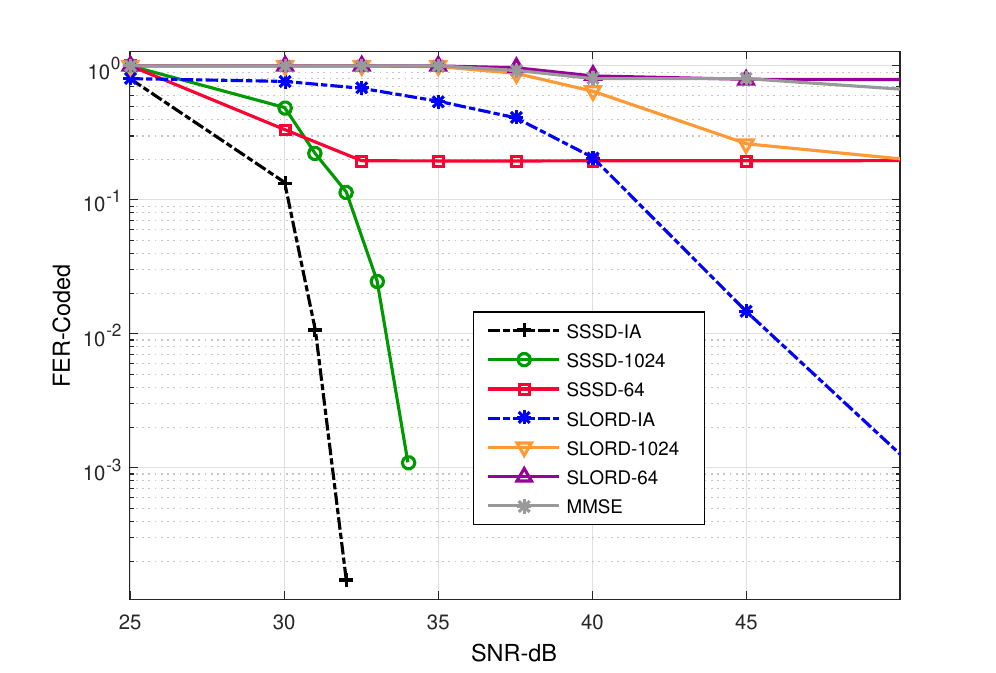}\vspace{-0.1in}
\caption{FER performance - correlated channels - 64-QAM on layer of interest.}
\label{f:ch9_plot3}
\end{figure}

Figure~\ref{f:ch9_plot1} shows that for uncorrelated channels, the best performance is achieved by Subspace-Log-MAP and LORD-Log-MAP classifiers, when the MTs on the remaining layers are assumed to be 1024-QAM. The Subspace-Max-Log-MAP and sub-optimal ALRT classifiers lag behind, but are also capable of achieving unity CCR at high SNR. However, assuming 64-QAMs instead of 1024-QAMs resulted in bad classification performance for both subspace-decomposition and LORD-based classifiers. The exhaustive Log-MAP classifier with $T=10$ observations only was added as a reference, and the much less complex proposed approaches outperformed it. Also, the reference feature based (\nth{4} order CCs) classifier performed very bad with an all-QAM set of hypotheses. Figure~\ref{f:ch9_plot2} then shows that for highly correlated channels, only Subspace-Log-MAP performs well at high SNR, approaching the upper Log-MAP bound.

The corresponding coded FER performance of the proposed detectors with high channel correlation is shown in Fig.~\ref{f:ch9_plot3}. The detectors were simulated assuming the layer of interest to use 64-QAM (following successful per-layer MC), while the MTs at the remaining layers were unknown, and randomly hopping over possible hypotheses. Both SLORD and SSSD were tested, assuming the remaining MTs to be 1024-QAM or 64-QAM. These detectors were compared to the regular MT-aware SLORD and SSSD that have perfect knowledge of MTs on all layers. While regular SSSD beats SLORD by more than $\unit[10]{dB}$, only assuming 1024-QAM in conjunction with SSSD was able to achieve near MT-aware performance. This declares the subspace-based classifiers winners in the context of joint MC and detection. \\

\textbf{Conclusion} \\

Low-complexity per-layer MC schemes have been proposed for MIMO systems, based on subspace decomposition. It has been shown that assuming the MT on all layers except the layer of interest to be a dense constellation results in good classification performance. This assumption has been proved to have a negligible performance degradation cost in SSSD, which made fully parallelizable efficient joint MC and data detection feasible.

%% file: Chapter10_MASSIVE.tex

\chapter{Massive MIMO 1-Bit Precoding}
\label{chapter:Massive}

In this chapter, the problem of efficient precoding in the downlink of massive multiple-input multiple-output systems is considered. We investigate the performance and complexity tradeoffs of search-based precoders in the context of 1-bit digital-to-analog converters. By adapting the procedures of popular search-based detection algorithms to 1-bit quantized precoding, two families of nonlinear precoders are proposed. The first employs QR-decomposition combined with tree-based search techniques, and the second uses Gibbs sampling for search enumerations without decomposing the channel. Simulations demonstrate that some of the proposed schemes outperform reference nonlinear precoders, both in performance and complexity with low order MIMO, and in performance with a graceful increase in computations in the context of massive MIMO with high order modulation types.

%

\section{System Model}\label{sec:ch10_sysmodel}
\begin{figure}[t!]
\centerline{
\includegraphics[width=5.5in,angle=0]{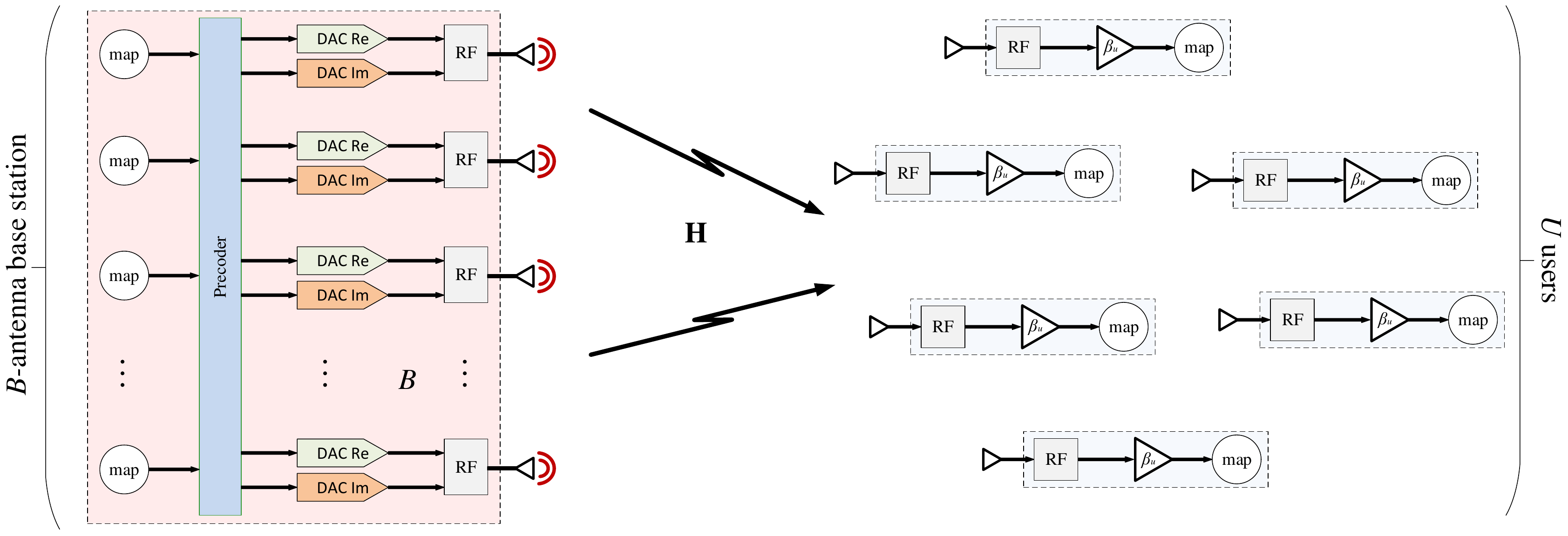}\vspace{-0.1in}
}
\caption{Massive MIMO downlink system model}
\label{f:plot0}
\end{figure}

We consider, as shown in Fig.\ref{f:plot0}, a downlink multiuser MIMO scenario in which a BS with $B$ antennas serves $U$ single-antenna UEs at the same time and frequency. While the RF chains are assumed ideal, the DACs at the BSs are subject to 1-bit quantization. The equivalent complex baseband input-output system relation is $\mbf{y} = \mbf{H}\mbf{x} + \mbf{n} $, where $\mbf{y}=[y_{1}\cdots y_{u}\cdots y_{U}^{}]^{T}\!\in\!\mathcal{C}^{U\times1}$ is the received vector, $\mbf{H}\!\in\!\mathcal{C}^{U\times B}$ is the downlink circularly-symmetric channel matrix with $\mathcal{CN}(0,1)$ entries, $\mbf{x}=[x_{1}\cdots x_{b}\cdots x_{B}^{}]^{T}\!\in\!\bar{\mathcal{L}}^{B\times1}$ is the precoded transmitted symbol vector, and $\mbf{n}\!\in\!\mathcal{C}^{U\times1}$ is the noise vector with $\mathcal{CN}(0,\sigma^2)$ entries. With finite-precision, the $\nth{b}$ symbol of $\mbf{x}$, $x_b\!=\! l_{\mathrm{R}}+jl_{\mathrm{I}}\in \bar{\mathcal{L}}$, has quantized in-phase and quadrature components, i.e., $l_{\mathrm{R}},l_{\mathrm{I}}\!\in\! \mathcal{L}$ where $\mathcal{L}=\{l_{0},l_{1},\cdots,l_{L-1}\}$ is the set of possible quantization labels and $\bar{\mathcal{L}}=\mathcal{L}\times\mathcal{L}$. For 1-bit quantization, we have $\bar{L}=\abs{\mathcal{L}}=2$.

Precoding is used to increase the array gain and reduce multi-user interference. Prior to precoding, the symbol vector $\mbf{s}\!=\![s_{1}\!\cdots\! s_{2}\!\cdots\! s_{U}^{}]^{T}\!\in\!\mathcal{M}^{U}$ is obtained by mapping the information bits to a QAM constellation $\mathcal{M}$. The BS then uses the knowledge of $\mbf{H}$ to precode $\mbf{s}$ into $\mbf{x}=\mathcal{P_r}(\mbf{s},\mbf{H})$, through a mapping function $\mathcal{P_r}(\cdot,\cdot)=\mathcal{M}^{U}\times\mathcal{C}^{U\times B}\rightarrow\bar{\mathcal{L}}^{B}$. The precoded vector satisfies the average power constraint $\mathsf{E}\left[\norm{\mbf{x}}^2\leq P\right]$ for some maximum transmit power $P$, and the SNR is consequently defined as $\mathsf{SNR}=P/\sigma^2$. We thus have $\bar{\mathcal{L}} = \{ \sqrt{P/2B}(\pm1 \pm\!j)\}$. At the receiver, we assume that the $\nth{u}$ UE is capable of accounting for array gain, by scaling the received symbol $y_u$ by a precoding factor $\beta\!\in\!\mathcal{R}$, that depends on a perfectly known $\mbf{H}$, to obtain an estimate $\hat{s}_u=\beta y_u$, which gets sliced over $\mathcal{M}$.

%
\section{Reference Precoders}
\label{sec:ch10_ref_precoding}

A popular approach for formulating the precoding optimization problem is minimizing the mean square error between the received signal and the transmitted symbols under the power constraint ~\cite{jacobsson2017quantized,Joham1468466}. With coarse quantization, an additional distortion due to finite precoder outputs exists. Since optimal precoding is exhaustive due to the cardinality of $\bar{\mathcal{L}}^{B}$, only LQPs and low-complexity NLQPs are feasible.

\subsection{Linear Quantized Precoders}
\label{sec:ch10_linear_precoding}

With LQPs, a precoding matrix $\mbf{P}\!\in\!\mathcal{C}^{B\times U}$ is designed based solely on $\mbf{H}$. The transmit vector is generated as a quantized version of $\mbf{P}\mbf{s}$, i.e., $\mbf{x} = \mathcal{Q}\left(\mbf{P}\mbf{s}\right)$, where $\mathcal{Q}(\cdot)\!:\!\mathcal{C}^{B}\!\rightarrow\!\bar{\mathcal{L}}^{B}$ is the quantizer-mapping function
\begin{equation}\label{eq:ch10_quantizer_map}
  \mathcal{Q}(\mbf{z}) = \sqrt{P/2B} \left(\sgn(\Re\{\mbf{z}\}) + j \sgn(\Im\{\mbf{z}\}) \right).
\end{equation}
The MMSE precoding problem is expressed as
\begin{align}\label{eq:ch10_LQP}
  & \min_{\mbf{P}\in\mathcal{C}^{B\times U},\ \beta\in\mathcal{R}} \ \ {\mathsf{E}_{\mbf{s}}[\norm{\mbf{s}-\beta\mbf{Hx}}^2] + \beta^2U\sigma^2} \\
   & \text{subject to} \ \ \mathsf{E}_{\mbf{s}}[\norm{\mbf{x}}^2] \leq P, \ \ \text{and} \ \ \beta>0.
\end{align}
Solving \eqref{eq:ch10_LQP} in closed form is challenging due to quantization. An alternative solution~\cite{jacobsson2017quantized} is to design linear precoders that assume infinite-resolution DACs at the BS, and then quantize the resulting precoded vector. For example, for the ZF precoder, we have
\begin{align}\label{eq:ch10_ZF}
  \mbf{P}^{\ZF} &= \frac{1}{\beta^{\ZF}}\mbf{H}^H(\mbf{HH}^H)^{-1} \\
  \beta^{ZF} &= \frac{1}{\sqrt{P}}\sqrt{\tr{((\mbf{HH}^H)^{-1})}}.
\end{align}
The resulting precoded vector is $\mbf{x}^{\ZF}=\mathcal{Q}(\mbf{P}^{\ZF}\mbf{s})$.

\subsection{Nonlinear Quantized Precoders}
\label{sec:ch10_nonlinear_precoding}

With NLQPs, $\mbf{x}$ is obtained as a function of both $\mbf{H}$ and the instantaneous $\mbf{s}$. The 1-bit NLQP problem is defined as
\begin{align}\label{eq:ch10_NLQP}
  & \min_{\mbf{x}\in\bar{\mathcal{L}}^B,\ \beta \in \mathcal{R}} \ \ {\norm{\mbf{s}-\beta\mbf{Hx}}^2 + \beta^2U\sigma^2} \\
   & \text{subject to} \ \ \beta>0.
\end{align}
Note that this resembles an $\text{L}_2$-norm regularized closest-vector problem (CVP), and solving it exhaustively requires $\abs{\bar{\mathcal{L}}^B} = 4^B$ Euclidean distance computations. The SD can solve the CVP exactly with less computations. To map the SD procedure to the SP at the transmitter side, it is noted in \cite{jacobsson2017quantized} that
\begin{align}\label{eq:ch10_SP1}
  \norm{\mbf{s}-\beta\mbf{Hx}}^2 + \beta^2U\sigma^2 &= \norm{\mbf{s}-\beta\mbf{Hx}}^2 + \beta^2 \frac{U\sigma^2}{P}\norm{\mbf{x}}^2 \\
   & = \norm{\bar{\mbf{s}}-\beta\bar{\mbf{H}}\mbf{x}}^2,
\end{align}
since $\norm{\mbf{x}}^2=P$ in the case of 1-bit quantization, and where $\bar{\mbf{s}} = [ \mbf{s}^T \ \mbf{0}_B^T]^T$ and $\bar{\mbf{H}} = [\mbf{H}^T \ \sqrt{U\sigma^2/P}\mbf{I}_B]^T$. The problem formulation in \eqref{eq:ch10_NLQP} can thus be rewritten as
\begin{equation}\label{eq:ch10_SP2}
  \min_{\mbf{x}\in\bar{\mathcal{L}}^{B}} \ \ \norm{\bar{\mbf{s}}-\beta\bar{\mbf{H}}\mbf{x}}^2,
\end{equation}
which can be transformed into a tree-search problem via QRD, where branches that exceed a specific radius constraint are pruned. However, the SP is still considered a brute-force solution in the context of massive MIMO. Hence, the most popular approximations to solve \eqref{eq:ch10_NLQP} are via semi-definite relaxation \cite{castaneda20171}, namely the SQUID precoder and its variants.

%
\section{Proposed QRD-Based Precoders}
\label{sec:ch10_prop_qrd}

Building on the construction in \eqref{eq:ch10_SP1} and applying QRD, we have $\bar{\mbf{H}}\!=\!\mbf{Q}\mbf{R}$, where $\mbf{Q}\!\in\!\mathcal{C}^{(B+U) \times B}$ has orthonormal columns, and $\mbf{R}\!\in\!\mathcal{C}^{B\times B}$ is a square UTM. Let $\tilde{\mbf{s}}\!=\![\tilde{s}_{1}\cdots \tilde{s}_{b}\cdots \tilde{s}_{B}^{}]\!=\!\mbf{Q}^H\bar{\mbf{s}}\!\in\!\mathcal{C}^{B\times1}$ and $\tilde{\mbf{R}}\!=\! \beta\mbf{R}\!=\![r_{bl}^{}]\!\in\!\mathcal{C}^{B\times B}$ with $r_{bb}\!\in\!\mathcal{R}^{+}$, the problem in \eqref{eq:ch10_SP2} can be expressed as
\begin{equation}\label{eq:ch10_QRD_precoding}
  \min_{\mbf{x}\in\bar{\mathcal{L}}^{B}} \ \ \norm{\tilde{\mbf{s}}-\tilde{\mbf{R}}\mbf{x}}^2.
\end{equation}
We next present low-complexity solutions to~\eqref{eq:ch10_QRD_precoding}. The first solution follows from the NC detector. We propose the NC precoder (NCP) in which successive back-substitution and slicing are applied to suppress co-antenna interference. Hence, $\mbf{x}^{\NCP}=[ x^{\NCP}_1\cdots x^{\NCP}_n\cdots x^{\NCP}_N ]$ is computed as
\begin{equation}\label{eq:ch10_SIC}
x^{\NCP}_b = \mathcal{Q}\left( \left(\tilde{s}_{b} - \sum_{l=b+1}^{B} r_{bl} x^{\NCP}_{l}\right)/r_{bb} \right),
\end{equation}
for $b = B,B-1,\cdots, 1$.

In the second solution, we map the search routine in the CD to a chase precoder (CP) at the transmitter. The CP proceeds by populating a list $\mathcal{S}(\tilde{\mbf{s}},\tilde{\mbf{R}})$ of candidate symbol vectors for final decision. It first partitions $\tilde{\mbf{s}}$, $\tilde{\mbf{R}}$, and $\mbf{x}$ as
\begin{equation}\label{eq:ch10_partitionCP}
    \tilde{\mbf{s}} =
        \begin{bmatrix}
            \tilde{\mbf{s}}_{1} \\
            \tilde{s}_{B}^{}
        \end{bmatrix}, \ \
    \tilde{\mbf{R}} =
        \begin{bmatrix}
            \mbf{A} & \mbf{b} \\
            \mbf{0} & c
        \end{bmatrix}, \ \
    \mbf{x} =
        \begin{bmatrix}
            \mbf{x}_{1} \\
            x_{B}^{}
        \end{bmatrix},
\end{equation}
where $\tilde{\mbf{s}}_{1} \!\in\! \mathcal{C}^{(B-1)\times1}$, $\tilde{s}_{N}^{} \!\in\! \mathcal{C}^{1\times1}$, $\mbf{A} \!\in\! \mathcal{C}^{(B-1)\times(B-1)}$, $\mbf{b}\!\in\! \mathcal{C}^{(B-1)\times1}$, $c \!\in\! \mathcal{R}^{1\times1}$, $\mbf{x}_{1} \!\in\! \bar{\mathcal{L}}^{B-1}$, $\mbf{0}$ is a $1\times(B-1)$ vector of zero-valued entries, and $x_{B}^{} \!\in\! \bar{\mathcal{L}}$. Then, for each $x_B$ at the root layer, a candidate vector is accumulated via SIC on upper layers using~\eqref{eq:ch10_SIC} and added to $\mathcal{S}$. The number of candidate vectors in $\mathcal{S}$ is $\abs{\bar{\mathcal{L}}}\!=\!2^{\bar{L}}\!=\!4$, and the transmit vector is chosen from $\mathcal{S}$ to be
\begin{equation}\label{eq:ch10_CP2}
  \mbf{x}^{\CP} = \argmin_{\mbf{x} \in \mathcal{S}} \norm{\tilde{\mbf{s}}\!-\!\tilde{\mbf{R}}\mbf{x}}^{2}.
\end{equation}

Furthermore, we propose a third precoder that is based on LORD, which we call the layered orthogonal lattice precoder (LORP). Instead of executing the CP routine once, LORP repeats chase precoding with different layer orderings, each time with a different layer as a root, by cyclically shifting the columns of $\bar{\mbf{H}}$. The best output from these trials is the final solution. Each permuted $\bar{\mbf{H}}$ at step $t$, $t\!=\!1,\!\cdots\!,B$, is QR-decomposed into $\mbf{Q}^{(t)}$ and $\mbf{R}^{(t)}$ and then partitioned as in~\eqref{eq:ch10_partitionCP}. Let $\mbf{x}_{(t)}^{\CP}$ denote the CP solution at step $t$. Then, the final solution $\mbf{x}^{\LORP}$ is $\mbf{x}_{(t_{min})}^{\CP}$, where

\begin{equation}\label{eq:ch10_LORPout}
    t_{min} = \argmin_{ t \in \{1,\cdots,B\} } \norm{\tilde{\mbf{s}} - \tilde{\mbf{R}}\mbf{x}_{(t)}^{\CP}}^{2}.
\end{equation}
Since distances are preserved under different layer orderings with QRD, the accumulated candidate vectors across different partitions form an ``extended'' candidate list, which results in an added gain with LORP compared to CP. Note that the decomposition steps are independent, and can hence be computed in parallel.

All these proposed precoders depend on the value of the parameter $\beta$. However, the optimal $\beta$ is unknown in practice. Since the objective function in \eqref{eq:ch10_NLQP} is quadratic in $\beta$, we have
\begin{align}\label{eq:ch10_beta_hat}
    \hat{\beta}(\mbf{x}) &= \frac{\Re\{\mbf{s}^H\mbf{Hx}\}}{\norm{\mbf{Hx}}^2 + U\sigma^2} \\
    & = \frac{\Re\{\mbf{s}^H\mbf{Hx}\}}{\mbf{x}^H(\mbf{H}^H\mbf{H}+\frac{U\sigma^2}{P}\mbf{I}_B)\mbf{x}}.
\end{align}
Hence, knowing $\mbf{x}$ suffices to compute $\beta$. The proposed precoders can then be implemented iteratively. At the first iteration $\tau\!=\!1$, the precoder, say NCP, is initialized with $\beta^{\ZF}$. Then, after obtaining the precoding vector $\mbf{x}_{(\tau)}^{\NCP}$ at an iteration $\tau\!>\!1$, we can compute $\beta_{\tau\!+\!1}^{\NCP} \!=\! \hat{\beta}(\mbf{x}_{(\tau)}^{\NCP})$ using~\eqref{eq:ch10_beta_hat}. This continues until convergence, or for a maximum of 10 iterations.

Moreover, since $\beta$ depends on the instantaneous vector $\mbf{s}$ with NLQP, it cannot be estimated at the receiver. As a solution, it is shown in~\cite{7869149Jacobsson} that it is sufficient to modify the precoding problem in~\eqref{eq:ch10_NLQP} such that a common $\beta$ is chosen for a block of transmit symbols. This allows the UEs to estimate $\beta$ through pilot transmissions or blind estimation techniques. However, in the specific case of the constant envelope QPSK modulation, which is capacity-achieving with 1-bit quantization, there is no need for scaling. Hence, computing $\beta$ at the receiver is redundant in this case, but including it as an extra degree of freedom in the iterative precoding problem at the transmitter will still lead to performance enhancement.

%
\section{Proposed Non-QRD-Based Precoders}
\label{sec:ch10_prop_nqrd}

Taking into consideration the fact that the QRD of a large matrix is computationally demanding, we propose alternative search-based precoders that exploit \eqref{eq:ch10_NLQP} directly, without decompositions. The resultant schemes are similar to precoding designs with finite candidates in the literature. We hereby study their applicability to 1-bit massive MIMO. The first precoder mimics the behaviour of Gibbs-sampling-based detectors for large MIMO, and it is hence called the Gibbs precoder (GP).

\begin{algorithm}[!t]
\caption{Gibbs precoder algorithm}\label{ch10_GPpalg}
\begin{algorithmic}[1]
\State $\mbf{x}_{(\tau=1)}^{\GP}\!=\!\mbf{x}^{\ZF}$, $\beta_{(\tau=1)}^{\GP}\!=\!\beta^{\ZF}$
\For{$\tau =2:10$}
	\State $\acute{\mbf{x}}\!=\!\mbf{x}_{(\tau-1)}^{\GP}$
	\For{$b=1:B$}
        \State $x^{\GP}_b(\tau)\!=\!\argmin_{\acute{x}_b \in \mathcal{X}} \norm{\mbf{s}-\beta_{(\tau-1)}^{\GP}\mbf{H}\acute{\mbf{x}}}_2^2 \!+\! {\beta_{(\tau-1)}^{\GP}}^2U\sigma^2$
	\EndFor
    \State $\beta^{\GP}_{(\tau)} = \Re\{\mbf{s}^\mathcal{H}\mbf{H}\mbf{x}^{\GP}_{(\tau)}\}/ \left(\norm{\mbf{H}\mbf{x}^{\GP}_{(\tau)}}_2^2 + U\sigma^2\right)$
    \If {$\mbf{x}^{\GP}_{(\tau)} == \mbf{x}^{\GP}_{(\tau-1)}$}
    	\State Break
    \EndIf
\EndFor
\end{algorithmic}
\end{algorithm}

Starting from the ZF solution at $\tau\!=\!1$, we have $\mbf{x}_{(\tau=1)}^{\GP}\!=\!\mbf{x}^{\ZF}$ and $\beta_{(\tau=1)}^{\GP}\!=\!\beta^{\ZF}$. At each iteration $\tau \!>\!1$, the GP routine begins by varying the symbol $\acute{x}_b$, $1\!<\!b\!<\!B$, in the vector $\acute{\mbf{x}}\!=\![\acute{x}_{1}\cdots \acute{x}_{b}\cdots \acute{x}_{B}^{}]^{T}\!=\!\mbf{x}_{(\tau-1)}^{\GP}$, while keeping the remaining elements of $\acute{\mbf{x}}$ intact. For each possible value of $\acute{x}_b$, the metric $d(\acute{\mbf{x}})=\norm{\mbf{s}-\beta_{(\tau-1)}^{\GP}\mbf{H}\acute{\mbf{x}}}_2^2 + {\beta_{(\tau-1)}^{\GP}}^2U\sigma^2$ gets computed and stored in memory, alongside its corresponding symbol flip. This repeats for all remaining symbols in $\acute{\mbf{x}}$, where every time a specific symbol is being explored, the remaining symbols retain their values from the initial solution $\mbf{x}_{(\tau-1)}^{\GP}$. After accumulating $B\times(2^{\bar{L}}-1)+1$ metric values, the symbol flip that best minimizes $d$ is applied to the initial solution to obtain $\mbf{x}_{(\tau)}^{\GP}$. The corresponding procedure is summarized in Algorithm \ref{ch10_GPpalg}. Let $\mathcal{G}$ be the set of $B\times(2^{\bar{L}}-1)+1$ tested candidate vectors, GP effectively solves
\begin{equation}\label{eq:GP}
  \mbf{x}_{(\tau)}^{\GP} = \argmin_{\acute{\mbf{x}} \in \mathcal{G}} \norm{\mbf{s}-\beta_{(\tau-1)}^{\GP}\mbf{H}\acute{\mbf{x}}}_2^2 + {\beta_{(\tau-1)}^{\GP}}^2U\sigma^2.
\end{equation}
This differs from conventional Gibbs sampling, where exploring a new symbol starts after updating the solution with the best value of the previous symbol. In the proposed GP all symbols can be explored in parallel. Furthermore, metric computations can be significantly reduced when only one symbol is modified compared to the initial solution. Computing $\beta_{(\tau)}^{\GP}$ using \eqref{eq:ch10_beta_hat} completes one iteration, and this gets repeated until convergence, or for a maximum of $T\!=\!10$ iterations.

The GP can be seen as the non-QRD version of LORP. We similarly propose the single-layer GP (SGP) as a non-QRD version of CP. With SGP, only one arbitrary symbol is varied in the search routine. Furthermore, we add as a reference the exhaustive ML precoder, that in each iteration $\tau$ exhaustively searches all possible values of $\acute{\mbf{x}}\!\in\!\bar{\mathcal{L}}^{B}$, to find the vector that best minimizes the distance metric $d(\acute{\mbf{x}})$. Since QRD does not modify distances, SP and ML precoding are identical.

%
\section{Complexity Analysis}
\label{sec:complexity}

\begin{figure}[t!]
\centering
\includegraphics[width=4in]{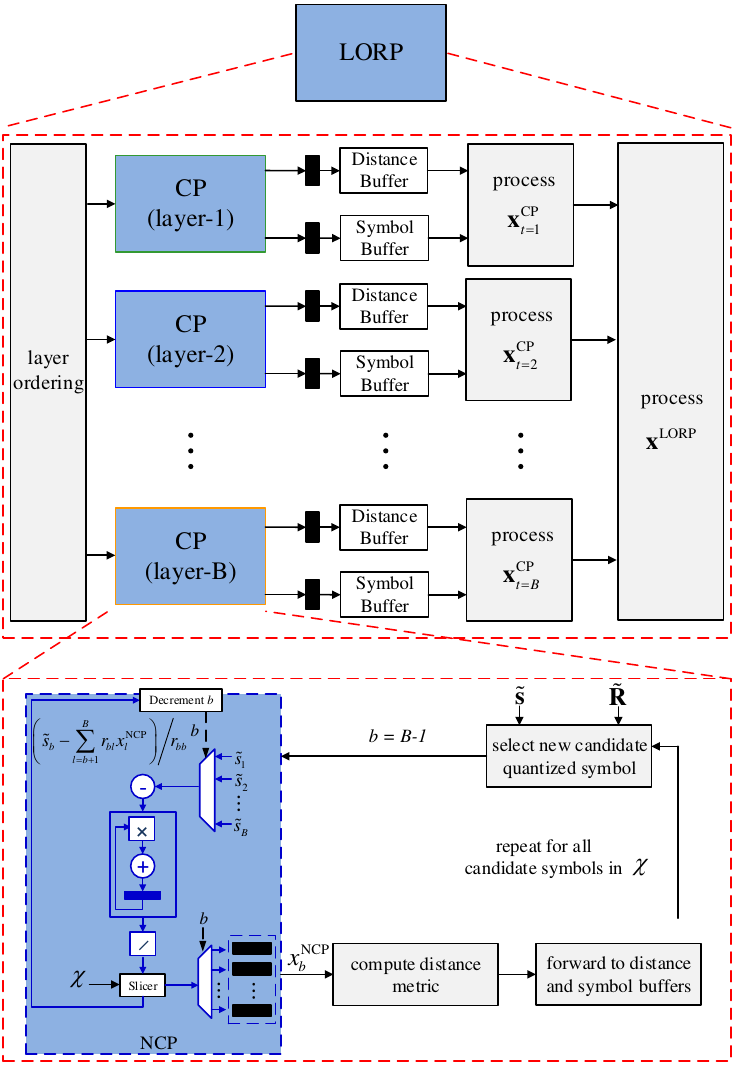}\vspace{-0.1in}
\caption{Modular architecture design for QRD-based precoders.}
\label{f:ch10_plot6}
\end{figure}

\begin{figure}[t!]
\centering
\includegraphics[width=4in]{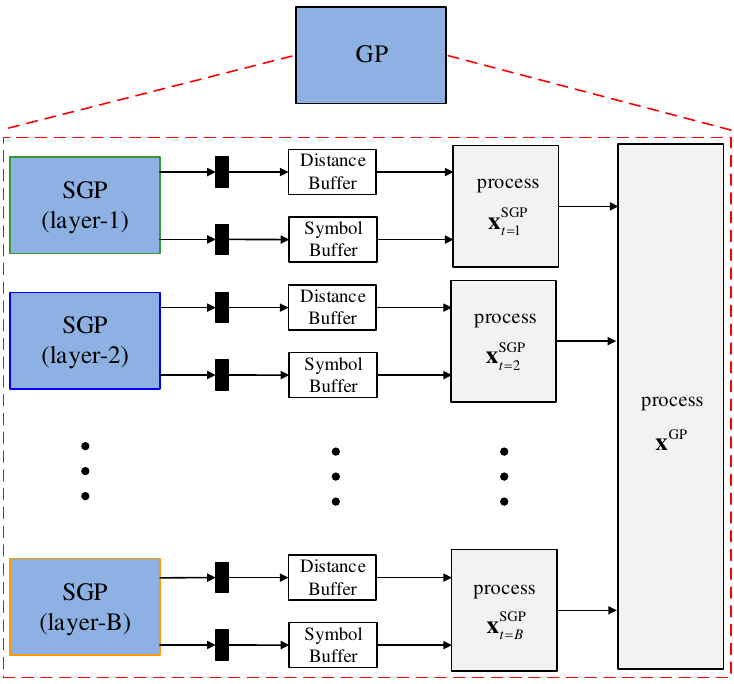}\vspace{-0.1in}
\caption{Modular architecture design for non-QRD-based precoders.}
\label{f:ch10_plot7}
\end{figure}

\begin{figure}[t!]
\centering
\includegraphics[width=5.83in]{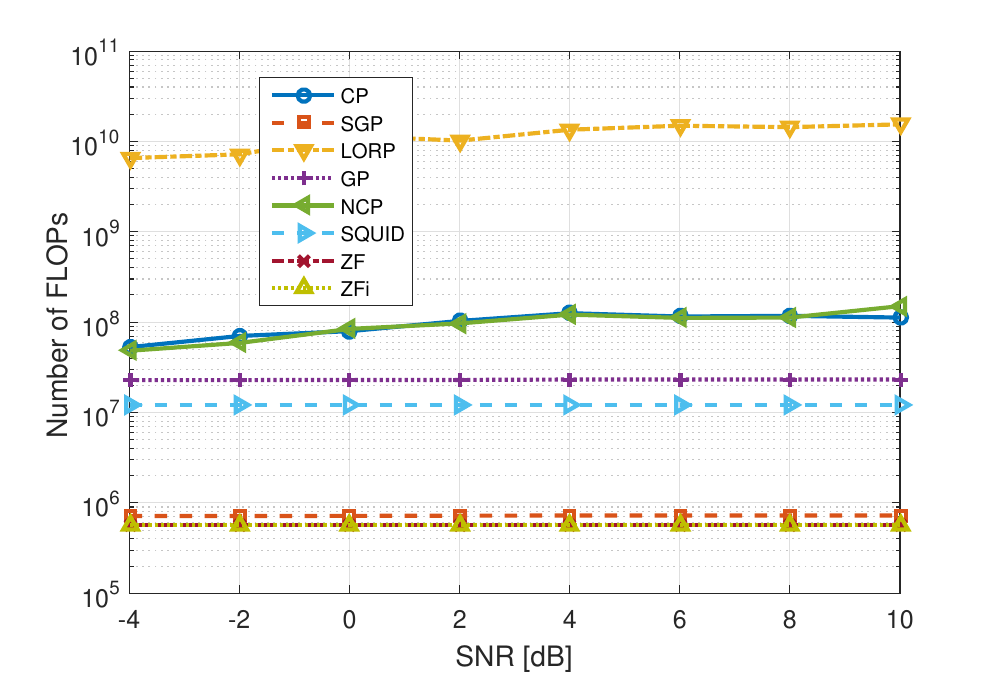}\vspace{-0.1in}
\caption{Complexity in number of FLOPs - 16$\times$128 MIMO.}
\label{f:ch10_plot4}
\end{figure}

\begin{figure}[t!]
\centering
\includegraphics[width=5.83in]{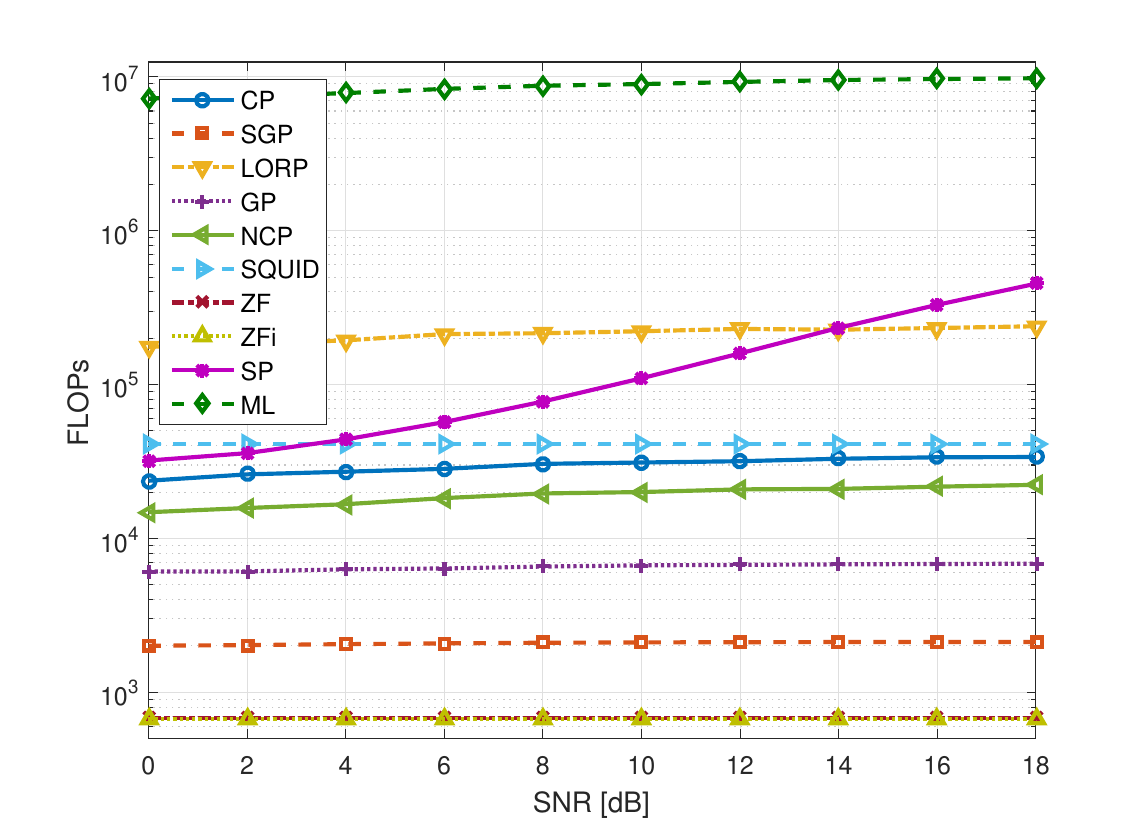}\vspace{-0.1in}
\caption{Complexity in number of FLOPs - 2$\times$8 MIMO.}
\label{f:ch10_plot5}
\end{figure}

A modular cost-efficient architecture that implements the QRD-based precoders is shown in Fig.~\ref{f:ch10_plot6}, and its non-QRD-based  counterpart is shown in Fig.~\ref{f:ch10_plot7}. The designs are hierarchical, showing LORD using CP building blocks, that themselves use NCP, while GP uses SGP blocks. With CP, the distances computed and their symbol vectors are directly forwarded for a processing unit at the corresponding layer of interest. Since the CP processes can run in parallel on all layers, the LORP precoding vector can be computed after a CP processing delay of one layer. The LORP architecture is thus partially parallelizable. A similar behavior can be noted with SGP and GP, where SGP is fully parallelizable but GP is not.

We analyze the complexity in terms of floating-point operations (FLOPs) based on real multiplications and additions. Real division and square-root operations are assumed equivalent to a real multiplication, while complex multiplication requires $4$ real multiplications and $2$ real additions, and complex addition requires $2$ real additions. The complexity plots are generated via simulations to take into account the fact that some of the proposed schemes converge faster than others. Note that since the search is for the best combination of quantization vectors, it can be argued that the complexity is almost independent of the true MT prior to quantization.

Figure~\ref{f:ch10_plot4} compares the average number of FLOPs per transmitted frame for the studied precoders versus SNR, in the context of massive $16\times128$ MIMO. Except for SP, the entailed complexity of the precoders is almost independent of SNR. While SGP is the least complex (near-ZF complexity), LORP is shown to be relatively exhaustive and impractical. In between, CP, NCP, GP, and SQUID are all feasible solutions. Compared to SQUID, GP is twice as complex, while CP and NCP result in a linear increase in complexity. Figure~\ref{f:ch10_plot5} then compares the complexity of the studied precoders in the case of $2\times8$ multiuser MIMO. Here again LORP is relatively complex, but it is much less complex than ML precoding, and even less complex than SP at high SNR. The remaining precoders, CP, NCP, GP, and SGP are all less complex than the reference SQUID. \\

Nevertheless, these complexity numbers can be significantly reduced, if real representations of the complex matrices were employed with 1-bit precoding. In this case, all multiplications with candidate precoding vectors can be conducted by simple sign flips. Finally, it can be argued that the complexity of the proposed precoders will only scale up marginally with higher order coarse quantizations.

%
\section{Simulation Results and Discussions}
\label{sec:simulation}

\begin{figure}[t!]
\centering
\includegraphics[width=5.83in]{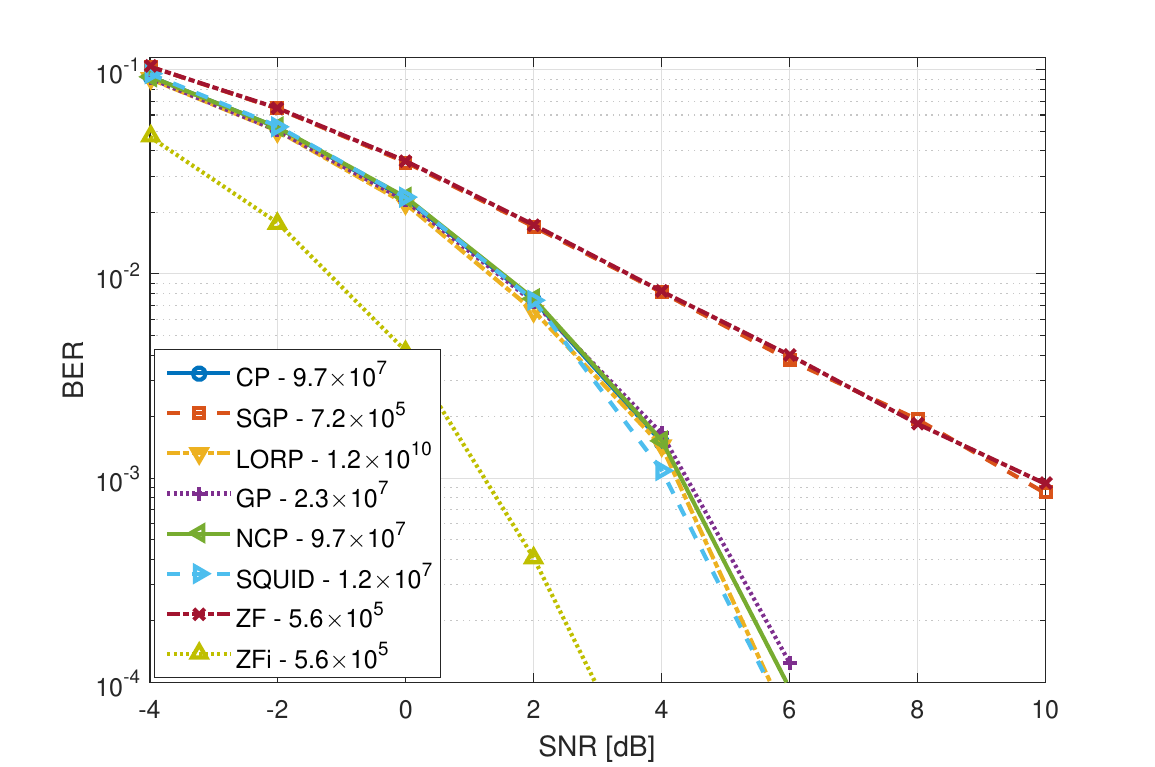}\vspace{-0.1in}
\caption{Uncoded BER performance - 16$\times$128 MIMO - QPSK.}
\label{f:ch10_plot1}
\end{figure}

The studied precoding schemes are simulated following the system model of Sec.~\ref{sec:ch10_sysmodel}. The simulation chain that includes reference precoding schemes is regenerated from~\cite{jacobsson2017quantized}. Quantized ZF and ZF with infinity resolution (ZFi) are simulated for reference, alongside SQUID and SP. Figure~\ref{f:ch10_plot1} shows the uncoded bit error rate (BER) performance with $16\times128$ massive MIMO, when $\mathcal{M}$ is QPSK. All proposed precoders achieve the same performance as SQUID, lagging behind ZFi by $\unit[3]{dB}$, except for SGP, which is as bad as ZF. Note that the average number of required FLOPs is added to the legends.

\begin{figure}[t!]
\centering
\includegraphics[width=5.83in]{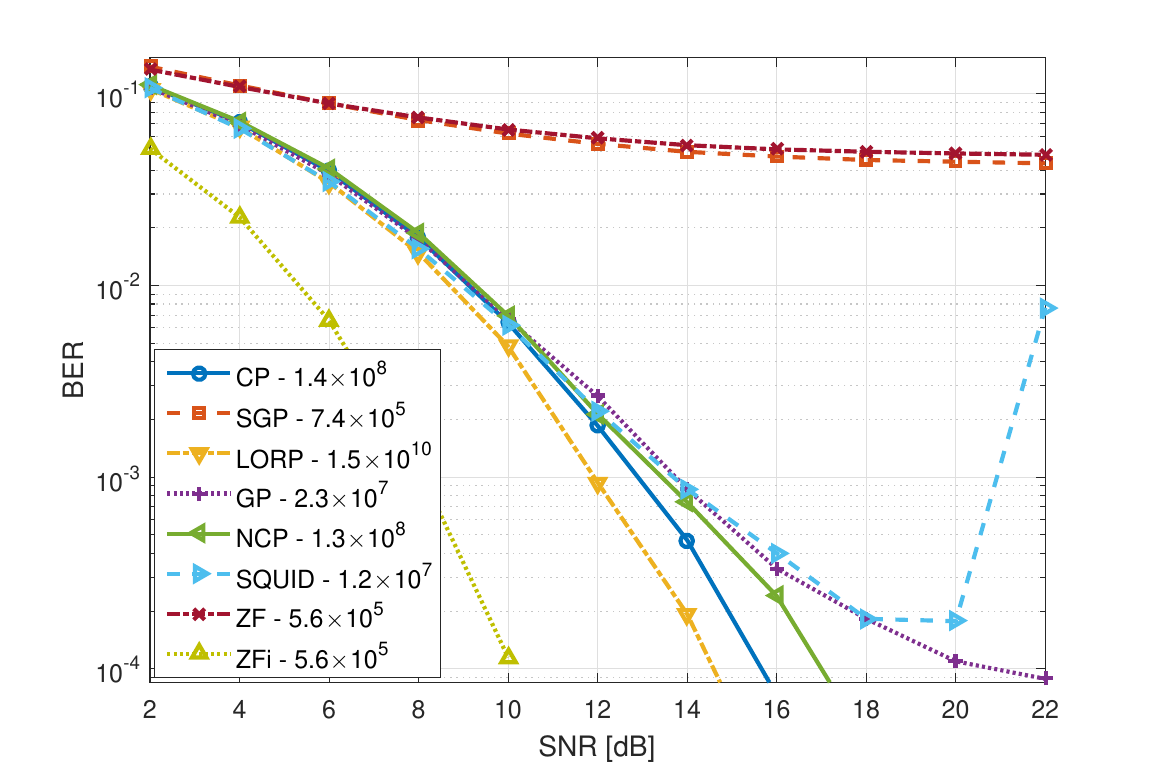}\vspace{-0.1in}
\caption{Uncoded BER performance - 16$\times$128 MIMO - 16-QAM.}
\label{f:ch10_plot2}
\end{figure}
\begin{figure}[t!]
\centering
\includegraphics[width=5.83in]{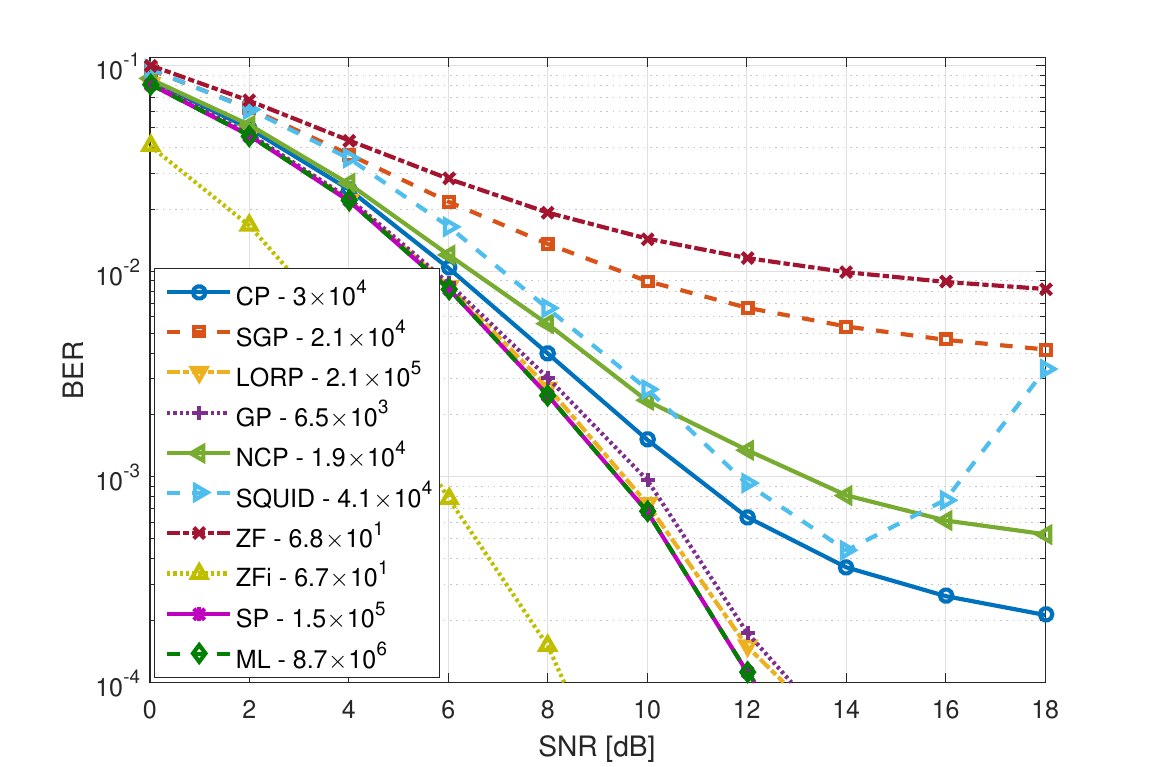}\vspace{-0.1in}
\caption{Uncoded BER performance - 2$\times$8 MIMO - QPSK.}
\label{f:ch10_plot3}
\end{figure}

Figure~\ref{f:ch10_plot2} then shows the results for the massive MIMO scenario with 16-QAM. The first thing to notice here is that SQUID is numerically unstable at high SNR with high order MTs. The best performing search-based precoder with high order constellations is LORP, which lags behind ZFi by $\unit[4]{dB}$, followed by CP, NCP, and GP, respectively.

As for the case of lower order $2\times8$ MIMO systems, Fig.~\ref{f:ch10_plot3} shows that even with QPSK, SQUID fails to converge at high SNR. The best performing precoders in this scenario are LORP and GP, which achieve near ML/SP performance. SGP introduced a slight improvement compared to ZF, but both are significantly outperformed by CP and NCP.

Taking both performance and complexity into consideration, it can be argued that CP, NCP, and GP are efficient candidates for massive MIMO precoding with  coarse quantization. They outmatch reference nonlinear precoders in the literature, in performance, complexity, or both, depending on the scenario. Table~\ref{table:ch10_summary} summarizes the results, and illustrates the corresponding scenario-dependent recommendations. The classification metrics are Inefficient, Feasible and Efficient.

Despite its low complexity, SGP does not perform well and is thus inefficient. On the other hand, despite good performance, LORP and SP are only applicable with low MIMO orders. SQUID is clearly the winner in the context of massive MIMO with QPSK. However, NCP, CP, and GP are scalable and efficient solutions for most scenarios. CP is the best solution at high SNR with massive MIMO and large MTs, while GP is the best solution with low-order MIMO, and a very efficient solution with massive MIMO as well.

Therefore, contrary to popular belief, the use of search-based precoding in massive MIMO systems is valid and efficient. This observation is emphasized in an adaptive setting (adaptive modulators and precoders), where the proposed modular architectures would result in significant simplifications. Since the proposed less complex precoders constitute their more complex extensions, an optimized hardware with simple switching circuitry can insure seamless hopping over different precoders depending on runtime channel conditions. \\

\textbf{Conclusion} \\

Two families of efficient nonlinear search-based precoders have been proposed, in the context of massive MIMO with 1-bit DACs, which are QRD-based and non-QRD-based. It has been shown through empirical simulations that the proposed precoders can achieve significant performance and complexity gains compared to reference nonlinear precoders with low order MIMO systems, as well as performance gains at the expense of graceful complexity costs with massive MIMO.

\renewcommand{\arraystretch}{1.8}

\begin{landscape}

\begin{table*}[t!]

\caption{\textbf{Summary of Precoders and Recommendations}}
\begin{center}
\begin{tabular}{|>{\centering\arraybackslash}p{1.8cm}|>{\centering\arraybackslash}p{1.8cm}|>{\centering\arraybackslash}p{1.8cm}|>{\centering\arraybackslash}p{1.8cm}|>{\centering\arraybackslash}p{1.8cm}|>{\centering\arraybackslash}p{1.8cm}|>{\centering\arraybackslash}p{1.8cm}|>{\centering\arraybackslash}p{1.8cm}|>{\centering\arraybackslash}p{1.8cm}|>{\centering\arraybackslash}p{1.8cm}|}

    \hline

    MIMO Order & MT Order & SNR Range & NCP & CP & LORP & SGP & GP & SP & SQUID \\
    \hline\hline

    \multirow{4}{*}{Massive}

    & \multirow{2}{*}{QPSK}
    & \multirow{1}{*}{Low SNR} & \cellcolor{blue!20}Feasible & \cellcolor{blue!20}Feasible & \cellcolor{red!20}Inefficient
    & \cellcolor{red!20}Inefficient & \cellcolor{green!20}Efficient & \cellcolor{red!20}Inefficient & \cellcolor{green!20}Efficient \\\cline{3-10}
    & & \multirow{1}{*}{High SNR} & \cellcolor{blue!20}Feasible & \cellcolor{blue!20}Feasible & \cellcolor{red!20}Inefficient
    & \cellcolor{red!20}Inefficient & \cellcolor{green!20}Efficien & \cellcolor{red!20}Inefficient & \cellcolor{green!20}Efficient \\ \cline{2-10}

    & \multirow{2}{*}{16-QAM}
    & \multirow{1}{*}{Low SNR} & \cellcolor{blue!20}Feasible & \cellcolor{blue!20}Feasible & \cellcolor{red!20}Inefficient & \cellcolor{red!20}Inefficient & \cellcolor{green!20}Efficient & \cellcolor{red!20}Inefficient & \cellcolor{green!20}Efficient \\\cline{3-10}
    & &\multirow{1}{*}{High SNR} & \cellcolor{green!20}Efficient & \cellcolor{green!20}Efficient & \cellcolor{red!20}Inefficient
    & \cellcolor{red!20}Inefficient & \cellcolor{blue!20}Feasible & \cellcolor{red!20}Inefficient & \cellcolor{red!20}Inefficient \\
    \hline

    \multirow{4}{*}{Low-Order}

    & \multirow{2}{*}{QPSK}
    & \multirow{1}{*}{Low SNR} & \cellcolor{blue!20}Feasible & \cellcolor{blue!20}Feasible & \cellcolor{blue!20}Feasible
    & \cellcolor{red!20}Inefficient & \cellcolor{green!20}Efficient & \cellcolor{blue!20}Feasible & \cellcolor{red!20}Inefficient \\\cline{3-10}
    & & \multirow{1}{*}{High SNR} & \cellcolor{red!20}Inefficient & \cellcolor{red!20}Inefficient & \cellcolor{blue!20}Feasible
    & \cellcolor{red!20}Inefficient & \cellcolor{green!20}Efficient & \cellcolor{red!20}Inefficient & \cellcolor{red!20}Inefficient \\ \cline{2-10}

    & \multirow{2}{*}{16-QAM}
    & \multirow{1}{*}{Low SNR} & \cellcolor{blue!20}Feasible & \cellcolor{blue!20}Feasible & \cellcolor{blue!20}Feasible & \cellcolor{red!20}Inefficient & \cellcolor{green!20}Efficient & \cellcolor{blue!20}Feasible & \cellcolor{red!20}Inefficient \\\cline{3-10}
    & &\multirow{1}{*}{High SNR} & \cellcolor{red!20}Inefficient & \cellcolor{red!20}Inefficient & \cellcolor{blue!20}Feasible
    & \cellcolor{red!20}Inefficient & \cellcolor{green!20}Efficient & \cellcolor{red!20}Inefficient & \cellcolor{red!20}Inefficient \\
    \hline

\end{tabular}
\end{center}
\label{table:ch10_summary}
\end{table*}

\end{landscape}

%% file: Chapter11_CONCLUSION.tex

\chapter{Conclusions and Future Work}
\label{chapter:Conclusion}

\textbf{Conclusions} \\

In the first part of this thesis, single-user MIMO systems have been addressed. Heuristic techniques have been first proposed to increase the efficiency of low order MIMO systems, by enhancing iterative detection/decoding receivers with high order MTs, as well as reducing the preprocessing channel matrix QRD overhead in popular detectors. Then, a family of low-complexity MIMO detectors that employ punctured QRD in lieu of regular QRD has been proposed and studied, both analytically, by deriving bounds on the capacity, diversity gains, and error probability, and empirically through simulations. The proposed HO detectors have been shown to achieve significant computational savings in the context of large MIMO systems, while at the same time achieving the same diversity gains as their QRD-based counterparts. Furthermore, significant performance gains have been observed with the proposed SO detectors, especially with highly correlated channels. An architectural design has been proposed, by using the detectors of lower complexity as building blocks in their more complex extensions, and it has been established that the proposed schemes scale up efficiently, both in the number of antennas and the constellation size. In particular, SO per-layer subspace detection has been shown to achieve a $\unit[2.5]{dB}$ SNR gain in $256$-QAM $16\!\times\!16$ MIMO, while saving $77\%$ of N/C computations.

In the second part of the thesis, several low-complexity MU-MIMO detectors have been proposed, that are based on joint detection and MC. The discussion was first motivated by studying dual-layer systems, and then extended to higher order systems. In particular, the MC complexity overhead in the context of joint MC and SSSD has been shown to reduce to $0.23\%$. It has been concluded that while assuming the MT of the interferer without estimation is sufficient with good channel conditions, MC is required at high SNR with correlated channels. An extension to this work was to consider MC in a single-user scenario with adaptive MTs. Per-layer MC has been achieved via subspace decomposition, and significant performance and complexity reduction gains have been noted.

It can be argued that the proposed schemes in this thesis are better candidates for large MIMO detection than reference detectors in the literature when complexity is taken into consideration. For example, detection based on local search \cite{Vardhan_2008,Li_2010} does not achieve near-ML diversity, nor does it scale up efficiently with high order modulation constellations. Similarly, heuristic tabu search algorithms \cite{Datta_2010,Srinidhi_2011} do not perform well with high order MTs, and their performance is hard to track analytically. Detectors based on message passing on graphical models \cite{Goldberger_2011,Narasimhan_2014} have been recently extended to support high order MTs; however, they are better suited for joint iterative detection and decoding schemes. Furthermore, with LR~\cite{Wubben_2011,Zhou_LR_2013}, processing large channel matrices using conventional reduction schemes is costly, and low complexity schemes such as element-based LR incur performance degradation. Nevertheless, LR can be implemented in combination with our proposed approaches, despite the fact that the resultant LLR computations are not straightforward. Finally, detection using Monte Carlo sampling \cite{Datta_Gibbs_2013} is clearly outperformed by the proposed schemes.

In the last part of this thesis, preliminary results on the scalability of nonlinear search-based precoding in the context of massive MIMO have been discussed.

\textbf{Future Work} \\

As a future work, we will mainly aim at investigating low-complexity nonlinear precoding and detection algorithms for massive MIMO, which can outperform conventional state-of-the-art linear solutions with graceful computational costs. Our focus would be on the case where neither the transmitter nor the receiver have any a priori CSI. This implies that the fading realizations have to be learned through pilot transmission followed by channel estimation at the receiver, based on coarsely quantized observations. For simplicity, we will assume single-cell operation at early stages of the work. We will try to exploit the extra degrees of freedom to combat quantization effects. The investigated algorithms are to be supported by theoretical performance analysis and practical architectural designs, in order to present well defined feasible end-to-end solutions for future 5G wireless systems. In particular, we can proceed as follows:

\begin{itemize}
  \item Investigate search-based  NLQP algorithms for the downlink. Explore space for performance and complexity tradeoffs. Extend the investigation to include the complexity of the UE into account, and explore space for tradeoffs between precoding at the BS and detection at the UEs.
  \item Analyse the performance of the proposed NLQPs theoretically, using diversity studies, bit error rate (BER) computations, and capacity analyses.
  \item Design practical architectures that realise the proposed NLQPs. Implement and test the proposed designs in hardware to obtain exact measures of the occupied area and achievable throughput.
  \item Investigate quantized detectors for the uplink and explore performance-complexity tradeoffs. Analyse the performance of the proposed detectors theoretically.
  \item Design practical architectures that can realise the proposed detectors. Implement and test the proposed designs in hardware.
  \item Investigate channel estimation schemes and analyse the impact on the overall system theoretically.
  \item Design practical architectures that can realise the proposed channel estimators, jointly with the detectors, or as separate entities. 
  \item Investigate alternative uses of the degrees of freedom that are provided by the large number of BS antennas. One approach is to design smart coding schemes across the extra dimensions. For instance, a group of antennas with 1-bit data converters can be jointly encoded to support a single data stream, as if they were a single antenna element with higher resolution. This can be thought of as layered quantization.\\
\end{itemize}
Moreover, the work in several sections in this thesis can be further extended. In what follows we highlight some examples in a random order of importance.

\begin{itemize}
  \item Implement the proposed efficient architectures in hardware.
  \item Further analyse the AIR-based detectors and compare them to our proposed WRD detectors, both in performance and complexity.
  \item Extend the capacity analysis to propose optimized resource allocation schemes that make best use of the proposed detectors.
  \item Derive the capacity bounds of the proposed shemes for very large MIMO.
  \item Conduct a theoretical study to identify at what precoding conditions is MC beneficial to mitigate intra-cell interference.
  \item Investigate the performance-complexity tradeoffs of solutions that trade precoding at the BS with MC followed by IA detection at the receiver.
\end{itemize}